\documentclass[11pt,twoside,a4paper]{article}

\usepackage[T1]{fontenc}
\usepackage[utf8]{inputenc}
\usepackage[ngerman,english]{babel}
\usepackage{csquotes}

\usepackage{amsmath}
	\numberwithin{equation}{subsection}
    \numberwithin{table}{section}
    \numberwithin{figure}{section}
\usepackage{mathtools}
\usepackage{amssymb}
\usepackage{mathrsfs}
\usepackage{dsfont}
\usepackage{bbm}
\usepackage{siunitx}
    \DeclareSIUnit \kpc {kpc}
    \DeclareSIUnit \Mpc {Mpc}
    \DeclareSIUnit \barn {b}
\usepackage[tableposition=top, format=plain, labelfont=bf]{caption} 
\usepackage[usenames, dvipsnames]{xcolor} 
    \definecolor{mygreen}{RGB}{56,87,35}
    \definecolor{myblue}{RGB}{0,107,146}
    \definecolor{mybrown}{RGB}{128,102,74}
    \definecolor{mymagenta}{RGB}{147,66,134}
    \definecolor{mycyan}{RGB}{0,153,125}
    \definecolor{myred}{RGB}{158,63,50}
\usepackage[colorlinks]{hyperref}
    \makeatletter
    \define@key{Hyp}{footnotecolor}{\HyColor@HyperrefColor{#1}\@footnotecolor}
    \patchcmd{\@footnotemark}{\hyper@linkstart{link}}{\hyper@linkstart{footnote}}{}{}
    \makeatother
    \hypersetup{linkcolor=myred,citecolor=myblue,urlcolor=mymagenta,footnotecolor=mybrown}
\usepackage[capitalize]{cleveref}
    \crefformat{appendix}{App.~#2#1#3}
    \makeatletter
    \creflabelformat{equation}{%
      \textup{%
        \hypersetup{
          linkcolor=mygreen,
          linkbordercolor=mygreen,
        }%
        #2#1#3%
      }%
    }
    \makeatother

\usepackage{fancyhdr}
    \fancyheadoffset[RO,LE]{0pt}
    \fancypagestyle{normal}{
	   \fancyfoot{}                        
	   \fancyhf{}
	   \fancyhead[LE]{\rightmark}
	   \fancyhead[RO]{\leftmark}
	   \fancyfoot[LE,RO]{\bfseries\thepage}
	   \setlength{\headheight}{13.6pt}
      \setlength{\parindent}{0.5cm}
    }
    \fancypagestyle{acronyms}{
    	\fancyfoot{}                        
    	\fancyhf{}
    	\fancyhead[LE]{ACRONYMS}
    	\fancyhead[RO]{ACRONYMS}
    	\fancyfoot[LE,RO]{\bfseries\thepage}
    	\setlength{\headheight}{13.6pt}
    }
    \fancypagestyle{abstract}{
    	\fancyfoot{}                        
    	\fancyhf{}
    	\fancyhead[LE]{ABSTRACT}
    	\fancyhead[RO]{ABSTRACT}
    	\fancyfoot[LE,RO]{\bfseries\thepage}
    	\setlength{\headheight}{13.6pt}
        \setlength{\parindent}{0.5cm}
    }
    \fancypagestyle{acknowledgements}{
    	\fancyfoot{}                        
    	\fancyhf{}
    	\fancyhead[LE]{ACKNOWLEDGEMENTS}
    	\fancyhead[RO]{ACKNOWLEDGEMENTS}
    	\fancyfoot[LE,RO]{\bfseries\thepage}
    	\setlength{\headheight}{13.6pt}
        \setlength{\parindent}{0.5cm}
    }
    \fancypagestyle{appendix}{
    	\fancyfoot{}                        
    	\fancyhf{}
    	\fancyhead[LE]{\rightmark}
    	\fancyhead[RO]{APPENDIX}
    	\fancyfoot[LE,RO]{\bfseries\thepage}
    	\setlength{\headheight}{13.6pt}
        \setlength{\parindent}{0.5cm}
    }
\usepackage{chngcntr} 
	\counterwithin{figure}{section}
\usepackage[toc,page]{appendix}
\usepackage[backend=bibtex,natbib=false,url=false,doi=false,isbn=false,bibencoding=ascii,style=numeric-comp,sorting=none,maxbibnames=5,giveninits=true]{biblatex}
\usepackage[acronym,nopostdot]{glossaries}
\usepackage{glossary-mcols}
\makeglossaries
\newacronymstyle{long-short-emph}%
{%
  \ifglshaslong{\glslabel}%
  {
    \ifdefempty\glscustomtext
    {%
      \ifglsused\glslabel
      {%
        \glsifplural
        {%
          \glscapscase
          {%
            \acronymfont{\glsentryshortpl{\glslabel}}\glsinsert
          }%
          {%
            \acronymfont{\Glsentryshortpl{\glslabel}}\glsinsert
          }%
          {%
            \mfirstucMakeUppercase
              {\acronymfont{\glsentryshortpl{\glslabel}}\glsinsert}%
          }%
        }%
        {%
          \glscapscase
          {%
            \acronymfont{\glsentryshort{\glslabel}}\glsinsert
          }%
          {%
            \acronymfont{\Glsentryshort{\glslabel}}\glsinsert
          }%
          {%
            \mfirstucMakeUppercase
              {\acronymfont{\glsentryshort{\glslabel}}\glsinsert}%
          }%
        }%
      }%
      {%
        \glsifplural
        {%
          \glscapscase
          {%
            \emph{\glsentrylongpl{\glslabel}\glsinsert}\space
            (\firstacronymfont{\glsentryshortpl{\glslabel}})%
          }%
          {%
            \emph{\Glsentrylongpl{\glslabel}{\glsinsert}}\space
            (\firstacronymfont{\glsentryshortpl{\glslabel}})%
          }%
          {%
            \mfirstucMakeUppercase
              {\emph{\glsentrylongpl{\glslabel}{\glsinsert}}\space
               (\firstacronymfont{\glsentryshortpl{\glslabel}})}%
          }%
        }%
        {%
          \glscapscase
          {%
            \emph{\glsentrylong{\glslabel}\glsinsert}\space
            (\firstacronymfont{\glsentryshort{\glslabel}})%
          }%
          {%
            \emph{\Glsentrylong{\glslabel}\glsinsert}\space
            (\firstacronymfont{\glsentryshort{\glslabel}})%
          }%
          {%
            \mfirstucMakeUppercase
             {\emph{\glsentrylong{\glslabel}\glsinsert}\space
             (\firstacronymfont{\glsentryshort{\glslabel}})}%
          }%
        }%
      }%
    }%
    {%
      \glscustomtext
    }%
  }%
  {
    \ifglsused{\glslabel}{\glsgenentryfmt}{\emph{\glsgenentryfmt}}%
  }%
}%
{%
  \GlsUseAcrStyleDefs{long-short}%
}
\setacronymstyle{long-short-emph}

\newacronym{bbn}{BBN}{Big Bang nucleosynthesis}
\newacronym{be}{BE}{Boltzmann equation}
\newacronym{bs}{BS}{bound states}
\newacronym{bsf}{BSF}{bound state formation}
\newacronym{bsm}{BSM}{beyond Standard Model}

\newacronym{com}{COM}{center of momentum}
\newacronym{cmb}{CMB}{cosmic microwave background}
\newacronym{cl}{CL}{confidence level}
\newacronym{cta}{CTA}{Cherenkov Telescope Array}

\newacronym{dm}{DM}{dark matter}
\newacronym{dof}{dof}{degrees of freedom}
\newacronym{df}{DF}{distribution function}
\newacronym[\glslongpluralkey={dwarf spheroidal galaxies}]{dsph}{dSph}{dwarf spheroidal galaxy}

\newacronym{edm}{EDM}{electric dipole moment}
\newacronym{eft}{EFT}{effective field theory}
\newacronym{ewsb}{EWSB}{electroweak symmetry breaking}

\newacronym{fermi}{Fermi-LAT}{Fermi Large Area Telescope}
\newacronym{fi}{FI}{freeze-in}
\newacronym{frw}{FRW}{Friedmann-Robertson-Walker}
\newacronym{fo}{FO}{freeze-out}

\newacronym{gc}{GC}{galactic center}

\newacronym{hl}{HL}{high-luminosity}

\newacronym{igm}{IGM}{intergalactic medium}

\newacronym{lcdm}{$\Lambda$CDM}{$\Lambda$-cold dark matter}
\newacronym{lhc}{LHC}{Large Hadron Collider}
\newacronym{lhs}{LHS}{left-hand side}
\newacronym{los}{LoS}{line of sight}
\newacronym{lsz}{LSZ}{Lehmann-Symanzik-Zimmermann}

\newacronym{macho}{MACHO}{massive astrophysical compact halo object}
\newacronym{mond}{MOND}{modified Newtonian dynamics}

\newacronym{nfw}{NFW}{Navarro-Frenk-White}
\newacronym{nreft}{NREFT}{non-relativistic EFT}

\newacronym{pnreft}{pNREFT}{potential NREFT}

\newacronym{qcd}{QCD}{quantum chromodynamics}
\newacronym{qed}{QED}{quantum electrodynamics}
\newacronym{qft}{QFT}{quantum field theory}

\newacronym{rhs}{RHS}{right-hand side}
\newacronym{rms}{rms}{root-mean-square}

\newacronym{se}{SE}{Sommerfeld enhancement}
\newacronym{sm}{SM}{Standard Model}
\newacronym{sw}{sW}{superWIMP}

\newacronym{ts}{TS}{test statistics}

\newacronym{uv}{UV}{ultraviolet}

\newacronym{vev}{vev}{vacuum expectation value}

\newacronym{wdm}{WDM}{warm dark matter}
\newacronym{wimp}{WIMP}{weakly interacting massive particle}

\usepackage{booktabs} 
\usepackage{graphicx} 

\newcommand{\quotes}[1]{``#1''}
\newcommand{\newp}{\\[1mm]\indent}
\newcommand{\newpp}{\\[1mm]}
\newcommand*{\cf}{cf.\ }
\newcommand*{\eg}{e.\,g.\ }
\newcommand*{\ie}{i.\,e.\ }
\newcommand*{\etal}{et al.\,}
\newcommand*{\app}{App.}
\newcommand*{\eref}{Ref.\,}
\newcommand*{\erefs}{Refs.\,}

\usepackage{geometry} 
\usepackage{todonotes}

\renewcommand{\vec}[1]{\boldsymbol{#1}}
\newcommand{\uvec}[1]{\hat{\vec{#1}}}
\newcommand{\abs}[1]{\left|#1\right|} 
\newcommand{\dd}[2][]{\,\text{d}^{#1}#2\,} 
\newcommand{\dv}[3][]{\frac{\text{d}^{#1}#2}{\text{d}{#3}^{#1}}} 
\newcommand{\pdv}[3][]{\frac{\partial^{#1}#2}{\partial{#3}^{#1}}} 
\newcommand{\expval}[1]{\left\langle #1 \right \rangle} 
\newcommand{\order}[1]{\mathcal{O}\left(#1\right)} 
\DeclareMathOperator{\diag}{diag} 
\DeclareMathOperator{\arccot}{arccot} 
\renewcommand{\Im}[1]{\text{Im}\left\{#1\right\}}
\renewcommand{\Re}[1]{\text{Re}\left\{#1\right\}}

\newcommand{\CS}[3]{\Gamma^{#1}_{#2#3}} 
\newcommand{\HGone}[3]{\hspace{-1mm}\hphantom{I}_1F_1\left(#1;#2;#3\right)} 
\newcommand{\HGtwo}[4]{\hspace{-1mm}\hphantom{I}_2F_1\left(#1,#2;#3;#4\right)} 
\newcommand{\hlone}{h_l^{(1)}} 
\newcommand{\hltwo}{h_l^{(2)}} 
\newcommand{\Jkn}[1]{\vec{\mathcal{J}}_{\vec{k},#1}} 
\newcommand{\Ykn}[1]{\vec{\mathcal{Y}}_{\vec{k},#1}} 
\newcommand{\Ikn}[1]{\mathcal{I}_{\vec{k},#1}} 
\newcommand{\Kkn}[1]{\mathcal{K}_{\vec{k},#1}} 
\newcommand{\jl}{j_l} 
\newcommand{\nl}{n_l} 
\newcommand{\kone}[1]{K_1\left(#1\right)} 
\newcommand{\ktwo}[1]{K_2\left(#1\right)} 
\newcommand{\Laguerre}[3]{L_{#1}^{(#2)}\left(#3\right)} 
\newcommand{\spec}[3]{^{#1}{#2}_{#3}} 

\newcommand{\geff}{g_{\text{eff}}} 
\newcommand{\heff}{h_{\text{eff}}}  
\newcommand{\gstar}{g_\star}  
\newcommand{\MPl}{M_{\text{\tiny Pl}}} 
\newcommand{\rhocrit}{\rho_{crit,0}} 
\newcommand{\mwdm}{m_{\text{\tiny{WDM}}}} 
\newcommand{\Tprod}{T_{\text{prod}}} 

\newcommand{\vmol}{v_{\text{M\o l}}} 
\newcommand{\vrel}{v_{\text{rel}}} 
\newcommand{\feq}[1]{f^{\text{eq}}_{#1}} 
\newcommand{\ndeq}[1]{n^{\text{eq}}_{#1}} 
\newcommand{\Cfionetwo}{\mathscr{C}^{\text{\tiny FI}}_{1\to 2}} 
\newcommand{\Cfitwotwo}{\mathscr{C}^{\text{\tiny FI}}_{2\to 2}}
\newcommand{\Cfotwotwo}{\mathscr{C}^{\text{\tiny FO}}_{2\to 2}}
\newcommand{\Yfionetwo}[1]{Y^{\text{\tiny FI}}_{#1,1\to 2}} 
\newcommand{\Yfitwotwo}[1]{Y^{\text{\tiny FI}}_{#1,2\to 2}} 
\newcommand{\Yfi}[1]{Y^{\text{\tiny FI}}_{#1}} 
\newcommand{\Yeq}[1]{Y^{\text{eq}}_{#1}} 

\newcommand{\mchi}{m_\chi} 
\newcommand{\mvphi}{m_\varphi} 
\newcommand{\Omegadm}{\Omega_{\text{\tiny DM}}} 
\newcommand{\Omegadmt}{\Omega_{\text{\tiny DM},0}} 
\newcommand{\sigmavmolavg}[2]{\expval{\sigma\vmol}_{#1\to #2}} 
\newcommand{\sigmaann}{\sigma_{\text{ann}}} 
\newcommand{\sigmabsf}{\sigma_{\text{\tiny{BSF}}}} 
\newcommand{\Gammadecn}{\Gamma^{(n)}_{\text{dec}}} 
\newcommand{\Gammadec}{\Gamma_{\text{dec}}} 
\newcommand{\Gammaion}{\Gamma_{\text{ion}}} 
\newcommand{\sigmavrelavg}{\expval{\sigma\vrel}} 
\newcommand{\sigmaannexp}{\expval{\sigmaann\vrel}} 
\newcommand{\sigmabsfexp}{\expval{\sigmabsf\vrel}} 
\newcommand{\sigmabsfeffexp}{\sigmabsfexp_{\text{eff}}} 
\newcommand{\sigmaanntotexp}{\sigmaannexp_{\text{tot}}} 
\newcommand{\Gammaionexp}{\expval{\Gammaion}} 
\newcommand{\Gammadecexp}{\expval{\Gammadec}} 

\newcommand{\Ek}{\mathcal{E}_{\vec{k}}} 
\newcommand{\Enl}{\mathcal{E}_{nl}} 
\newcommand{\En}{\mathcal{E}_{n}} 
\newcommand{\Ekn}[1]{\Delta\mathcal{E}_{#1}^{\vec{k}}} 
\newcommand{\gammanl}{\gamma_{nl}} 
\newcommand{\phik}{\phi_{\vec{k}}} 
\newcommand{\psinlm}{\psi_{nlm}} 
\newcommand{\psin}{\psi_{n}} 
\newcommand{\phiq}[1]{\phi_{\vec{#1}}} 
\newcommand{\chikl}{\chi_{\vec{k},l}} 
\newcommand{\chitkl}{\tilde{\chi}_{\vec{k},l}} 
\newcommand{\chiqs}[2]{\chi_{\vec{#1},#2}} 
\newcommand{\chinl}{\chi_{nl}} 
\newcommand{\chitnl}{\tilde{\chi}_{nl}} 
\newcommand{\chiij}[1]{\chi_{#1}} 
\newcommand{\ukl}{u_{\vec{k},l}} 

\newcommand{\Sann}{S_{\text{ann},l}} 
\newcommand{\Sannl}[1]{S_{\text{ann},#1}} 
\newcommand{\Sdec}{S_{\text{dec},l}^{(n)}} 
\newcommand{\Sdecnl}[2]{S_{\text{dec},#2}^{#1}} 
\newcommand{\SBSF}{S_{\text{\tiny{BSF}}}} 

\newcommand{\nry}{\text{NRY}_{\gamma_5}} 
\newcommand{\pnry}{\text{pNRY}_{\gamma_5}} 
\newcommand{\phius}{\left|\phi_{\text{us}}\right\rangle} 

\newcommand{\SSQ}[2]{\left|\mathcal{U}_{\vec{#1},\vec{#2}}\right\rangle} 
\newcommand{\BSQ}[1]{\left|\mathcal{B}_{\vec{#1},n}\right\rangle} 
\newcommand{\SSQcc}[2]{\left\langle\mathcal{U}_{\vec{#1},\vec{#2}}\right|} 
\newcommand{\BSQcc}[1]{\left\langle\mathcal{B}_{\vec{#1},n}\right|} 
\newcommand{\Vac}{\left|\Omega\right\rangle} 
\newcommand{\Vaccc}{\left\langle\Omega\right|} 
\newcommand{\ESS}[2]{\omega_{\vec{#1},\vec{#2}}} 
\newcommand{\ESSdiffq}[2]{\varepsilon_{\vec{#1},\vec{#2}}} 
\newcommand{\EBS}[1]{\omega_{\vec{#1},n}} 
\newcommand{\Ephi}{\omega_\varphi(\vec{P}_\varphi)} 
\newcommand{\PhiQBS}[2]{\Phi_{\vec{#1},\vec{#2}}} 
\newcommand{\PsiQBS}[1]{\Psi_{\vec{#1},n}} 
\newcommand{\PhiQBSt}[2]{\tilde{\Phi}_{\vec{#1},\vec{#2}}} 
\newcommand{\PsiQBSt}[1]{\tilde{\Psi}_{\vec{#1},n}} 
\newcommand{\phiPphi}{\left|\varphi_{\vec{P}_\varphi}\right\rangle} 
\newcommand{\Pphi}{P_{\varphi}} 
\newcommand{\NnormBS}[1]{\mathcal{N}_{\vec{#1}}} 
\newcommand{\SEwavefQBSt}[1]{\tilde{\psi}_{\vec{#1},n}} 
\newcommand{\SEwavefQSSt}[2]{\tilde{\phi}_{\vec{#1},\vec{#2}}} 
\newcommand{\SEwavefBSt}{\tilde{\psi}_{n}}
\newcommand{\SEwavefSSt}[1]{\tilde{\phi}_{\vec{#1}}} 
\newcommand{\Gfour}{G^{(4)}} 
\newcommand{\Gfourt}{\tilde{G}^{(4)}} 
\newcommand{\Gfive}{G^{(5)}} 
\newcommand{\Gfivet}{\tilde{G}^{(5)}} 
\newcommand{\Afive}{\mathcal{A}^{(5)}} 
\newcommand{\Cfive}{\mathcal{C}^{(5)}} 
\newcommand{\Cfiveamp}[1]{\mathcal{C}_{#1-\text{amp}}^{(5)}} 
\newcommand{\Mkn}{\mathcal{M}_{\vec{k}\to n}} 
\newcommand{\Mknvec}{\vec{\mathcal{M}}_{\vec{k}\to n}} 
\newcommand{\Mtrans}{\mathcal{M}_{\text{trans}}} 
\newcommand{\Mtransvec}{\vec{\mathcal{M}}_{\text{trans}}} 

\newcommand{\gsbsf}{g_s^{\text{\tiny BSF}}} 
\newcommand{\gsna}{g_s^{\text{\tiny NA}}} 
\newcommand{\alphaem}{\alpha_{\text{em}}} 
\newcommand{\alphasann}{\alpha_{s,\text{ann}}} 
\newcommand{\alphasS}{\alpha^{\scriptscriptstyle S}_s} 
\newcommand{\alphagSs}{\alpha^{\scriptscriptstyle S}_{g,\scriptscriptstyle{[\mathbf{1}]}}} 
\newcommand{\alphagSo}{\alpha^{\scriptscriptstyle S}_{g,\scriptscriptstyle{[\mathbf{8}]}}} 
\newcommand{\alphasBs}{\alpha^{\scriptscriptstyle B}_{s,\scriptscriptstyle{[\mathbf{1}]}}} 
\newcommand{\alphagBs}{\alpha^{\scriptscriptstyle B}_{g,\scriptscriptstyle{[\mathbf{1}]}}} 
\newcommand{\alphagB}{\alpha^{\scriptscriptstyle B}_{g}} 
\newcommand{\alphasBSFs}{\alpha^{\text{\tiny BSF}}_{s,\scriptscriptstyle{[\mathbf{1}]}}}
\newcommand{\alphasNA}{\alpha^{\text{\tiny NA}}_{s,\scriptscriptstyle{[\mathbf{1}]}}} 
\newcommand{\ttilde}{\tilde{t}} 
\newcommand{\mttilde}{m_{\tilde{t}}} 
\newcommand{\tautilde}{\tilde{\tau}} 
\newcommand{\mtautilde}{m_{\tautilde}} 
\newcommand{\lchi}{\lambda_\chi} 
\newcommand{\lH}{\lambda_H} 
\newcommand{\YB}{Y_{B}} 
\newcommand{\Yttbar}{Y_{\ttilde\ttilde^*}} 
\newcommand{\Ysw}[1]{Y^{\text{\tiny sW}}_{#1}} 
\newcommand{\Ytot}{Y_{\text{tot}}} 
\newcommand{\zetas}{\zeta_{\scriptscriptstyle S}} 
\newcommand{\zetab}{\zeta_{\scriptscriptstyle B}} 
\newcommand{\zetagamma}{\zeta_\gamma} 

\newcommand{\mphi}{m_{\phi}} 
\newcommand{\muphih}{\mu_{\phi h}} 
\newcommand{\rvir}{r_{\text{vir}}} 
\newcommand{\Mvir}{M_{\text{vir}}} 
\newcommand{\rhot}{\tilde{\rho}_{\chi}} 
\newcommand{\psit}{\tilde{\psi}} 
\newcommand{\et}{\tilde{\epsilon}} 
\newcommand{\vt}{\tilde{v}} 
\newcommand{\vrms}{v_{\text{rms}}} 
\newcommand{\vesc}{v_{\text{esc}}} 
\newcommand{\vesct}{\tilde{v}_{\text{esc}}} 
\newcommand{\gNFW}{g_{\text{\tiny{NFW}}}} 
\newcommand{\gEin}{g_{\text{\tiny{Ein}}}}
\newcommand{\fchit}{\tilde{f}_\chi} 
\newcommand{\vcm}{v_{\text{cm}}} 
\newcommand{\vcmt}{\tilde{v}_{\text{cm}}} 
\newcommand{\vrelt}{\tilde{v}_{\text{rel}}} 
\newcommand{\Pxrel}{P_{x,\text{rel}}} 
\newcommand{\Jsann}[1]{J_{\text{ann},#1}^{(\alpha)}} 
\newcommand{\Jsbsf}{J_{\text{\tiny{BSF}}}^{(\alpha)}} 

\begin{document}

\pagenumbering{roman}
\begin{titlepage}
\AddToHookNext{shipout/background}{
  \begin{tikzpicture}[remember picture,overlay, opacity=0.15]
  \node at (15cm,-25cm) {
    \includegraphics[width=40cm]{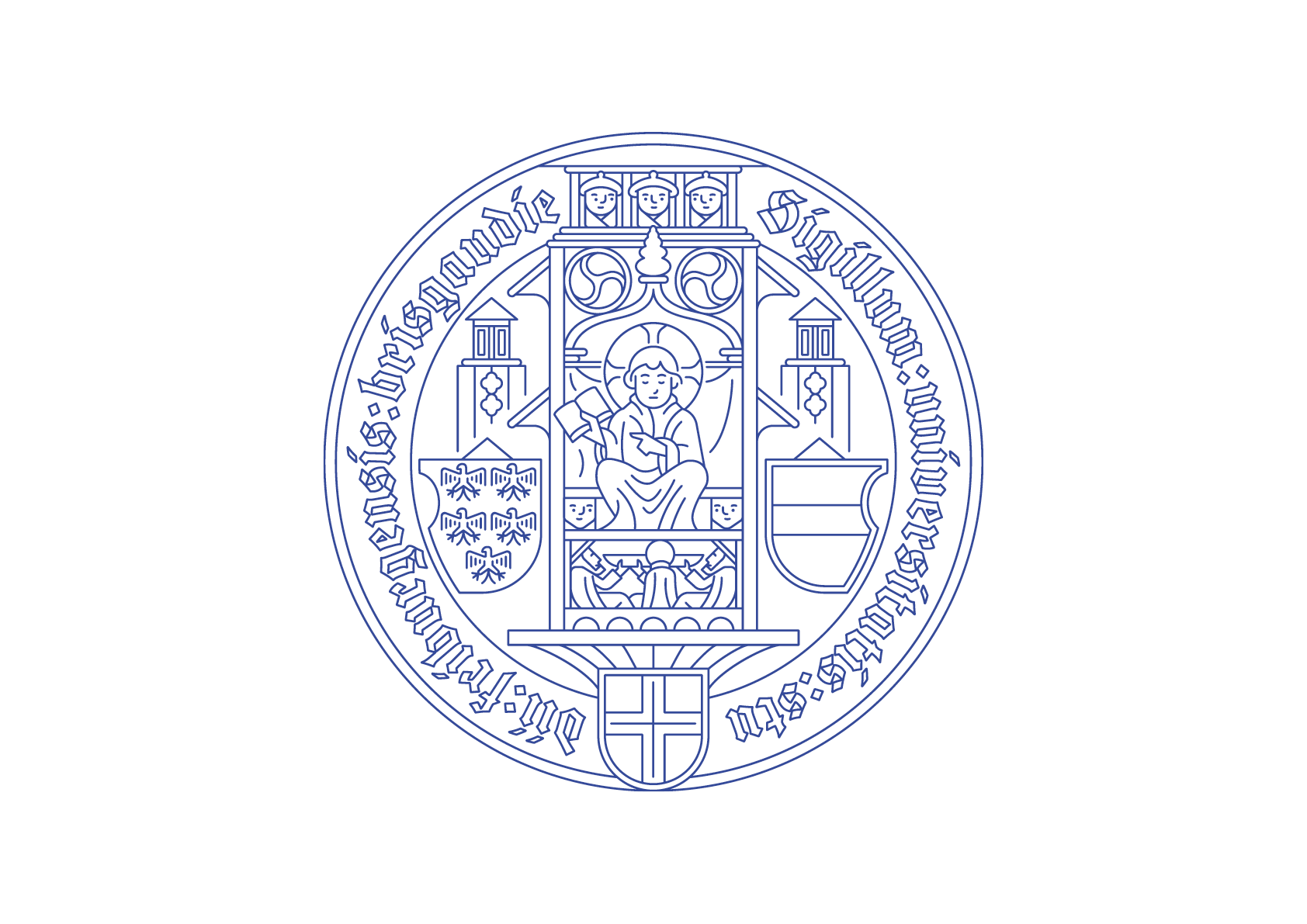}
  };
  \end{tikzpicture}
}

\textcolor{white}{h}
\vspace{10mm}
\begin{center}

\scalebox{1.25}{\hspace{-20mm}\parbox{\textwidth}{{\centering \LARGE  \bfseries \hspace{4mm}The impact of non-perturbative effects \\[0.5cm]\hspace{20.5mm}in dark matter production and detection}}}
\\[0.7cm]
\vspace{35mm}
\selectlanguage{ngerman}
{\large \bfseries Dissertation}\\[1.5cm]
zur Erlangung des Doktorgrades\\ [0.2cm]
der Fakultät für Mathematik und Physik der\\ [0.9cm]
{\large \bfseries Albert-Ludwigs-Universität}\\[0.2cm]
{\large \bfseries Freiburg im Breisgau}\\ [2.cm]
\vspace{5mm}
 vorgelegt von \\ [0.3cm]
\vspace{5mm}
{\large \bfseries Julian Bollig}\\

 \vspace{20mm}
{\large \bfseries  Mai 2024} \\
\end{center}

\end{titlepage}

\newpage
\thispagestyle{empty}
\phantom{!}
\vfill
\parindent=0pt
Dekan: Prof.\,Dr.\,Michael Růžička\\[1mm]
Referent: JProf.\,Dr.\,Stefan Vogl\\[1mm]
Korreferent: Prof.\,Dr.\,Heidi Rzehak\\[1mm]
Tag der mündlichen Prüfung: 25.07.2024
\selectlanguage{english}
\newpage

\thispagestyle{abstract}
Dark matter stands out as one of the most important unsolved mysteries in particle physics and cosmology. We investigate the influence of non-perturbative effects on the production of dark matter as well as their impact on direct and indirect detection to attain insight on its fundamental nature. Specifically, we focus on Sommerfeld enhancement and the formation of dark matter bound states both of which can alter the total dark matter annihilation cross section and thus modify dark matter observables in the early and late Universe. To this end, we conducted two studies focusing on different dark matter observables, which are likely to be probed by upcoming experiments.\newp
In the first study, we investigate the production properties of a dark matter candidate with feeble couplings to Standard Model fermions mediated by a dark scalar in the $\si{\TeV}$ mass range. The scalar possesses Standard Model quantum numbers, allowing for a sizable new physics production cross-section at the LHC. By analyzing dark matter production through freeze-in and the superWIMP mechanisms, we discover a significant suppression of the dark matter yield in the latter due to non-perturbative effects, which favors higher masses for the scalar mediator. Thus, we conclude that testing this scenario, even after the HL-LHC upgrade, poses a greater challenge than anticipated.\newp
For our second study, we focus on indirect detection properties of a dark matter candidate in the $\si{\TeV}$ mass range. It possesses scalar- and pseudo-scalar couplings to a massive dark mediator, which connects to the Standard model. Scalar interactions induce an attractive Yukawa potential in the dark sector, which facilitates non-perturbative effects. We discuss their impact on the relic density of dark matter and their effect on indirect detection. Regarding the latter, we examine current limits on indirect detection signals by Planck and Fermi-LAT as well as prospects from CTA. We find that large portions of the model's parameter space are testable in the near future.

\newpage
\thispagestyle{empty}
\mbox{}
\newpage

\thispagestyle{abstract}
\selectlanguage{ngerman}
Dunkle Materie ist eines der wichtigsten ungelösten Rätsel in der Teilchenphysik und Kosmologie unserer Zeit. Wir untersuchen den Einfluss von nicht-perturbativen Effekten auf die Produktion und den direkten wie indirekten Nachweis von Dunkler Materie, um Einblicke in ihre Natur zu erhalten. Insbesondere konzentrieren wir uns auf das Phänomen der Sommerfeld-Verstärkung und die Möglichkeit, Bindungszustände Dunkler Materie zu bilden. Beide Effekte können den Wirkungsquerschnitt der Annihilation von Dunkler Materie verändern und somit Dunkle Materie Observablen im frühen und späten Universum modifizieren. Zu diesem Zweck haben wir zwei Studien durchgeführt, die Fokus auf Observablen Dunkler Materie legen, welche aller Voraussicht nach durch bevorstehende Experimente getestet werden können.\newp
In der ersten Studie untersuchen wir die Produktionseigenschaften eines Dunkle Materie Kandidaten mit sehr schwachen Kopplungen an Standardmodell-Fermionen, die durch einen dunklen Skalar im $\si{\TeV}$-Massenbereich vermittelt werden. Der Skalar besitzt Standardmodell-Quantenzahlen, was einen beträchtlichen Wirkungsquerschnitt für die Produktion neuer Physik am LHC ermöglicht. Bei der Analyse der Produktion von Dunkler Materie über den Freeze-In- und SuperWIMP-Mechanismus entdecken wir eine signifikante Unterdrückung der Ausbeute an Dunkler Materie durch nicht-perturbative Effekte, die höhere Massen für den skalaren Mediator begünstigen. Wir kommen daher zu dem Schluss, dass die Testung dieses Szenarios, selbst nach dem HL-LHC-Upgrade, eine größere Herausforderung darstellt als vorher angenommen wurde.\newp
In unserer zweiten Studie konzentrieren wir uns auf die indirekten Nachweis-Eigenschaften eines Dunkle Materie Kandidaten im $\si{\TeV}$-Massenbereich. Dieser besitzt skalare und pseudoskalare Kopplungen zu einem massiven dunklen Mediator, welcher die Verbindung zum Standardmodell herstellt. Skalare Wechselwirkungen induzieren ein attraktives Yukawa-Potential im dunklen Sektor, das nicht-perturbative Effekte ermöglicht. Wir diskutieren ihre Auswirkungen auf die Reliktdichte Dunkler Materie und ihren Einfluss auf den indirekten Nachweis. In Bezug auf Letzteres untersuchen wir die aktuellen Grenzen für indirekte Nachweissignale von Planck und Fermi-LAT sowie die Aussichten für CTA. Wir stellen fest, dass große Teile des Parameterraums des Modells in naher Zukunft getestet werden können.

\selectlanguage{english}
\newpage
\thispagestyle{empty}
\mbox{}
\newpage

\thispagestyle{acknowledgements}
There are many people who have helped, supported and believed in me on this journey throughout the last couple of years. Words cannot express how grateful I am to have crossed paths with each and every one of you in one way or the other. However, to quote here a probably famous (or not so famous) person: Let me at least try!\newp
First and foremost, I deeply want to thank my supervisor, JProf.\,Dr.\,Stefan Vogl, who has not only given me this chance in the first place but also courageously supported me at each step along the way. His deep understanding of the subject matter that he shared with me on numerous occasions and the advice he gave me, substantially let me grow as a scientist. On this note, I also thank Dr.\,Simone Biondini for a fruitful collaboration over the past years.\newp
Further, I want to give credit to my research training group, in particular Prof.\,Dr.\,Markus Schumacher and my second supervisor, Prof.\,Dr.\,Stefan Dittmaier, not only for granting me funding and enabling me to enjoy conferences and workshops around the world but especially for maintaining a scientific community of fellow PhD students and Postdocs in Freiburg, I could share my experiences with.\newp
Scientific work usually does not happen in a void, and it needs good colleagues to foster an (occasionally) productive environment. Therefore, my special thanks go to María Dias and Selina Machnitzky for cheering up my everyday and also bearing my grumpy moments. Moreover, I'd like to thank the PhD students and Postdocs from the Dittmaier working group, in particular Jan Schwarz, Yann Stoll, Tim Engel, Sebastian Schuhmacher, Jonas Rehberg, Max Reyer and Jóse Lüis Hernando, for many joyful moments and memorable discussions. Out of the ones mentioned above, a shout-out goes to María Dias and Yann Stoll for accepting the ingrateful task of proofreading my first draft, I do owe you one!\newp
Outside of the scientific context, I would like to thank my friends, which supported me in the last years and which I hold very dear. The deep connections I am grateful to have made with you, are my pillars in life. Sebastian Ochs, Elias Eulig, Felix Wolf, Thomas Pfeil, Felix Bitzer, Alexander Prill, Elisabeth Alexander, Florian Staudt and Alexander Holler: Thank you very much for the past years and the years to come.\newp
Finally, I want to thank my family, in particular my parents and my grandmother. The unconditional love and support you gave me helped me to overcome all the hurdles life has confronted me with so far. I could have really not done it without you!
\newpage
\thispagestyle{empty}
\mbox{}
\newpage

{\hypersetup{hidelinks}\tableofcontents}
\newpage
{\hypersetup{hidelinks}\listoffigures\listoftables}
\newpage
\pagestyle{acronyms}
\printglossary[style=mcolindex,type=\acronymtype,nonumberlist]
\newpage
\thispagestyle{empty}
\mbox{}
\newpage
\pagestyle{normal}
\pagenumbering{arabic}

\section{Preface}
\label{sec:preface}

The existence of dark matter is one of the most pressing problems of modern physics. Despite our precise knowledge regarding its gravitational properties and abundance in the Universe, its fundamental particle nature remains elusive. Consequently, it also represents a severe shortcoming in the otherwise highly successful Standard Model of particle physics, lacking to provide us with a viable dark matter candidate. The specifics of its spin, interactions with itself, the Standard Model, or potential new particles in an extended dark sector remain unknown, and its possible mass ranges over fifty orders of magnitude \cite{Bertone:2004pz,Bertone:2010zza,Bauer:2017qwy}.\newp
Over recent decades, myriad efforts have been dedicated to shedding light on dark matter, both theoretically and experimentally. With numerous theoretical models proposing various dark matter signatures across a multitude of experimental setups, the experimental quest nowadays resembles searching for a needle in a haystack. Although searches in some piles of the hay have become quite advanced, so far no clear evidence for the needle has been found. Going forward with this strategy, it is essential that the straw that has been meticulously trawled through, can be thrown away, \ie it is imperative to discard theories contradicting experimental observations. This necessitates robust predictions of dark matter observables from theoretical models when confronted with experimental data. Yet, due to the highly complex environments in which dark matter interacts across the cosmic epochs, accurate predictions still pose serious challenges in dark matter model building. When first proposing a viable parameter space for a certain model to be tested, typically a variety of different assumptions have to be imposed to enable analytical calculations or reach numerical conclusions with limited computing power. This inevitably increases the risk that significant effects, which could substantially influence the model's parameter space, are inadvertently being overlooked, because they are mistakenly estimated to be subleading.\newp
In recent years, a particular set of previously overlooked effects has received a considerable amount of attention among the dark matter community (see \eref\cite{Tulin:2017ara} for a comprehensive overview and the references therein). This thesis aims to explore certain aspects of these effects in dark matter modelling and detection. The effects in question stem from non-perturbative self-interactions among dark matter particles in the non-relativistic regime, mediated by light bosonic particles with strong couplings to the dark matter candidates. Alongside Sommerfeld enhancement, our focus lies on the intriguing prospect of dark matter particles forming metastable bound states, potentially catalyzing dark matter annihilations and significantly influencing dark matter observables. These effects not only impact the overall dark matter abundance by altering production rates in the early Universe but also affect potential dark matter signals at late times.\newp
To this end, we conducted two studies examining these effects within a class of dark matter models. These models were chosen for their sensitivity to these effects and their testability by upcoming experiments, rendering them particularly intriguing subjects for investigation. Both studies outlined in this thesis have been published and are accessible in \erefs\cite{Bollig:2021psb, Biondini:2023ksj}. The first study delves into the influence of non-perturbative effects on dark matter production rates in the early Universe, focusing on a non-thermal dark matter model with feeble couplings to the Standard Model heat bath. The second study also considers these effects on dark matter production but concentrates on their implications for indirect detection. This study employs a model featuring thermal dark matter with no direct couplings to the Standard Model, but rather interactions with a thermalized dark mediator.\newp
We start with a concise overview on the principles of cosmology in \cref{sec:introductiontocosmology}, encompassing a short history of the Universe, its main properties at its earliest stages, and its thermal evolution. Moving forward to \cref{sec:dmintro}, we introduce the concept of dark matter, including experimental evidence, its primary characteristics, potential dark matter candidates, and methods for dark matter detection. \Cref{sec:DMproduction} elaborates on quantifying the evolution of dark matter and introduces fundamental concepts for calculating its present-day observed abundance. In \cref{sec:non-perturbative-effects}, we present two approaches to address the non-perturbative effects examined in this thesis. \Cref{sec:impactonnonthermalDMproduction,sec:indirectdetection} encompass two studies conducted by the author, focusing on non-perturbative effects in dark matter production and detection, which form the core of this thesis. We wrap up with concluding remarks in \cref{sec:conclusion}.

\paragraph{Conventions and notation:}$~$\newpp
In this thesis, we will adopt natural units, denoted as $c=\hbar= k_B\equiv 1$, unless explicitly stated otherwise. Additionally, 4-vector operations will adhere to the metric signature $(+,-,-,-)$. Greek lower case letters, such as $\mu, \nu = 0,1,2,3$, will serve as indices for 4-vectors, while Latin lower case letters, like $i,j=1,2,3$, will denote spatial indices. Bold variables, such as $\vec{x}$ ($\uvec{x}$), will represent 3-dimensional (unit-)vectors, while non-bold vector quantities will either denote 4-vectors $x\equiv x^\mu$ or absolute values of 3-vectors $x\equiv \abs{\vec{x}}$, as clear from the context. Time derivatives will be indicated by a dot, \ie $\dot{y}(t,x)$, while other derivatives will be denoted as $y'(x)$ for brevity. Regarding spherical coordinates, $\Omega\equiv (\theta,\phi)$ will represent the solid angle with the measure $\dd{\Omega}=\dd{\!\cos\theta}\dd{\phi}$, where $\phi\in[0,2\pi)$ denotes the azimuthal angle and $\theta\in[0,\pi)$ represents the polar angle. All values for physical constants and particle physics properties used in this work, have been extracted from the latest \textit{Particle Data Group} review \cite{Workman:2022ynf}, if not referenced otherwise.
\section{Introduction to cosmology}%
\label{sec:introductiontocosmology}

Starting with an introduction to cosmology, our aim is not to comprehensively cover the $13.8$ billion years \cite{Planck:2018vyg} of cosmological evolution. Instead, we aim to provide the reader with an overview of key concepts and terminologies relevant to this work. To compile this section, we have drawn inspiration primarily from \erefs\cite{Baumann:2022mni,Dodelson:2003ft}, although similar information can be found in numerous other books on modern cosmology.\newp
We will commence with a concise overview of significant cosmological milestones in the Universe's history in \cref{subsec:historyoftheuniverse}. Transitioning to \cref{subsec:FRWuniverse}, our focus will shift to elucidate the fundamental equations that govern the early Universe, providing a quantitative understanding of its expansion dynamics. Subsequently, in \cref{subsec:thermalhistory}, we will introduce key variables to delineate the thermal evolution of the early Universe.


\subsection{A very short history of the Universe}
\label{subsec:historyoftheuniverse}

The \textit{Big Bang} is defined as the time singularity of the metric used to describe the early Universe. It will serve as our temporal reference point. In its earliest stage, the Universe likely underwent a period of exceedingly rapid expansion, commonly referred to as \textit{inflation} \cite{Guth:1980zm,Baumann:2009ds}. This inflationary epoch effectively smoothed out any pre-existing spatial curvature and erased information regarding the initial particle composition. After the inflationary phase terminated, the Universe transitioned into a stage of \textit{reheating} \cite{Kofman:1994rk,Kofman:1997yn}. During this phase, a substantial amount of energy was injected into the \gls{sm} sector, leading to the formation of an intensely hot and dense plasma consisting of \gls{sm} and potentially \gls{bsm} particles. This marks the starting point for our subsequent discussion. Although the temperature of the \gls{sm} plasma after reheating can be as low as $T_{\text{rh}}\gtrsim \SI{4}{\MeV}$ \cite{Hannestad:2004px}, in the remainder of this thesis, we will presume it to be well above the $\si{\TeV}$ scale. This assumption is made without specific considerations about the underlying details of the inflationary model.\newp
As the temperature of the \gls{sm} bath cools down, particles undergo a process of decoupling from the thermal plasma. This typically occurs when the expansion rate of the Universe surpasses the interaction rates of the particles in question with the surrounding plasma. As an example, neutrinos decouple at around $\SI{1}{\MeV}$. Subsequently, they become free streaming, meaning they propagate through the Universe without further scattering, and thereby constitute the \textit{cosmic neutrino background} \cite{Lopez:1998aq}. Detecting this background remains challenging up to present day due to the exceedingly weak interactions of neutrinos with the rest of the \gls{sm}. The earliest remnants from the primordial Universe which have been observed so far, lie in the abundances of light elements, forming once the temperature drops below the binding energy of nucleons. This phenomenon, known as \gls{bbn}, initiates at temperatures around $T\lesssim \SI{1}{\MeV}$ \cite{Sarkar:1995dd, Fields:2019pfx}, corresponding to a cosmic age of approximately $\SI{3}{\min}$. Presently, constraints derived from these observations play a crucial role in establishing stringent bounds on the cosmological evolution of particles beyond the \gls{sm}.\newp
At the scale of approximately a few hundred $\si{\keV}$, the thermal bath retains only a significant abundance of protons, electrons, and photons. The enduring strong coupling between electrons and photons at these energies primarily arises from Thomson scattering ($e^-+\gamma\to e^- +\gamma$). As the temperature drops below the binding energy of hydrogen at a few $\si{\eV}$, neutral hydrogen atoms begin to form during a phase called \textit{recombination} \cite{Zeldovich:1969ff,Peebles:1968ja}, occurring $260,000-380,000$ years after the Big Bang. Recombination induces a sharp drop in the free electron density, rendering Thomson scattering inefficient, and thereby making the Universe transparent to photons for the first time. Subsequently, the photons from the \gls{sm} bath decouple and have since traversed freely. They constitute the \gls{cmb} \cite{Penzias:1965wn,Hu:2001bc}, a relic of this cosmological epoch occurring approximately $380,000$ years following the Big Bang, which has been observed with remarkable precision. Its examination has yielded profound insights into the underlying physics governing cosmological evolution.\newp
The great abundance of non-relativistic matter initiated a slow process of gravitational collapse, giving rise to the formation of the first stars, galaxies, and the emergence of large-scale structures. These transformative processes span durations extending from hundreds of millions to a few billion years and persist through to the present day.


\subsection{The Friedmann-Robertson-Walker Universe}
\label{subsec:FRWuniverse}

\begin{figure}[ht]
    \centering
    \includegraphics[width=\textwidth]{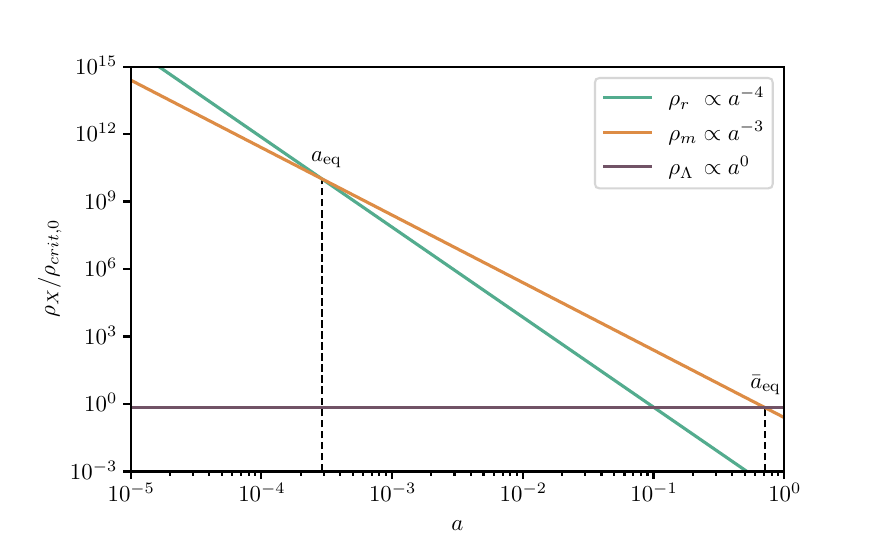}
    \caption[The scaling of the energy density with the scale factor $a$.]{The scaling of the energy density (normalized by $\rhocrit$) with the scale factor $a$. Displayed are three different components: radiation (r), matter (m) and dark energy in form of a cosmological constant ($\Lambda$). Also shown are the points of matter radiation equality $a_{\text{eq}}$ as well as dark energy matter equality $\bar{a}_{\text{eq}}$. The scale factor today is chosen to be $a(t_0)\equiv 1$.}
    \label{fig:scalingofscalefactor}
\end{figure}

On cosmological scales, the Universe exhibits isotropy, meaning that its properties appear uniform in every direction. This observation leads to the inference of homogeneity, suggesting that the Universe is uniform not only as seen from Earth but at any point in space. Since the structures we see today have grown with time, cosmologists believe that at early stages, the Universe was nearly perfectly homogeneous and isotropic (apart from quantum fluctuations). This proposition is encapsulated in the \textit{cosmological principle}, which imposes stringent constraints on the nature of space-time anticipated in the early Universe. In fact, there is just one (non-equivalent) metric which aligns with these criteria: the \gls{frw} metric, parametrized by
\begin{equation}
    \dd{s}^2=\dd{t}^2-a(t)^2\left[\frac{\dd{r}^2}{1-\kappa^2r^2}+r^2\dd{\Omega}^2\right]
\end{equation}
in spherical space coordinates with $\dd{\Omega}\equiv \dd{\!\cos\theta}\dd{\phi}$ and $\kappa$ a curvature parameter, which we subsequently set to zero in accordance with current experimental data \cite{Planck:2018vyg}.\newp
The parameter $a(t)$ is the \textit{scale factor} of the Universe, describing its expansion. The properties of particles in the Universe scale differently with $a(t)$. Obviously, distances grow proportional to $a(t)$, while particle momenta $p$ scale with $1/a(t)$, which can be seen by employing the \textit{geodesic equation} using an \gls{frw} metric. Likewise, the wavelength of massless particles like the photon $\lambda=2\pi/p$ will scale with $a(t)$. This leads to the interesting concept of \textit{redshift}, meaning that light emitted with a wavelength $\lambda_1$ at some point in the cosmological history will have a larger wavelength $\lambda_0=a(t_0)/a(t_1)\lambda_1>\lambda_1$ when it reaches us today. The redshift is usually parametrized by a parameter $z$, which is defined by 
\begin{equation}
    \frac{\lambda_0}{\lambda_1}\equiv 1+z =\frac{a(t_0)}{a(t_1)}\,,
\end{equation}
and can be used as an alternative measure of time to describe cosmological evolution. As a reference point, one usually sets $a(t_0)\equiv 1$ with $t_0$ denoting the age of the Universe today.\newp
The dynamics of the Universe is governed by the \textit{Einstein field equations}
\begin{equation}
    G_{\mu\nu}=8\pi G\, T_{\mu\nu}\,,
\end{equation}
which relate the space-time curvature of the Universe encoded in the Einstein tensor $G_{\mu\nu}$ with its energy content parametrized by the energy-momentum tensor $T_{\mu\nu}$. Its proportionality is given by the gravitational constant $G$. The Einstein tensor is defined by
\begin{equation}
    G_{\mu\nu}\equiv R_{\mu\nu}-\frac{1}{2}R\,g_{\mu\nu}\,,
\end{equation}
where 
\begin{equation}
    R_{\mu\nu}\equiv \partial_\lambda\CS{\lambda}{\mu}{\nu}-\partial_\nu\CS{\lambda}{\mu}{\lambda}+\CS{\lambda}{\lambda}{\rho}\CS{\rho}{\mu}{\nu}-\CS{\rho}{\mu}{\lambda}\CS{\lambda}{\nu}{\rho}\,,\quad R\equiv {R^\mu}_\mu=g^{\mu\nu}R_{\mu\nu}\,,
\end{equation}
are, respectively, the Ricci tensor and scalar, and $g_{\mu\nu}\equiv\diag{(1,-a(t)^2,-a(t)^2,-a(t)^2)}$ denotes the \gls{frw} metric (in Cartesian spatial coordinates). The $\CS{\mu}{\alpha}{\beta}$ are called Christoffel symbols and describe a \textit{metric connection}. They are given by
\begin{equation}
    \CS{\mu}{\alpha}{\beta}\equiv \frac{1}{2}g^{\mu\lambda}\left(\partial_\alpha g_{\beta\lambda}+\partial_\beta g_{\alpha\lambda}-\partial_\lambda g_{\alpha\beta}\right)
\end{equation}
and have to be calculated using the corresponding metric. The non-vanishing components of the Einstein tensor ${G^\mu}_\nu=g^{\mu\lambda}G_{\lambda\nu}$ for the \gls{frw} metric yield
\begin{equation}
    {G^0}_0=3\left(\frac{\dot{a}}{a}\right)^2, \quad {G^i}_j=\left[2\frac{\ddot{a}}{a}+\left(\frac{\dot{a}}{a}\right)^2\right]\delta^i_j\,.
\end{equation}\newp
From homogeneity and isotropy arguments one can argue that the energy-momentum tensor of the system must equal the one of a perfect fluid ${T^\mu}_\nu=\diag{(\rho,-P,-P,-P)}$ in its rest frame, where the energy density $\rho$ and pressure $P$ are solely functions of time. They are linked through energy-momentum conservation
\begin{equation}
     {T^\mu}_{\nu;\mu}\equiv \partial_\mu {T^\mu}_\nu + \CS{\mu}{\mu}{\lambda}{T^\lambda}_\nu-\CS{\lambda}{\mu}{\nu}{T^\mu}_\lambda=0\,,
\end{equation}
where the $\nu=0$ component defines a continuity equation, given for the \gls{frw} metric by
\begin{equation}
    \label{eq:continuityequation}
    \dot{\rho}+3\frac{\dot{a}}{a}\left(\rho+P\right)=0\,.
\end{equation}
The Universe is in general composed of three idealized components, which can be classified by their relation between energy density and pressure. Their difference in scaling with $a(t)$ can be deduced from \cref{eq:continuityequation}. The first component is \textit{matter} for which $\abs{P}\ll \rho$ and therefore $\rho\propto a(t)^{-3}$, \ie the dilution of energy density mirrors the expansion of the volume $V\propto a(t)^3$. It is comprised of all non-relativistic particle species which are part of the \gls{sm} and beyond at a given time. The second component is \textit{radiation}, for which $P=\rho/3$ and thus $\rho\propto a(t)^{-4}$. For this type also the redshift of the energy $E\propto a(t)^{-1}$ has been included in the dilution. All relativistic particles such as photons and neutrinos as well as gravitational waves fall under this category. The third component is referred to as \textit{dark energy} and has negative pressure $P\approx-\rho$ \cite{Planck:2018vyg}, which results in a constant energy density $\rho\propto a^0$. Not much is known about dark energy, except that it is needed to explain the accelerated expansion of the Universe we observe nowadays \cite{SupernovaSearchTeam:1998fmf,SupernovaCosmologyProject:1998vns}. The easiest way to include such a component into the cosmological picture, which is consistent with observations, is to add a cosmological constant $\Lambda\,g_{\mu\nu}$ to the Einstein field equations (which would exactly yield $P=-\rho$). Due to the different scaling of the energy density with $a(t)$, each of the three components has dominated the Universe for a certain period of time, as we can see in \cref{fig:scalingofscalefactor}. In the beginning, the Universe has been dominated by radiation, the era we predominantly work in for the remainder of this thesis. At a redshift of $z\sim 3400$, or equivalently $a_{\text{eq}}\sim \SI{2.9e-4}{}$, we reach matter radiation equality and the matter component takes over, approximately $60,000$ years after the Big Bang. Within this era, most structure formation processes take place. Only very recently on cosmological time scales, at a redshift of $z\sim 0.4$ (or $\bar{a}_{\text{eq}}\sim 0.71$, ca.\,$9$ billion years after the Big Bang), we have reached equality between matter and dark energy, such that today the latter is starting to dominate the overall energy density.\newp
Combining our results for the Einstein and the energy-momentum tensor results in the \textit{Friedmann equations}
\begin{align}
    \label{eq:Friedmann1}
    \left(\frac{\dot{a}}{a}\right)^2&=\frac{8\pi G}{3} \rho\,,\\
    \frac{\ddot{a}}{a} &= -\frac{4\pi G}{3}\left(\rho+3P\right)\,,
\end{align}
where $\rho$ and $P$ are understood as the respective sum over the individual energy density and pressure components mentioned above. The first Friedmann equation as well as many other cosmological properties are usually written in terms of the \textit{Hubble parameter} $H\equiv \dot{a}/a$ instead of $a$. This is especially convenient, because $H$ is the expansion rate of the Universe at a given time. The first Friedmann equation can thus be recasted into the form 
\begin{equation}
    H^2(a)=H_0^2\left[\Omega_{r,0}\left(\frac{a_0}{a}\right)^4+\Omega_{m,0}\left(\frac{a_0}{a}\right)^3+\Omega_{\Lambda,0}\right]\,,
\end{equation}
where $H_0$ is the \textit{Hubble constant}, \ie the value of the Hubble parameter measured today, $a_0\equiv a(t_0)\equiv 1$ and the $\Omega_{X,0}=\rho_{X,0}/\rhocrit$ denote the dimensionless density parameters (or \textit{abundances}) of the different energy density components at present time, with $\rhocrit=3H_0^2/(8\pi G)$ the \textit{critical energy density} today. Neglecting an additional component, which would arise from a non-zero curvature, $\Omega_r+\Omega_m+\Omega_\Lambda=1$ is true at all times. The best fit values for the abundances today are $\Omega_{r,0}=\SI{5.38(15)e-5}{}$, $\Omega_{m,0}=\SI{0.315(7)}{}$, $\Omega_{\Lambda,0}=\SI{0.685(7)}{}$ with $H_0=100\,h\,\si{\km\,\s^{-1}\Mpc^{-1}}$ and $\rhocrit=\SI{2.8e11}{}\,h^2\,M_{\odot}\,\si{\Mpc^{-3}}$, where $M_{\odot}=\SI{1.98841(4)e30}{\kg}$ denotes the solar mass and $h=0.674(5)$ \cite{Planck:2018vyg}.


\subsection{The thermal history of the early Universe}
\label{subsec:thermalhistory}

From the perfect black-body spectrum of the \gls{cmb} \cite{Planck:2018vyg}, we have sufficient evidence that the \gls{sm} sector in the early Universe was in local \textit{thermal equilibrium}. Thermal equilibrium (\ie no net heat transfer) implies that two or more systems share the same temperature, which is true for the \gls{sm} plasma. This allows us to use equilibrium thermodynamics in order to quantify the Universe at this stage.\newp
We start by defining a distribution function $f(\vec{x},\vec{p},t)$ which quantifies how the particles and their momentum are distributed in a certain volume over time. As a consequence of homogeneity and isotropy in the early Universe, the distribution function of a bath particle should be independent of $\vec{x}$ and only depend on the magnitude of the momentum $p=\abs{\vec{p}}$. Leaving the time dependence implicit, we will work with $f(\vec{x},\vec{p},t)=f(p)$ in the following. \textit{Kinetic equilibrium} is a statement about how efficiently energy between two or more systems can be exchanged on cosmological scales. If it is maintained, the distribution function of a particle species $a$ with energy $E_a(p)=\sqrt{p^2+m_a^2}$ equals a \textit{Bose-Einstein} or \textit{Fermi-Dirac} distribution
\begin{equation}
    f_a(p)\equiv\frac{1}{\exp\left\{\frac{E_a(p)-\mu_a(T_a)}{T_a}\right\}\pm 1}\,,
\end{equation}
depending on if the particles in question are bosons ($-$) or fermions ($+$). As we can see, the distribution function of a particle species in kinetic equilibrium depends on two macroscopic variables, its temperature $T_a$ and its chemical potential $\mu_a(T_a)$ the latter of which characterizes the response of a system to a change in the particle number. For two or more systems in \textit{chemical equilibrium}, there is no change in the concentration of the particles between the systems, implicating that the total chemical potential vanishes. Systems in chemical and kinetic equilibrium are also in thermal equilibrium and vice versa (see \eg\eref\cite{Kolb:1990vq}). Therefore, we can replace the distribution functions for \gls{sm} particles with their equilibrium values 
\begin{equation}
    \label{eq:feq}
    \feq{a}(p)\equiv\frac{1}{e^{E_a(p)/T}\pm 1}
\end{equation}
at a common temperature $T$, where we can safely assume all chemical potentials to be negligible at early times. For non-relativistic particles, we can reduce their distribution functions to a \textit{Maxwell-Boltzmann} form $\feq{a}(p)\approx e^{-E_a(p)/T}$.\newp
The \textit{number density} of a particle species $a$, \ie the number of particles of type $a$ within a certain volume $V$, is defined as
\begin{equation}
    \label{eq:numberdensity}
	n_a\equiv g_a\int \frac{\dd{p_a^3}}{(2\pi)^3} f_a(p)\,,
\end{equation}
where $g_a$ denotes their number of internal \gls{dof}. The energy density and pressure of a species $a$ can be defined likewise
\begin{align}
    \rho_a \equiv & ~g_a\int\frac{\dd[3]{p}}{(2\pi)^3} f_a(p) E_a(p)\,,\\
    P_a \equiv & ~g_a\int\frac{\dd[3]{p}}{(2\pi)^3} f_a(p) \frac{p^2}{3E_a(p)}\,.
\end{align}
Another important quantity for cosmological evolution is the entropy $S$ of the Universe, which is conserved in thermal equilibrium. As we can neglect the insignificant enhancement of entropy from non-equilibrium processes at later stages, we can treat the expansion of the Universe as adiabatic, leading to a conservation of entropy throughout its evolution. It will be in the following more convenient to work with the \textit{entropy density} $s\equiv S/V$ which can be related to the other thermodynamic variables via 
\begin{equation}
    s=\frac{\rho+P}{T}\quad\text{for}\quad \mu=0\,,
\end{equation}
where a summation over all species is implicit. Its value today yields $s_0=\SI{2891}{\cm^{-3}}$.\newp
\begin{figure}[ht]
    \centering
    \includegraphics[width=\textwidth]{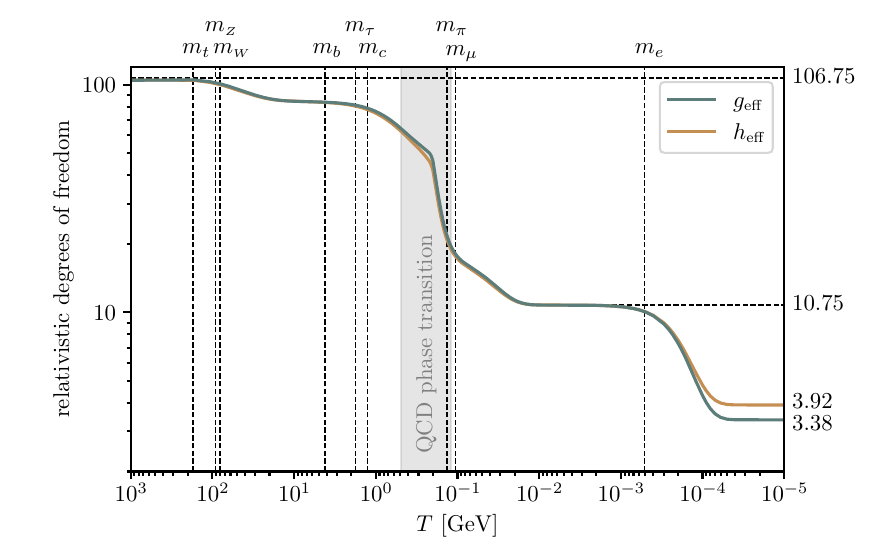}
    \caption[The evolution of the effective number of relativistic energy and entropy degrees of freedom.]{The evolution of the effective number of relativistic energy ($\geff$) and entropy ($\heff$) \gls{dof} is depicted. Vertical lines correspond to masses of \gls{sm} particles, while horizontal lines indicate important benchmark values. The temperature range at which the \acrshort{qcd} phase transition occurs is shaded in gray. Adapted from \eref\cite{Baumann:2022mni} with data taken from \eref\cite{Laine:2015kra}.}
    \label{fig:relativisticdofs}
\end{figure}
For particles in thermal equilibrium, their number density can be determined via
\begin{equation}
    \label{eq:neq}
    \ndeq{a}(T)=\frac{g_a}{2\pi^2}T^3 \int_0^\infty \dd{u}\frac{u^2}{e^{\sqrt{u^2+(m/T)^2}}\pm 1}\,,
\end{equation}
which is solely a function of $T$ (as well as the particles mass) and has no analytical solution. However, we can find analytic approximations in the relativistic regime (where $m/T\ll 1$) and by using the Maxwell-Boltzmann approximation for $\feq{a}$.\footnote{In the subsequent sections, the Maxwell-Boltzmann ansatz will be used even in situations where particles are not necessarily non-relativistic. However, as long as the particles are not strictly massless, the differences with respect to the full relativistic treatment are expected to be small.} They read
\begin{equation}
    \label{eq:neqapprox}
    \ndeq{a,\text{\tiny MB}}(T)= \frac{g_a}{2\pi^2}\,m^2\,T\,K_2\left(\frac{m}{T}\right),\qquad 
    \ndeq{a,\text{rel}}(T)=\frac{\zeta(3)}{\pi^2} g T^3
    \begin{cases}
        1 & \text{bosons} \\
        \frac{3}{4} & \text{fermions}
    \end{cases}\,,
\end{equation}
where $K_2(x)$ is the modified Bessel function of second order defined in \cref{eq:BesselKdef} and $\zeta(3)\approx 1.202$. Collecting contributions from all species (assuming a common temperature), the total energy and entropy density can be written as 
\begin{align}
    \label{eq:energydensity}
    \rho(T) &= \sum_j \rho_j(T)= \frac{\pi^2}{30}\geff(T) T^4\,,\\
    \label{eq:entropydensity}
	s(T)& = \sum_j \frac{\rho_j(T)+P_j(T)}{T} = \frac{2\pi^2}{45}\heff(T)T^3\,,
\end{align}
where $\geff(T)$ and $\heff(T)$ are labelled the relativistic energy and entropy \gls{dof} (see \eref\cite{Husdal:2016haj} for a detailed overview). Their name arises because they effectively count how many particles are still relativistic at a given temperature $T$ and thus contribute to the total energy and entropy density in a radiation-dominated Universe. The evolution of $\geff(T)$ and $\heff(T)$ with the \gls{sm} bath temperature is depicted in \cref{fig:relativisticdofs}. At very high temperatures, when all \gls{sm} bath particles are relativistic, they take on a value of $106.75$, which is simply the sum of all bosonic \gls{dof} ($g_b=28$) plus the sum of all fermionic \gls{dof} ($g_f=90$), where the latter is multiplied by a factor of $7/8$. When a particle species $a$ becomes non-relativistic around $m_a\gtrsim T$, its contribution to $\geff(T)$ and $\heff(T)$ gets suppressed by a factor $e^{-m_a/T}$. A severe drop in the relativistic \gls{dof} happens around the confinement scale of \gls{qcd}, $\Lambda_{\text{\tiny QCD}}\sim\SI{400}{\MeV}$, when the quarks left in the plasma combine into hadrons, which are all non-relativistic (except pions) below the temperature of the \gls{qcd} phase transition. After the decoupling of pions and muons, we are left with photons, electrons, and neutrinos in the \gls{sm} bath, constituting a value of $10.75$ at the onset of \gls{bbn} around $\SI{1}{\MeV}$. The gap between $\geff(T)$ and $\heff(T)$ at very low temperatures is a result of their different scaling with $T_\nu/T$, where $T_\nu$ denotes the neutrino temperature after decoupling.\newp
We can combine \cref{eq:Friedmann1,eq:energydensity} to obtain an expression for the Hubble rate during the radiation dominated era, which is then given by
\begin{equation}
    \label{eq:Hubble}
	H(T)= \sqrt{\frac{4\pi^3}{45\MPl^2}}\geff^{1/2}(T)T^2\,,
\end{equation}
where we substituted the gravitational constant with the Planck mass via $\MPl=G^{-1/2}$.

\section{Introduction to dark matter}
\label{sec:dmintro}

In the era of precision cosmology, it is possible to measure the baryonic content of the Universe, \ie the amount of visible non-relativistic matter like hydrogen, helium etc. to percent level accuracy. Its measured value is $\Omega_{b,0}=0.0493(6)$ \cite{Planck:2018vyg}, which constitutes about a fifth of the total energy density allocated to matter, today (\cf\cref{subsec:FRWuniverse}). The immediate question regarding the nature of the remaining matter content, which is not contained in visible objects, has puzzled cosmologists and particle physicists for almost a century. It has been given the name \gls{dm} but up to the present day, it remains unclear, what this additional and very abundant matter contribution actually is. \newp
While the structure and content of this section is based on \erefs\cite{Bertone:2004pz,Bauer:2017qwy}, the information presented about \gls{dm} is also available in various other references. We will begin by providing a brief historical overview on the discovery of dark matter as well as its experimental evidence up to present day in \cref{subsec:dmexpevidence}. In \cref{subsec:dmcharacteristicsandcandidates}, we will elaborate on the known properties of \gls{dm} and classify different \gls{dm} candidates. Finally, we will discuss the different detection methods for particle \gls{dm} in \cref{subsec:detectionmethods}.


\subsection{Discovery and experimental evidence}
\label{subsec:dmexpevidence}

\begin{figure}
    \centering
    \includegraphics[width=0.8\textwidth]{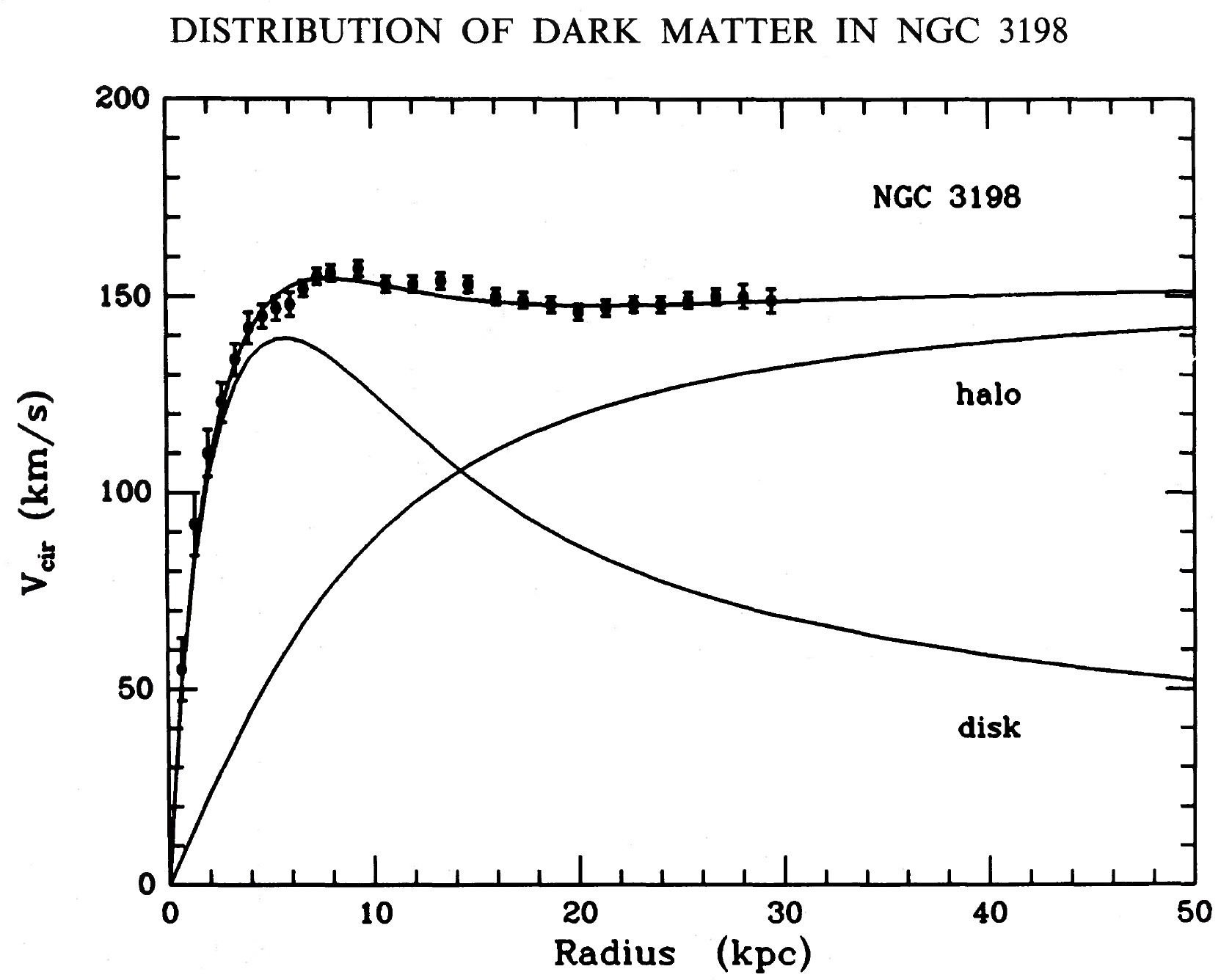}
    \caption[Circular velocity of an exemplary galaxy as a function of the radius.]{Circular velocity of an exemplary galaxy as a function of the radius. Displayed are the theoretical expectations from a luminous galactic disk, a \gls{dm} halo and a combination of both. Experimental data is shown as dots with respective error bars. Taken from \eref\cite{vanAlbada:1984js}.}
    \label{fig:rotationcurves}
\end{figure}

The first evidence of \gls{dm} has been observed in galaxy clusters. In 1933, the astronomer Fritz Zwicky studied the redshift of various galaxy clusters \cite{Zwicky:1933gu} using an analysis, which had been published two years prior by Edwin Hubble and Milton Humason \cite{Hubble:1931zz}. He took particular interest in the Coma Cluster, for which he calculated the average velocity dispersion of the $\approx 800$ galaxies within it by applying the \textit{virial theorem}.\footnote{The virial theorem connects the average kinetic energy of a particle system bound by a conservative force (like gravity) with its average potential energy. It can be applied to galaxy clusters, since the diameter of the galaxies within the cluster ($\sim\SI{100}{\kpc}$) is much smaller than the diameter of the cluster itself ($\sim\SI{10}{\Mpc}$).} By making additional assumptions about the average masses of the galaxies obtained from visible matter sources as well as the physical size of the gravitationally bound system, he obtained an average velocity dispersion of $\sim\SI{80}{\km/\s}$, which was in great contrast to the averaged observed dispersion of $\sim\SI{1000}{\km/\s}$ along the \gls{los}. He deduced from this comparison a large abundance of non-luminous matter within the cluster.\newp
Although similar observations were made in other clusters in the following years, it would take over three decades, until the idea of missing mass in the Universe had been established among cosmologists. A game changer, which convinced many physicists that the \gls{dm} problem was real, had been measurements of galaxy rotation curves performed by Vera Rubin and Kent Ford in the 1970's \cite{Rubin:1970zza,Rubin:1978kmz}. Ken Freeman soon realized that the shape of these rotation curves cannot be explained by luminous matter alone \cite{Freeman:1970mx}, an argument which was later extended and linked to the missing mass problem in galaxy clusters by Jaan Einasto, Jim Peebles and others \cite{Einasto:1974,Ostriker:1974}. The circular velocity of a galactic object is given by 
\begin{equation}
    \label{eq:circularvelocity}
    v_{\text{cir}}^2(r)=\frac{GM(r)}{r}\,,
\end{equation}
with $M(r)$ being the enclosed mass in a spherical volume with radius $r$. As sketched in \cref{fig:rotationcurves} for an exemplary galaxy, the circular velocity of a galactic disk of luminous matter is expected to rise linearly $v(r)\propto r$ in the inner part, where the matter density is approximately constant. This rise would then be followed by a $v(r)\propto r^{-1/2}$ decay in the outer parts of the galaxy, where luminous matter becomes dilute. However, the measurements strongly indicate a plateau outside of the galactic disk, hinting towards a spherical and invisible matter halo with a density $\rho(r) \propto r^{-2}$ accompanying it and thus greatly extending the size of the galaxy.\newp
\begin{figure}[ht!]
    \centering
    \includegraphics[width=0.9\textwidth]{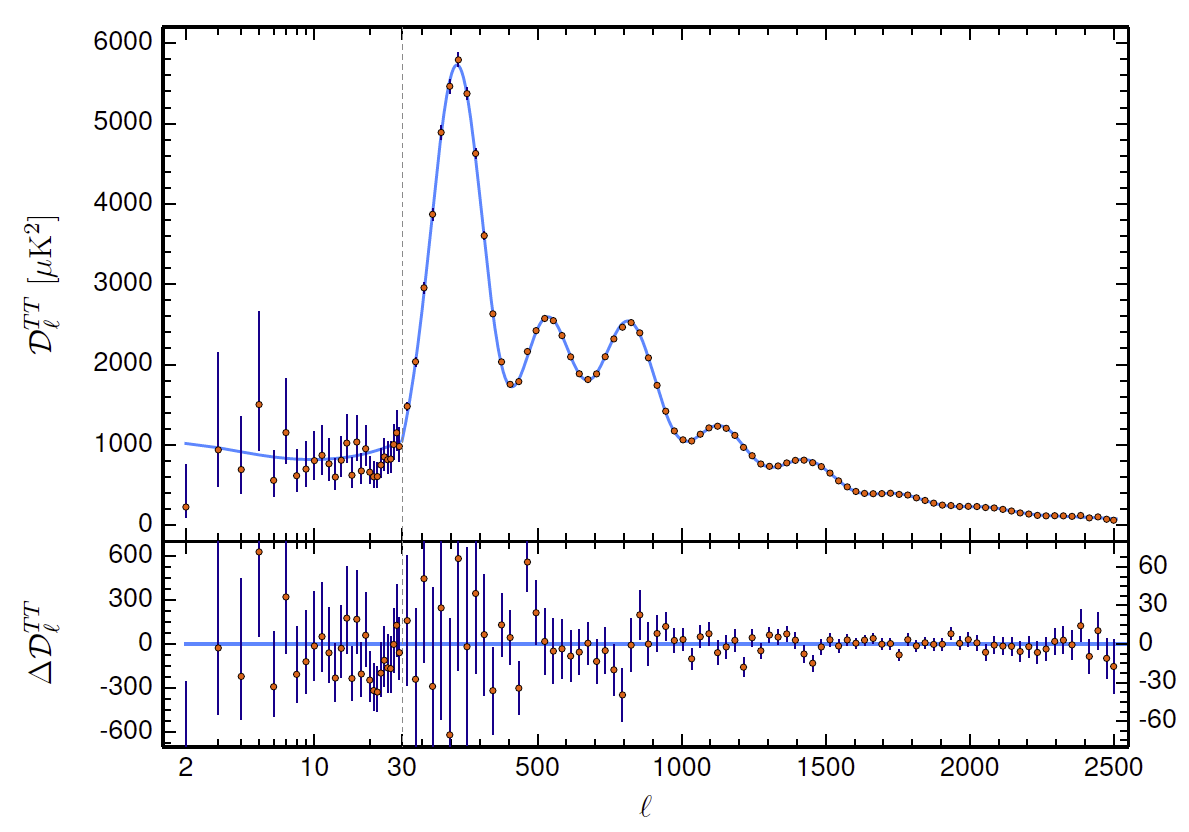}
    \caption[Temperature power spectrum of the CMB.]{Temperature fluctuations of the \gls{cmb} in terms of the two-point correlation functions expanded in spherical harmonics with angular modes $l$ (also sometimes called \textit{\gls{cmb} power spectrum}). The continuous line represent the best-fit function from the \gls{lcdm} model. Taken from \eref\cite{Planck:2018vyg}.}
    \label{fig:cmbpowerspectrum}
\end{figure}
Another strong evidence for \gls{dm}, which is particularly important for a better understanding of its nature, has been observed in a collision of two galaxy clusters, nowadays famously known as the \textit{bullet cluster} \cite{Clowe:2006eq}. The ionized gas within the two colliding systems interacts through \gls{sm} processes and thus experiences a drag force. It would remain behind as the accompanying \gls{dm} halos passed through each other without collision. Exactly this behaviour has been observed in the bullet cluster. The the location of the heated gas was measured there by X-ray surveys, whereas the \gls{dm} concentration could be determined through \textit{gravitational lensing}.\footnote{Gravitational lensing is a powerful method to observe mass concentrations, in particular of matter which does not emit photons and, thus, cannot be easily be detected otherwise. It is agnostic to possible \gls{sm} interactions of the matter in question, as it relies on the bending of light through massive objects within the \gls{los}.}\newp
The most precise measurements of the different energy components in the Universe, including the matter content observed today, stem from the cosmological model describing the temperature fluctuations of the \gls{cmb} \cite{Planck:2018vyg}.\footnote{Other important observations to constrain the cosmological abundances of certain energy components, on which we will not comment on further, come from supernovae measurements \cite{SDSS:2014iwm} and \textit{baryon acoustic oscillations} \cite{BOSS:2012tck}.} The so-called \gls{lcdm} model, represented by the continuous line in \cref{fig:cmbpowerspectrum}, has been a huge success in describing these primordial deviations. It requires a total of six input parameters, which have been fitted to the data collected by various surveys over the last decades. The figure itself contains measurements of the two-point correlation function of the temperature fluctuations expanded in spherical harmonics. The height and width of the peaks of the function contain information about the values of the different abundances which have been stated in the beginning of this section as well as in \cref{subsec:FRWuniverse}. Its prediction for the present day \gls{dm} abundance yields $\Omegadmt=0.265(7)$ \cite{Planck:2018vyg}. The \gls{lcdm} model, also known as \textit{Standard Model of Cosmology}, conflates our modern understanding of the Universe with \gls{dm} as a necessary but yet poorly understood component.


\subsection{Characteristics and possible candidates}
\label{subsec:dmcharacteristicsandcandidates}

Over the years, it has been possible to derive certain characteristics, \gls{dm} should possess. To sum it up, a viable \gls{dm} candidate needs to be dark, cold, stable, collisionless, and non-baryonic. What is meant by these terms will be discussed in the following. First and foremost, it should be \textit{dark}, meaning that its couplings to \gls{sm} particles should be small. Particularly, this implies that \gls{dm} must be (almost) electrically neutral. This is necessary to avoid detection and to allow it to form halos around galaxies without collapsing into disk-like structures, as is true for baryonic matter (see, for example, \eref\cite{McDermott:2010pa}). Further, \gls{dm} must be sufficiently \textit{cold}, \ie it must be non-relativistic during the epoch of structure formation. This constraint comes from the observation of large scale structures, which prefer theoretical results computed using cold (non-relativistic) over hot (relativistic) \gls{dm} already in the regime, where linear perturbation theory can be applied. These observations are reflected in \gls{cmb} data, which requires non-relativistic \gls{dm} to account for the growth of small perturbations in the power spectrum. At smaller scales, probed usually by galaxy surveys, numerical N-body simulations are needed to accurately describe the formation of structure. Here, the mismatch for hot \gls{dm} becomes even more apparent (see \eg\eref\cite{Peebles:1980yev}), leading to robust constraints on the velocity of \gls{dm} at late times. For thermally produced \gls{dm} (\cf\cref{sec:DMproduction}), it is also possible to put a stringent constraint of $\gtrsim \order{\si{\keV}}$ on its mass (see \eg\eref\cite{Irsic:2017ixq}). Since the effects of \gls{dm} have been observed on a variety of different time scales (from the \gls{cmb} up to present day), it must be sufficiently \textit{stable}. This means that \gls{dm} annihilation or conversion processes into \gls{sm} (and potential \gls{bsm}) particles had to become inefficient at early times and \gls{dm} decay rates must be very slow on cosmological time scales. Stringent limits on metastable \gls{dm} are very model dependent. As mentioned in the discussion of the bullet cluster in \cref{subsec:dmexpevidence}, \gls{dm} needs also to be \textit{collisionless}, meaning that its self-interactions and the dissipation of kinetic energy into other particles must be small. This also ensures, that \gls{dm} halos are almost spherical. Current bounds on the \gls{dm} self-interaction cross section are $\sigma/m_{\text{\tiny DM}}\lesssim \SI{1}{\cm^2/g}$ \cite{Tulin:2017ara}. Finally, \gls{dm} must be of \textit{non-baryonic} nature. This requirement is clear from the previous discussion of \gls{cmb} constraints but it is also imposed by \gls{bbn}, as \gls{dm} cannot play a measurable role in the formation of light elements.\newp
In the discussion above, we have implicitly assumed that \gls{dm} has a particle nature, \ie that it consists of one (or more) elementary or composite particles, which we have not yet discovered. Although this possibility is by far the most accepted within the cosmology and (astro-)particle physics community (and also the one we will consider throughout this thesis), we will briefly highlight two other options, which have gained some attention. The first one is known under the label of \gls{mond} \cite{Milgrom:1983ca,Milgrom:1983pn,Milgrom:1983zz}. The original idea is that in opposition to Newtons second law, the gravitational force scales as $F=ma^2/a_0$ in the limit of very low accelerations ($a\ll a_0\sim\SI{1.2e-10}{\m/\s^2}$). This would give an explanation for the observed motion of stars and gas within a galaxy without additional matter, meaning that this theory gets by without the need of \gls{dm}. While successful in predicting the correct rotation curves of galaxies, no implementation of \gls{mond} to date can fully account for all the phenomena caused by \gls{dm} on different cosmological scales. Even the so far most elaborate theory of \gls{mond}, known as \textit{Tensor-Vector-Scalar gravity} \cite{Bekenstein:2004ne} has yet to explain the observations of \gls{dm} made in the bullet cluster. Although not entirely impossible, this explanation for \gls{dm} seems highly unlikely, as it requires modifying the well-tested theory of \textit{general relativity}, a challenge in itself.\newp
The second alternative to particle \gls{dm} has been historically the first plausible candidate for \gls{dm} entertained among the astrophysics community. It comprises \glspl{macho} such as planets, brown dwarfs, red dwarfs, white dwarfs, neutron stars and black holes, \ie astronomical objects which are much fainter than ordinary stars. As these are made of baryonic matter, it has been ruled out by \gls{cmb} and also \gls{bbn} measurements alongside surveys which measured \textit{microlensing} events in galaxies (see \eg\eref\cite{Becker:2004ni}), that \glspl{macho} which formed during structure formation could account for a considerable amount of the mass missing in galaxies. There is however one caveat which still allows for one specific type of these objects, namely black holes which have formed prior to \gls{bbn}, so-called \textit{primordial black holes} \cite{Carr:1974nx,Carr:2020xqk}. If light enough, they can evade bounds from microlensing surveys and would not be considered as classical baryonic matter, which clustered into macroscopic objects at a much later stage. Up to present day, there is an active community exploring primordial black holes as \gls{dm} candidates as well as their implications on the physics of the early Universe.\newp
As mentioned above, for the remainder of this thesis, we will assume that \gls{dm} has a particle nature. Armed with knowledge about the criteria that \gls{dm} candidates must meet, we can start to categorize them. As there are numerous possibilities from a model building perspective, we will restrict ourselves here to the most prominent solutions to the \gls{dm} problem (see \eg\erefs\cite{Bertone:2004pz,Feng:2010gw} for a broader overview). The most intuitive approach to start with is to look for a \gls{dm} candidate within the \gls{sm} of particle physics. The only viable option at first glance are \gls{sm} neutrinos, as they are stable, electrically neutral and weakly interacting. However, it has been shown, that \gls{sm} neutrinos cannot account for a considerable amount of \gls{dm} as they are too light and therefore too hot to explain the large scale structures in the Universe. Nevertheless, if there were an additional neutrino species with a mass in the $\si{\keV}$ range or higher, it could realistically make up for \gls{dm}. This viable \gls{dm} candidate offers a rich phenomenology and has become widely known as \textit{sterile neutrino} \cite{Dodelson:1993je,Boyarsky:2018tvu,Dasgupta:2021ies}.\newp
Another possibility to introduce a new particle which can account for \gls{dm} emerged from the hunt for solutions to the \textit{strong CP-problem} of \gls{qcd}. At heart, this problem is related to a CP-violating term in the \gls{qcd} Lagrangian. The relevance of this term is related to a quantity $\bar{\theta}$, which can, in principle, be of $\order{1}$ but has been found to be smaller than $\lesssim\SI{2e-10}{}$ from measurements of the electric dipole moment of the neutron \cite{Seng:2014lea}. A popular theoretical explanation for this \quotes{unnatural} small value, is the proposal of a spontaneously broken global $U(1)$ symmetry (the \textit{Peccei–Quinn symmetry}), which drives $\bar{\theta}$ dynamically towards zero \cite{Peccei:1977a,Peccei:1977b}. This introduces a Nambu-Goldstone boson, the so-called \textit{axion}, which acquires a small mass through an explicit breaking of the symmetry \cite{Weinberg:1977ma,Wilczek:1977pj}. Within a mass range of $m_a\sim 10^{-6}-10^{-4}\,\si{\eV}$, those metastable particles could have been produced in the early Universe through the \textit{vacuum misalignment} mechanism in a sufficient amount to constitute for \gls{dm}. A relaxation of the requirement to solve the strong CP-problem, led to an even richer phenomenology of \textit{axion-like particles} (ALPs), which together with the original \gls{qcd}-axion have become one of the most studied \gls{dm} candidates so far \cite{Jaeckel:2010ni,Choi:2020rgn}.\newp
In the 1990s it had been realized, that an electrically neutral self-annihilating particle with a cross section of $\sim\SI{e-26}{\cm^3/\s}$ undergoing thermal freeze-out (\cf\cref{sec:DMproduction}) can precisely account for the \gls{dm} content observed today. As this cross section is very close to cross sections arising in weak interactions, it was assumed that \gls{dm} could consist of new stable particles with masses in the $\si{\MeV}-\si{\TeV}$ regime that interact on the electroweak scale. These \glspl{wimp} have become subject to countless theoretical studies and experimental searches over the last few decades \cite{Steigman:1984ac,Arcadi:2017kky}. The term \gls{wimp} itself is quite generic and encompasses a broad range of more refined theories. In many studies, \glspl{wimp} have been implemented into theoretically complete extensions of the \gls{sm} such as \textit{supersymmetric theories}, where the lightest supersymmetric partners are \gls{wimp} \gls{dm} candidates \cite{Jungman:1995df}. In the era of particle colliders which can probe physics up to the $\si{\TeV}$ scale, it will become clear whether the \gls{wimp} paradigm proves successful, requires modification, or declines in favor of other \gls{dm} models.\newp
With so many different options to choose from, the model building community has in recent years shifted from \gls{dm} candidates embedded in \gls{uv} complete theories like supersymmetry, towards \textit{simplified models} \cite{Abdallah:2015ter,Albert:2016osu,Morgante:2018tiq}. Within these models it has become easier to study certain characteristics of \gls{dm} candidates, like their self-interacting properties as well as a variety of different couplings to \gls{sm} particles that \gls{dm} might possess. Also extensions to the dark sector which comprise more than one new particle can now be studied with considerably less effort. From the cosmological evolution of these particles as well as their detection signatures, one can draw conclusions about the nature of particle \gls{dm} without relying on a specific theory. In \cref{sec:impactonnonthermalDMproduction,sec:indirectdetection} we will employ simplified models to study the impact of non-perturbative effects on \gls{dm} observables.


\subsection{Detection methods}
\label{subsec:detectionmethods}

The quest for \gls{dm} continues as there has been, so far, no clear evidence for a signal which can be traced back to particles beyond the \gls{sm}. Considering a thermally or non-thermally (\cf\cref{sec:DMproduction}) produced \gls{wimp}-like particle in the $\si{\MeV}-\si{\TeV}$ mass range, there are three fundamentally different ways to search for \gls{dm}, namely collider signals, direct and indirect detection.\footnote{There is the additional option to search for \gls{dm} through its gravitational interactions. However, as valuable as they have been to explore general characteristics of \gls{dm}, these experiments will usually not reveal information about its particle content. As an exception, there are very recent proposals to search for \gls{dm} using \textit{gravitational wave} signals (see \eg\eref\cite{Bringmann:2023iuz}).} All of them will be used in the following to constrain the phenomenology of our simplified models (\cf\cref{sec:impactonnonthermalDMproduction,sec:indirectdetection}), which is why we will comment on them briefly in the subsequent discussion (see \eg\erefs\cite{MarrodanUndagoitia:2015veg,Klasen:2015uma,Gaskins:2016cha,Arcadi:2017kky,Kahlhoefer:2017dnp} for more comprehensive overviews).\newp
If \gls{dm} or other particles in the dark sector possess non-negligible couplings with \gls{sm} particles, they are produced in collider experiments provided that they are light enough. The most prominent collider experiments at the moment to search for \gls{bsm} particles are ATLAS \cite{ATLAS:2008xda} and CMS \cite{CMS:2008xjf} at the \gls{lhc}. Both are able to probe new physics up to the $\si{\TeV}$ scale. How to search for a new particle in collider experiments highly depends on its properties, \ie what its interactions with \gls{sm} particles are, what the expected mass range is, which charges and spin it possesses and if it is stable on collider scales. A collection of various searches for \gls{wimp}-like \gls{dm} and \gls{bsm} particles with different properties conducted by both experiments can be found in \erefs\cite{Buchmueller:2017qhf,Boveia:2018yeb}. For our study in \cref{sec:impactonnonthermalDMproduction}, we will restrict ourselves to charged scalars in the $\si{\TeV}$ mass range, which do not decay inside of the experiment.\newp
There is a large class of \gls{dm} candidates (\eg\glspl{wimp}) which can undergo scattering with atomic nuclei or their accompanying electrons via either spin-independent or spin-dependent couplings. One can look for these scattering processes in \textit{direct detection} experiments \cite{MarrodanUndagoitia:2015veg,Schumann:2019eaa}. Although there are many different working principles of direct detection experiments, there are also commonalities among most of them: Firstly, they are \quotes{passive} experiments, \ie they take advantage of the large abundance of \gls{dm} around us and wait, until a \gls{dm} particle scatters with the detector material. The energy deposited in the material can then be detected. From the absence of a signal, they can put upper limits on the \gls{dm}-nucleus cross section. In order to detect these extremely rare scattering events, direct detection experiments require an extremely low background. They are typically situated deep underground to shield them from cosmic radiation, and they use highly purified materials for both the detector medium and supporting scaffold. The three leading direct detection experiments today for \gls{dm} masses above $\SI{1}{\GeV}$ all operate with liquid xenon. They are called PandaX \cite{PandaX:2014mem}, XENONnT \cite{XENON:2024wpa}, and LZ \cite{LZ:2019sgr}. We will employ direct detection constraints in \cref{sec:indirectdetection} to determine the maximally allowed coupling of dark sector particles to the \gls{sm}.\newp
The third possibility to search for \gls{dm} is through \textit{indirect detection} \cite{Cirelli:2010xx,Slatyer:2017sev,Slatyer:2021qgc}. This type of experiments can probe \gls{dm} which annihilates or decays into \gls{sm} particles. The ways to search for indirect detection signals are quite diverse. One can, for example look for effects on the anisotropies in the \gls{cmb} caused by late time energy injection from the dark sector into the \gls{sm}. This can be achieved with microwave telescopes. Notably, for the upcoming discussion, emphasis will be placed on the results from the \textit{Planck satellite} \cite{Planck:2006aa}. Another possibility is to search for cosmic signals from decaying or annihilating \gls{dm} in overdense regions.\footnote{These annihilations or decays must be inefficient enough not to alter the \gls{dm} abundance observed today in a detectable amount to match the stability criterion.} Prominent candidates to detect by these experiments are cosmic rays (highly energetic ionized particles like antiprotons \cite{AMS:2015tnn,Cuoco:2016eej,Heisig:2020nse} or positrons \cite{PAMELA:2008gwm,AMS:2019rhg,John:2021ugy}), X- or gamma-rays (highly energetic photons) \cite{Fermi-LAT:2015att,Fermi-LAT:2016uux,Fermi-LAT:2017opo,HESS:2016mib,VERITAS:2017tif,Hoof:2018hyn}, and neutrinos \cite{ANTARES:2015vis,IceCube:2022clp,IceCube:2021xzo}. Searching for each of these remnants possesses unique strengths and limitations. The relative simplicity of detection, along with the unimpeded travel of photons from their source to Earth, renders gamma-rays particularly intriguing. Concretely, we will be interested in results from surveys of \gls{dm} dominated satellite galaxies of the Milky Way and the Andromeda Galaxy \cite{Strigari:2018utn}. Of great importance in the following will therefore be the \gls{fermi}, which is installed on the Fermi gamma-ray Space Telescope \cite{Fermi-LAT:2009ihh}, and the \gls{cta} \cite{CTAConsortium:2017dvg}, a next-generation ground-based gamma-ray experiment, which is currently under construction and will be commissioned in the near future. We will use existing gamma-ray data on dwarf spheroidal galaxies from \gls{fermi} as well as prospect data on the galactic center of the Milky Way provided by the \gls{cta} collaboration, alongside with Planck data on \gls{cmb} anisotropies, for our phenomenological discussion in \cref{sec:indirectdetection}.
\section{Dark matter production in the early Universe}
\label{sec:DMproduction}
\fancyhead[RO]{\uppercase{DM production in the early Universe}}

The change in the abundance of \gls{dm} in the early Universe depends on the initial \gls{dm} abundance after reheating as well as its interactions with the \gls{sm} bath. The latter also determine whether \gls{dm} is initially in thermal equilibrium with the \gls{sm} heat bath and how long chemical and kinetic equilibrium with the \gls{sm} can be maintained (\cf\cref{subsec:thermalhistory}).\newp
We will focus in the following on two scenarios which are important for the subsequent discussion, namely \gls{fo} and \gls{fi}. In \gls{fo}, \gls{dm} particles are initially in thermal equilibrium with the \gls{sm} heat bath. When annihilation or, more generally, conversion processes into \gls{sm} particles become slow compared to the expansion rate of the Universe, \ie the conversion rate $\Gamma_{\text{conv}}\ll H$, they drop out of chemical equilibrium and the \gls{dm} abundance becomes constant. This commonly happens when \gls{dm} particles become non-relativistic ($m_{\text{\tiny DM}}\gg T$). Further assuming no (asymmetric) energy injection into the two now chemically decoupled sectors, they will maintain the same temperature until kinetic decoupling, which typically happens much later. In \gls{fi}, \gls{dm} never reaches thermal equilibrium with the \gls{sm} and the initial \gls{dm} abundance is assumed to be negligible. \gls{dm} gets produced through \gls{sm} conversion processes, which typically leave the momentum distributions of the particles involved thermal. Therefore, also here, a common temperature is approximately retained. When the \gls{dm} production processes become inefficient, the \gls{dm} abundance again levels off.\newp
We will start by introducing the set of equations that, for our purpose, sufficiently describe the evolution of \gls{dm} in the early Universe in \cref{subsec:BE}. In \cref{subsubsec:Liouvilleoperator,subsubsec:integratedcollisionterm}, we calculate both sides of the equations separately and merge them for typical \gls{fi} and \gls{fo} scenarios in \cref{subsec:BEfiandfo}. \Cref{subsec:thermallyaveragedcrossection} deals with the calculation of the thermally averaged cross section, a quantity needed to describe the $2\to 2$ processes governing \gls{fo} and \gls{fi} for our scenarios of interest.


\subsection{The Boltzmann equation -- a general discussion}
\label{subsec:BE}

The evolution of \gls{dm} in the early Universe is commonly described using a semi-classical approach, known as the \gls{be}, which models \gls{dm} as a homogeneous and isotropic particle fluid in an expanding background. This approach is applicable to particle species with de Broglie wavelengths much shorter than the size of the Universe.\footnote{This condition holds for most particle \gls{dm} models, particularly for non-relativistic \gls{dm} with masses well above the $\si{\keV}$ range, which is our main focus.} For our subsequent calculation of the \gls{dm} number density to describe its evolution, we will follow the discussion in Kolb and Turner's book \cite{Kolb:1990vq} together with the improved analysis by Gondolo and Gelmini \cite{Gondolo:1990dk}. In doing so, we will adopt the \gls{be} as it offers a convenient framework. A detailed derivation of the \gls{be} using kinetic theory can be found, for example, in \eref\cite{Bernstein:1988bw}.\newp
The \gls{be} for a species $a$ with a distribution function $f_a$ in its most general form can be written as 
\begin{equation}
	\hat{L}[f_a]=C[f_a]\,,
\end{equation}
where $\hat{L}$ is the \textit{Liouville operator} encapsulating the space-time effects of the background metric and $C[f_a]$ denotes the \textit{collision term}, which specifies all interactions of the species with the other particles present in the plasma. Note that, in principle, the \gls{rhs} of the \gls{be} contains all possible kinematic processes of species $a$ involving an arbitrary number of particles. However, we will focus in the following on $1\to 2$ and $2\to 2$ processes only. Since we are usually interested in the whole species $a$ at a certain time $t$, we integrate over the full Lorentz invariant momentum space 
\begin{equation}
\label{eq:integrated_BE}
	g_a \int \frac{\dd[3]{p_a}}{(2\pi)^3E_a}\hat{L}[f_a]= 2g_a\int \dd{\Pi_a}C[f_a]\equiv\sum_i\mathscr{C}_i\,,
\end{equation}
where we introduced the Lorentz-invariant phase-space measure as
\begin{equation}
	\dd{\Pi_a}\equiv\frac{\dd[3]{p_a}}{(2\pi)^32E_a}\,,
\end{equation}
with $g_a$ being the internal degrees of freedom of species $a$ and $E_a=\sqrt{p_a^2+m_a^2}$ its energy given by the usual energy momentum relation. The \textit{integrated collision term} $\sum_i\mathscr{C}_i$ contains all interaction processes $i$ involving the target species. We will calculate both sides of \cref{eq:integrated_BE} for our models of interest in the following.


\subsubsection{The Liouville operator in the FRW Universe}
\label{subsubsec:Liouvilleoperator}

The Liouville operator is defined as 
\begin{equation}
	\hat{L}[f_a]\equiv p_a^\alpha\pdv{f_a}{x_a^\alpha}-\CS{\alpha}{\beta}{\gamma}p_a^\beta p_a^\gamma\pdv{f_a}{p_a^\alpha}\,,
\end{equation}
where the Christoffel symbols $\CS{\alpha}{\beta}{\gamma}$ in our case are calculated using the \gls{frw} metric (\cf\cref{subsec:FRWuniverse}). Relabelling $x_a^0=t$, $p_a^0=E_a$, $\abs{\vec{p}_a}=p_a$ and accounting for homogeneity ($\partial f_a /\partial x_a^i=0$) and isotropy ($\vec{p}_a\cdot\vec{\nabla} f_a=0$), \ie $f_a=f_a(p_a,t)=f(E_a,t)$, the Liouville operator yields
\begin{equation}
	\hat{L}[f_a]=E_a\pdv{f_a}{t}-Hp_a^2\pdv{f_a}{E_a}\,,
\end{equation}
with $H$ the Hubble parameter defined in \cref{eq:Hubble}. Recalling the definition of the number density $n_a$ of a species $a$ from \cref{eq:numberdensity} and inserting $\hat{L}$ into the \gls{lhs} of \cref{eq:integrated_BE}, we obtain 
\begin{equation}
    \label{eq:LHS1}
	\text{(LHS)}=\pdv{}{t}\left(g_a\int \frac{\dd[3]{p_a}}{(2\pi)^3}f_a(p_a,t)\right)-H g_a\int\frac{\dd[3]{p_a}}{(2\pi)^3}\frac{p_a^2}{E_a}\pdv{f_a}{E_a} = \dot{n}_a+3Hn_a\,,
\end{equation} 
where we used $\partial E_a/\partial p_a= E_a/p_a$ to rewrite the respective derivative of $f_a$ and then partially integrated the second term.\newp
It is favorable to re-express the \gls{lhs} in terms of the yield $Y_a\equiv n_a/s$, a quantity which is (unlike $n_a\propto a(t)^{-3}$) independent of the expansion of the Universe. This can be seen easily because the total entropy of a unit volume $S=s(t)a(t)^3$ in the Universe is approximately conserved throughout its history (\cf\cref{subsec:thermalhistory}) and thus the entropy density $s(t)\propto a(t)^{-3}$ at all times. Assuming entropy conservation, we can use $\dd{s}/\dd{t}= -3Hs$ to check that 
\begin{equation}
    \label{eq:yieldtimederivative}
	s\dv{Y_a}{t}=s\dv{(n_a/s)}{t}=\dv{n_a}{t}-\frac{n_a}{s}\dv{s}{t}=\dot{n}_a+3Hn_a\,.
\end{equation}
We also want to switch from a time to a (thermal bath) temperature dependence of $Y_a$, since it is the more natural variable to work with in this context. The relation between time and temperature can also be obtained from entropy conservation. Since $s\propto \heff(T) T^3$ (\cf\cref{eq:entropydensity}), from $\dd{S}/\dd{t}=0$ we obtain \cite{Srednicki:1988ce}
\begin{equation}
    \label{eq:timetemperaturerelation}
	\dv{T}{t}=-HT\left(1+\frac{1}{3}\frac{T}{\heff}\dv{\heff}{T}\right)^{-1}=-HT\frac{\heff}{\geff^{1/2}}\gstar^{-1/2}\,,
\end{equation}
where we have defined the modified relativistic effective \gls{dof}
\begin{equation}
	\gstar^{1/2}\equiv\frac{\heff}{\geff^{1/2}}\left(1+\frac{1}{3}\frac{T}{\heff}\dv{\heff}{T}\right)
\end{equation}
with respect to the relativistic energy and entropy \gls{dof} (\cf\cref{fig:relativisticdofs}) for later convenience.\footnote{Note, however, that for most bath temperatures $T$, the physical values of $\geff$, $\heff$ and $\gstar$ will be almost the same.} Combining \cref{eq:LHS1,eq:yieldtimederivative,eq:timetemperaturerelation}, the \gls{lhs} of the integrated \gls{be} yields
\begin{equation}
	\label{eq:LHS_BE}	
	\text{(LHS)}=s\dv{Y_a}{t}=s\dv{T}{t}\dv{Y_a}{T}=-sHT\frac{\heff}{\geff^{1/2}}\gstar^{-1/2}\dv{Y_a}{T}.
\end{equation}


\subsubsection{The integrated collision term}
\label{subsubsec:integratedcollisionterm}

The \gls{rhs} of \cref{eq:integrated_BE} highly depends on the form of interactions the target species participates in. In all generality, for a process $a+b+\ldots \to i + j + \ldots$ the integrated collision term is given by
\begin{align}
    \label{eq:integratedcollisionterm}
    \mathscr{C}_{a + b + \ldots \to i + j + \ldots}=-\sum_{\text{dof}}\int & \dd{\Pi_a}\dd{\Pi_b}\ldots\dd{\Pi_i}\dd{\Pi_j}\ldots\nonumber\\
    &(2\pi)^4\delta^{(4)}(p_a+p_b+\ldots-p_i-p_j-\ldots)\nonumber\\
	&~~\left[\abs{\mathcal{M}}^2_{a+b+\ldots\to i+j+\ldots} f_af_b\ldots(1\pm f_i)(1\pm f_j)\ldots\right.\nonumber\\
    &~~\left.\,-\abs{\mathcal{M}}^2_{i+j+\ldots\to a+b+\ldots} f_if_j\ldots(1\pm f_a)(1\pm f_b)\ldots\right].
\end{align}
The $\abs{\mathcal{M}}^2_{a+b+\ldots\leftrightarrow i+j+\ldots}$ denote the squared matrix elements of the process and its respective inverse, the $f_a, f_b, \ldots$ describe the distribution functions of the particles involved and the $(1\pm f_a)(1\pm f_b)\ldots$ are quantum statistical factors. The sum for a single process is taken over the internal \gls{dof} of all initial and final state particles. Interactions that do not change the particle species with up to four particles involved (\ie scattering processes) vanish upon integration of the \gls{be}. The remaining processes under consideration are $1\to 2$ and $2\to 2$ \gls{fi} as well as $2\to 2$ \gls{fo} processes, denoted as $\Cfionetwo$, $\Cfitwotwo$, and $\Cfotwotwo$, respectively, which we will calculate in the following. We want to emphasize here that for both \gls{fi} processes, we will derive the integrated collision term with respect to a particle species $a$ that is not the \gls{dm} candidate. The reason for this is that the \gls{dm} candidate will only appear as a final state in \gls{fi} processes, which would make a direct calculation inconvenient. However, we will make contact with the equations describing the evolution of \gls{dm} in \cref{subsec:BEfiandfo}.

\paragraph{The $\mathbf{1\to 2}$ freeze-in processes}$~$\newpp
For a particle decay process $a\to b+\chi$, where $\chi$ denotes the \gls{dm} candidate, the \gls{rhs} of the integrated \gls{be} reduces to 
\begin{align}
	\Cfionetwo=-\sum_{\text{dof}}\int & \dd{\Pi_a}\dd{\Pi_b}\dd{\Pi_\chi}(2\pi)^4\delta^{(4)}(p_a-p_b-p_\chi)\nonumber\\
	&\left[\abs{\mathcal{M}}^2_{a\to b\chi} f_a(1\pm f_b)(1\pm f_\chi)-\abs{\mathcal{M}}^2_{b\chi\to a} (1\pm f_a)f_b f_\chi\right],
\end{align}
where for the \gls{fi} scenario we can ignore the second term because we expect $f_\chi\approx 0$. Furthermore, we assume for simplicity a Maxwell-Boltzmann statistic for all other particles and neglect the statistical factors $(1\pm f_X)\approx 1$. Absorbing the sum over all \gls{dof} as well as a factor $1/g_a$ of the initial state in the summed and averaged squared matrix element $\overline{\abs{\mathcal{M}}^2}_{a\to b\chi}$, the term can be simplified into 
\begin{equation}
	\Cfionetwo=-\int \dd{\Pi_a}g_af_a \underbrace{\int\dd{\Pi_b}\dd{\Pi_\chi}(2\pi)^4\delta^{(4)}(p_a-p_b-p_\chi) \overline{\abs{\mathcal{M}}^2}_{a\to b\chi}}_{(I)}.
\end{equation}
The integral over the final state particles yields $(I)=2m_a \Gamma_{a\to b\chi}$, where 
\begin{equation}
    \label{eq:1to2particledecayrate}
	\Gamma_{a\to b\chi}\equiv\frac{f_s}{32\pi^2m_a^2}\,p^*\int \dd{\Omega}\overline{\abs{\mathcal{M}}^2}_{a\to b\chi}
\end{equation}
denotes the usual definition of the (partial) decay rate of a particle $a$ with $f_s = 1/2$ ($1$) for (non-)identical final states. In this context, $p^*=|\vec{p}_b|=|\vec{p}_\chi|=\sqrt{\lambda(m_a^2,m_b^2,m_\chi^2)}/(2m_a)$ yields the absolute momentum of the final state particles in the rest frame of $a$ with $\lambda(x,y,z)\equiv(x-(\sqrt{y}+\sqrt{z})^2)(x-(\sqrt{y}-\sqrt{z})^2)$ being the Källén function. Thus, the integrated collision term can be written as
\begin{equation}
    \label{eq:Cfionetwo}
	\Cfionetwo=-\Gamma_{a\to b\chi} \int \frac{\dd[3]{p_a}}{(2\pi)^3}g_af_a\frac{m_a}{E_a}=-\ndeq{a}\Gamma_{a\to b\chi}\expval{\frac{m_a}{E_a}}\equiv-\ndeq{a}\Gammadecexp_{a\to b\chi},
\end{equation}
where we assumed $f_a=\feq{a}\propto e^{-E_a/T}$ to follow a Maxwell-Boltzmann distribution in equilibrium (with $\mu_a\approx0$). Further, 
\begin{align}
	\expval{\frac{m_a}{E_a}}\equiv\frac{\int \dd{n_a}\frac{m_a}{E_a}}{\int \dd{n_a}}=\frac{\int_0^\infty \dd{p_a}p_a^2e^{-E_a/T}\frac{m_a}{E_a}}{\int_0^\infty \dd{p_a}p_a^2e^{-E_a/T}}=\frac{\int_1^\infty \dd{y}\sqrt{y^2-1}e^{-xy}}{\int_1^\infty \dd{y}\sqrt{y^2-1}ye^{-xy}}=\frac{\kone{x}}{\ktwo{x}}
\end{align}
denotes the average of the corresponding quantity with respect to the number density measure $\dd{n_a}\equiv g_af_a\dd[3]{p_a}/(2\pi)^3$ of species $a$, where we have substituted $y\equiv E_a/m_a$ and $x\equiv m_a/T$. In the last step we formulated the integrals in terms of modified Bessel functions of second kind (\cf\cref{eq:BesselKdef}). Together with $\Gamma_{a\to b\chi}$ they define a thermally averaged decay rate
\begin{equation}
    \label{eq:Gammadecexp}
    \Gammadecexp_{a\to b\chi}\equiv\frac{\kone{x}}{\ktwo{x}}\Gamma_{a\to b\chi}\,.
\end{equation}
We will briefly discuss the two limiting cases from the definitions of the modified Bessel functions given in \cref{eq:BesselKdef,eq:BesselKexplargearguments}. At later times, for large $x\gg 1$, $\kone{x}/\ktwo{x}\approx 1$, such that $\Gammadecexp_{a\to b\chi} \approx \Gamma_{a\to b\chi}$ for the thermal average. For $x\ll1$ at early times, especially relevant for \gls{fi}, the thermal average suppresses the decay rate via $\kone{x}/\ktwo{x}= x/2 +\order{x^2}$, such that in this regime, $2\to2$ processes are expected to dominate (\cf\cref{fig:fi_example} for a concrete example).

\paragraph{The $\mathbf{2\to 2}$ freeze-in processes}$~$\newpp
Following similar steps as for the $1\to 2$ case (also assuming $f_\chi\approx 0$, $(1\pm f_X)\approx 1$), the collision term of an interaction $a+b\to c+\chi$ producing \gls{dm} in a \gls{fi} scenario can be written as
\begin{align}
	\label{eq:CollisionTermBE2to2}
	\Cfitwotwo &=-\sum_{\text{dof}}\int \dd{\Pi_a}\dd{\Pi_b}\dd{\Pi_c}\dd{\Pi_\chi}(2\pi)^4\delta^{(4)}(p_a+p_b-p_c-p_\chi)\nonumber\\
	&\hphantom{\sum_{\substack{\text{initial and}\\ \text{final dof's}}}\int\quad}\left[\abs{\mathcal{M}}^2_{ab\to c\chi} f_a f_b(1\pm f_c)(1\pm f_\chi)-\abs{\mathcal{M}}^2_{c\chi\to ab} (1\pm f_a)(1\pm f_b)f_c f_\chi\right]\nonumber\\
	&=-\int \dd{\Pi_a}\dd{\Pi_b}g_af_a g_bf_b \underbrace{\int\dd{\Pi_c}\dd{\Pi_\chi}(2\pi)^4\delta^{(4)}(p_a+p_b-p_c-p_\chi) \overline{\abs{\mathcal{M}}^2}_{ab\to c\chi}}_{(II)}\,,
\end{align}
where the integral $(II)=4F(s)\sigma_{ab\to c\chi}$ matches the common definition of the $2\to2$ cross section of the process times a prefactor $F(s)\equiv\sqrt{(p_a\cdot p_b)^2-m_a^2m_b^2}=\sqrt{\lambda(s,m_a^2,m_b^2)}/2$, which is a function of the squared \gls{com} energy $s=(p_a+p_b)^2$ alone. A more convenient way to write $F(s)=\vmol E_aE_b$ is using the M\o ller velocity, which has the useful property that $\vmol n_an_b$ is invariant under Lorentz transformations. In terms of the individual particle velocities $\vec{v}_a=\vec{p}_a/E_a$ and $\vec{v}_b=\vec{p}_b/E_b$, the M\o ller velocity is defined as 
\begin{equation}
	\vmol=\sqrt{\abs{\vec{v}_a-\vec{v}_b}^2-\abs{\vec{v}_a\times\vec{v}_b}^2}
\end{equation}
and reduces to the relative velocity $\vrel\equiv\abs{\vec{v}_a-\vec{v}_b}$ in the \gls{com} frame (where $\vec{v}_a\parallel\vec{v}_b$). The integrated \gls{be} then simplifies to 
\begin{equation}
    \label{eq:Cfitwotwo}
	\Cfitwotwo=-\int\frac{\dd[3]{p_a}}{(2\pi)^3}\frac{\dd[3]{p_b}}{(2\pi)^3}g_af_ag_bf_b\sigma_{ab\to c\chi}\vmol = -\ndeq{a}\ndeq{b}\sigmavmolavg{ab}{c\chi},
\end{equation}
where we have again assumed that $f_{a,b}^{\,}=\feq{a,b}$ follow equilibrium Maxwell-Boltzmann statistics. We have further defined the thermally averaged cross section
\begin{equation}
    \label{eq:sigmavrelmolavg}
	\sigmavmolavg{ab}{c\chi}\equiv\frac{\int \dd{n_a} \dd{n_b}\sigma_{ab\to c\chi}\vmol}{\int \dd{n_a}\int \dd{n_b}}=\frac{\int \dd{\ndeq{a}} \dd{\ndeq{b}}\sigma_{ab\to c\chi}\vmol}{\int \dd{\ndeq{a}}\int \dd{\ndeq{b}}},
\end{equation}
which will be calculated in \cref{subsec:thermallyaveragedcrossection}. We want to emphasize here that the last equality of \cref{eq:sigmavrelmolavg} also holds for distributions outside of chemical equilibrium, since the momentum-independent chemical potentials cancel.

\paragraph{The $\mathbf{2\to 2}$ freeze-out processes}$~$\newpp
The \gls{fo} scenario for a particle species $a$ considering a $2\to 2$ process $a+b\to c+d$ starts equivalently to the \gls{fi} case with \gls{sm} particles in the final state. In models with only one additional particle accounting for \gls{dm} (\eg classical WIMP models), the \gls{dm} candidate $\chi$ coincides with the species $a$ in our discussion. However, in the following we will employ exclusively models with more than one particle in the dark sector. Within those, the species $a$ undergoing \gls{fo} can also refer to another particle in the dark sector and not the \gls{dm} candidate directly, which is why we keep the label $a$ for generality. We still assume Maxwell-Boltzmann statistics for all particle species and neglect statistical factors, but we can neither disregard any distribution functions nor assume all initial state particles to be in thermal equilibrium. However, since $2\to 2$ processes need to efficiently exchange energy between the \gls{sm} and the dark sector to be relevant for \gls{fo}, we can assume the final state particles to be part of, and therefore in thermal equilibrium with, the SM bath, \ie $f_cf_d=\feq{c}\feq{d}$, and that the \textit{principle of detailed balance} holds, implying $\feq{c}\feq{d}=\feq{a}\feq{b}$ (see \eg\eref\cite{Kolb:1990vq}). Further, we require the system to exhibit a $CP$ or $T$ symmetry, such that $\overline{\abs{\mathcal{M}}^2}_{ab\to cd}=\overline{\abs{\mathcal{M}}^2}_{cd\to ab}$. Going through similar steps as before, we arrive at
\begin{align}
    \label{eq:Cfotwotwo}
	\Cfotwotwo &=-\int\frac{\dd[3]{p_a}}{(2\pi)^3}\frac{\dd[3]{p_b}}{(2\pi)^3}g_ag_b\left(f_af_b-\feq{a}\feq{b}\right)\sigma_{ab\to cd}\vmol \nonumber\\
	&=- \sigmavmolavg{ab}{cd}\left(n_an_b-\ndeq{a}\ndeq{b}\right).
\end{align}


\subsection{The Boltzmann equations for freeze-in and freeze-out}
\label{subsec:BEfiandfo}

\begin{figure}[ht]
    \centering
    \includegraphics[width=\textwidth]{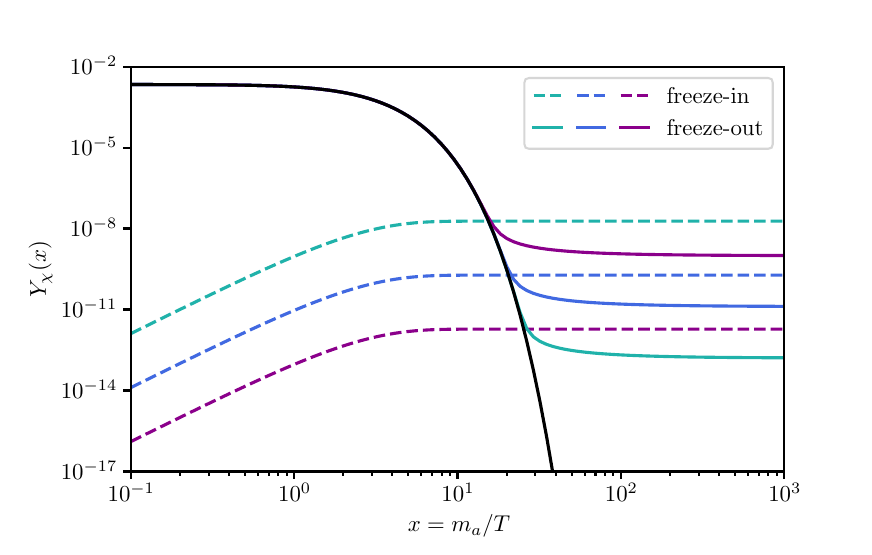}
    \caption[The yield of a particle species $\chi$ evolving over time for a typical freeze-in and freeze-out scenario.]{The yield of a particle species $\chi$ evolving over time as a function of $x\equiv m_a/T$ for a typical \gls{fi} (dashed) and \gls{fo} (solid) scenario, where $m_a$ is either the mass of the particle $\chi$ itself with $a=\chi$ (for \gls{fo}) or the mass of a particle $a$ involved in the production of $\chi$ (for \gls{fi}), respectively. Lighter colors correspond to higher values of the interaction rates which govern the \gls{fi} and \gls{fo} processes. The solid black line refers to the equilibrium yield $\Yeq{a}$ of species $a$.}
    \label{fig:FI_FO_example}
\end{figure}

With the information gathered above, we are now well-equipped to determine the \glspl{be} for typical \gls{fi} and \gls{fo} scenarios. We employ for this purpose a dimensionless quantity $x=m_*/T$ with $m_*$ being a mass scale commonly set to the mass $m_a$ of a species in the initial state of the interaction processes (but not necessarily). Furthermore, we define (\cf\cref{eq:neqapprox,eq:entropydensity})
\begin{equation}
    \label{eq:equilibriumyield}
	\Yeq{i}(x)\equiv\frac{\ndeq{i}(x)}{s(x)}=\frac{45g_i \eta(m_i,x)}{2\pi^4m_*^3\heff(x)},\quad \eta(m_i,x)=
\begin{cases}
   \frac{1}{2}m_*m_i^2x^2\ktwo{\frac{m_i}{m_*}x}\quad&\text{for}\quad m_i > 0\\
   \zeta(3)\,m_*^3\quad&\text{for}\quad m_i=0
\end{cases}
\end{equation}
to be the equilibrium yield of any particle species $i$.\footnote{Note that we have taken for a massless (massive) bosonic particle the relativistic (Maxwell-Boltzmann) approximation of the number density. If one considers a massless fermion, a factor of $7/8$ must be added.} Starting with \gls{fi}, we can combine \cref{eq:Cfionetwo,eq:Cfitwotwo} with \cref{eq:LHS_BE} and use the relation $\dd{\Yfi{a}}/\dd{x}=-\dd{\Yfi{\chi}}/\dd{x}$, to arrive at the corresponding \glspl{be} for $\chi$ considering $1\to 2$ and $2\to 2$ processes
\begin{align}
    \label{eq:dYFIdx12}
	\dv{\Yfionetwo{\chi}}{x}&=\sqrt{\frac{45}{4\pi^3}}\frac{\MPl}{m_*^2}\frac{\gstar^{1/2}}{\heff}\Gammadecexp_{a\to b\chi}\Yeq{a},\\
    \label{eq:dYFIdx22}
    \dv{\Yfitwotwo{\chi}}{x}&=\sqrt{\frac{\pi}{45}}\frac{\MPl m_*\gstar^{1/2}}{x^2}\sigmavmolavg{ab}{c\chi}\Yeq{a}\Yeq{b},
\end{align}
where we note that the $K_n$ within the definitions of the thermal averages are in general functions of $(m_a/m_*)x$ rather than $x$ alone. To obtain the total \gls{dm} yield $\Yfi{\chi}$ from these equations, we can simply add up the yields from the individual contributions, which can be extracted by integrating the equations above from $0$ to $x$ (assuming $\Yfi{\chi}(0)=0$). We have sketched $\Yfi{\chi}$ for a typical \gls{fi} process from particle decay of a species $a$ in \cref{fig:FI_FO_example} (dashed lines) alongside with the equilibrium yield $\Yeq{a}$ (black solid line), where we set the mass scale $m_*=m_a$. It is apparent from \cref{eq:dYFIdx12} that the \gls{dm} yield scales linearly with the decay rate, which in turn depends on the coupling strength between \gls{dm} and the decaying species. The efficiency of this process drastically decreases when the species $a$ becomes non-relativistic ($x\gg 1$) due to an exponential suppression of the equilibrium yield 
\begin{equation}
    \label{eq:equilibriumyieldapprox}
	\Yeq{a}(x)\xrightarrow{x\gg 1} \frac{45g_a}{4\sqrt{2}\pi^{7/2}\heff(x)}x^{3/2}e^{-x},
\end{equation}
rendering $\Yfi{\chi}$ constant for $x\gtrsim 10$. Including other kinematic processes (like $2\to 2$ production) will complicate the \gls{fi} dynamics but the general picture of \cref{fig:FI_FO_example} together with the efficiency statements will remain.\newp
To calculate the yield of a species $a$ in a \gls{fo} process, we first set $m_*=m_a$ and drop the label \gls{fo} for convenience. We then combine \cref{eq:Cfotwotwo,eq:LHS_BE}, which results in a first order differential equation
\begin{equation}
    \label{eq:freeze-out}
    \dv{Y_a}{x}=-\sqrt{\frac{\pi}{45}}\frac{\MPl m_a\gstar^{1/2}}{x^2}\sigmavmolavg{ab}{cd}\left(Y_a Y_b-\Yeq{a} \Yeq{b}\right).
\end{equation}
This differential equation usually has to be solved numerically assuming thermal equilibrium at the starting point, \ie $Y_a(x_s)=\Yeq{a}(x_s)$. We depicted a typical \gls{fo} process governed by \cref{eq:freeze-out} in \cref{fig:FI_FO_example} (solid lines) as well, assuming a constant thermally averaged cross section and setting $Y_a\equiv Y_b$.\footnote{This is the case, when either species $a$ and $b$ are the same, or particle $b$ is the antiparticle of $a$ in a theory where processes that could create an asymmetry between $a$ and $b$ are absent.} We observe that, at early times until $x\gtrsim 15$, the yield of species $a$ closely follows its equilibrium value. This can be explained by the efficiency of the exchange processes encapsulated in $\sigmavmolavg{ab}{cd}/x^2$, which quickly drive the system back to $\Yeq{a}$ whenever it experiences a small departure. As the efficiency of these processes decreases for $x\gg 1$ and $\Yeq{a}$ becomes exponentially suppressed, the \gls{rhs} of \cref{eq:freeze-out} vanishes and the yield becomes constant.\footnote{Defining the rate of the $2\to 2$ exchange processes as $\Gamma_a\sim \ndeq{a} \sigmavmolavg{ab}{cd}$, one can cast the \gls{fo} equation into the form $\dd{Y_a}/\dd{x}\propto -\Gamma_a/H(Y^2_a/{\Yeq{a}}^2-1)$. Here, it is apparent that \gls{fo} happens when $\Gamma_a\ll H$ as expected from the na\"\i ve picture.} The larger the cross sections of the processes under consideration, the longer equilibrium will be maintained and the smaller the final \gls{fo} yield will be (\cf\cref{fig:FI_FO_example}). Typical values for the \gls{fo} temperature, \ie the temperature at which $Y_a$ departs from equilibrium, are $x_{\text{\tiny{FO}}}\sim 25$.\newp
The produced \gls{dm} abundance $\Omegadmt$ from either of the two mechanisms up to the present day ($t_0$) can be calculated via
\begin{equation}
    \label{eq:DMabundance}
    \Omegadmt\equiv\Omegadm(t_0)=\frac{\rho_\chi(t_0)}{\rhocrit}=\frac{\mchi s_0 Y_{\chi,\infty}}{\rhocrit},
\end{equation}
where we used the approximation $\rho_\chi(t)=\mchi n_\chi(t)$ for a non-relativistic particle species and $Y_{\chi,\infty}\equiv Y_\chi(x\to\infty)$ denotes the \gls{dm} yield after number changing processes have stopped.


\subsection{The thermally averaged cross section}
\label{subsec:thermallyaveragedcrossection}

The cross section itself has to be computed individually for each particle physics process, however, the kinematics of the thermal average can be derived on a more generic level. Assuming Maxwell-Boltzmann statistics for all incoming particles\footnote{This is a simplification made in order to obtain analytic results. Being justified in the non-relativistic limit, for relativistic initial state particles (\eg massless gauge bosons), one, in principle, needs to employ the proper statistical distributions. However, as these corrections are usually small, we neglect them in the following.}, the thermally averaged cross section, which is only a function of the (squared) \gls{com} energy $s$, yields 
\begin{equation}
    \label{eq:thermallyaveragedcrosssectionI}
	\sigmavmolavg{ab}{cd}=\frac{\int \dd{\ndeq{a}} \dd{\ndeq{b}}\sigma_{ab\to cd}(s)\vmol}{\int \dd{\ndeq{a}}\int \dd{\ndeq{b}}}=\frac{\int \dd[3]{p_a} \dd[3]{p_b}e^{-E_a/T}e^{-E_b/T}\sigma_{ab\to cd}(s)\vmol}{\int \dd[3]{p_a}e^{-E_a/T}\int \dd[3]{p_b}e^{-E_b/T}}.
\end{equation}
To simplify the calculation, we switch to a more convenient set of variables comprising \cite{Gondolo:1998ah}
\begin{equation}
	E_+ = E_a+E_b,\quad E_- = E_a-E_b,\quad s = m_a^2+m_b^2+2E_aE_b-2p_ap_b\cos\theta,
\end{equation}
with $\theta$ denoting the angle between the initial state particles $a$ and $b$, and $p_i=\abs{\vec{p}_i}$. With $\det J=1/(4p_ap_b)$, the differentials in the new parametrization (for unpolarized initial states) are given by
\begin{equation}
	\dd[3]{p_a}\dd[3]{p_b} = 4\pi p_aE_a\dd{E_a}4\pi p_bE_b\dd{E_b}\frac{1}{2}\dd{\!\cos\theta}= 2\pi^2 E_aE_b\dd{E_+}\dd{E_-}\dd{s}\,,
\end{equation}
and the integration boundaries change from $\{ E_a\geq m_a, E_b\geq m_b, \abs{\cos\theta}\leq 1\}$ to 
\begin{equation}
	\left\{ s\geq (m_a+m_b)^2,~ E_+\geq \sqrt{s},~ \abs{E_-+E_+\frac{m_a^2-m_b^2}{s}}\leq \frac{2F(s)}{s}\sqrt{E_+^2-s}\right\}\,.
\end{equation}
Using $\int \dd[3]{p_i}e^{-E_i/T}=8\pi x^{-3}\eta(m_i,x)$ to calculate the denominator of \cref{eq:thermallyaveragedcrosssectionI} ($x=m_*/T$) and inserting the definition of the M\o ller velocity $\vmol=F(s)/(E_aE_b)$, the thermally averaged cross section reads
\begin{equation}
	\sigmavmolavg{ab}{cd}=\frac{x^6}{32 \eta(m_a,x) \eta(m_b,x)}\int\dd{s}F(s)\sigma_{ab\to c\chi}(s)\int\dd{E_+}e^{-E_+/T}\int\dd{E_-}\,.
\end{equation}
Since $\sigma_{ab\to c\chi}(s)$ is a function of $s$ alone, we can perform the integrals over $E_-$ and $E_+$ directly
\begin{equation}
	\int\dd{E_-}=4\frac{F(s)}{s}\sqrt{E_+^2-s^2},\quad\int\dd{E_+}e^{-E_+/T}\sqrt{E_+^2-s}=\sqrt{s}\frac{m_*}{x}\kone{\sqrt{s}\frac{x}{m_*}}\,.
\end{equation}
This leaves us with only one integral
\begin{equation}
	\label{eq:thermallyaveragedcrosssectionII}
	\sigmavmolavg{ab}{cd}=\frac{m_*x^5}{32\eta(m_a,x) \eta(m_b,x)}\int_{s_{\text{min}}}^\infty\dd{s}\frac{\lambda(s,m_a^2,m_b^2)}{\sqrt{s}}\kone{\sqrt{s}\frac{x}{m_*}}\sigma_{ab\to cd}(s)\,,
\end{equation}
where $s_{\text{min}}=\max\{(m_a+m_b)^2,(m_c+m_d)^2\}$ denotes the minimal possible energy transfer. The cutoff for processes where $(m_c+m_d)^2>(m_a+m_b)^2$ is also encoded in the kinematic phase space of the cross section. Therefore, we use $s_{\text{min}}=(m_a+m_b)^2$ for further calculations. The $2\to 2$ cross section in the \gls{com} frame as an integral over the solid angle is given by
\begin{equation}
	\label{eq:generalcrosssection}
	\sigma_{ab\to cd}=\frac{f_s}{64\pi^2 s}\frac{p_c^{*}}{p_a^{*}}\int\dd{\Omega}\overline{\abs{\mathcal{M}}^2}_{ab\to cd}\,,
\end{equation}
with $f_s=1/2\,(1)$ for (non-)identical final states. The initial and final state momenta are defined here as 
\begin{equation}
    p_a^*\equiv\sqrt{\frac{\lambda(s,m_a^2,m_b^2)}{4s}}\,,\quad p_c^*\equiv\sqrt{\frac{\lambda(s,m_c^2,m_d^2)}{4s}}\,.
\end{equation}\newp
The thermally averaged cross section for a specific process calculated with \cref{eq:thermallyaveragedcrosssectionII} exhibits in general a complicated dependence on $x$. For a \gls{fo} scenario, we are interested in processes happening at $x\sim 20$ or higher, such that it is usually sufficient to expand the thermally averaged cross section with respect to large $x$ and only consider the leading order term. Although confusing, in the literature this treatment is referred to as a \textit{partial wave expansion}.\footnote{Let us emphasise here, that an actual expansion in partial waves happens at the level of the matrix element, where its angular dependence is expanded in Legendre polynomials $P_l(\cos\theta)$, which are then matched to an expansion of the scattering amplitude that one would obtain from a non-relativistic approach using the Schrödinger equations. See \cref{app:partial_wave_analysis} and \cref{app:SEderivation} for a detailed discussion.} Accordingly, we will denote the $x^0$ term as the \textit{s-wave} contribution, the $x^{-1}$ term as the \textit{p-wave} contribution and so on. For massive initial particles, we start by expressing $\sigma\vrel$ in terms of $\vrel$ (instead of $s$) and perform a \textit{velocity expansion} of the cross section around $\vrel=0$, yielding
\begin{equation}
    \label{eq:crosssectionvrelexpansion}
	\sigma\vrel=\sum_{l=0}^\infty \sigma_l\,\vrel^{2l}
\end{equation}
for an analytic function with expansion coefficients $\sigma_l$. Applying \cref{eq:BesselKexplargearguments} for large arguments to $K_n(x)$ in \cref{eq:thermallyaveragedcrosssectionII}, we can, in fact, expand the whole expression for $x\to \infty$. After doing so, we can integrate analytically order by order over $\vrel$ and group the contributions in powers of $x^{-n}$. Only considering the s-wave contribution yields $\sigmavmolavg{ab}{cd}\approx \sigma_0$ for all initial mass configurations considered in the following.
\section{Non-perturbative effects in dark matter interactions}
\label{sec:non-perturbative-effects}
\fancyhead[RO]{\uppercase{Non-perturbative effects in DM interactions}}

\begin{figure}
    \centering
    \includegraphics[width=\textwidth]{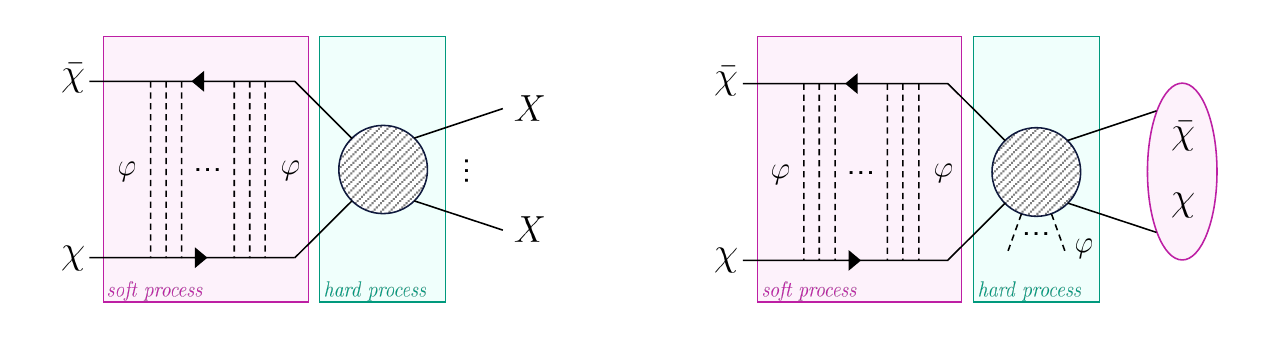}
    \caption[Feynman-like diagrams for particle antiparticle annihilation and a BSF process of an interacting $\bar{\chi}\chi$ pair with SE present.]{Feynman-like diagrams for particle antiparticle annihilation (left) and a \gls{bsf} process (right) of an interacting $\bar{\chi}\chi$ pair. The influence of the potential inducing \gls{se} is pictured by infinite exchanges of the mediator $\varphi$. Processes featuring high momentum transfer are shaded in cyan while processes of continuous low momentum transfer (including \gls{bs}) are colored in magenta.}
    \label{fig:SE_BSF_Feynman}
\end{figure}

In \gls{dm} models with two or more particles populating the dark sector, new interactions of the dark particles with the \gls{sm} and among themselves can occur, which are typically less constrained by current observations. Consequently, macroscopic observables such as the total \gls{dm} abundance or \gls{dm} distributions in galaxies might differ if these interactions are taken into account properly, possibly also changing the valid parameter space of the models under consideration when confronted with experimental data. This work is dedicated to study the impact of effects arising from interactions within the dark sector on cosmological and astrophysical observables in a variety of different models. More concretely, we will focus on two effects, namely \gls{se} \cite{Sommerfeld:1931qaf} and the presence of \gls{bs} within the dark sector. Both concepts alone have been known for a very long time in the \gls{sm} \cite{Sakharov:1948plh,Strassler:1990nw,Fadin:1990wx,Sumino:1992ai,Fadin:1994pm,Jezabek:1998pj,Grinstein:2007iv,Oliveira:2010uv,Mohorovicic:1934,Deutsch:1951zza,E598:1974sol,SLAC-SP-017:1974ind,E288:1977xhf,Brambilla:2010cs} but only recently, they have been applied in the context of \gls{dm} models as well \cite{Hisano:2002fk,Hisano:2003ec,Hisano:2004ds,Arkani-Hamed:2008hhe,Iengo:2009ni,Cassel:2009wt,ElHedri:2016onc,Blum:2016nrz,vonHarling:2014kha,Petraki:2015hla,Ellis:2015vna,Liew:2016hqo,Beneke:2016ync,Cirelli:2016rnw,Beneke:2016jpw,Mitridate:2017izz,Harz:2018csl,Biondini:2018pwp,Oncala:2019yvj,Binder:2019erp,Oncala:2021tkz,Biondini:2021ycj,Garny:2021qsr,Binder:2021vfo}. These effects are referred to as non-perturbative because they cannot be fully calculated in the framework of a perturbative \gls{qft}. However, since both effects occur when the particles are non-relativistic, they can be met with a quantum mechanical treatment. For our derivation of cross sections, decay rates etc. including the aforementioned effects, switching between both pictures will be advantageous in order to address the high energy (\textit{hard}) and low energy (\textit{soft}) momentum transfer processes within the interactions.\newp
\gls{se} occurs when a heavy particle antiparticle pair $\bar{\chi}\chi$ interacts with a light bosonic mediator $\varphi$, which has a Compton wavelength $\propto \mvphi^{-1}$ longer than the Bohr radius of the particle pair $\propto (\alpha \mchi)^{-1}$ (see \eg\eref\cite{Arkani-Hamed:2008hhe}), where $\alpha$ denotes the strength of the interaction.\footnote{\gls{se} can also occur for two particles with non-identical masses. This general case, however, will not be important for the subsequent discussion.} This is equivalent to the statement that the potential which is generated by $\varphi$ and acting on the $\bar{\chi}\chi$ pair is sufficiently \textit{long-ranged}. The condition above is naturally fulfilled for massless mediators, whereas for massive mediators it highly depends on the mass ratio between the mediator and the interacting pair. If in processes like particle antiparticle annihilation, the interaction length of the process $\propto (\vrel \mchi)^{-1}$ is much larger than the Bohr radius, the incoming particle wave functions cannot be considered as free anymore at the origin and the corresponding annihilation amplitude will receive corrections. These corrections will be larger, the smaller the relative velocity between the annihilating particles is compared to the strength of the potential ($\vrel \ll \alpha$). Moreover, if the potential generated by $\varphi$ is attractive, \gls{bs} can form.\newp
For illustration purposes, we have sketched a typical particle antiparticle annihilation process with \gls{se} as well as a \gls{bsf} process in \cref{fig:SE_BSF_Feynman}. The cross section of the hard annihilation process of a $\bar{\chi}\chi$ pair into arbitrary final states in the left diagram, depicted as a hatched circle, can be calculated with typical methods of perturbative \gls{qft}, where the results depend on the underlying model. The corrections due to \gls{se}, which can be described in the \gls{qft} framework as a resummation of soft $\varphi$ exchanges, are more model-independent due to their non-relativistic nature and solely depend on the form of the potential generated by $\varphi$ (given that the potential strength $\alpha$ has already been extracted from the \gls{qft} Lagrangian). Thus, they can be derived using a quantum mechanical approach by utilizing the scattering wave functions of the corresponding Schrödinger equation. We will assume in the following that the soft and hard processes of the interaction are well separable, \ie that we can compute the hard annihilation cross section and the \gls{se} factor independently from each other.\footnote{This standard treatment can, however, fail in certain regimes, \eg when \gls{bs} with zero energy appear in the Schrödinger equation or when the hard cross section is already so large that adding the \gls{se} factor would violate unitarity bounds at finite velocity. We direct the reader to \eref\cite{Blum:2016nrz} for an in-depth discussion of these cases.} In the right diagram of \cref{fig:SE_BSF_Feynman}, a \gls{bsf} process $\bar{\chi}+\chi\to \mathscr{B}(\bar{\chi}\chi) + n\cdot\varphi$ is sketched ($n=1,2,...$). The hard process can again be calculated with methods of perturbative \gls{qft} but this time a clear separation from the soft processes in order to calculate the \gls{bsf} cross section cannot be reached due to the off-shellness of the outgoing states. The \gls{se} correction to the \gls{bs} decay rate (not depicted) can in turn be calculated in analogy to the annihilation cross section.\newp
In \cref{subsec:SEannihilationBSdecay}, we will start by establishing a quantum mechanical treatment of \gls{se} corrections within particle antiparticle annihilations and \gls{bs} decay. We will further give insight on how to calculate the \gls{se} factors numerically in case an analytic treatment is not possible. Moving on, in \cref{subsec:nonperturbativeeffectsfromQFT} we introduce an approach to calculate the \gls{bsf} rate by leveraging methods from \gls{qft} within the non-relativistic limit. Finally, we present a different approach resorting to \gls{eft} in order to recalculate all observables in \cref{subsec:nonperturbativeeffectsfromEFT}. We discuss the merits of each method and compare their advantages.


\subsection{Sommerfeld enhancement in particle annihilation and bound state decay}
\label{subsec:SEannihilationBSdecay}

We will now calculate the \gls{se} factor for the most typical case of an annihilating $\bar{\chi}\chi$ pair alongside with the \gls{se} factor for the decay rate of a $\mathscr{B}(\bar{\chi}\chi)$ \gls{bs} into two final states. As mentioned earlier, we will assume the \gls{se} factor to be separable from the hard matrix element, such that the cross section and decay rate can be written as \cite{Iengo:2009ni,Cassel:2009wt,Petraki:2015hla}
\begin{align}
    \label{eq:SEsplittingann}
    \sigmaann\vrel&=\sum_{l=0}^{\infty}(\sigmaann\vrel)_l=\sum_{l=0}^{\infty} \sigma_l\, \vrel^{2l}\,\Sann\,,\\
    \label{eq:SEsplittingdec}
    \Gammadecn&=\sum_{l=0}^{\infty}\Gamma^{(n)}_{\text{dec},l}=\sum_{l=0}^{\infty}\sigma_{l}\,\Sdec\,,
\end{align}
with $l$ indicating an expansion in partial waves (\cf\cref{app:partial_wave_analysis}) where $\Sannl{0}$ denotes the s-wave enhancement factor, $\Sannl{1}$ the p-wave etc., and analogously for $\Sdec$, where $(n)\equiv\{nl^{(\prime)}m\}$ covers all quantum numbers of the bound state.\footnote{We want to emphasize here that the partial wave expansion number $l$ needs, in principle, to be distinguished from the orbital quantum number $l'$ of the bound state for $\Sdec$. However, we will see that an $\{nl'm\}$ bound state will only exhibit an $l$-wave contribution to the \gls{se} factor of the \gls{bs}, such that we will drop the distinguishing mark unless we want to emphasize it explicitly.} The $\sigma_l$ encapsulate the $l$-wave contributions to the hard cross section within this expansion where the partial matrix elements are stripped from their angular dependencies. Their definition together with a careful derivation of the equations for \gls{se}, which are stated in the following subsections, can be found in \cref{app:SEderivation}. For now, we will continue by setting up the Schrödinger equations for the scattering and \gls{bs} wave functions in \cref{subsubsec:Schroedingerequations}, which are needed to derive the corresponding \gls{se} factors in \cref{subsubsec:SEannihilations,subsubsec:SEboundstatedecays} for the potentials created by a massless and massive mediator $\varphi$.


\subsubsection{Schrödinger equations for scattering and bound states}
\label{subsubsec:Schroedingerequations}

The Schrödinger equations for a two-particle scattering state $\phik(\vec{r})$ and a \gls{bs} $\psinlm(\vec{r})$ in a spherically symmetric potential $V(\vec{r})=V(\abs{\vec{r}})$ take the form (see \ie\eref\cite{Landau:1991wop})
\begin{align}
	\label{eq:scatteringSchroedinger}
	\left[-\frac{1}{2\mu}\nabla^2+V(\vec{r})\right]\phik(\vec{r}) & =\Ek\phik(\vec{r})\,,\\
	\label{eq:boundstateSchroedinger}
	\left[-\frac{1}{2\mu}\nabla^2+V(\vec{r})\right]\psinlm(\vec{r}) &=\Enl\psinlm(\vec{r})\,,
\end{align}
with $\Ek=k^2/(2\mu)$, $\Enl=-\gammanl^2\kappa^2/(2\mu)$ the scattering and \gls{bs} energies. The scale $k=\mu \vrel$ hereby denotes the typical momentum transfer between the scattering particles (inversely proportional to their interaction length) and $\kappa=\mu \alpha $ is the Bohr momentum of the \gls{bs}, where $\mu=\mchi/2$ indicates the reduced mass of the system. While $\phik$ depends on the particles momentum transfer $\vec{k}$, $\psinlm$ is parametrized by the main and orbital quantum numbers $n$ and $l,m$ of the corresponding \gls{bs}, respectively. They obey the normalization 
\begin{align}
    \label{eq:scatteringwavefunctionnorm}
    \int\dd[3]{r}\phiq{q}^*(\vec{r})\phik(\vec{r}) & = (2\pi)^3\delta^{(3)}\left(\vec{q}-\vec{k}\right)\,,\\
    \label{eq:boundstatewavefunctionnorm}
    \int\dd[3]{r}\psinlm^*(\vec{r})\psi_{n'l'm'}(\vec{r}) &= \delta_{nn'}\delta_{ll'}\delta_{mm'}\,,
\end{align}
and completeness conditions
\begin{align}
    \label{eq:scatteringwavefunctioncomplete}
    \int\dd[3]{k}\phik^*(\vec{r})\phik(\vec{r}') & = (2\pi)^3\delta^{(3)}\left(\vec{r}-\vec{r}'\right)\,,\\
    \label{eq:boundstatewavefunctioncomplete}
    \sum_{n,l,m}\psinlm^*(\vec{r})\psi_{nlm}(\vec{r}') &= \delta^{(3)}\left(\vec{r}-\vec{r}'\right)\,.
\end{align}
The function $\gamma_{nl}$ parameterizes the energy splitting between different \gls{bs} on which we will comment on later. We want to emphasize here that the spin structure of the interacting particles plays no role for the subsequent discussion, and it is thus irrelevant if the particles involved are bosons or fermions. What will be important in the following, however, is if the mediator creating the potential $V(\vec{r})$ has a mass. Therefore, we will distinguish between a \textit{Yukawa potential} $V_{\scriptscriptstyle Y}(\vec{r})=-\alpha/r \exp(-\mvphi r)$ for a massive $\varphi$ and a \textit{Coulomb potential} $V_{\scriptscriptstyle C}(\vec{r})=-\alpha/r$ for the massless case.\newp
Due to the spherical symmetry of the potentials under consideration, we can expand $\phik$ in partial waves (see \cref{app:partial_wave_analysis}) and write $\psinlm$ in terms of spherical harmonics $Y_{lm}(\Omega_{\vec{r}})$ (see \cref{app:special_functions}). The wave functions then yield
\begin{align}
    \label{eq:scatteringwavefunctiondec}
	\phik(\vec{r})&=\sum_{l=0}^\infty (2l+1)\left[\frac{\chikl(\kappa r)}{\kappa r}\right] P_l(\vec{\hat{k}}\cdot\vec{\hat{r}})\,,\\
    \label{eq:boundstatewavefunctiondec}
	\psinlm(\vec{r})&=\kappa^{3/2}\left[\frac{\chinl(\kappa r)}{\kappa r}\right]Y_{lm}(\Omega_{\vec{r}})\,.
\end{align}
The radial parts of the scattering and \gls{bs} wave functions $\chikl$ and $\chinl$ depend on the exact form of the potential. Starting with a Yukawa potential, they are solutions to the radial Schrödinger equations
\begin{align}
    \label{eq:scatteringSchroedingerRadial}
    \chikl''(x)+\left(\frac{1}{\zeta^2}-\frac{l(l+1)}{x^2}+\frac{2}{x}e^{-\frac{x}{\xi}}\right) \chikl(x)&=0\,,\\
    \label{eq:boundstateSchroedingerRadial}
    \chinl''(x)+\left(-\gammanl^2(\xi)-\frac{l(l+1)}{x^2}+\frac{2}{x}e^{-\frac{x}{\xi}}\right)\chinl(x)&=0\,,
\end{align}
which can be obtained by inserting \cref{eq:scatteringwavefunctiondec,eq:boundstatewavefunctiondec} into \cref{eq:scatteringSchroedinger,eq:boundstateSchroedinger} and employing \cref{eq:LegendreDEQ,eq:sphericalharmonicsDEQ} to cancel the angular dependencies. The equations above were reparameterized in terms of $x\equiv \kappa r$ and we introduced the dimensionless variables
\begin{equation}
    \label{eq:zetaandxi}
    \zeta \equiv \frac{\kappa}{k}=\frac{\alpha}{\vrel},\qquad \xi\equiv \frac{\kappa}{\mvphi}=\frac{\alpha\mchi}{2\mvphi},
\end{equation}
which we will use throughout this work. This choice is convenient because the previous discussions regarding the significance of the \gls{se} and its relative importance as a correction to the annihilation cross-section can be easily expressed by requiring $\xi\gg 1$ and $\zeta\gg 1$, respectively. The Coulomb case can be obtained from \cref{eq:scatteringSchroedingerRadial,eq:boundstateSchroedingerRadial} by taking the asymptotic limit $\xi\to\infty$ (or $\mvphi\to 0$). Utilizing \cref{eq:Legendresum,eq:Legendreint,eq:sphericalharmonicsadditiontheorems,eq:sphericalharmonicsnorm,eq:deltaspherical}, the normalization and completeness conditions for the radial wave functions can be derived from the full expressions. They are given by 
\begin{align}
    \label{eq:scatteringwavefunctionnormRadial}
    \frac{4}{\kappa^3}\int_0^\infty \dd{x}\chiqs{q}{l}^*(x)\chikl(x) &= \frac{(2\pi)}{k^2}\delta(q-k)\,,\\
    \label{eq:boundstatewavefunctionnormRadial}
    \int_0^\infty \dd{x} \chinl^*(x)\chiij{n'l}(x)&=\delta_{nn'}\,,\\
    \label{eq:scatteringwavefunctioncompleteRadial}
    \frac{4}{\kappa^3}\int_0^\infty \dd{k} k^2\chikl^*(x)\chikl(x') &= (2\pi)\delta(x-x')\,,\\
    \label{eq:boundstatewavefunctioncompleteRadial}
    \sum_n\chinl^*(x)\chinl(x')&=\delta(x-x')\,.
\end{align}
Their form as a solution to the radial Schrödinger equations are well known for the Coulomb limit and yield in this parametrization (see \eg\app\,A of \eref\cite{Garny:2021qsr})
\begin{align}
    \label{eq:ColumbscatteringwavefunctionRadial}
    \chikl(x)=&\frac{\abs{\Gamma(1+l-i\zeta)}e^{\pi\zeta/2}e^{i\delta_l}}{(2l+1)!}\left(\frac{2ix}{\zeta}\right)^lxe^{-ix/\zeta}\HGone{1+l+i\zeta}{2l+2}{\frac{2ix}{\zeta}}\\
    \label{eq:ColumbboundstatewavefunctionRadial}
    \chinl(x)=&\frac{2}{n^2}\sqrt{\frac{(n-l-1)!}{(n+l)!}}\left(\frac{2x}{n}\right)^{l}x e^{-x/n}\Laguerre{n-l-1}{2l+1}{\frac{2x}{n}}
\end{align}
with $\delta_l=\arg(\Gamma(1+l-i\zeta))$ and where $\Gamma(x)$, $\HGone{a}{c}{x}$ and $\Laguerre{n}{\alpha}{x}$ denote the gamma function, the confluent hypergeometric function and the associated Laguerre polynomial, respectively (see \cref{app:special_functions}). The solutions for radial wave equations using a Yukawa potential have to be found numerically. Methods for this, which we will use in the following, were discussed in \erefs\cite{Iengo:2009ni,Petraki:2016cnz}.


\subsubsection{Sommerfeld enhancement in particle annihilations}
\label{subsubsec:SEannihilations}

The $l$-wave \gls{se} factor for particle annihilation is given by \cite{Cassel:2009wt} (\cf also \cref{app:SEderivation})
\begin{equation}
    \label{eq:SEannihilation}
    \Sann=\left[\frac{(2l+1)!!}{l!}\zeta^l\dv[l]{}{x}\left(\frac{\chikl(x)}{x}\right)\right]_{x=0}^2.
\end{equation}
For a massless mediator creating a Coulomb potential, we can simply plug in \cref{eq:ColumbscatteringwavefunctionRadial} and use \cref{eq:GammaS0prod,eq:HGonesumdef} to arrive at
\begin{equation}
    \label{eq:SEfactorannCoulomb}
    \Sann(\zeta)=\Sannl{0}(\zeta)\prod_{s=1}^l\left(1+\frac{\zeta^2}{s^2}\right),\quad\text{with}\quad\Sannl{0}(\zeta)=\frac{2\pi\zeta}{1-e^{-2\pi\zeta}}.
\end{equation}
We can easily see that the \gls{se} vanishes, \ie $\Sann(\zeta)\to 1$, for high relative velocities of the interacting particles and/or low couplings ($\zeta\to 0$), as expected. For low velocities, $\Sann(\zeta)\propto 1/\vrel^{1+2l}$, such that the cross section will be $\propto 1/\vrel$ and, therefore, diverges for $\vrel\to 0$. Usually this does not pose a problem since only the thermally averaged quantities are physical observables but it can lead to violation of unitarity for certain coupling strengths as explored \eg in \erefs\cite{Griest:1989wd,vonHarling:2014kha,Baldes:2017gzw}. We will implicitly check for these limitations in the considered models but will not discuss them further as they turn out to play no role for the subsequent analyses in our parameter regimes of interest.\newp
In case of a massive mediator, corresponding to a Yukawa potential in the Schrödinger equations, the \gls{se} factor can only be extracted numerically \cite{Iengo:2009ni}.
Without loss of generality, we can take $\chikl$ to be real because the phase of the scattering wave function does not enter \cref{eq:SEannihilation}. The normalization chosen corresponds at leading order to an asymptotic behavior of (\cf\cref{eq:scatteringwavefunctionsineapprox})
\begin{equation}
	\chitkl(x)\xrightarrow{x\to\infty}C_{s,l}\sin\left(\frac{x}{\zeta}-\frac{l\pi}{2}+\delta_l\right)\,,
\end{equation}
where we have defined $\chikl(x)=\zeta/C_{s,l}\,\chitkl(x)$ for later convenience, with $C_{s,l}$ being the difference in amplitude between the wave function with and without \gls{se}. Requiring the solution to \cref{eq:scatteringSchroedingerRadial} as well as its first derivative to be regular at the origin, we can define the initial conditions as 
\begin{equation}
    \chitkl(x)\xrightarrow{x\to 0}\left(\frac{x}{\zeta}\right)^{l+1},\qquad  \chitkl'(x)\xrightarrow{x\to 0}\frac{l+1}{\zeta}\left(\frac{x}{\zeta}\right)^{l}-\frac{1}{l+1}\left(\frac{x}{\zeta}\right)^{l+1}.
\end{equation}
These choices significantly simplify \cref{eq:SEannihilation} to
\begin{equation}
    \label{eq:SEannYukawa}
    \Sann=\left|\frac{(2l+1)!!}{C_{s,l}}\right|^2.
\end{equation}
The  constant $C_{s,l}$ for an s-wave ($l=0$) can be extracted best by defining the function $F_0(y)\equiv\tilde{\chi}_{\vec{k},0}(y)^2+\tilde{\chi}_{\vec{k},0}(y-\pi/2)^2$ which yields $\lim_{y\to\infty}F_0(y)=C_{s,0}^2$ at sufficiently high values for $y\equiv x/\zeta$. While this procedure works perfectly fine for $l=0$, for $l=1$ it is necessary to increase the precision of the asymptotic expansion up to sub-leading order to obtain a good result. Starting with \cite{Iengo:2009ni}
\begin{equation}
	\tilde{\chi}_{\vec{k},1}(y)\xrightarrow{y\to \infty}C_{s,1}\left(\sin\left(y-\frac{\pi}{2}+\delta_1\right)+\frac{\cos\left(y-\frac{\pi}{2}+\delta_1\right)}{y}\right)\,,
\end{equation}
we can obtain $C_{s,1}^2$ by defining $F_1(y)\equiv\tilde{\chi}_{\vec{k},1}(y)$, 
\begin{align}
	k(y)&\equiv \frac{(\pi^2-16y^2)^2}{8\pi(\pi^2-16y^2)}\left(F_1(y+\pi/4)^2-F_1(y-\pi/4)^2\right)\,,\\
	h(y)&\equiv (\pi^3-4\pi y^2)(k(y+\pi/4)+k(y-\pi/4))\,,\\
	j(y)&\equiv -8\frac{(h(y+\pi/4)+h(y-\pi/4))}{(\pi^2-16y^2)(3\pi^2+16(1+y^2))}\,,
\end{align}
and taking $\lim_{y\to\infty}j(y)=C_{s,1}^2$. Numerically, we implemented the second method with a relative error of $\varepsilon_{\text{rel}}<10^{-3}$ as a breaking condition for the convergence algorithm.


\subsubsection{Sommerfeld enhancement in bound state decays}
\label{subsubsec:SEboundstatedecays}

Bound states can only form when the corresponding potential is sufficiently long-ranged.  This is always the case for potentials generated by massless particles (\eg Coulomb potentials). For Yukawa potentials, which are created by massive mediators, this statement is less obvious. A rough estimate for the existence of a \gls{bs} is $\xi\gtrsim n^2$ but one might want to do better. Considering \gls{bsf} by radiating off one massive mediator, a stronger criterion to create a bound state is $\Ek-\Enl>\mvphi$ which translates in dimensionless coordinates to 
\begin{equation}
    \label{eq:bsfexistencecriterium}
    \xi >\frac{2}{\alpha(\gammanl^2(\xi)+1/\zeta^2)}\xrightarrow{\zeta\to \infty}\frac{2}{\alpha\gammanl^2(\xi)}\gg n^2
\end{equation}
in the static limit ($\vrel\to 0$), if the function $\gammanl(\xi)$ is known. \gls{bsf} will become effective, once the Yukawa potential can get resolved by the wavelength of the incoming state, \ie $k\gtrsim\mvphi$ or $\xi\gtrsim \zeta$. A criterion to determine, if a potential is long-ranged enough, is then to demand that the \textit{screening length} $D$ which corresponds to $\mvphi^{-1}$ for a Yukawa potential should be above a critical value $D_0$ \cite{Rogers:1970}. For a bound state to exist, we demand $\xi>\xi_c=0.8399$ for $n=1$ ($l=0$), which will be reflected in the Schrödinger equation (\cf\cref{eq:boundstateSchroedingerRadial}) by $\gammanl^2(\xi)\to 0$ as $\xi\to\xi_c$.\footnote{The critical value $\xi_c$ for the $n=1$ and other bound states can be extracted from Table 4 in \eref\cite{Rogers:1970}.}. In the Coulomb limit ($\xi\to\infty$) there is no screening and $\gammanl^2(\xi)\to 1/n^2$, where $n$ states of different $l$ are degenerate.\newp
The $l'$-wave \gls{se} factor for bound state decay into two final states is given by \eref\cite{Petraki:2015hla} (\cf also \cref{app:SEderivation})
\begin{equation}
    \label{eq:SEdecay}
    \Sdec=\frac{\kappa^3}{4\pi}\frac{\delta_{ll'}\delta_{m0}}{2l+1}\left[\frac{(2l+1)!!}{l!}\alpha^l\dv[l]{}{x}\left(\frac{\chinl(x)}{x}\right)\right]_{x=0}^2\,,
\end{equation}
where $\delta_{ll'}$ projects out the correct angular momentum state and $\delta_{m0}$ results from the azimuthal symmetry of the potential. Using \cref{eq:ColumbboundstatewavefunctionRadial} for a Coulomb potential, the \gls{se} factor for \gls{bs} decay simplifies to
\begin{equation}
    \Sdec=\frac{\kappa^{3+2l}}{\pi\mu^{2l}}\frac{\delta_{ll'}\delta_{m0}}{2l+1}\frac{(n+l)!}{n^{2(l+2)}(n-l-1)!(l!)^2}\,.
\end{equation}
In order to find the correct bound state wave function for a given $\xi$ in a Yukawa potential, we follow the procedure in \eref\cite{Petraki:2016cnz} where we start by substituting $\rho\equiv 2x\gammanl(\xi)$ as well as $\chinl(\rho)\equiv \chitnl(\rho)/C_{b,nl} $ in \cref{eq:boundstateSchroedingerRadial} yielding
\begin{equation}
	\chitnl''(\rho)+\left(-\frac{1}{4}-\frac{l(l+1)}{\rho^2}+\frac{e^{-\rho/(2\xi\gammanl(\xi))}}{\rho\gammanl(\xi)}\right)\chitnl(\rho)=0\,,
\end{equation}
with $C_{b,nl}$ being the \gls{bs} equivalent to $C_{s,l}$ for scattering state solutions. Apart from determining regular solutions to this equation using the corresponding initial conditions 
\begin{equation}
    \label{eq:boundstateinitialconditions}
    \chitnl(\rho)\xrightarrow{\rho\to 0} \left(\frac{\rho}{2\gammanl(\xi)}\right)^{l+1},\qquad \chitnl'(\rho)\xrightarrow{\rho\to 0} \frac{(l+1)}{2\gammanl(\xi)} \left(\frac{\rho}{2\gammanl(\xi)}\right)^{l},
\end{equation}
we also have to resolve $\gammanl(\xi)$ by requiring that the \gls{bs} solutions have to be localized wave packages, \ie
\begin{equation}
    \lim_{\rho\to \infty}\chitnl(\rho)=0.
\end{equation}
For this purpose we guess an initial $\gammanl^{(0)}$ using the approximate formula $\gammanl^{(0)}\simeq 1/n (1-n^2\xi_c/\xi)^b$ (where the values for $b$ for a given $\{n,l\}$ pair can be extracted from \eref\cite{Petraki:2016cnz}) and perform an iterative shooting method demanding $\chitnl(\rho_\infty)=0$, with a finite $\rho_\infty$ for numerical feasibility ($\rho_\infty\simeq 60$ was checked to be sufficient for our purpose). After we have determined the correct $\gammanl(\xi)$ up to the desired precision and changed the variables back to $x$, we can use the normalization condition (\cf\cref{eq:boundstatewavefunctionnormRadial})
\begin{equation}
    \int_0^\infty \dd{x} \abs{\chitnl(x)}^2 = C_{b,nl}^2
\end{equation}
to determine $C_{b,nl}$ and thus the full \gls{bs} wave function $\chinl(x)$. Plugging in \cref{eq:boundstateinitialconditions} into \cref{eq:SEdecay}, the \gls{se} factor for \gls{bs} decay into two final states within this parametrization yields
\begin{equation}
    \Sdec=\frac{\kappa^{3+2l}}{\pi\mu^{2l}}\frac{\delta_{ll'}\delta_{m0}}{2l+1}\left
    (\frac{(2l+1)!!}{2C_{b,nl}}\right)^2.
\end{equation}


\subsection{Non-perturbative effects from the Bethe-Salpeter approach}
\label{subsec:nonperturbativeeffectsfromQFT}

Now, we know how to explicitly calculate the \gls{se} factors for particle antiparticle annihilation and \gls{bs} decay utilizing the Schrödinger equations for non-relativistic scattering and \gls{bs}. This was made possible because we could clearly separate soft and hard momentum exchange processes within these interactions. By inserting the corrected rates into a set of Boltzmann equations as described in \cref{sec:DMproduction}, treating the free $\bar{\chi}\chi$ pair as well as each $\mathscr{B}(\bar{\chi}\chi)$ bound state as a separate species, we can track the evolution of their yield, which eventually leaves us with a corrected \gls{dm} abundance. This, we can then compare to the abundance we would have obtained without taking them into account in order to quantify their relative importance. However, so far one important link is still missing: we have to find a way to describe the production of \gls{bs}, \ie we need a suitable method to calculate processes like the one pictured in \cref{fig:SE_BSF_Feynman} (right). This task is significantly more challenging because the aforementioned na\"ive separation between soft and hard scales cannot be applied anymore. In this and in the next subsection, we will present two independent methods to achieve this goal, each of them having different advantages but an equivalent outcome. The first one has been adapted to the \gls{dm} problem initially in \eref\cite{Petraki:2015hla} and was later extended in various ways (see \eg\erefs\cite{Petraki:2016cnz,Cirelli:2016rnw,Harz:2017dlj,Harz:2018csl,Oncala:2018bvl,Harz:2019rro,Binder:2019erp,Oncala:2019yvj,Oncala:2021tkz,Oncala:2021swy,Garny:2021qsr}).\newp
The main idea starts by utilizing \textit{Bethe-Salpeter wave equations}, which we rederive and connect to the Schrödinger equations in \cref{subsubsec:BetheSalpeter} to build up the scaffold and nomenclature of the method. Using common techniques from \gls{qft} as well as appropriate approximations for the non-relativistic limit, we will derive the desired \gls{bsf} cross section in \cref{subsubsec:BSFfromBetheSalpeter}. Note that we will highlight in the following only the pivotal steps and most important derivations from \eref\cite{Petraki:2015hla}. For a detailed and in-depth discussion, the reader is referred to the original paper and the references therein (most importantly \erefs\cite{Itzykson:1980rh,Silagadze:1998ri}).


\subsubsection{The Bethe-Salpeter wave equation}
\label{subsubsec:BetheSalpeter}

We first want to derive the Bethe-Salpeter wave equations, which in simplified terms can be seen as a relativistic extension of the Schrödinger equations. They can be utilized to characterize a \gls{bs} of a two-particle quantum system, incorporating concepts such as spin, which the Schrödinger equations do not encompass. For simplicity, we employ a general system of two scalar particles $\chi_1$ and $\chi_2$ with momenta $\vec{q}_{1,2}$ and (not necessarily equal) masses $m_{1,2}$ interacting via a mediator $\varphi$ being either a scalar or vector boson. We will later discuss what changes in the derivation if $\chi_1$ and $\chi_2$ were fermions instead.

\paragraph{The Bethe-Salpeter wave functions}$~$\newpp
We define $\SSQ{Q}{q}$ to be a two-particle $\chi_1\chi_2$ scattering state with total and relative momentum $\vec{Q}=\vec{q}_1+\vec{q}_2$ and $\vec{q}=\eta_2\vec{q}_1-\eta_1\vec{q}_2=\mu\vec{\vrel}$, respectively, where $m=m_1+m_2$, $\mu=m_1m_2/m$ denote the total and reduced mass of the system, and $\eta_{1,2}=m_{1,2}/m$ (with $\eta_1+\eta_2=1$).\footnote{Unless stated otherwise, capital and small unindexed roman letters appearing in the subsequent discussion will always refer to total and relative 3- or 4-momenta of a two-body system as defined above.} The total energy of the scattering state is then given by $\ESS{Q}{q}$. Further, we denote the one-particle \gls{bs} as $\BSQ{Q}$ with energy $\EBS{Q}$, where $n\equiv\{nlm\}$ takes into account all its quantum numbers. Within this notation, we can define the scattering and \gls{bs} \textit{Bethe-Salpeter wave functions} (first introduced in \eref\cite{Salpeter:1951sz})
\begin{align}
    \label{eq:BetheSalpeterSS}
    \PhiQBS{Q}{q}(x_1,x_2)&\equiv\Vaccc T\chi_1(x_1)\chi_2(x_2)\SSQ{Q}{q}\,,\\
    \label{eq:BetheSalpeterBS}
    \PsiQBS{Q}(x_1,x_2)&\equiv\Vaccc T\chi_1(x_1)\chi_2(x_2)\BSQ{Q}\,,
\end{align}
which represent, roughly speaking, a covariant extension of the Schrödinger wave functions into the relativistic regime, with $T$ denoting the time-ordering operator and $\Vac$ represents the vacuum state of the interacting theory.\footnote{For a particle antiparticle pair as often used throughout this work, we can substitute $\chi_1(x_1)\to\chi(x_1)$ and $\chi_2(x_2)\to\chi^\dagger(x_2)$.} Further, we define
\begin{align}
    \label{eq:BetheSalpeterSSstar}
    \PhiQBS{Q}{q}^\star(x_1,x_2)&\equiv\Vaccc \bar{T}\chi_1(x_1)\chi_2(x_2)\SSQ{Q}{q}^*=\SSQcc{Q}{q}T\chi_1^\dagger(x_1)\chi_2^\dagger(x_2)\Vac\,,\\
    \label{eq:BetheSalpeterBSstar}
    \PsiQBS{Q}^\star(x_1,x_2)&\equiv\Vaccc \bar{T}\chi_1(x_1)\chi_2(x_2)\BSQ{Q}^*=\BSQcc{Q}T\chi_1^\dagger(x_1)\chi_2^\dagger(x_2)\Vac\,,
\end{align}
with $\bar{T}$ the anti-time-ordering operator and asterix denoting the complex conjugate of a quantity. 
We will use these functions in the following to derive the \gls{bsf} cross section as well as the \gls{se} factors used in \cref{subsubsec:SEannihilations,subsubsec:SEboundstatedecays} by connecting them to the usual Schrödinger wave functions in the non-relativistic limit.\newp
For later convenience, we want to separate the \gls{com} from the relative motion of the particles. To do so, we first switch to a more suitable coordinate system in position space $X=\eta_1x_1+\eta_2 x_2$ and $x=x_1-x_2$. Using the 4-momentum operator $\hat{P}$ to rewrite
\begin{equation}
    \chi_1(x_1)=e^{i\hat{P}X}\chi_1(\eta_2x)e^{-i\hat{P}X},\quad \chi_2(x_2)=e^{i\hat{P}X}\chi_2(-\eta_1 x)e^{-i\hat{P}X}\,,
\end{equation}
we obtain after a few steps
\begin{equation}
    \PhiQBS{Q}{q}(x_1,x_2)=e^{-iQX}\Vaccc T\chi_1(\eta_2x)\chi_2(-\eta_1x)\SSQ{Q}{q}\equiv e^{-iQX}\PhiQBS{Q}{q}(x)\,,
\end{equation}
where $\PhiQBS{Q}{q}(x)$ only depends on the distance between the two particles, and analogously for the other wave functions. Their Fourier transforms can then be defined accordingly
\begin{equation}
    \label{eq:BSwavefunctionsFT}
    \PhiQBSt{Q}{q}^{(\star)}(p)\equiv \int\dd[4]{x}\PhiQBS{Q}{q}^{(\star)}(x)e^{\pm ipx},\quad \PsiQBSt{Q}^{(\star)}(p)\equiv \int\dd[4]{x}\PsiQBS{Q}^{(\star)}(x)e^{\pm ipx}\,,
\end{equation}
with $p$ denoting a relative momentum coordinate defined analogously to $q$.\footnote{More concretely we define $P\equiv p_1+p_2$, $p\equiv\eta_2p_1-\eta_1p_2$ such that the Fourier transform $\PhiQBSt{Q}{q}(p_1,p_2)=\int\dd[4]{x_1}\dd[4]{x_2}\PhiQBS{Q}{q}(x_1,x_2)e^{ip_1x_1}e^{ip_2x_2}=\delta^{(4)}(P-Q)\PhiQBSt{Q}{q}(p)$ reflects the desired momentum separation.}

\paragraph{The 4-point Green's function}$~$\newpp
The 4-point Green's (or correlation) function including the mediator exchange between $\chi_1$ and $\chi_2$ is defined as 
\begin{equation}
    \Gfour(x_1,x_2;y_1,y_2)\equiv\Vaccc T\chi_1(y_1)\chi_2(y_2)\chi_1^\dagger(x_1)\chi_2^\dagger(x_2)\Vac\,,
\end{equation}
evaluated at arbitrary space-time points $x_{1,2}$ and $y_{1,2}$. It satisfies the \textit{Dyson-Schwinger equation} \cite{Schwinger:1948yk,Schwinger:1951ex,Schwinger:1951hq}
\begin{align}
    \Gfour(x_1,x_2;y_1,y_2)=\,&S_1(x_1-y_1)S_2(x_2-y_2)+\int \dd[4]{z_1}\dd[4]{z_2}\dd[4]{z_1'}\dd[4]{z_2'}\nonumber\\
    &S_1(x_1-z_1)S_2(x_2-z_2)W(z_1,z_2;z_1',z_2')\Gfour(z_1',z_2',y_1,y_2)\,,
\end{align}
with $S_{1,2}$ being the full $\chi_{1,2}$ propagators and $W$ describes the perturbative 4-point interaction kernel between $\chi_1$ and $\chi_2$.\footnote{The kernel $W$ involves all truncated 4-fermion diagrams that cannot be disconnected by cutting two fermion lines (see \eg\eref\cite{Itzykson:1980rh}).} Switching to \gls{com} coordinates does neither alter $\Gfour$ nor $W$ due to their translational invariance. Their Fourier transforms can thus be defined as
\begin{align}
    \Gfourt(p,p';Q)&\equiv\int\dd[4]{x}\dd[4]{y}\dd[4]{(X-Y)}\Gfour(x,y;X-Y)e^{i(px-p'y)}e^{iQ(X-Y)},\\
    \tilde{W}(p,p';Q)&\equiv\int\dd[4]{x}\dd[4]{y}\dd[4]{(X-Y)}W(x,y;X-Y)e^{i(px-p'y)}e^{iQ(X-Y)},
\end{align}
which we can use to recast the Dyson-Schwinger equation in momentum space
\begin{equation}
    \label{eq:DysonSchwingerFT}
     \Gfourt(p,p';Q)=(2\pi)^4\delta^{(4)}(p-p')S(p;Q)+S(p;Q)\int\frac{\dd[4]{k}}{(2\pi)^4}\tilde{W}(p,k;Q)\Gfourt(k,p';Q),
\end{equation}
where we further defined for convenience
\begin{equation}
    \label{eq:combinedpropagator}
    S(p;Q)\equiv\tilde{S}_1(\eta_1Q+p)\tilde{S}_2(\eta_2Q-p) \quad\text{with} \quad \tilde{S_j}(p)\equiv\int\dd[4]{z}S_j(z)e^{ipz}. 
\end{equation}\newp 
We will employ \cref{eq:DysonSchwingerFT} later to derive the wave equations for the Bethe-Salpeter wave functions.
Before we can do this, we need to decompose the 4-point Green's function into contributions from bound and scattering states, such that $\Gfour$ can be linked to $\PhiQBS{Q}{q}$ and $\PsiQBS{Q}$, separately. This can be attained by applying the one- and two-particle completeness relation 
\begin{equation}
    \sum_{n} \int \frac{\dd[3]{Q}}{(2\pi)^3 2\EBS{Q}}\BSQ{Q}\BSQcc{Q}+\int\frac{\dd[3]{Q}}{(2\pi)^3 2\ESS{Q}{q}}\frac{\dd[3]{q}}{(2\pi)^32\ESSdiffq{Q}{q}}\SSQ{Q}{q}\SSQcc{Q}{q}=\mathds{1}
\end{equation}
to $\Gfour$ directly, with $\ESSdiffq{Q}{q}$ denoting the energy of $q$. This enables us to split up
\begin{align}
    \Gfour(x,y;X-Y)&=\sum_n \Gfour_n(x,y;X-Y)+\Gfour_{\mathcal{U}}(x,y;X-Y)\,,\\
    \Gfourt(p,p';Q)&=\sum_n \Gfourt_n(p,p';Q)+\Gfourt_{\mathcal{U}}(p,p';Q)\,,
\end{align}
with $G_n^{(4)}$ and $G_{\mathcal{U}}^{(4)}$ denoting the separated Green's functions for the bound and scattering state, respectively.\footnote{We used here the standard normalization of one-particle momentum eigenstates, namely $\langle\vec{p}_j|\vec{k}_j\rangle=2E_j(\vec{p};\vec{Q})(2\pi)^3\delta^{(3)}(\vec{p}_j-\vec{k}_j)$, with $E_j(\vec{p};\vec{Q})$ denoting the energy of state $|\vec{p}_j\rangle$. To leading order in the interaction strength, we can assume the two-particle state $\SSQ{Q}{q}$ to yield just the outer product of the two composing one-particle states. Thus, $2\ESS{Q}{q}2\ESSdiffq{Q}{q}\simeq 2E_1(\vec{q};\vec{Q})E_2(\vec{q};\vec{Q})$.} After a few steps, we arrive at
\begin{align}
    \label{eq:Greens4FTBS}
    \Gfourt_n(p,p';Q)=&i\int\dd[4]{x}\dd[4]{y}e^{i(px-p'y)}\frac{\PsiQBS{Q}(x)\PsiQBS{Q}^{\star}(y)e^{-i[Q^0-\EBS{Q}][h_-(x^0)-h_+(y^0)]}}{2\EBS{Q}[Q^0-\EBS{Q}+i\epsilon]}\\
    \label{eq:Greens4FTBSpole}
    &\xrightarrow{Q^0\to\,\EBS{Q}}\frac{i\PsiQBSt{Q}(p)\PsiQBSt{Q}^{\star}(p')}{2\EBS{Q}[Q^0-\EBS{Q}+i\epsilon]}\,,\\
    \label{eq:Greens4FTSS}
    \Gfourt_{\mathcal{U}}(p,p';Q)=&i\int\frac{\dd[3]{q}}{(2\pi)^3}\int\dd[4]{x}\dd[4]{y}e^{i(px-p'y)}\frac{\PhiQBS{Q}{q}(x)\PhiQBS{Q}{q}^{\star}(y)e^{-i[Q^0-\ESS{Q}{q}][h_-(x^0)-h_+(y^0)]}}{2\ESS{Q}{q}2\ESSdiffq{Q}{q}[Q^0-\ESS{Q}{q}+i\epsilon]}\\
    \label{eq:Greens4FTSSpole}
    &\xrightarrow{Q^0\to\,\ESS{Q}{q}}\int\frac{\dd[3]{q}}{(2\pi)^3}\frac{i\PhiQBSt{Q}{q}(p)\PhiQBSt{Q}{q}^{\star}(p')}{2\ESS{Q}{q}2\ESSdiffq{Q}{q}[Q^0-\ESS{Q}{q}+i\epsilon]}\,,
\end{align}
for the Fourier transformed entities, where \cref{eq:Greens4FTBSpole,eq:Greens4FTSSpole} describe their behavior close to their poles. The functions
\begin{equation}
    h_\pm(z^0)\equiv\frac{1}{2}(\eta_2-\eta_1)z^0\pm\frac{1}{2}\abs{z^0}
\end{equation}
are introduced through time ordering and exponentiated after employing the integral representation of the Heavyside $\Theta$ function (\cf\cref{eq:HeavysideIntegral}). A detailed calculation as well as the corresponding expressions for $\Gfour_{n/\mathcal{U}}(x,y;X-Y)$ can be found in \eref\cite{Petraki:2015hla}.

\paragraph{The Bethe-Salpeter wave equation}$~$\newpp
With the definition of the operator 
\begin{equation}
    A(p,p';Q)\equiv(2\pi)^4\delta^{(4)}(p-p')S^{-1}(p;Q)-\tilde{W}(p,p';Q)\,,
\end{equation}
we can recast the Dyson Schwinger equation (\cf\cref{eq:DysonSchwingerFT}) into 
\begin{equation}
    \int\frac{\dd[4]{k}}{(2\pi)^4}A(p,k;Q)\Gfourt(k,p';Q)=(2\pi)^4\delta^{(4)}(p-p')\,.
\end{equation}
In this form, we can see that the 4-point Green's function possesses the general solution
\begin{equation}
    \Gfourt(p,p';Q)=\sum_n c_n^{-1}(Q)C_n(p;Q)C_n^\dagger(p';Q)+\int\frac{\dd[3]{q}
    }{(2\pi)^3} f^{-1}_{\vec{q}}(Q) F_{\vec{q}}(p;Q)F^\dagger_{\vec{q}}(p';Q)\,,
\end{equation}
where $C_n$ and $F_{\vec{q}}$ denote here the discrete and continuous eigenfunctions of $A$ with eigenvalues $c_n$ and $f_{\vec{q}}$ left to be determined. Comparing the expression above to the separated Green's functions, given in \cref{eq:Greens4FTBS,eq:Greens4FTSS}, we can easily extract their structure\footnote{The following relations are fixed since the eigenvalues must not depend on the momenta $p,p'$ and we demand that the eigenfunctions are free of singularities.}
\begin{align}
    C_n(p;Q)\propto &\int\dd[4]{x}\PsiQBS{Q}(x)e^{ipx}e^{-i[Q^0-\EBS{Q}]h_-(x^0)}\,,\\
    C^\dagger_n(p';Q)\propto &\int\dd[4]{y}\PsiQBS{Q}^{\star}(y)e^{-ip'y}e^{i[Q^0-\EBS{Q}]h_+(y^0)}\,,\\
    F_{\vec{q}}(p;Q)\propto &\int\dd[4]{x}\PhiQBS{Q}{q}(x)e^{ipx}e^{-i[Q^0-\ESS{Q}{q}]h_-(x^0)}\,,\\
    F^\dagger_{\vec{q}}(p';Q)\propto &\int\dd[4]{y}\PhiQBS{Q}{q}^{\star}(y)e^{-ip'y}e^{i[Q^0-\ESS{Q}{q}]h_+(y^0)}\,,\\
    c_n(Q)\propto &~1-\EBS{Q}/Q^0,\qquad f_{\vec{q}}(Q)\propto\,1-\ESS{Q}{q}/Q^0.
\end{align}
Inserting these quantities into the eigenvalue equations
\begin{align}
    \int\frac{\dd[4]{k}}{(2\pi)^4}A(p,k;Q)C_n(k;Q)=&c_n(Q)C_n(p;Q)\,,\\
    \int\frac{\dd[4]{k}}{(2\pi)^4}A(p,k;Q)F_{\vec{q}}(k;Q)=&f_{\vec{q}}(Q)F_{\vec{q}}(p;Q)\,,
\end{align}
and putting them on-shell (\ie $Q^0\to\EBS{Q}$ and $Q^0\to\ESS{Q}{q}$, respectively), we arrive at the Bethe-Salpeter wave equations
\begin{align}
    \label{eq:BSequationBS}
    \PsiQBSt{Q}(p)=S(p;Q)\int\frac{\dd[4]{k}}{(2\pi)^4}\tilde{W}(p,k;Q)\PsiQBSt{Q}(k)\,,\\
    \label{eq:BSequationSS}
    \PhiQBSt{Q}{q}(p)=S(p;Q)\int\frac{\dd[4]{k}}{(2\pi)^4}\tilde{W}(p,k;Q)\PhiQBSt{Q}{q}(k)\,,
\end{align}
which aim to describe scattering and bound states in a relativistic and covariant formalism.\footnote{The homogeneous Bethe-Salpeter wave equations do not contain information about the normalization of the wave functions. For a detailed discussion about their derivation, the reader is referred to \eref\cite{Petraki:2015hla}.} By taking appropriate limits and approximations, we can revive the usual Schrödinger equation from them, as we will see in the next paragraph.

\paragraph{The Schrödinger equations from the instantaneous approximation}$~$\newpp
The Bethe-Salpeter wave equations, given in \cref{eq:BSequationBS,eq:BSequationSS}, lack exact solutions in their general form. Nevertheless, through specific assumptions, one can simplify these equations in order to obtain analytic results. One prevalent simplification method is the \textit{instantaneous approximation} (see \eg\eref\cite{Lucha:2012ky}). It is used in the non-relativistic regime, where the energy exchange between two bound or unbound particles is $q^0\sim \abs{\vec{q}}^2/(2\mu)\ll \abs{\vec{q}}$. Under these circumstances, we can ignore the $p^0$ and $p^{\prime 0}$ dependence of the interaction kernel $\tilde{W}(p,p';Q)$ to leading order in $\alpha$ and $\vrel$. Moreover, for all cases of interest in the following, $\tilde{W}$ will only depend on $\abs{\vec{p}-\vec{p}'}$ and neither on $\vec{p},\vec{p}'$ nor $Q$ alone.  Therefore, we will assume $\tilde{W}(p,p';Q)\simeq \mathcal{W}(\abs{\vec{p}-\vec{p}'})$ throughout this work. By doing so, we can see from the Bethe-Salpeter wave equations that the quantities $S^{-1}(p;Q)\PsiQBSt{Q}(p)$ and $S^{-1}(p;Q)\PhiQBSt{Q}{q}(p)$ do not depend on $p^0$ any longer. We will account for that by redefining 
\begin{align}
    \label{eq:SchroedingerSEconnectionBS}
    \SEwavefQBSt{Q}(\vec{p})&\equiv\sqrt{2\NnormBS{Q}(\vec{p})}\mathcal{S}_0(\vec{p};Q)S^{-1}(p;Q)\PsiQBSt{Q}(p)\,,\\
    \label{eq:SchroedingerSEconnectionSS}
    \SEwavefQSSt{Q}{q}(\vec{p})&\equiv\sqrt{\frac{2\NnormBS{Q}(\vec{p})}{2\ESSdiffq{Q}{q}}}\mathcal{S}_0(\vec{p};Q)S^{-1}(p;Q)\PhiQBSt{Q}{q}(p)\,,
\end{align}
with $\mathcal{S}_0(\vec{p};Q)\equiv\int\frac{\dd{p^0}}{2\pi}S(p;Q)$ and $\NnormBS{Q}(\vec{p})$ a normalization factor. The $\SEwavefQBSt{Q}(\vec{p})$ and $\SEwavefQSSt{Q}{q}(\vec{p})$ can then be associated with the Fourier transforms of the scattering and bound state Schrödinger wave functions as defined in \cref{subsubsec:Schroedingerequations}.\footnote{One can show that in position space, the Schrödinger wave functions are connected to the Bethe-Salpeter wave functions at equal times (\ie $x_1^0=x_2^0$, \cf\cref{eq:BetheSalpeterSS,eq:BetheSalpeterBS}). Therefore, $\SEwavefQBSt{Q}^{\star}(\vec{p})=\SEwavefQBSt{Q}^*(\vec{p})$ and analogously for $\SEwavefQSSt{Q}{q}$.} Choosing
\begin{equation}
    \label{eq:SEnormalization}
    \NnormBS{Q}(\vec{p})\equiv\frac{E_1(\vec{p};\vec{Q})E_2(\vec{p};\vec{Q})}{E_1(\vec{p};\vec{Q})+E_2(\vec{p};\vec{Q})}
\end{equation}
will recover the usual normalization conditions for the Schrödinger wave functions as given in \cref{eq:scatteringwavefunctionnorm,eq:boundstatewavefunctionnorm} (see \eref\cite{Petraki:2015hla} for further insight). Employing \cref{eq:SchroedingerSEconnectionBS,eq:SchroedingerSEconnectionSS} to the Bethe-Salpeter wave equations, given in \cref{eq:BSequationBS,eq:BSequationSS}, yields
\begin{align}
    \label{eq:BSequationInstAppBS}
    \frac{\mathcal{S}_0^{-1}(\vec{p};Q)}{\sqrt{2\NnormBS{Q}(\vec{p})}}\SEwavefQBSt{Q}(\vec{p})&=\int\frac{\dd[3]{k}}{(2\pi)^3}\frac{\mathcal{W}(\abs{\vec{p}-\vec{k}})}{\sqrt{2\NnormBS{Q}(\vec{k})}}\SEwavefQBSt{Q}(\vec{k})\quad \text{with}\quad Q^0=\EBS{Q}\,,\\
    \label{eq:BSequationInstAppSS}
    \frac{\mathcal{S}^{-1}_0(\vec{p};Q)}{\sqrt{2\NnormBS{Q}(\vec{p})}}\SEwavefQSSt{Q}{q}(\vec{p})&=\int\frac{\dd[3]{k}}{(2\pi)^3}\frac{\mathcal{W}(\abs{\vec{p}-\vec{k}})}{\sqrt{2\NnormBS{Q}(\vec{k})}}\SEwavefQSSt{Q}{q}(\vec{k})\quad \text{with}\quad Q^0=\ESS{Q}{q}.
\end{align}
Since we are already in the non-relativistic regime, we might as well continue by making further non-relativistic approximations. Assuming $\abs{\vec{P}},\abs{\vec{p}}\ll P^0,m_1,m_2$, we expand 
\begin{align}
    \label{eq:E1approx}
    E_1(\vec{p},\vec{P})&\simeq \eta_1\left(m+\frac{\vec{P}^2}{2m}\right)+\frac{\vec{P}\cdot\vec{p}}{m}+\frac{\vec{p}^2}{2m_1}\,,\\
    \label{eq:E2approx}
    E_2(\vec{p},\vec{P})&\simeq \eta_2\left(m+\frac{\vec{P}^2}{2m}\right)-\frac{\vec{P}\cdot\vec{p}}{m}+\frac{\vec{p}^2}{2m_1}\,,\\
    \label{eq:Esumapprox}
    E_1(\vec{p},\vec{P})+E_2(\vec{p},\vec{P})&\simeq m+\frac{\vec{P}^2}{2m}+\frac{\vec{p}^2}{2\mu}\,,
\end{align}
up to second order in the momenta. This enables us to also approximate (\cf\app\,C of \eref\cite{Petraki:2015hla} for the derivation)
\begin{align}
    \mathcal{S}_0^{-1}(\vec{p};Q)&=-\frac{i2E_1(\vec{p},\vec{Q})E_2(\vec{p},\vec{Q})\left[(Q^0)^2-(E_1(\vec{p},\vec{Q})+E_2(\vec{p},\vec{Q}))^2\right]}{E_1(\vec{p},\vec{Q})+E_2(\vec{p},\vec{Q})}\\
    &\simeq -i4m\mu\left(Q^0-m-\frac{\vec{P}^2}{2m}-\frac{\vec{p}^2}{2\mu}\right) = -i4m\mu\left(\mathcal{E}-\frac{\vec{p}^2}{2\mu}\right)
\end{align}
to the same order, where we conveniently defined 
\begin{equation}
    \label{eq:totalenergydefinition}
    Q^0\equiv m+\frac{\vec{Q}^2}{2m}+\mathcal{E}\,,
\end{equation}
with $\mathcal{E}\in\{\mathcal{E}_{n},\mathcal{E}_{\vec{q}}\}$ denoting the \gls{bs} and scattering state energies for $Q^0=\EBS{Q}$ and $Q^0=\ESS{Q}{q}$, respectively. The normalization coefficients cancel on both sides if expanded to leading order in $\vec{p}$ and $\vec{k}$, such that we can now write \cref{eq:BSequationInstAppBS,eq:BSequationInstAppSS} as 
\begin{align}
    \label{eq:SchroedingerequationBSmomentumspace}
    \left(-\frac{\vec{p}^2}{2\mu}+\En\right)\SEwavefBSt(\vec{p})=-\frac{1}{i4m\mu}\int\frac{\dd[3]{k}}{(2\pi)^3}\mathcal{W}(\abs{\vec{p}-\vec{k}})\SEwavefBSt(\vec{k})\,,\\
    \label{eq:SchroedingerequationSSmomentumspace}
    \left(-\frac{\vec{p}^2}{2\mu}+\mathcal{E}_{\vec{q}}\right)\SEwavefSSt{q}(\vec{p})=-\frac{1}{i4m\mu}\int\frac{\dd[3]{k}}{(2\pi)^3}\mathcal{W}(\abs{\vec{p}-\vec{k}})\SEwavefSSt{q}(\vec{k})\,.
\end{align}
Note that we have dropped the index $\vec{Q}$ in the labelling of the wave functions since we have eliminated its explicit dependence in the equations. Fourier transforming the quantities above, we arrive at the usual Schrödinger equations as defined in \cref{eq:scatteringSchroedinger,eq:boundstateSchroedinger} with 
\begin{equation}
    \label{eq:potentialfrominteractionkernel}
    V(\vec{r})\equiv-\frac{1}{i4m\mu}\int\frac{\dd[3]{p}}{(2\pi)^3}\mathcal{W}(\abs{\vec{p}})e^{i\vec{p}\cdot\vec{r}}
\end{equation}
being the non-relativistic potential, which we can now derive from the interaction kernel of the theory. For the models under consideration, this has been done in \cref{app:potential_for_different_mediators}.\newp
For the case that $\chi_{1,2}$ are fermions, the derivation of the Bethe-Salpeter equations (and its solutions) becomes much more involved due to the spin structure of the particles. A full treatment, which can be reviewed \eg in \eref\cite{Carbonell:2010zw}, goes beyond the scope of this work. However, as we are only interested in a leading order approximation in the involved momenta, it suffices in the following to use the equations for scalar particles with only a few modifications. First, we have to account for the difference in mass dimensionality of the fermions between relativistic spinors with mass dimension $3/2$ and non-relativistic fields, which are agnostic to the spin properties and have a mass dimension of $1$. Thus, we introduce a mapping factor for the Schrödinger wave functions $\SEwavefBSt^f(\vec{p})\to \sqrt{2m_1}\sqrt{2m_2}\SEwavefBSt(\vec{p})$ and analogously for $\SEwavefSSt{q}^f(\vec{p})$, which can easily be extracted from the normalization of a single spinor in the Dirac representation $\sqrt{E_i(\vec{p};\vec{Q})+m_i}\simeq \sqrt{2m_i}$. This additional factor has no direct impact on the Schrödinger equations derived from \cref{eq:SchroedingerequationBSmomentumspace,eq:SchroedingerequationSSmomentumspace} due to cancellation on both sides but it will become important in the next section. Moreover, a fermionic propagator in the non-relativistic regime differs by a factor $2m_i$ from a scalar propagator, such that we further take $\tilde{S}^f_i(p_i)\to 2m_i\tilde{S}_i(p_i)$ in this limit. The additional factor $S^f(p;Q)\to 4m\mu S(p;Q)$ from the Bethe-Salpeter wave equations then cancels the corresponding factor in the definition of the non-relativistic potential $V(\vec{r})$ (\cf\cref{eq:potentialfrominteractionkernel}) for a fermionic particle antiparticle pair.


\subsubsection{The bound state formation cross section}
\label{subsubsec:BSFfromBetheSalpeter}
We are now ready to calculate the cross section for the formation of a \gls{bs} as sketched in \cref{fig:SE_BSF_Feynman} (right). Using the framework we employed in the previous section, to leading order in the coupling $\alpha$ this process can be sufficiently described by 
\begin{equation}
    \chi_1(k_1)+\chi_2(k_2) \to \mathscr{B}(\chi_1(p_1)\chi_2(p_2))+\varphi(P_\varphi),
\end{equation}
where $\mathscr{B}(\chi_1\chi_2)$ denotes a bound state with quantum numbers $n\equiv\{nlm\}$. The S-matrix element we want to calculate is, thus, 
\begin{equation}
	\label{eq:SmatrixElementBSF}
	\prescript{}{\text{out}}{\left\langle\mathcal{B}_{\vec{P},n};\varphi_{\vec{P}_\varphi}\middle|\mathcal{U}_{\vec{K},{\vec{k}}}\right\rangle_{\text{in}}}=\left\langle\mathcal{B}_{\vec{P},n};\varphi_{\vec{P}_\varphi}\middle|\mathsf{S}\middle|\mathcal{U}_{\vec{K},{\vec{k}}}\right\rangle\,,
\end{equation}
with $\vec{K}, \vec{P}$ the total and $\vec{k},\vec{p}$ the relative 3-momenta of the scattering and \gls{bs} defined in analogy to \cref{subsubsec:BetheSalpeter}. The state of the mediator which gets radiated off during a \gls{bsf} process has 4-momentum $\Pphi=K-P$ and is denoted in the following by $\phiPphi$.

\paragraph{The 5-point Green's function}$~$\newpp
\begin{figure}
    \centering
    \includegraphics[width=\textwidth]{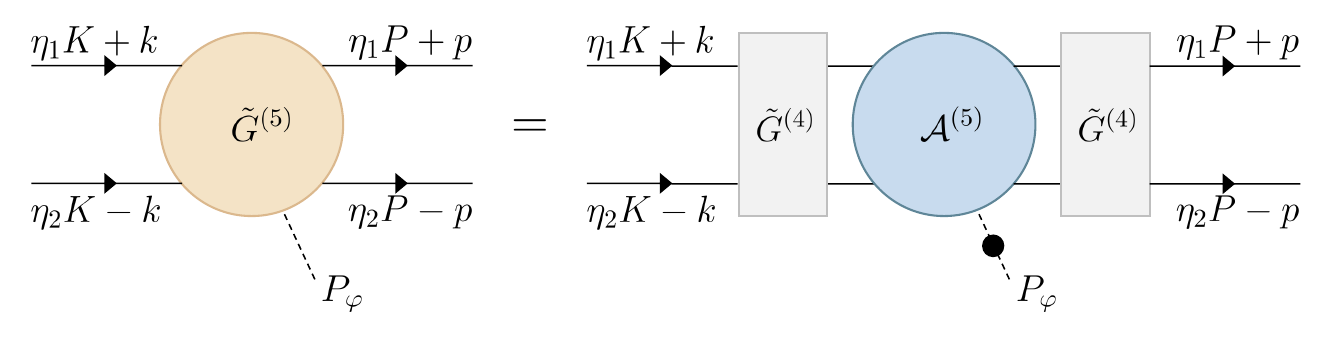}
    \caption[Decomposition of the 5-point Green's function.]{Decomposition of the 5-point Green's function $\Gfivet$ of a $\bar{\chi}\chi$ pair and a bosonic mediator $\varphi$ according to \cref{eq:5Greensdecomposition}. The black filled dot refers to the full propagator of the bosonic mediator $\varphi$, whereas the gray boxes on the right denote the 4-point Green's function $\Gfourt$. The momenta of the incoming (outgoing) particles are also displayed in terms of total and relative momenta of the system. A definition of $\Afive$ is given in the text. Adapted from \eref\cite{Petraki:2015hla}.}
    \label{fig:Greensfunctiondecomposition}
\end{figure}
As a first step to determine the S-matrix element for \gls{bsf} we consider the 5-point Green's function, which is given by 
\begin{equation}
	\Gfive(x_1,x_2;y_1,y_2,X_\varphi)\equiv\Vaccc T\varphi(X_\varphi)\chi_1(y_1)\chi_2(y_2)\chi_1^\dagger(x_1)\chi_2^\dagger(x_2)\Vac.
\end{equation}
Due to its translational invariance, we can switch to \gls{com} coordinates $X,Y$ and $x,y$, which are equivalent to the ones used for the 4-point Green's function (\cf\cref{subsubsec:BetheSalpeter}), and define the Fourier transform as
\begin{align}
    \Gfivet(\eta_1K+k,&\eta_2K-k;\eta_1P+p,\eta_2P-p,\Pphi)\equiv\int \dd[4]{X_\varphi}\dd[4]{X}\dd[4]{Y}\dd[4]{x}\dd[4]{y}\nonumber\\
	&e^{i(KX+kx-PY-py+\Pphi X_\varphi)}\Gfive(X+\eta_2x,X-\eta_1x;Y+\eta_2y,Y-\eta_1y,X_\varphi)\,,
\end{align}
where the conjugate momenta of $X,x$ and $Y,y$ are $K,k$ and $P,p$, respectively. As pictured in \cref{fig:Greensfunctiondecomposition}, the 5-point Green's function in momentum space can be decomposed into 
\begin{align}
    \label{eq:5Greensdecomposition}
    \Gfivet(&\eta_1K+k,\eta_2K-k;\eta_1P+p,\eta_2P-p,\Pphi)=\tilde{S}_\varphi(\Pphi)\int\frac{\dd[4]{p'}}{(2\pi)^4}\frac{\dd[4]{k'}}{(2\pi)^4} \Gfourt(k',k;K)\nonumber \\
    \Gfourt&(p,p';P)(2\pi)^4\delta^{(4)}(K-P-\Pphi) i\Afive(\eta_1K+k',\eta_2K-k';\eta_1P+p',\eta_2P-p',\Pphi)\,,
\end{align}
where $\Afive$ can be defined via $\Cfive$, the sum of all connected diagrams\footnote{We want to emphasize here that $\Cfive$ does also contain diagrams which are not fully connected. Otherwise, $\Afive$ would just be the sum of all fully connected and amputated diagrams.} 
\begin{equation}
    i\Cfive(k_1,k_2;p_1,p_2,\Pphi)\equiv\tilde{S}_\varphi(\Pphi)\tilde{S}_1(k_1)\tilde{S}_2(k_2)\tilde{S}_1(p_1)\tilde{S}_2(p_2)i\Afive(k_1,k_2;p_1,p_2,\Pphi)\,,
\end{equation}
and $\tilde{S}_\varphi(\Pphi)$ denotes the propagator of a scalar $\varphi$, which is given by 
\begin{equation}
	\label{eq:fullphipropagator}
	\tilde{S}_{\varphi}(\Pphi)\equiv\frac{i Z_\varphi(\vec{P}_\varphi)}{\Pphi^2-m_\varphi^2+i\varepsilon}\,,\quad \text{with}\quad Z_\varphi(\vec{P}_\varphi)=\abs{\left\langle\Omega\middle|\varphi(0)\middle|\varphi_{\vec{P}_\varphi}\right\rangle}^2
\end{equation}
its field strength renormalization. For later convenience, we also define $\Cfiveamp{\varphi}$, the sum of all connected and $\varphi$-amputated diagrams, via
\begin{equation}
	i\Cfive(k_1,k_2;p_1,p_2,\Pphi)\equiv\tilde{S}_\varphi(\Pphi)i\Cfiveamp{\varphi}(k_1,k_2;p_1,p_2,\Pphi)\,,
\end{equation}
such that we can cast $\Afive$ in a more convenient form
\begin{align}
	i\Afive(\eta_1K+k,&\eta_2K-k;\eta_1P+p,\eta_2P-p,\Pphi)=\nonumber\\
	&S^{-1}(k;K)S^{-1}(p;P)i\Cfiveamp{\varphi}(\eta_1K+k,\eta_2K-k;\eta_1P+p,\eta_2P-p,\Pphi)
\end{align}
with $S(p;P)$ defined as in \cref{eq:combinedpropagator}. 

\paragraph{The \gls{bsf} matrix element}$~$\newpp
We now want to link the 5-point Green's function to the S-matrix element. This can be achieved by employing the \gls{lsz} reduction formula (see \eg\eref \cite{Peskin:1995ev}). Close to on-shell, \ie $\Pphi^0\to \omega_\varphi(\vec{P}_\varphi)$, $P^0\to\EBS{P}$, $K^0\to \ESS{K}{k}$, the \gls{lsz} formula is given by
\begin{align}
	\int \dd[4]{X_\varphi}e^{i\Pphi X_\varphi}&\!\int\dd[4]{X}e^{iKX}\!\int\dd[4]{Y}e^{-iPY}\Gfive(X+\eta_2x,X-\eta_1x;Y+\eta_2y,Y-\eta_1y,X_\varphi)\nonumber\\
	\sim&\Bigg[\frac{i\left\langle\Omega\middle|\varphi(0)\middle|\varphi_{\vec{P}_\varphi}\right\rangle}{2\Ephi\left(\Pphi^0-\Ephi+i\varepsilon\right)}\Bigg]\Bigg[\frac{i\Vaccc T\chi_1(\eta_2x)\chi_2(-\eta_1x)\BSQ{P}}{2\EBS{P}\left(P^0-\EBS{P}+i\varepsilon\right)}\Bigg]\nonumber\\
	&\int\frac{\dd[3]{k'}}{(2\pi)^32\ESSdiffq{K}{k'}}\frac{i\SSQcc{K}{k'}T\chi^\dagger_1(\eta_2y)\chi^\dagger_2(-\eta_1y)\Vac}{2\ESS{K}{k'}\left(K^0-\ESS{K}{k'}+i\varepsilon\right)}\left\langle\mathcal{B}_{\vec{P},n};\varphi_{\vec{P}_\varphi}\middle|\mathsf{S}\middle|\mathcal{U}_{\vec{K},{\vec{k'}}}\right\rangle\,,
\end{align}
where $\sim$ means that both sides share the same singularities in this limit. In order to equate them, we need to extract the residues of the singularities from this relation. We start by Fourier transforming both sides with respect to $x$ and $y$. This yields on the \gls{lhs} just the definition of $\Gfivet$. On the \gls{rhs} it enables us to exchange the two-point correlator functions with the Bethe-Salpeter wave functions in momentum space (\cf\cref{eq:BetheSalpeterBS,eq:BetheSalpeterSSstar,eq:BSwavefunctionsFT}). Hence, we arrive at 
\begin{align}
    \label{eq:LSZcompare1}
	\Gfivet(\eta_1K+q,\eta_2K-q;&\eta_1P+p,\eta_2P-p,\Pphi)\nonumber\\
    \sim&\Bigg[\frac{i\sqrt{Z_\varphi(\vec{P}_\varphi)}}{{2\Ephi\left(\Pphi^0-\Ephi+i\varepsilon\right)}}\Bigg]
    \Bigg[\frac{i\PsiQBSt{P}(p)}{2\EBS{P}\left(P^0-\EBS{P}+i\varepsilon\right)}\Bigg]\nonumber\\
    &\int\frac{\dd[3]{k'}}{(2\pi)^3 2\ESSdiffq{K}{k'}}\frac{i\PhiQBSt{K}{k'}^{\star}(q)}{2\ESS{K}{k'}\left(K^0-\ESS{K}{k'}+i\varepsilon\right)}\left\langle\mathcal{B}_{\vec{P},n};\varphi_{\vec{P}_\varphi}\middle|\mathsf{S}\middle|\mathcal{U}_{\vec{K},{\vec{k'}}}\right\rangle\,,
\end{align}
which we can compare with \cref{eq:5Greensdecomposition} in the on-shell limit (\cf\cref{eq:Greens4FTBSpole,eq:Greens4FTSSpole})
\begin{align}
    \Gfivet(&\eta_1K+q,\eta_2K-q;\eta_1P+p,\eta_2P-p,\Pphi)\sim \frac{i Z_\varphi(\vec{P}_\varphi)}{\Pphi^2-m_\varphi^2+i\varepsilon}\nonumber\\
    &\int\frac{\dd[4]{p'}}{(2\pi)^4}\frac{\dd[4]{q'}}{(2\pi)^4} \frac{i\PsiQBSt{P}(p)\PsiQBSt{P}^{\star}(p')}{2\EBS{P}\left(P^0-\EBS{P}+i\epsilon\right)}\int\frac{\dd[3]{k'}}{(2\pi)^32\ESS{K}{k'}}\frac{i\PhiQBSt{K}{k'}(q')\PhiQBSt{K}{k'}^{\star}(q)}{2\ESSdiffq{K}{k'}\left(K^0-\ESS{K}{k'}+i\epsilon\right)}\nonumber \\
    &(2\pi)^4\delta^{(4)}(K-P-\Pphi) i\Afive(\eta_1K+q',\eta_2K-q';\eta_1P+p',\eta_2P-p',\Pphi)\,.
\end{align}
We can extract the residues from $\Pphi^0\to \omega_\varphi(\vec{P}_\varphi)$ and $P^0\to\EBS{P}$ right away. To determine the leading singularity in $K^0\to \ESS{K}{k}$ a few more steps are in order. First, we multiply both sides by $\tilde{N}_{\vec{k}}(q,q'';\vec{K})\PhiQBSt{K}{k}(q'')$ with $\tilde{N}_{\vec{k}}$ being the normalization of $\PhiQBSt{K}{k}$. In a second step, we integrate over $q$ and $q''$ and use the normalization condition (\cf\eref\cite{Petraki:2015hla})
\begin{equation}
	\int \frac{\dd[4]{q}}{(2\pi)^4}\frac{\dd[4]{q''}}{(2\pi)^4}\PhiQBSt{K}{k'}^{\star}(q)\tilde{N}_{\vec{k}'}(q,q'';K)\PhiQBSt{K}{k}(q'')=2\ESS{K}{k}2\ESSdiffq{K}{k}(2\pi)^3\delta^{(3)}(\vec{k}-\vec{k}')
\end{equation}
to match the singularity. Relabelling $p'\to p$, $q'\to q$ we can eventually equate
\begin{align}
    \left\langle\mathcal{B}_{\vec{P},n};\varphi_{\vec{P}_\varphi}\middle|\mathsf{S}\middle|\mathcal{U}_{\vec{K},{\vec{k'}}}\right\rangle=&\sqrt{Z_\varphi(\vec{P}_\varphi)}\int\frac{\dd[4]{p}}{(2\pi)^4}\frac{\dd[4]{q}}{(2\pi)^4}\PsiQBSt{P}^{\star}(p)\PhiQBSt{K}{k}(q) (2\pi)^4\delta^{(4)}(K-P-\Pphi)\nonumber \\ &i\Afive(\eta_1K+q,\eta_2K-q;\eta_1P+p,\eta_2P-p,\Pphi). 
\end{align}
Conventionally defining $\left\langle\mathcal{B}_{\vec{P},n};\varphi_{\vec{P}_\varphi}\middle|\mathsf{S}\middle|\mathcal{U}_{\vec{K},{\vec{k'}}}\right\rangle\equiv(2\pi)^4\delta^{(4)}(K-P-\Pphi)\Mkn$ and switching to $\Cfiveamp{\varphi}$, the matrix element of the \gls{bsf} cross section is given by
\begin{align}
    \label{eq:MktongeneralBS}
	\Mkn = \sqrt{Z_\varphi(\vec{P}_\varphi)}\int\frac{\dd[4]{p}}{(2\pi)^4}\frac{\dd[4]{q}}{(2\pi)^4}S^{-1}(p,P)\PsiQBSt{P}^{\star}(p)S^{-1}(q,K)\PhiQBSt{K}{k}(q)\Cfiveamp{\varphi}
\end{align}
for a scalar mediator. For a vector mediator, one would redefine $\Mkn=\varepsilon_\mu\Mkn^\mu$ and $\Cfiveamp{\varphi}=\varepsilon_\mu(\Cfiveamp{\varphi})^\mu$ with $\varepsilon_\mu$ the polarization vector of $\varphi$.

\paragraph{The \gls{bsf} cross section in the non-relativistic limit}$~$\newpp
Processes which enable the formation of \gls{bs} happen deep in the non-relativistic regime with momentum exchanges typically of the order of the binding energy $\En \propto \mu \alpha^2$. Thus, it is justified to employ the instantaneous approximation in order to rewrite \cref{eq:MktongeneralBS} in terms of the Schrödinger wave functions (\cf\cref{eq:SchroedingerSEconnectionBS,eq:SchroedingerSEconnectionSS})
\begin{equation}
	\label{eq:MkngeneralSE}
	\Mkn\simeq\sqrt{2\ESSdiffq{K}{k}}\int\frac{\dd[3]{p}}{(2\pi)^3}\frac{\dd[3]{q}}{(2\pi)^3}\frac{\SEwavefQBSt{P}^*(\vec{p})\SEwavefQSSt{K}{k}(\vec{q})}{\sqrt{2\mathcal{N}_{\vec{P}}(\vec{p})2\mathcal{N}_{\vec{K}}(\vec{q})}}\Mtrans(\vec{q},\vec{p})\,,
\end{equation}
where we assumed $Z_\varphi(\vec{P}_\varphi)\simeq 1$ to leading order in $\alpha$.
The transition matrix element on the \gls{rhs} of the equation has been defined as
\begin{align}
	\label{eq:Mtrans}
	\Mtrans(\vec{p},\vec{q})\equiv &~\mathcal{S}^{-1}_0(\vec{q},K)\mathcal{S}^{-1}_0(\vec{p},P)\nonumber\\
    &\int\frac{\dd{p^0}}{2\pi}\frac{\dd{q^0}}{2\pi}\Cfiveamp{\varphi}(\eta_1K+q,\eta_2K-q;\eta_1P+p,\eta_2P-p,\Pphi).
\end{align}
Making use of \cref{eq:SEnormalization,eq:E1approx,eq:E2approx}, we can approximate 
\begin{equation}
\label{eq:normalizationapprox}
	\frac{1}{\sqrt{2\mathcal{N}_{\vec{P}}(\vec{p})2\mathcal{N}_{\vec{K}}(\vec{q})}}\simeq \frac{1}{2\mu}\left(1-\frac{\vec{p}^2+\vec{q}^2}{4\mu^2}\left(1-\frac{3\mu}{m}\right)\right)
\end{equation}
to first order in $\vec{p}^2, \vec{q}^2$ which will be proportional to $\alpha^2$ and $\vrel^2$, respectively. Consequently, we also need to expand the Schrödinger wave functions as well as $\Mtrans(\vec{p},\vec{q})$ to the same order when computing the matrix element. Corrections from the total initial and final state momenta $\vec{K}$ and $\vec{P}$ are of $\order{\alpha^2+\vrel^2}$ and can thus be neglected (\ie we will only expand them to leading order). Further, we take $\ESSdiffq{K}{k}\simeq \mu$ to leading order in $\vec{k}^2\propto \vrel^2$ for the same reasons. Dropping all indices of the total momenta we arrive at 
\begin{equation}
	\label{eq:MknSEapprox}
	\Mkn\simeq\frac{1}{\sqrt{2\mu}}\int\frac{\dd[3]{p}}{(2\pi)^3}\frac{\dd[3]{q}}{(2\pi)^3}\left(1-\frac{\vec{p}^2+\vec{q}^2}{4\mu^2}\left(1-\frac{3\mu}{m}\right)\right)\SEwavefBSt^*(\vec{p})\SEwavefSSt{k}(\vec{q})\Mtrans(\vec{q},\vec{p})
\end{equation}
for the \gls{bsf} matrix element, where the concrete dependence on the particle physics model is hidden in $\Mtrans$.\newp
The differential \gls{bsf} cross section into the $n$-th \gls{bs} in the \gls{com} frame (\ie $\vec{K}=0$, $\vec{P}=-\vec{P}_\varphi$) is given by 
\begin{equation}
    \dv{\sigmabsf^{(n)}}{\Omega}=\frac{1}{2\sqrt{\lambda(s,m_1^2,m_2^2)}}\frac{\abs{\vec{P}}}{16\pi^2\sqrt{s}}\overline{\abs{\Mkn}^2}\,,
\end{equation}
in accordance with \cref{eq:generalcrosssection}. Using $s=\omega_{\vec{K}=0,\vec{k}}^2=(m+\Ek)^2$ (see \cref{eq:totalenergydefinition}) as well as $M_n=m+\En$ for the bound state mass, to leading order in $\alpha^2$, $\vrel^2$ we obtain
\begin{equation}
    \abs{\vec{P}}=\sqrt{\frac{\lambda(s,M_n^2,\mvphi^2)}{4s}}\simeq (\Ek-\En)\left(1-\frac{\mvphi^2}{(\Ek-\En)^2}\right)^{1/2}\,,
\end{equation}
where we also assumed $\mvphi\ll m$. Further, $2\sqrt{\lambda(s,m_1^2,m_2^2)}\simeq 4mk=4m\mu\vrel$ to leading order in $k$, such that the \gls{bsf} cross section can be written as 
\begin{equation}
    \label{eq:generalBSFcrosssection}
    \sigmabsf^{(n)}\vrel = \frac{(\Ek-\En)}{64\pi^2m^2\mu}\left(1-\frac{\mvphi^2}{(\Ek-\En)^2}\right)^{1/2} \int\dd{\Omega}\overline{\abs{\Mkn}^2}.
\end{equation}\newp
A full derivation of the matrix element for the \gls{bsf} cross section considering a fermionic particle antiparticle system is beyond the scope of this work. However, at leading order in the corresponding momenta we can still use \cref{eq:MknSEapprox} together with the appropriate non-relativistic modifications for fermionic entities (\cf\cref{subsubsec:BetheSalpeter}) to obtain the correct result.


\subsection{Non-perturbative effects from EFT}
\label{subsec:nonperturbativeeffectsfromEFT}

A complementary approach exists for describing \gls{se} and the presence of \gls{bs} within a theoretical model. While this method has long been established for computing the effects of non-perturbative \gls{sm} processes \cite{Luke:1997ys,Bodwin:1994jh,Caswell:1985ui,Pineda:1997bj,Pineda:1998kn,Brambilla:2002nu,Brambilla:2004jw}, its application to the context of \gls{dm} is relatively recent \cite{Biondini:2018ovz,Biondini:2021ccr,Biondini:2021ycj,Biondini:2023zcz,Binder:2023ckj}. This approach leverages the framework of \gls{eft}, which naturally facilitates the separation of scales required to compute these phenomena. Analogous to the Bethe-Salpeter approach outlined in \cref{subsec:nonperturbativeeffectsfromQFT}, this method heavily depends on the symmetries and particle content of the \gls{uv} Lagrangian to which the \gls{eft} is matched.\footnote{In the following, we refer to the theory applicable at the highest energy scales considered as the \quotes{\gls{uv} Lagrangian} or \quotes{full Lagrangian}. This theory may itself be an \gls{eft}, but it must be power-counting renormalizable, \ie all operators must have mass dimension $d\leq 4$.} The computations required to determine the necessary observables are generally more extensive than in the \gls{uv} theory, as the \gls{eft} Lagrangian incorporates a greater number of interaction terms that must be considered. However, once computed, many of the obtained results can be applied to different contexts, as \glspl{eft} at lower energies are often more analogous to each other than their \gls{uv} counterparts. Additionally, one enjoys better control over the size of corrections at higher orders due to the clear hierarchy of scales, making it more straightforward to include thermal corrections (or justify their neglect). For these reasons, we will employ this method to ascertain the leading-order processes for the model presented in \cref{sec:indirectdetection}, which features a much more intricate structure than the one discussed in \cref{sec:impactonnonthermalDMproduction}, for which the approach outlined in the previous section suffices. The calculations of the non-perturbative observables for the model in \cref{sec:indirectdetection} will also be conducted within the \gls{eft} framework and later cross-checked using the Bethe-Salpeter approach discussed earlier.\newp
Within systems which experience \gls{se} and the presence of \gls{bs}, there are usually three important energy scales to consider. The first one is the \textit{hard} scale, which is set to the mass of the interacting (anti)particle $\mchi$. The second one, labelled as the \textit{soft} scale, is proportional to the relative momentum exchange $p\propto \mchi\vrel$ of the interacting $\bar{\chi}\chi$ pair. As non-perturbative effects become important in the regime where $\vrel \sim \alpha$, the soft scale is commonly set to $\mchi\alpha$. The third scale is denoted as \textit{ultra-soft} and lives at the binding energies of the \gls{bs} or the kinetic energy of the scattering state, respectively. It is thus conveniently set to $\mchi\alpha^2$. These three energy scales are hierarchically ordered as follows: $\mchi\gg\mchi\alpha\gg\mchi\alpha^2$. In principle, more energy scales can be relevant, like the mass of the mediator particle $\mvphi$ or thermal scales which can arise from interactions of the particles in question with a surrounding plasma. The construction of the \gls{eft} for the problem at hand highly depends on where these additional scales are situated with respect to the hard, soft and ultra-soft scales. We will assume in the following that the mass of the mediator particle is at most as large as the soft scale and treat the temperature scale as if it always lives below the ultra-soft scale.\footnote{The latter statement is not true in general when describing an evolving Universe, as the temperature runs over all scales of the \glspl{eft}. A justification for this choice is given when applying the approach to the model in \cref{subsec:dmmodelid}.}\newp
The method to construct the desired \gls{eft} is a two-step procedure \cite{Caswell:1985ui,Pineda:1997bj,Pineda:1998kn}. Starting from the full Lagrangian, first all energy and momentum modes of the order of the hard scale are integrated out, resulting in a so-called \gls{nreft}. Within this theory, the soft and ultra-soft modes are still intertwined, which is why in a second step, the soft modes are integrated out. Apart from the soft momentum exchanges between the non-relativistic particle pairs, this also includes the mediator mass from our previous assumptions. This procedure yields the \gls{pnreft}, which we will use to calculate the relevant cross sections and decay widths. We will highlight in the following how such a \gls{pnreft} is constructed using the \gls{uv} Lagrangian of \cref{sec:indirectdetection} as a concrete example. The derivation of the corresponding \glspl{eft} can be found in \erefs\cite{Biondini:2021ycj,Luke:1997ys,Biondini:2021ccr,Luke:1996hj,Biondini:2023zcz} in greater detail. 


\subsubsection{Non-relativistic EFT}
\label{subsubsec:NREFT}

We will consider in the following a $\bar{\chi}\chi$ (Dirac) fermion pair which has scalar and pseudo-scalar interactions with a massive real scalar mediator $\phi$. The corresponding couplings are denoted as $g$ and $g_5$ where we assume that $g\gg g_5$ (see \cref{eq:LmodID} for the full Lagrangian). This assumption ensures that $\alpha\equiv g^2/(4\pi)$ sets the dynamical scales of the \glspl{eft}, while $\alpha_5\equiv g_5^2/(4\pi)$ remains subleading. After integrating out the hard scale modes, we are left with the \gls{nreft} labelled as $\nry$, which contains only non-relativistic fermions and antifermions apart from the scalars which have energy and momenta below $\mchi$. The idea behind the construction of the $\nry$ Lagrangian (as for any \gls{nreft}) is to use Pauli spinor fields $\varsigma,\eta$ representing the non-relativistic fermionic fields, where $\varsigma$ annihilates a fermion and $\eta^\dagger$ an antifermion, whereas the scalar field $\phi$ is retained from the \gls{uv} Lagrangian. The effective Lagrangian is then given as an expansion in terms of the inverse of the hard scale $1/\mchi$ reading
\begin{equation}
    \mathcal{L}_{\nry} \equiv \sum_{i,n} c_i\frac{\mathcal{O}^{(d)}_i}{\mchi^n}.
\end{equation}
The effective operators $\mathcal{O}^{(d)}_i$ with mass dimension $d=4+n$ can be constructed from the present fields $\varsigma,\eta$ and $\phi$ along with time and space differentials $\partial_0$, $\vec{\nabla}$ as well as the three vector of Pauli matrices $\vec{\sigma}$ to account for spin flipping interactions. Viable operators have to respect certain properties of the effective Lagrangian, which are in our case only $C$ and $CPT$ invariance, rotational as well as translational invariance and hermiticity.\footnote{Depending on the underlying full Lagrangian, also other symmetries have to be respected such as gauge invariance or invariance with respect to other discrete symmetries like $P$ and $T$.} Using field redefinitions and other \gls{eft} methods, at each order in $1/\mchi$ we are able to reduce the amount of effective operators to a minimal and unique number (whereas the operators itself are, in general, not unique). The matching coefficients $c_i$ can then be determined partly using the Lorentz invariance of the full theory and through explicit matching calculations. For examples on how to construct an \gls{nreft}, the reader is directed to earlier works on the non-relativistic effective \gls{sm} theories NRQED and NRQCD \cite{Bodwin:1994jh,Kinoshita:1995mt,Manohar:1997qy,Paz:2015uga}.\newp
The contributions to the $\nry$ Lagrangian which will become important in the following can be split up into four sectors, reading
\begin{align}
    \label{eq:NRY_scheme}
    \mathcal{L}_{\nry} \supset \mathcal{L}_{\text{bilinear}}^{\varsigma} + \mathcal{L}_{\text{bilinear}}^{\eta} + \mathcal{L}_{\text{4-fermions}} + \mathcal{L}_{\text{scalar}} \,.
\end{align}
The scalar sector of the effective Lagrangian remains formally unaltered when compared to the full one, since it comprises only of mediator interactions which are at scales lower than the hard scale.\footnote{We will ignore in the following the scalar Higgs portal couplings present in \cref{eq:LmodID}, as they are irrelevant for the subsequent discussion due to their small values.} The interaction vertices between the (anti)fermion and the mediator are contained in the bilinear Lagrangians. With the matching coefficients set to their tree level values, the bilinear parts up to $\order{1/\mchi}$ yield \cite{Biondini:2021ycj}
\begin{align}
    \label{eq:bilinear_varsigma}
    \mathcal{L}^{\text{bilinear}}_{\varsigma} =\varsigma^\dagger \left( i \partial_0  - \,  g\phi +  g_5 \frac{\vec{\sigma} \cdot [\vec{\nabla} \phi]}{2\mchi} - g_5^2 \frac{\phi^2}{2\mchi} + \frac{\vec{\nabla}^2}{2\mchi} \right) \varsigma \,,\\
    \label{eq:bilinear_eta}
    \mathcal{L}^{\text{bilinear}}_{\eta} = \eta^\dagger \left( i \partial_0  + \,  g\phi +  g_5 \frac{\vec{\sigma} \cdot [\vec{\nabla} \phi]}{2\mchi} + g_5^2 \frac{\phi^2}{2\mchi}  - \frac{\vec{\nabla}^2}{2\mchi} \right) \eta \,,
\end{align}
(see \app\,A of \eref\cite{Biondini:2021ccr} for different construction methods) where $[\vec{\nabla} \phi]$ indicates that the derivative acts solely on the scalar field. We can see that the leading order (and parity violating) pseudo-scalar interaction is accompanied by a factor $p/\mchi$, which for the present soft and ultra-soft momenta leads to a suppression of at least $\alpha$ when compared to the purely scalar interactions.\newp
\begin{figure}
    \centering
    \includegraphics[width=\textwidth]{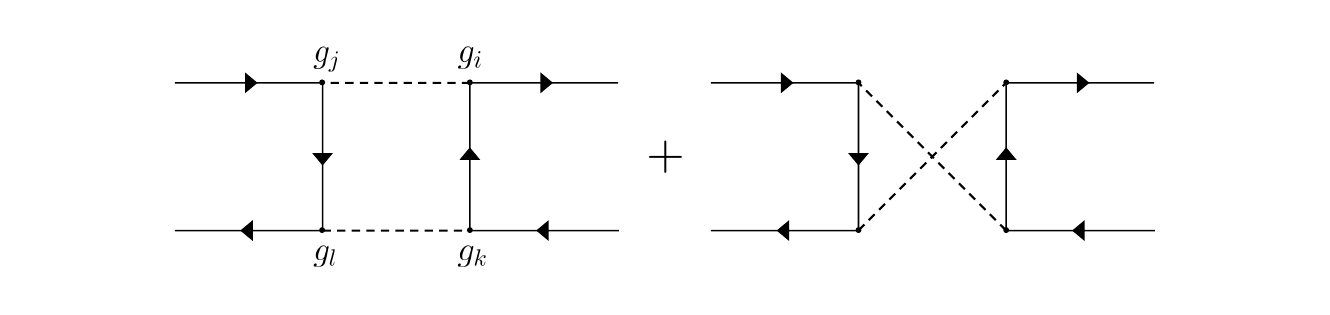}
    \caption[The diagrams contributing to the imaginary part of the 4-point Green's function for EFT matching.]{The diagrams contributing to the imaginary part of the 4-point Green's function, which can be matched to the $\nry$. The couplings $g_{i,j,k,l}$ between the \gls{dm} fermion and the scalar mediator can either be $g$ or $g_5$.}
    \label{fig:boxdiagrams}
\end{figure}
The 4-fermion Lagrangian includes annihilations of fermion antifermion pairs into light mediators \cite{Bodwin:1994jh}. These processes have been integrated out in the \gls{nreft} because they happen at a scale $2\mchi$. Employing the optical theorem, the pair annihilation process $\bar{\chi}+\chi\to 2\phi$ corresponds to the leading order term in the imaginary part of the 4-point Green's function, which is used for the matching procedure (see \eref\cite{Biondini:2023zcz}). The latter arises from the box diagrams of the \gls{uv} Lagrangian displayed in \cref{fig:boxdiagrams}. It will be most convenient in the following to adopt the notation from NRQED and NRQCD \cite{Caswell:1985ui,Bodwin:1994jh} to classify the different 4-fermion operators. The labelling of the operators and matching coefficients is therefore spectroscopic, \ie they inherit the structure $\spec{2S+1}{L}{J}$ with $S$ the total spin of the annihilating pair and $L,J$ their relative and total angular momentum. This will turn out to be particularly useful when assigning the \gls{se} factors. At dimension $6$, there are just two independent effective operators \cite{Bodwin:1994jh}
\begin{equation}
    \label{eq:dimension_6_operators}
    \mathcal{O}(\spec{1}{S}{0})\equiv\varsigma^\dagger \eta \, \eta^\dagger \varsigma \, , \quad \mathcal{O}(\spec{3}{S}{1})\equiv  \varsigma^\dagger \, \vec{\sigma} \, \eta \cdot \eta^\dagger \, \vec{\sigma} \, \varsigma \,,
\end{equation}
which contribute to the 4-fermion Lagrangian 
\begin{equation}
    \label{eq:dimension_6_lag}
    (\mathcal{L}_{\text{4-fermions}} )_{d=6} = \frac{f(\spec{1}{S}{0})}{\mchi^2} \, \mathcal{O}(\spec{1}{S}{0}) + \frac{f(\spec{3}{S}{1})}{\mchi^2} \, \mathcal{O}(\spec{3}{S}{1}) \,,
\end{equation}
with $f(\spec{1}{S}{0})$ and $f(\spec{3}{S}{1})$ the spin singlet and triplet matching coefficients. As apparent from the operator structure, they take part in the velocity independent s-wave contributions of the annihilation cross section. At dimension $8$, one finds $7$ independent effective operators \cite{Bodwin:1994jh}
\begin{align}
    \mathcal{O}(\spec{1}{P}{1}) &\equiv \varsigma^\dagger \left( - \frac{i}{2} \overset{\leftrightarrow}{\vec{\nabla}} \right) \eta \cdot \eta^\dagger \left( -\frac{i}{2} \overset{\leftrightarrow}{\vec{\nabla}} \right) \varsigma \, ,  
    \\
    \mathcal{O}(\spec{3}{P}{0}) &\equiv \frac{1}{3} \varsigma^\dagger \left( -\frac{i}{2} \overset{\leftrightarrow}{\vec{\nabla}} \cdot \vec{\sigma} \right) \eta\,  \eta^\dagger \left( -\frac{i}{2} \overset{\leftrightarrow}{\vec{\nabla}} \cdot \vec{\sigma} \right) \varsigma\, , 
    \\
    \mathcal{O}(\spec{3}{P}{1}) &\equiv \frac{1}{3} \varsigma^\dagger \left( -\frac{i}{2} \overset{\leftrightarrow}{\vec{\nabla}} \times \vec{\sigma} \right) \eta  \cdot  \eta^\dagger \left( -\frac{i}{2} \overset{\leftrightarrow}{\vec{\nabla}} \times \vec{\sigma} \right) \varsigma\, , 
    \\
    \mathcal{O}(\spec{3}{P}{2}) &\equiv  \varsigma^\dagger \left( -\frac{i}{2} \overset{\leftrightarrow}{\nabla}  \phantom{s}^{(i} \sigma^{j)} \right) \eta\,  \eta^\dagger \left( -\frac{i}{2} \overset{\leftrightarrow}{\nabla}  \phantom{s}^{(i} \sigma^{j)} \right) \varsigma \, ,
    \\
    \mathcal{P}(\spec{1}{S}{0}) &\equiv \frac{1}{2} \left[ \varsigma^\dagger \eta \, \eta^\dagger \left( \frac{i}{2} \overset{\leftrightarrow}{\vec{\nabla}} \right)^2 \varsigma + \text{h.c.} \right] \, ,
    \\
    \mathcal{P}(\spec{3}{S}{1}) &\equiv \frac{1}{2} \left[ \varsigma^\dagger \vec{\sigma} \eta \, \cdot \eta^\dagger \vec{\sigma} \left( \frac{i}{2} \overset{\leftrightarrow}{\vec{\nabla}} \right)^2 \varsigma + \text{h.c.} \right] \, ,
    \\
    \mathcal{P}(\spec{3}{S}{1},\spec{3}{D}{1}) &\equiv \frac{1}{2} \left[ \varsigma^\dagger \sigma^i \eta \, \cdot \eta^\dagger \sigma^j  \left( \frac{i}{2} \right)^2 \overset{\leftrightarrow}{\nabla}  \phantom{x}^{(i} \overset{\leftrightarrow}{\nabla}\,^{j)}  \varsigma + \text{h.c.} \right] \, ,
\end{align}
contributing to the 4-fermion Lagrangian
\begin{align}
    \label{eq:dimension_8_lag}
    (\mathcal{L}_{\text{4-fermions}} )_{d=8} =~& \frac{f(\spec{1}{P}{1})}{\mchi^4} \mathcal{O}(\spec{1}{P}{1}) + \frac{f(\spec{3}{P}{0})}{\mchi^4} \mathcal{O}(\spec{3}{P}{0}) + \frac{f(\spec{3}{P}{1})}{\mchi^4} \mathcal{O}(\spec{3}{P}{1}) + \frac{f(\spec{3}{P}{2})}{\mchi^4} \mathcal{O}(\spec{3}{P}{2})\nonumber \\
    &  + \frac{g(\spec{1}{S}{0})}{\mchi^4} \mathcal{P}(\spec{1}{S}{0}) + \frac{g(\spec{3}{S}{1})}{\mchi^4} \mathcal{P}(\spec{3}{S}{1}) + \frac{g(\spec{3}{S}{1},\spec{3}{D}{1})}{\mchi^4} \mathcal{P}(\spec{3}{S}{1},\spec{3}{D}{1})\, ,
\end{align}
where $\eta^\dagger \overset{\leftrightarrow}{\vec{\nabla}} \varsigma \equiv \eta^\dagger (\vec{\nabla} \varsigma)- (\vec{\nabla} \eta)^\dagger \varsigma$ and the notation $T^{(ij)}$ for a rank 2 tensor represents its traceless symmetric components, \ie $T^{(ij)}\equiv(T^{ij}+T^{ji})/2-T^{kk}\delta^{ij}/3$. The dimension $8$ operators all contain spatial derivatives which contribute to the velocity dependent parts of the annihilation cross section. We denote in \cref{eq:dimension_8_lag} the leading order p-wave operators and matching coefficients with $\mathcal{O}$ and $f$, respectively, whereas the $\vrel^2$ contributions to the s-wave are labelled with $\mathcal{P}$ and $g$.\newp
The calculations of the matching coefficients are performed while maintaining the dependence of the results on the mass ratio $\tilde{R}\equiv \mphi/\mchi$.\footnote{This generalizes the results of \eref\cite{Biondini:2021ycj}. Although the effect is modest, because the mediator mass is at most of the order of $\mchi\alpha$, the result is useful for future studies, if one considers higher mediator masses. Note however, that in this case the construction of the \gls{nreft} would change, as the mediator scale has to be integrated out together with the fermion mass at the hard scale.} At leading order in both couplings, the imaginary parts of the matching coefficients yield \cite{Biondini:2023ksj}
\begin{align}
    \label{eq:IMmatch_coeff_1}
    \Im{f(\spec{1}{S}{0})} &= 2  \pi \alpha \alpha_5 \,\mathcal{F}(\tilde{R}) \, , \\
    \label{eq:IMmatch_coeff_2}
    \Im{g(\spec{1}{S}{0})} &= - \frac{8 \pi \alpha \alpha_5}{3} \frac{\left( 1-\frac{13}{8}\tilde{R}^2+\frac{5}{8}\tilde{R}^4 - \frac{3}{32} \tilde{R}^6\right)}{\big(1-\tilde{R}^2/2\big)^2\big(1-\tilde{R}^2\big)} \,\mathcal{F}(\tilde{R})\, , \\
    \label{eq:IMmatch_coeff_3}
    \Im{f(\spec{3}{P}{0})} &= \frac{\pi}{6} \left[ 3 \alpha \left( 2 - \frac{\mathcal{G}(\tilde{R})}{3} \right) - \alpha_5 \,\mathcal{G}(\tilde{R})  \right]^2  \mathcal{F}(R)  \, ,  \\
    \label{eq:IMmatch_coeff_4}
    \Im{f(\spec{3}{P}{2})} &= \frac{\pi}{15} (\alpha+\alpha_5)^2 \,\mathcal{G}(\tilde{R})^2 \,\mathcal{F}(\tilde{R}) \, ,
\end{align}
with contributions from $f(\spec{3}{S}{1})$, $f(\spec{1}{P}{1})$, $f(\spec{3}{P}{1})$, $g(\spec{3}{S}{1})$ and $g(\spec{3}{S}{1},\spec{3}{D}{1})$ vanishing. We introduced in the equations above the auxiliary functions 
\begin{equation}
    \mathcal{F}(\tilde{R}) = \frac{\sqrt{1-\tilde{R}^2}}{\big( 1-\tilde{R}^2/2 \big)^2}  \, , \quad  \mathcal{G}(\tilde{R}) =  \frac{1-\tilde{R}^2}{1-\tilde{R}^2/2}  \, ,
\end{equation}
for a more compact notation. The matching coefficients are calculated by computing the matrix elements of the \gls{uv} box diagrams in \cref{fig:boxdiagrams}, splitting up the 4-fermion spinors into their Pauli spinor components
\begin{equation}
    u(\vec{p})\equiv\sqrt{\frac{E_{\vec{p}}+\mchi}{2E_{\vec{p}}}}\begin{pmatrix}\varsigma \\ \frac{\vec{p}\cdot\vec{\sigma}}{E_{\vec{p}}+\mchi} \varsigma\end{pmatrix},\qquad v(\vec{p})\equiv\sqrt{\frac{E_{\vec{p}}+\mchi}{2E_{\vec{p}}}}\begin{pmatrix}\frac{\vec{p}\cdot\vec{\sigma}}{E_{\vec{p}}+\mchi}\eta \\ \eta\end{pmatrix}\,,
\end{equation}
and expanding around $\vec{p}/\mchi$ to (next to) leading order to compare their structure with the dimension 6 (8) operators of the \gls{nreft}. The fermion propagators are expanded as well, whereas the scalar mediator propagators can be put on-shell, since we are only interested in the imaginary parts. We can explain the vanishing $\spec{3}{S}{1}$ and $\spec{1}{P}{1}$ components using C symmetry arguments. The fermion antifermion pair has $C=(-1)^{L+S}$ whereas $C=1^n$ for $n$ scalar final states. Therefore, only $C=1$ states are allowed for the initial pair which selects $\spec{1}{S}{0}$ and $\spec{3}{P}{J}$ with $J=0,1,2$.


\subsubsection{Potential non-relativistic EFT}
\label{subsubsec:pNREFT}

We can use \gls{nreft} to calculate the annihilation cross sections but it is not sufficient, yet, to describe \gls{se} or properly account for \gls{bs} dynamics. We strive for a theory, where the equations of motion have the form of a Schrödinger equation and the \gls{dof} are bound and scattering states rather than fermions and antifermions. Moreover, we want a unique power-counting, as the soft and ultra-soft scales in the \gls{nreft} are still intertwined. To achieve this, we switch to another effective theory by integrating out the soft scale $\mchi\alpha$, which corresponds to the inverse relative distance $1/r$ between a $\bar{\chi}\chi$ pair. In our case, the mediator mass $\mphi$ also needs to be integrated out simultaneously, as it is allowed to be of the order of the soft scale, whereas thermal scales can be set to zero due to the assumption $T\lesssim \mchi\alpha^2$. The resulting theory is a \gls{pnreft}, which for our example case is labelled $\pnry$. It contains as \gls{dof} only $\bar{\chi}\chi$ pairs as well as mediators $\phi$ of energies at the order of the ultra-soft scale $\mchi\alpha^2$.\footnote{Note that the fermion antifermion pair can still have momenta up to the soft scale, as they appear only as $\vec{p}^2/\mchi\sim \mchi\alpha^2$ in the observables, whereas the mediator momenta must be ultra-soft.} These fermion pairs can easily translate to the scattering and \gls{bs} governed by a Schrödinger equation in the quantum mechanical picture and the power counting is unique as there is only one dynamical scale present.\newp
The Lagrangian of a \gls{pnreft} is organized as an expansion in $1/\mchi$, $\vec{p}$ and $\alpha$ which is inherited from the \gls{nreft} together with an expansion in $r\sim (\mchi\alpha)^{-1}$ specific to the \gls{pnreft}.\footnote{In general, there is an additional expansion in $\alpha$ at the soft scale due to its running. However, we will neglect this for now and justify our decision in \cref{subsec:dmproductionpnreft} for the specific cases of $\nry$ and $\pnry$.} The matching coefficients of a \gls{pnreft}, which are functions of $r$, are called \textit{potentials} for reasons which will become clear in a moment. The first thing to notice is that the structure of the Lagrangians of the scalar and bilinear sectors does not change when moving from $\nry$ to $\pnry$ because they do not depend on $1/r$ and are therefore insensitive to the scale which is integrated out, \ie
\begin{align}
    \mathcal{L}_{\text{scalar}}^{\pnry}&=\mathcal{L}_{\text{scalar}}^{\nry}\,,\\
    \mathcal{L}_{\text{bilinear}}^{\pnry}&=\mathcal{L}_{\text{bilinear}}^{\varsigma, \nry} + \mathcal{L}_{\text{bilinear}}^{\eta, \nry}\,.
\end{align}
However, the actual size of each term is not yet explicit from the $\nry$ Lagrangian and it is also not clear, that the mediator fields are only dependent on ultra-soft scales. Therefore, we project the $\pnry$ Hamiltonian onto the Fock subspace
\begin{equation}
    \label{eq:projectionoperator}
    \int\dd[3]{x_1}\dd[3]{x_2}\varphi_{ij}(\vec{x_1},\vec{x_2},t)\varsigma_i^\dagger(\vec{x_1},t)\eta_j(\vec{x_2},t)\phius\,,
\end{equation}
where the state $\phius$ contains an arbitrary number of scalar mediators with energies much smaller than the soft scale and no heavy (anti-)fermions, \ie $\varsigma(\vec{x})\phius=\eta^\dagger(\vec{x})\phius=0$. The scattering and bound states of the $\bar{\chi}\chi$ pair are represented by the bilocal wave function field $\varphi_{ij}(\vec{x_1},\vec{x_2},t)$, where $i,j$ are Pauli spinor indices, which we will suppressed in the following. The Pauli spinor fields satisfy the equal time anti-commutation relations
\begin{equation}
    \left\{\varsigma_i(\vec{x}),\varsigma^\dagger_j(\vec{y})\right\}=\delta_{ij}\,\delta^{(3)}(\vec{x}-\vec{y})\,,\qquad\left\{\eta^\dagger_i(\vec{x}),\eta_j(\vec{y})\right\}=\delta_{ij}\,\delta^{(3)}(\vec{x}-\vec{y})\,,
\end{equation}
with all other combinations vanishing. Performing this projection in the bilinear sector yields \cite{Biondini:2021ccr,Biondini:2021ycj}
\begin{align}
    \label{eq:pNRY}
    \mathcal{L}_{\text{bilinear}}^{\pnry}  \supset \int \dd[3]{r} &\, \varphi^\dagger(\vec{r},\vec{R},t) \left\{   i \partial_0 +\frac{\vec{\nabla}^2_{\vec{r}}}{\mchi} +\frac{\vec{\nabla}^2_{\vec{R}}}{4\mchi} + \frac{\vec{\nabla}^4_{\vec{r}}}{4 \mchi^3}\right. \nonumber \\
    & \left. - 2 g \phi(\vec{R}) -g\frac{ r^i r^j }{4}  \left[ \nabla_{\vec{R}}^i \nabla_{\vec{R}}^j \, \phi (\vec{R})   \right] -   g \phi(\vec{R}) \frac{\vec{\nabla}^2_{\vec{r}}}{\mchi^2}    \right\} \varphi(\vec{r},\vec{R},t),
 \end{align}
where we also switched to \gls{com} coordinates $\vec{r}\equiv\vec{x_1}-\vec{x_2}$, $\vec{R}\equiv(\vec{x_1}+\vec{x_2})/2$ and only kept terms up to order $\mchi\alpha^4$ in the power counting of $\pnry$ for reasons which will become apparent later.\footnote{Note that the integration over $\vec{R}$ is missing since the displayed quantity is actually the Lagrangian density. The spatial derivatives in the new coordinate system yield $\vec{\nabla}_{\vec{x_{1,2}}}=\vec{\nabla}_{\vec{R}}/2\pm\vec{\nabla}_{\vec{r}}$.} To see the scaling of the different terms, however, we need to know the power counting rules for the different quantities in the equation \cite{Luke:1996hj,Luke:1997ys,Biondini:2021ccr}. We already know that inverse relative distance together with the corresponding derivative scale as $r^{-1},\vec{\nabla}_{\vec{r}}\sim \mchi\alpha$. The time derivative, the scalar field and the inverse \gls{com} coordinate as well as its derivative scale as $\partial_0, g\phi, R^{-1}, \vec{\nabla}_{\vec{R}} \sim \mchi\alpha^2 (\sim T)$, which also self-consistently leads to $r\ll R$ as needed for the expansion. The first line of \cref{eq:pNRY} can then be determined easily, where the first two terms are of $\mchi\alpha^2$ and the last two follow an $\mchi\alpha^4$ scaling behaviour. Note that the last term had to be added from the $\vec{\nabla}^4/(8\mchi^3)$ contributions at $\order{1/\mchi^3}$ in the mass expansion (not displayed in \cref{eq:bilinear_varsigma,eq:bilinear_eta}) due to these power counting rules. The terms in the second line emerge after a \textit{multipole expansion} of the mediator field in terms of $\vec{r}$ 
\begin{equation}
    \phi(\vec{x_{1,2}})=\phi\left(\vec{R}\pm\frac{\vec{r}}{2}\right)=\phi(\vec{R})\pm\frac{\vec{r}}{2}\cdot\vec{\nabla}_{\vec{R}}\,\phi(\vec{R})+\frac{1}{8}r^ir^j\nabla^i_{\vec{R}}\nabla^j_{\vec{R}}\,\phi(\vec{R})+\order{r^3}
\end{equation}
in order to disentangle the different scales induced by $\vec{r}$ and $\vec{R}$. We can see that the first two terms in the second line of \cref{eq:pNRY} correspond to the monopole and quadrupole of the expansion, whereas the dipole terms cancel. Note that the quadrupole term is needed in the following to capture the leading non-trivial dynamics among the scattering and bound states. The last term is added from the $\pm g\{\vec{\nabla},\{\vec{\nabla},\phi\}\}/(8\mchi^2)$ contributions to the bilinear Lagrangians at $\order{1/\mchi^2}$ in the mass expansion, as it is of the same order as the quadrupole term. In the purely scalar part of the $\pnry$ Lagrangian, we only keep the $\phi(\vec{R})$ terms of the multipole expansion. The resulting structure is the same as for the underlying full theory and therefore not displayed.\newp
\begin{figure}
    \centering
    \includegraphics[width=\textwidth]{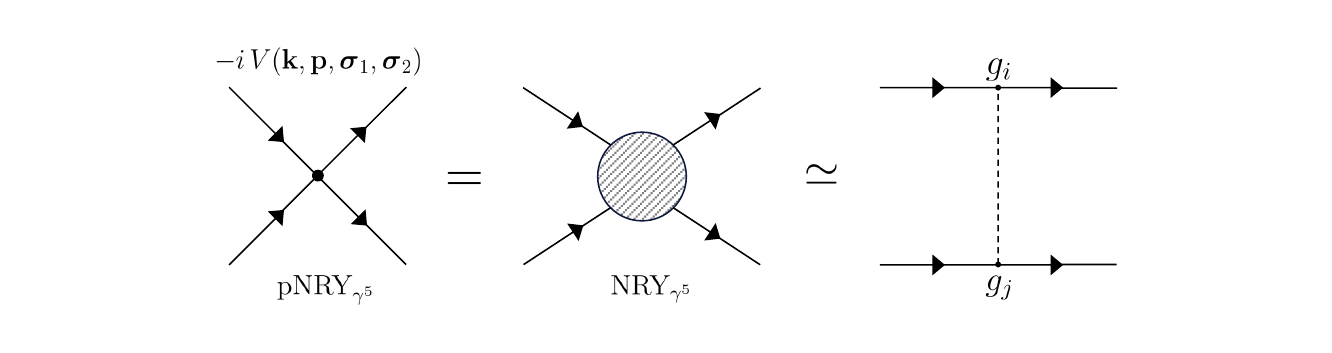}
    \caption[Matching of the 4-fermion Lagrangians between the $\nry$ and $\pnry$ theory via the off-shell 4-point Green's functions.]{Matching of the 4-fermion Lagrangians between the $\nry$ and $\pnry$ theory via the off-shell 4-point Green's functions. Solid lines denote (anti)fermions, whereas dashed lines refer to scalar mediator propagators. At leading order, the 4-point Green's function of $\nry$ consists only of one-boson exchange diagrams with vertex couplings $g_{i,j}\in\{g,g_5\}$. The matching coefficients of the $\pnry$ are contained in $V(\vec{k},\vec{p},\vec{\sigma}_1,\vec{\sigma}_2)$ (in momentum space), where $\vec{k}$ is the momentum transfer between the $\bar{\chi}\chi$ pair and thus also the momentum of the mediator propagator in the leading order diagrams.}
    \label{fig:potentialmatching}
\end{figure}
The 4-fermion sector of the $\pnry$ Lagrangian features new terms when compared to the $\nry$. After projection onto the low energy subspace using \cref{eq:projectionoperator}, its Lagrangian reads
\begin{equation}
    \mathcal{L}_{\text{4-fermion}}^{\pnry} = -\int \dd[3]{r} \, \varphi^\dagger(\vec{r},\vec{R},t)V(\vec{r},\vec{p},\vec{\sigma}_1,\vec{\sigma}_2)\varphi(\vec{r},\vec{R},t)\,,
\end{equation}
with matching coefficients $V(\vec{r},\vec{p},\vec{\sigma}_1,\vec{\sigma}_2)$, where the $\vec{\sigma}_1 (\vec{\sigma}_2)$ denotes the spin matrix of the (anti)fermion field in the bilinear $\nry$ Lagrangian. This \textit{potential} comprises a real and imaginary part, where the latter again corresponds to the annihilation process $\bar{\chi}+\chi\to 2\phi$. The potential, given by
\begin{equation}
    V(\vec{r},\vec{p},\vec{\sigma}_1,\vec{\sigma}_2)=V^{(0)}+\frac{V^{(1)}}{\mchi}+\frac{V^{(2)}}{\mchi^2}\,,
\end{equation}
is organized as an expansion in $\alpha$, $1/r$, and $1/\mchi$ and can be calculated by matching the off-shell 4-point Green's function of $\nry$ onto $\pnry$ (see \cref{fig:potentialmatching}). At leading order, the $\nry$ Green's function corresponds to one boson exchange diagrams. Calculating the purely scalar contribution gives rise to $V(\vec{k},\vec{p},\vec{\sigma}_1,\vec{\sigma}_2)|_{g^2}=-4\pi\alpha/(\vec{k}^2-\mphi^2)$ which after a Fourier transform yields the static Yukawa potential, matching
\begin{equation}
    V^{(0)}=-\frac{\alpha}{r}e^{-\mphi r}
\end{equation}
in the expansion of the potential. From the mixed scalar-pseudo-scalar and pure pseudo-scalar diagrams one obtains \cite{Biondini:2021ycj}
\begin{align}
    V(\vec{r},\vec{p},\vec{\sigma}_1,\vec{\sigma}_2)|_{g g_5^{},g_5^2}=&-\frac{g g_5}{4\pi\mchi}\frac{\vec{r}\cdot\vec{\sigma}_1-\vec{r}\cdot\vec{\sigma}_2}{r^3}\nonumber\\
    &+\frac{\pi\alpha_5}{\mchi^2}\left[-\frac{\vec{\sigma}_1\cdot\vec{\sigma}_2}{3}\delta^{(3)}(\vec{r})+\frac{3\vec{\sigma}_1\cdot\uvec{r}\,\vec{\sigma}_2\cdot\uvec{r}-\vec{\sigma}_1\cdot\vec{\sigma}_2}{4\pi r^3}\right]\,,
\end{align}
which enter the $\mchi$ suppressed $V^{(1)}$ and $V^{(2)}$ terms. From power counting arguments, where $\delta^{(3)}(\vec{r})\sim (\mchi\alpha)^3$, we can see that these terms are of order $(g/g_5)\mchi\alpha^3$ and $(g/g_5)^2\mchi\alpha^4$, respectively, and thus suppressed compared to the purely scalar term, which scales as $\mchi\alpha^2$. As we will work later with moderate couplings $\alpha<0.25$ and already assumed $g_5<g$, we can therefore neglect mixed and pure pseudo-scalar contributions and only retain the Yukawa term $V^{(0)}$. Collecting all leading $\mchi\alpha^2$ terms in the $\pnry$ Lagrangian, we are left with 
\begin{equation}
    \mathcal{L}^{\pnry} \supset \int \dd[3]{r} \, \varphi^\dagger(\vec{r},\vec{R},t) \left[ i \partial_0 +\frac{\vec{\nabla}^2_{\vec{r}}}{\mchi}-V^{(0)}-2g\phi(\vec{R})\right]\varphi(\vec{r},\vec{R},t).
\end{equation}
The corresponding equation of motion for $\varphi(\vec{r},\vec{R},t)$ yields a Schrödinger equation
\begin{equation}
    \left(-\frac{\vec{\nabla}^2_{\vec{r}}}{\mchi}+V^{(0)}+2g\phi(\vec{R})\right)\varphi(\vec{r},\vec{R},t)=i \partial_0 \,\varphi(\vec{r},\vec{R},t)
\end{equation}
featuring a Yukawa potential.\footnote{Since $T\lesssim \mchi\alpha^2$, no thermal contributions will enter the potential at any order.} Assuming the equation to be independent of $\vec{R}$ with no constant background field (\ie $\phi(\vec{R})\to 0$) and the Hamiltonian to be independent of time, we obtain the stationary Schrödinger equations given in \cref{eq:scatteringSchroedinger,eq:boundstateSchroedinger}. This also relates the bilocal field to the non-relativistic scattering and bound state wave functions $\phik$ and $\psinlm$, which can be computed with the methods displayed in \cref{subsubsec:Schroedingerequations}.\newp
As mentioned before, pair annihilations into scalar mediators are encoded in the imaginary part of the matching coefficients in $\pnry$. The Lagrangian of these terms does not discriminate if the bilocal fields $\varphi(\vec{r},\vec{R},t)$ represent scattering or \gls{bs}. Therefore, in $\pnry$, \gls{bs} decays can be computed from the same Lagrangian as annihilations cross sections. In $\nry$, annihilations are described by local 4-fermion operators. The same is true for $\pnry$, where the 4-fermion operators of $\nry$ create local terms in the imaginary part of the potential $V(\vec{k},\vec{p},\vec{\sigma}_1,\vec{\sigma}_2)$. The relevant terms for annihilation, including s-wave and p-wave contributions up to $\vrel^2$, are given by \cite{Biondini:2021ycj,Brambilla:2002nu,Brambilla:2004jw}
\begin{align}
    \label{eq:LannpNRY}
    \mathcal{L}_{\text{ann}}^{\pnry}=& \frac{i}{\mchi^2} \, \int\dd[3]{r} \varphi^\dagger (\vec{r}) \delta^{(3)}(\vec{r}) \left[ 2 \Im{f(\spec{1}{S}{0})} \!- \!\vec{S}^2 \!\left(\Im{f(\spec{1}{S}{0})} \!- \! \Im{f(\spec{3}{S}{0})}\right) \right] \varphi (\vec{r}) \nonumber\\
    &-\frac{i}{\mchi^4} \, \int \dd[3]{r} \varphi^\dagger (\vec{r}) \mathcal{T}_{SJ}^{ij} \nabla_{\vec{r}}^i \delta^{(3)}(\vec{r})  \nabla_{\vec{r}}^j \,  \Im{f(\spec{2S+1}{P}{J})}\varphi \, (\vec{r}) \nonumber\\
    &-\frac{i}{2\mchi^4} \, \int \dd[3]{r}\varphi^\dagger (\vec{r}) \,  \Omega_{SJ}^{ij} \left\{ \delta^{(3)}(\vec{r}),\nabla_{\vec{r}}^i \nabla_{\vec{r}}^j  \right\}  \Im{g(\spec{2S+1}{S}{J})} \varphi \, (\vec{r}) \, ,
\end{align}
where $\vec{S}=(\vec{\sigma}_1+\vec{\sigma}_2)/2$ denotes the total spin matrix of the pair\footnote{Note that $\vec{S}$ is an operator which is properly defined as $\vec{S}=\vec{\sigma}_1/2\otimes \mathds{1}_2 + \mathds{1}_1\otimes\vec{\sigma}_2/2$ and acts on the $\mathbf{1/2}\otimes\mathbf{1/2}$ spin space (where $\mathds{1}=\mathds{1}_1\otimes\mathds{1}_2$ is the unity operator). We will mostly project it on total spin states, where $\vec{S}^2|S,m_S\rangle=S(S+1)|S,m_S\rangle$ with $S=0~ (S=1)$ for a total spin singlet (triplet).} and $\mathcal{T}^{ij}_{SJ},\Omega^{ij}_{SJ}$ are spin projection operators, which read \cite{Brambilla:2004jw}
\begin{align}
    \label{eq:pNRY_ann_lag}
    \mathcal{T}^{ij}_{01}&=\delta^{ij}(2-\vec{S}^2),\quad \mathcal{T}^{ij}_{10}=\frac{1}{3}S^iS^j,\quad \mathcal{T}^{ij}_{11}=\frac{1}{2}\varepsilon^{kil}\varepsilon^{kjm}S^lS^m,\nonumber\\
    \mathcal{T}^{ij}_{12}&=\left(\frac{\delta^{ik}S^l+\delta^{il}S^k}{2}-\frac{\delta^{kl}S^i}{3}\right)\left(\frac{\delta^{jk}S^l+\delta^{jl}S^k}{2}-\frac{\delta^{kl}S^j}{3}\right)\\
    \Omega^{ij}_{00}&=\delta^{ij}(2-\vec{S}^2),\quad \Omega^{ij}_{11}=\delta^{ij}\vec{S}^2.
\end{align}
Note that we have suppressed the $\vec{R}$ and $t$ dependence of the $\varphi(\vec{r},\vec{R},t)$ to favor a more compact notation, whereas the $\vec{r}$ dependence contributes only at $\vec{r}=0$ due to the contact nature of the 4-fermion interactions already at $\nry$. The $\nry$ matching coefficients in \cref{eq:pNRY_ann_lag} are given in \cref{eq:IMmatch_coeff_1,eq:IMmatch_coeff_2,eq:IMmatch_coeff_3,eq:IMmatch_coeff_4}.\newp
When projected onto a total spin state, we can define the expectation value of the bilocal field as 
\begin{align}
    \label{eq:bilocalfielddef}
    \varphi^{(S)}_{ij}(t,\vec{r},\vec{R})\equiv & ~\langle S,m_S |\varphi_{ij}(t,\vec{r},\vec{R})|S,m_S\rangle\nonumber\\
    =&\int\frac{\dd[3]{P}}{(2\pi)^3}\left[\sum_n e^{-i\mathcal{E}_nt+i\vec{P}\cdot\vec{R}}\,\psin(\vec{r})\,\tilde{S}^{(S)}_{ij}\, a_n(\vec{P})\right.\nonumber\\
    &\hspace{1.9cm}+\left.\int\frac{\dd[3]{p}}{(2\pi)^3}e^{-i\mathcal{E}_{\vec{p}}t+i\vec{P}\cdot\vec{R}}\,\phiq{p}(\vec{r})\,\tilde{S}^{(S)}_{ij}\,a_{\vec{p}}(\vec{P})\right],
\end{align}
with $\tilde{S}^{(0)}_{ij}=\delta_{ij}/\sqrt{2}$ the spin matrix elements for a singlet ($S=0$) and $\tilde{S}^{(1)}_{ij}=(\vec{\sigma}\cdot\vec{\epsilon})_{ij}/\sqrt{2}$ for a triplet state ($S=1$), where $\vec{\epsilon}$ denotes the spin polarization vector. The matrix elements are already normalized, such that $(\tilde{S}^{(S)}_{ji})^*\tilde{S}^{(S)}_{ij}=1$ and $(\tilde{S}^{(0)}_{ji})^*\tilde{S}^{(1)}_{ij}=0$. The $a_{\vec{p}}(\vec{P})$ and $a_n(\vec{P})$ are the ladder operators of the scattering and \gls{bs} of the $\bar{\chi}\chi$ pair, respectively, where $|\vec{p},\vec{P}\rangle = a_{\vec{p}}^\dagger(\vec{P})\Vac$ creates a scattering and $|n,\vec{P}\rangle = a_n^\dagger(\vec{P})\Vac$ a bound state. They fulfill the commutation relations
\begin{align}
    \left[a_n(\vec{P}),a_{n'}^\dagger(\vec{P}')\right]&=(2\pi)^3 \delta_{nn'}\delta^{(3)}(\vec{P}-\vec{P}'),\\
    \left[a_{\vec{p}}(\vec{P}),a_{\vec{p}'}^\dagger(\vec{P}')\right]&=(2\pi)^6\delta^{(3)}(\vec{p}-\vec{p}')\delta^{(3)}(\vec{P}-\vec{P}'),
\end{align}
where we note that $n$ comprises all quantum numbers of the bound state.\newp
The two approaches outlined in \cref{subsec:nonperturbativeeffectsfromQFT,subsec:nonperturbativeeffectsfromEFT} together with the treatment of the \gls{se} factors discussed in \cref{subsec:SEannihilationBSdecay} will be extensively used in the following two sections to account for non-perturbative effects in the \gls{dm} observables for the explicit models under consideration. Whereas the Bethe-Salpeter method is sufficient to describe non-perturbative effects in \cref{sec:impactonnonthermalDMproduction}, in \cref{sec:indirectdetection} we will resort to the \gls{eft} approach for a better handling of the various scales involved, and use the Bethe-Salpeter method only for cross checking.
\newpage
\thispagestyle{empty}
\mbox{}
\newpage
\section{Non-perturbative effects in non-thermal dark matter production}
\label{sec:impactonnonthermalDMproduction}
\fancyhead[RO]{\uppercase{Non-perturbative effects in non-thermal DM production}}

Now that we have gained an understanding of the theoretical and experimental challenges posed by \gls{dm}, have developed tools to delineate its evolutionary dynamics and address non-perturbative effects in \gls{dm} interactions, our focus shifts towards practical application. In the following two sections, we explore two distinct theoretical models of \gls{dm}, each heavily influenced by non-perturbative effects. The primary research topic of this thesis revolves around conducting a comprehensive investigation into the phenomenology of \gls{dm} within these models, with a particular emphasize on \gls{se} and the existence of \gls{bs}.\newp
Examining the production of \gls{dm} in the early Universe offers a path towards a clearer understanding of its fundamental nature. Within this area of research, theoretical predictions of experimental signals for \gls{dm} candidates produced by thermal \gls{fo} attained a high degree of refinement within the past decades (see \eg\erefs\cite{Bertone:2004pz,Arcadi:2017kky} and references therein). Non-thermal production mechanisms, such as \gls{fi}, which involve \gls{dm} that does not thermalize with the \gls{sm} heat bath (\cf\cref{sec:DMproduction}), have received significantly less recognition to date \cite{Covi:1999ty,Feng:2003xh,Hall:2009bx}. Nevertheless, these mechanisms, by generating potentially long-lived states within the dark sector, can lead to markedly different collider signatures, rendering them intriguing subjects for study.\newp
We will consider in this section a minimal class of \gls{dm} models which allow for non-thermal production and can be tested with collider experiments. More concretely, we focus on so-called $t$-channel mediator models with feeble couplings to the \gls{sm}, as they have been adopted frequently in \gls{dm} searches at the LHC \cite{Chang:2013oia,An:2013xka,DiFranzo:2013vra,Papucci:2014iwa,Garny:2014waa,Ibarra:2015nca,Kahlhoefer:2017dnp,Arina:2020tuw,Arcadi:2021glq}. These models usually comprise a very weakly interacting \gls{dm} candidate as well as a mediator with \gls{sm} quantum numbers. Besides a testable production rate of the mediator particle at colliders for mediator masses below $\order{\SI{1}{\TeV}}$, the mediator interactions with \gls{sm} gauge bosons give rise to potentially large non-perturbative effects as discussed in \cref{sec:non-perturbative-effects}. We compute in the following the two non-thermal production mechanisms relevant for the parameter choices of our model, namely \gls{fi} and \gls{sw}, with a particular emphasize on \gls{se} and \gls{bsf}. The results will be compared to collider and astrophysical bounds to determine the viable parameter space of the model.\newp
We start by introducing our $t$-channel mediator model(s) in \cref{subsec:tchannelmediatormodel}. In \cref{subsec:nonthermalprodmechanisms}, we explain the two non-thermal production mechanisms mentioned above and set up the \glspl{be} for our specific applications. Subsequently, in \cref{subsec:DMabundancesW}, we modify the \glspl{be} to include non-perturbative effects after computing their impact on the corresponding production rates in \cref{subsec:nonpertubativeeffectsdmproduction}. Then, in \cref{subsec:constraintsnth}, we discuss various astrophysical, cosmological, and collider constraints, which we combine to obtain a complete picture of the open parameter space in \cref{subsec:parameterspacenthprod}.\newp
The work presented in this section has been published in \eref\cite{Bollig:2021psb}. It appeared as a preprint on the arXiv around the same time as \eref\cite{Decant:2021mhj}, which also studied the impact of bound state effects on the \gls{sw} mechanism in the same mediator model but focused more on the astrophysical limits of non-thermal \gls{dm}. The latter is hence complementary to our analysis. A later paper by Binder \etal\cite{Binder:2023ckj} provided a more comprehensive study on \gls{bsf} within this class of models by taking into account a large number of higher \gls{bs} as well as \gls{bs} transitions. Their results will not be considered in the following analysis.


\subsection[The \texorpdfstring{$t$}{???}-channel mediator model with feeble couplings]{\protect\boldmath The \texorpdfstring{$t$}{???}-channel mediator model with feeble couplings}
\label{subsec:tchannelmediatormodel}

The simplified model under consideration consists of two new particles in the dark sector. First, a \gls{dm} candidate $\chi$, which is taken to be a singlet under the \gls{sm} gauge group to allow for non-thermal production. In order to also avoid a direct coupling to the \gls{sm} via a Higgs portal, we assume it to be of fermionic nature. More concretely, we (arbitrarily) choose $\chi$ to be a Majorana fermion, but the outcome for a Dirac fermion is comparable. Furthermore, we assume a $\mathbb{Z}_2$ symmetry under which all (non-)SM particles are considered even (odd). This guarantees the stability of the \gls{dm} candidate and prevents a direct coupling to the \gls{sm} via the neutrino portal. Secondly, we introduce a scalar mediator $\varphi$ in the $\order{\si{\TeV}}$ mass regime, eventually linking $\chi$ to the \gls{sm} via a Yukawa-type interaction $\mathcal{L} \subset \lchi \bar{f} \varphi \chi$, with $f$ a \gls{sm} fermion and $\lchi$ the coupling which gatekeeps the connection between the \gls{dm} candidate and the \gls{sm} heat bath. Due to the gauge invariance of the Lagrangian, the quantum numbers of the mediator must match the quantum numbers of the \gls{sm} fermion, but conceptually, interactions with all \gls{sm} fermions are possible. To simplify things, we assume the mediator to be a singlet under the weak gauge group. Therefore, only interactions with right-handed quarks or leptons are allowed. Disregarding further differences in flavor because they will not alter the features of our model on a qualitative level, we will consider only interactions with the heaviest quarks and leptons. This leaves us with two possible choices: 1) a color-charged \textit{top-philic} mediator denoted by $\ttilde$, which interacts with the right-handed top quark, and 2) a \textit{lepto-philic} mediator $\tautilde$, carrying only hypercharge while interacting with the right-handed $\tau$-lepton.\footnote{The notation is borrowed from supersymmetry, but we want to emphasize that the gauge theory employed is not supersymmetric, and neither are the introduced fields.} An overview of the introduced dark sector particles is given in \cref{tab:particleoverview}.\newp
For the top-philic mediator $\ttilde$, the Lagrangian reads
\begin{eqnarray}
	\mathcal{L}^{\ttilde}_{\text{DS}}&=& \frac{i}{2}\bar{\chi}\gamma^\mu\partial_\mu\chi-\frac{1}{2}\mchi\bar{\chi}\chi-\mttilde^2\,\ttilde^*\ttilde\,,\\
	\mathcal{L}^{\ttilde}_{\text{int}}&=&\abs{D_\mu \ttilde}^2+\lchi \bar{t}_R\,\ttilde\,\chi+\lH \,\ttilde^*\ttilde\,(H^\dagger H)+h.c.\,,
\end{eqnarray}
and likewise for $\tautilde$ (exchanging $\bar{t}_R$ with $\tau_R^+$). Here, $D_\mu$ denotes the covariant derivative, and $H$ is the \gls{sm} Higgs doublet present due to an additional Higgs portal of the mediator. Together with the \gls{dm} and mediator masses $\mchi$, $\mttilde$ ($\mtautilde$) as well as the \gls{dm} mediator coupling $\lchi$, its coupling strength $\lH$ represents one of the four free parameters of our theory. We assume in the following $m_{\tilde{t}}$ ($\mtautilde$) $\geq m_\chi$, since an inverted mass hierarchy would lead to a stable (color-)charged relic and thus a vastly different phenomenology. For readability, we will continue our discussion using only the top-philic mediator model. Calculations for the lepto-philic mediator are analogous unless stated otherwise. 
\begin{table}
    \centering
    \caption[Summary of all dark sector fields introduced in the simplified models considered in \cref{sec:impactonnonthermalDMproduction}.]{Summary of all dark sector fields introduced in the simplified models considered in \cref{sec:impactonnonthermalDMproduction}. Note that all of these particles are odd under an additional $\mathbb{Z}_2$ symmetry. \label{tab:particleoverview}}
    \begin{tabular}{cccc}
    \toprule
        \vspace{0.05cm}
		new particles & type & $SU(3)_c\times SU(2)_L\times U(1)_Y$ \\
		\midrule
		$\ttilde$ & bosonic scalar & $(\mathbf{3},\mathbf{1},4/3)$  \\[0.1cm]
		$\tautilde$ & bosonic scalar & $(\mathbf{1},\mathbf{1},-2)$  \\[0.1cm]
		$\chi$ & Majorana fermion & $(\mathbf{1},\mathbf{1},0)$ \\[0.1cm]
		\bottomrule
    \end{tabular}
\end{table}


\subsection{Non-thermal production mechanisms}
\label{subsec:nonthermalprodmechanisms}

\begin{figure}
    \centering 
    \includegraphics[width=\textwidth]{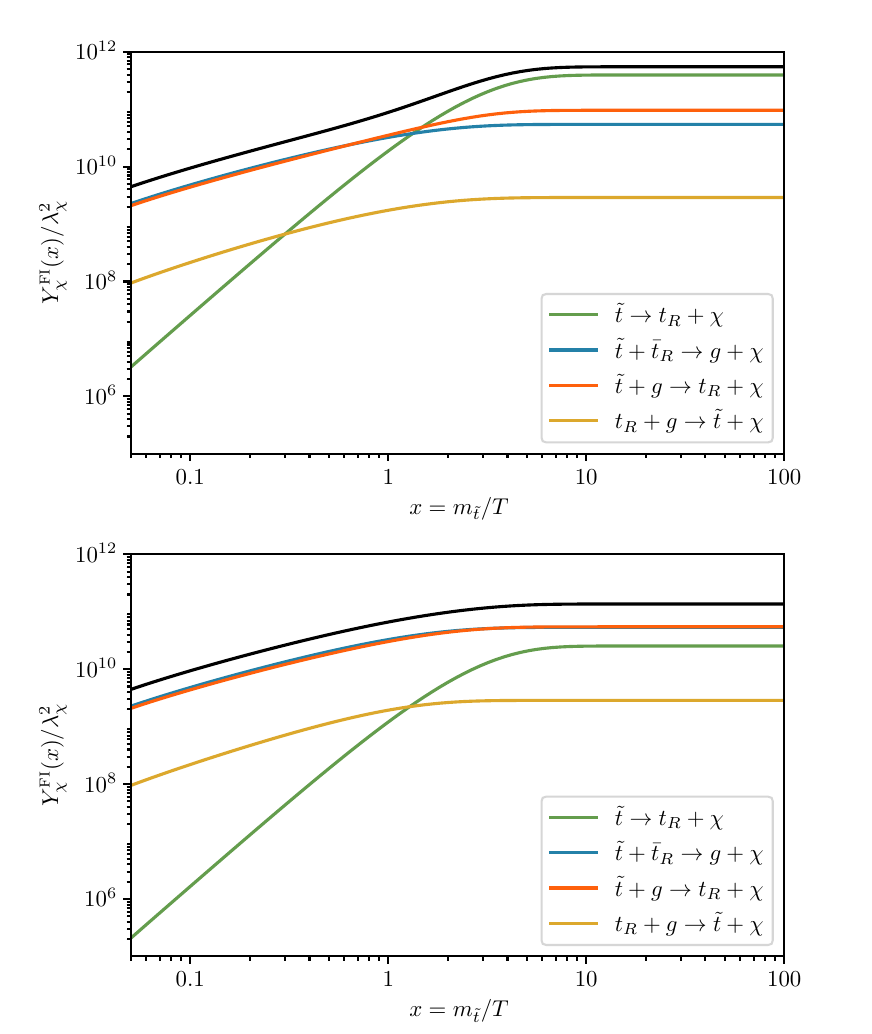}
    \caption[Contributions to the FI yield of DM from the most relevant production processes in the top-philic mediator model.]{Contributions to the \gls{fi} yield of \gls{dm} from the most relevant production processes in the top-philic mediator model as a function of $x\equiv\mttilde/T$. The yields on the top (bottom) are stripped of their $\lchi^2$ dependence and calculated for exemplary masses of $\mttilde=\SI{1}{\TeV}$, $\mchi=\SI{100}{\GeV}$ ($\mchi=\SI{800}{\GeV}$). The black line depicts the total \gls{dm} yield from \gls{fi} production.}
    \label{fig:fi_example}
\end{figure}

As already discussed in \cref{sec:DMproduction} for the case of \gls{fi}, we require for non-thermal production the initial number density of \gls{dm} after reheating to be negligible. Therefore, we set $n_\chi(T_{\text{rh}})= 0$ in the following. Moreover, \gls{dm} shall never reach thermal equilibrium with the \gls{sm}, which is why we further demand the processes driving the production of \gls{dm} to be slow on cosmological time-scales throughout the entire evolution of the Universe. This means that
\begin{equation}
\label{eq:estimate_non_thermal}
\left.\frac{\langle \Gamma \rangle_{\text{tot}}}{H}\right|_{T_{\text{max}}} \ll 1 \,,
\end{equation}
with $\langle \Gamma \rangle_{\text{tot}}$ being the total thermally averaged production rate and $T_{\text{max}}$ the temperature at which the production of $\chi$ reaches its maximum. Since all \gls{dm} production processes relevant for \gls{fi} within our model scale $\propto \lchi^2$, \cref{eq:estimate_non_thermal} imposes tight upper limits on the \gls{dm} mediator coupling. If the two-particle decay of the mediator is kinematically allowed, we can roughly estimate the constraint on $\lchi$ by taking $\langle \Gamma \rangle_{\text{tot}} \approx \Gamma_{\ttilde\to t_R\chi}$ with 
\begin{equation}
    \label{eq:mediatordecayrate1to2}
    \Gamma_{\ttilde\to t_R\chi}=\lchi^2\frac{\sqrt{\lambda(m_{\tilde{t}}^2,m_t^2,m_\chi^2)} (\mttilde^2-\mchi^2-m_t^2)}{16\pi\mttilde^3}\,,
\end{equation}
and $T_{\text{max}}\approx 0.3 \, \mttilde$.\footnote{For the top-philic mediator, contributions from $2\to2$ processes can overtake the production rate from $1\to 2$ mediator decay for certain mass configurations as we will see in the following (\cf\cref{fig:fi_example}). In these cases, a more detailed analysis is in order.} Given a strong mass hierarchy between $\mchi$ and $\mttilde$, \cref{eq:estimate_non_thermal} yields $\lambda_\chi \ll 2 \times 10^{-9} \sqrt{m_{\tilde{t}}/\si{\GeV}}$ for $\mttilde \gtrsim \SI{500}{\GeV}$. For $T \gtrsim \mttilde$, the mediator is in thermal equilibrium with the \gls{sm} bath due to its gauge couplings with \gls{sm} bosons, which results in production and annihilation rates much faster than the expansion of the Universe.\newp
The evolution of \gls{dm} and the mediator can in general be described by a system of coupled \glspl{be} \cite{Garny:2017rxs}. For \gls{fi}, the complexity of the system can be drastically reduced (\cf\cref{subsec:BEfiandfo}), resulting in a \gls{dm} yield given by 
\begin{equation}
    \label{eq:Yfreezein}
	\Yfi{\chi}(x)=2\left[\Yfi{\chi,1\to 2}(x)+\sum_{2\to 2} \Yfi{\chi,2\to 2}(x)\right]\,,
\end{equation}
where $x\equiv\mttilde/T$, and an additional factor $2$ arises from the summation over equal contributions from $\ttilde$ and $\ttilde^*$. \Cref{eq:Yfreezein} can be derived by integrating over \cref{eq:dYFIdx12,eq:dYFIdx22} with 
\begin{align}
    \label{eq:Yfi1to2}
	\Yfi{\chi,1\to 2}(x)\equiv &~\xi_1^{\text{FI}}\int_0^x\dd{x'}\frac{\gstar^{1/2}(x')}{\heff(x')}\Yeq{\ttilde}(x') x'\Gammadecexp_{\ttilde\to t_R\chi}\,,\\
    \label{eq:Yfi2to2}
	\Yfi{\chi,2\to 2}(x)\equiv &~\xi_2^{\text{FI}}\int_0^x \dd{x'}\gstar^{1/2}(x')\frac{\Yeq{a}(x')\Yeq{b}(x')}{x'^2}\expval{\sigma \vrel}_{ab\to c\chi}\,,
\end{align}
where the equilibrium yields $\Yeq{i}$ are given in \cref{eq:equilibriumyield}, and the prefactors are defined as $\xi_1^{\text{FI}}\equiv\sqrt{45/(4\pi^3)}\MPl/\mttilde^2$ and $\xi_2^{\text{FI}}\equiv\sqrt{\pi/45}\MPl\mttilde$. In the color-charged mediator model we take into account the $1\to 2$ decay process $\ttilde\to t_R+\chi$ alongside the $2\to2$ processes $\ttilde+\bar{t}_R\to g+\chi$, $\ttilde+g\to t_R+\chi$, and $t_R+g\to \ttilde+\chi$, which are also relevant due to the relative size of the strong coupling $\alpha_s(T)$. For the lepto-philic mediator model, only the $\tautilde\to \tau_R^-+\chi$ decay process needs to be considered, because $2\to2$ processes are suppressed by the coupling strength of electromagnetic interactions.\newp
The thermally averaged matrix elements for the $2\to 2$ processes $\expval{\sigma \vrel}_{ab\to c\chi}$ can be calculated numerically using \cref{eq:thermallyaveragedcrosssectionII}. In the course of the procedure, the integration over $s$ in the $\ttilde + g \to t_R + \chi$ channel requires some additional care due to its infrared divergence at tree level. Calculating the loop diagrams cancelling this divergence goes beyond the scope of this work, which is why we follow \eref\cite{Garny:2018ali} and set a minimal bound of $\sqrt{s_{\text{min}}}=(1+\epsilon) \mttilde$ with $\epsilon=0.1$ instead. Analogously, one could also introduce a thermal mass for the gluon. We checked that the results do not change with the cutoff scheme and that they also remain constant upon variation of $\epsilon$. Further, we neglect thermal corrections to the \gls{dm} yield, which are estimated to give an $\mathcal{O}(10\%)$ correction to the relic density \cite{Biondini:2020ric}.\newp
Contributions from different processes to the \gls{fi} yield in the top-philic mediator model are depicted in \cref{fig:fi_example} for two different choices of the mediator and \gls{dm} mass. We can see that \gls{dm} production through \gls{fi} is most efficient for $T\sim\mttilde$ (or $x\sim 1$) and levels off swiftly for $T\ll\mttilde$ due to the exponential suppression by the equilibrium yields (\cf\cref{eq:equilibriumyieldapprox}). Additionally, we notice that the $2\to 2$ processes become important at early times and for mass configurations where the mediator decay is phase-space suppressed. For $T\lesssim \mttilde/10$ (or $x\gtrsim 10$), the \gls{fi} contributions become negligible, allowing us to safely consider $\Yfi{\chi,\infty} \equiv \Yfi{\chi}(x=100)$ as the final yield from this production mechanism. If the Yukawa coupling $\lchi$ is very small, the amount of \gls{dm} produced from \gls{fi} (which can be calculated via \cref{eq:DMabundance}) is insufficient to account for the complete amount of \gls{dm} observed today. In this regime, another production mechanism becomes important, and as $\lchi$ decreases, it becomes dominant: the so-called superWIMP mechanism.\newp
Up to $T\gtrsim\mttilde/10$ (or $x\lesssim 10$), the mediator maintains thermal equilibrium with the \gls{sm} heat bath through sufficiently fast interactions with the \gls{sm} gauge bosons. However, for $T\ll \mttilde$ (or $x\gg 1$) these processes become ineffective, and the mediator eventually freezes out (\cf\cref{subsec:BEfiandfo}). Assuming further mediator decays to proceed long after the \gls{fo} process has been completed, the \gls{be} for a $\ttilde~\ttilde^*$ pair is described by \cref{eq:freeze-out} in this regime and reads 
\begin{equation}
	\label{eq:MFOwoSEwoBSF}
	\dv{\Yttbar}{x}=-\frac{1}{2}\xi_1(x)\sigmaannexp_{0}\left(\Yttbar^2-{\Yeq{\tilde{t}\tilde{t}^*}}^2\right)\,,
\end{equation}
where $\Yttbar=Y_{\ttilde}+Y_{\ttilde^*}$ (resulting in the additional factor of $1/2$), and the prefactor is given by $\xi_1(x)\equiv\sqrt{\pi/45}\MPl\mttilde\gstar^{1/2}(x) x^{-2}$. The thermally averaged annihilation cross section $\sigmaannexp_{0}$ incorporates all relevant $\ttilde+\ttilde^*\to \text{SM}+\text{SM}$ depletion processes of the mediator and the index $0$ indicates that we have not taken into account non-perturbative corrections yet. For the top-philic mediator, assuming $\lH=0$, the s-wave annihilation cross section reads 
\begin{equation}
	\label{eq:swaveanncrosssectionsttilde}
	\sigmaannexp_{0}^{\ttilde}\approx \sigma_0^{\ttilde}=\frac{14\pi \alphasann^2}{27\mttilde^2}+\frac{64\pi \alphaem\alphasann}{81\mttilde^2}\left[1+\left(1-\frac{m_Z^2}{4\mttilde^2}\right)\tan^2\theta_W\right]\,,
\end{equation}
with contributions from $g+g$, $g+\gamma$ and $g+Z$ final states.\footnote{Contributions from \eg $\gamma+\gamma$ or $Z+Z$ (for $\lH=0$) have been neglected since they are suppressed by $\alphaem^2$.} Allowing for $\lH>0$, the annihilation channels $t+\bar{t}$, $W^++W^-$ and $h+h$ can become sizeable. We have taken here the electromagnetic coupling $\alphaem\equiv\alphaem(m_Z)\approx 1/128$ measured at the $Z$ pole and evaluate the strong coupling $\alphasann\equiv\alpha_s(\mttilde)$ at the mediator mass scale (assuming the annihilating particles to be non-relativistic). In the lepto-philic mediator model, the dominant final states for $\lH=0$ are $\gamma+\gamma$, $\gamma+Z$ and $Z+Z$, resulting in an s-wave cross section of 
\begin{align}
    \label{eq:swaveanncrosssectiontautilde}
	\sigmaannexp_{0}^{\tautilde}\approx \sigma_0^{\tautilde}=&\frac{2\pi\alphaem^2}{\mtautilde	^2}+\left(1-\frac{m_Z^2}{4\mtautilde^2}\right)\frac{4\pi\alphaem^2}{\mtautilde^2}\tan^2\theta_W\nonumber\\
	&+\sqrt{1-\frac{m_Z^2}{\mtautilde^2}}\frac{8\mtautilde^4-8\mtautilde^2m_Z^2+3m_Z^4}{(m_Z^2-2\mtautilde^2)^2}\frac{\pi\alphaem^2}{\mtautilde^2}\tan^4\theta_W.
\end{align} 
For sizable $\lH>0$, the channels $W^++W^-$, $h+h$, and $t+\bar{t}$ become accessible at tree-level, with the first two potentially dominating. The co-annihilation processes which have been important for \gls{fi} production of \gls{dm}, do not play a role here because they are suppressed by the very small coupling $\lchi$.\newp
After \gls{fo}, the relic abundance of the mediator has to be depleted to avoid constraints from \gls{bbn}. For a Yukawa coupling $\lchi>0$, this is achieved through its decay, opening up another production mode for \gls{dm}, known as superWIMP mechanism \cite{Covi:1999ty,Feng:2003xh,Feng:2003uy}. In this mechanism, one \gls{dm} particle $\chi$ is produced for each frozen out mediator, resulting in a total dark matter yield given by $\Ysw{\chi,\infty}\equiv \Yttbar(x_p)$, where $x_p$ refers to a temperature regime after \gls{fo} where the mediator yield is stable and before decay processes take place. As long as the mediator abundance is fully depleted before \gls{bbn}, this mechanism is agnostic to the size of the coupling $\lchi$.\newp 
The final \gls{dm} yield is then given by a combination of \gls{fi} and \gls{sw} production, $Y_{\chi,\infty}=\Yfi{\chi,\infty}+\Ysw{\chi,\infty}$, where the relative importance of the two mechanisms is moderated by the value of the Yukawa coupling. When establishing the valid parameter space for our model in \cref{subsec:parameterspacenthprod}, this will enable us to fix one of the four free model parameters by comparing the calculated abundance (\cf\cref{eq:DMabundance}) to its experimental value observed today. Before proceeding, we investigate the influence of non-perturbative effects, which can affect the \gls{fo} process of the mediator and thus influence the \gls{sw} contribution to the overall dark matter abundance.


\subsection{Non-perturbative effects in dark matter production}
\label{subsec:nonpertubativeeffectsdmproduction}

The long-ranged interactions of the mediator with the \gls{sm} gauge fields give rise to non-perturbative effects in \gls{dm} production (\cf\cref{sec:non-perturbative-effects}). Since \gls{fo} typically happens in the non-relativistic regime where $\zeta\gg 1$, these effects can lead to sizeable corrections to the mediator yield after \gls{fo} (and therefore to the \gls{dm} abundance). Following the discussion in \cref{app:potential_for_different_mediators}, the potentials entering the Schrödinger equations (\cf\cref{eq:scatteringSchroedinger,eq:boundstateSchroedinger})
\begin{align*}
	\left[-\frac{1}{2\mu}\nabla^2+V(\vec{r})\right]\phik(\vec{r}) & =\Ek\phik(\vec{r})\,,\\
	\left[-\frac{1}{2\mu}\nabla^2+V(\vec{r})\right]\psinlm(\vec{r}) &=\Enl\psinlm(\vec{r})\,,
\end{align*}
are given by
\begin{equation}
    \label{eq:ttautildepotential}
    V_{\ttilde}(\vec{r})=-\frac{\alpha_g}{r}\,,\qquad V_{\tautilde}(\vec{r})=-\frac{1}{r}\left(\alpha_\gamma+\alpha_Z e^{-m_Z r}+\alpha_He^{-m_H r}\right)\,,
\end{equation}
for both mediator models. The form of the individual contributions and their couplings can be extracted from \cref{tab:potentials}. For the top-philic mediator model, we will only consider the dominating gluon potential, which is Coulomb-like in the perturbative regime, for both scattering and bound states \cite{Fischler:1977yf}. In the lepto-philic mediator model, we will also employ the $Z$ boson and Higgs potential for the scattering states as they can be sizeable for certain mass configurations of the \gls{dm} and mediator particle. For the \gls{bs} Schrödinger equation within this model, only the Coulomb potential is taken into account, which we justify in \cref{app:potential_for_different_mediators}.\newp
The energy eigenstates of the Schrödinger wave functions are given by 
$\Ek=k^2/(2\mu)$ for scattering and $\En=-\kappa^2/(2\mu n^2)$ for \gls{bs}, where $n$ denotes the principal quantum number and $l$ is degenerate. The reduced mass is given by $\mu=\mttilde/2$ and $k\equiv\abs{\vec{k}}=\mu\vrel$ denotes the average momentum transfer in scattering processes. The Bohr momentum of the bound state is defined as $\kappa=\mu \alphagB$, where $\alphagB$ denotes the interaction strength of the gluon potential (\cf\cref{tab:potentials}) evaluated at $\kappa$.\footnote{In case of the lepto-philic mediator the running of $\alphaem$ can be neglected.} Starting with a $\ttilde~\ttilde^*$ pair in the (anti)fundamental representation, from group theoretical arguments ($\mathbf{3}\otimes\mathbf{\bar{3}}=\mathbf{8}\oplus\mathbf{1}$) we can see that the resulting final state can either be in a color singlet or octet configuration. For a scattering state, both representations are possible, while for a \gls{bs} only the singlet state is populated since the color octet creates a repulsive potential, making it impossible for \gls{bs} to form. Therefore, $\alphagB\equiv \alphagBs=4/3\alphasBs$ with $\alphasBs\equiv \alpha_s(\mu \alphagBs)$.\newp
In the top-philic mediator model, the \gls{se} factors and the \gls{bsf} cross section can be calculated using analytical solutions for $\phik(\vec{r})$ and $\psinlm(\vec{r})$ as given in \cref{eq:scatteringwavefunctiondec,eq:boundstatewavefunctiondec,eq:ColumbscatteringwavefunctionRadial,eq:ColumbboundstatewavefunctionRadial}. For the lepto-philic mediator model, this is also possible for the \gls{bsf} cross section (as we neglect the $Z$ and Higgs contributions), for the \gls{se} factors, however, a numerical solution has to be employed (\cf\cref{subsubsec:SEannihilations,subsubsec:SEboundstatedecays}). Since this is a qualitative study on the impact of non-perturbative effects, we will restrict ourselves calculating only the dominating s-wave contributions to the corresponding quantities. With respect to \gls{bs}, we will for simplicity take into account only the ground state ($\{nlm\}\to\{100\}$) and neglect excited states as well as bound-to-bound transitions. For a quantitative study on higher bound state effects, the reader is referred to \eref\cite{Binder:2023ckj}.\newp
By taking into account these non-perturbative effects, the \gls{be}, given in \cref{eq:MFOwoSEwoBSF}, for mediator \gls{fo} is modified in two ways. Firstly, the thermally averaged annihilation cross section $\sigmaannexp_{0}\to\sigmaannexp$ receives corrections from \gls{se}. Secondly, we employ a separate \gls{be} for each bound state and couple it to \cref{eq:MFOwoSEwoBSF} through \gls{bsf} and ionization processes. The new system of \glspl{be} is then given by (\cf\eg\eref\cite{Garny:2021qsr})
\begin{align}
    \label{eq:BoltzmannYttbarnonpert}
    \dv{\Yttbar}{x}&=-\frac{1}{2}\xi_1(x)\sigmaannexp\left(\Yttbar^2-{\Yeq{\tilde{t}\tilde{t}^*}}^2\right)-\frac{1}{2}\xi_1(x)	\sigmabsfexp\Yttbar^2+2\xi_2(x)\Gammaionexp\YB\,,\\
    \label{eq:BoltzmannYBnonpert}
	\dv{\YB}{x}&=-\xi_2(x)\Gammadecexp\left(\YB-\Yeq{B}\right)+\frac{1}{4}\xi_1(x)\sigmabsfexp\Yttbar^2-\xi_2(x)\Gammaionexp\YB\,,
\end{align}
with $\YB$ the bound state yield of the ground state and $\xi_2(x)\equiv\xi_1(x)/s(x)$ denoting a second prefactor. The equilibrium yields in the non-relativistic regime ($x\gg 1$) are given by (\cf\cref{eq:equilibriumyieldapprox})
\begin{align}
	\Yeq{\tilde{t}\tilde{t}^*}(x)&=\frac{45g_{\tilde{t}}}{2\sqrt{2}\pi^{7/2}\heff(x)}x^{3/2}e^{-x}\,,\\
	\label{eq:YBeq}
	\Yeq{B}(x)&=\frac{g_B}{2g_{\tilde{t}}}\frac{\heff\left(\frac{m_B}{\mttilde}x\right)}{\heff(x)}\Yeq{\tilde{t}\tilde{t}^*}\left(\frac{m_B}{\mttilde}x\right)\,,
\end{align}
where $m_B=2\mttilde+\mathcal{E}_1$ is the mass of the ground state and $\mathcal{E}_1=-4/9\,\mttilde(\alphasBs)^2$ denotes its binding energy. The internal \gls{dof} are $g_{\ttilde}=3$ ($g_{\tautilde}=1$) for the individual mediator particles and $g_B=1$ for the bound ground state. In the regime where $x\gtrsim1$, we can set $\heff(m_B/\mttilde~x)/\heff(x)\approx 1$. In the following subsections, we will calculate $\sigmaannexp$ as well as the thermally averaged \gls{bsf} cross section $\sigmabsfexp$, the bound state ionization rate $\Gammaionexp$ and its decay rate $\Gammadecexp$ for the top- and lepto-philic mediator model.


\subsubsection{Dark matter annihilation cross section and bound state decay rate}
\label{subsubsec:sWSEann+decay}

Following the decomposition in \cref{eq:SEsplittingann}, the \gls{se} corrected s-wave contribution to the thermally averaged annihilation cross section in the top-philic mediator model is given by
\begin{equation}
    \sigmaannexp\approx \sum_i \sigma_{0}^i \expval{\Sannl{0}^{\scriptscriptstyle [\hat{\mathbf{R}}_i]}}\,,
\end{equation}
where $\sigma_{0}^i$ are the s-wave contributions to the annihilation cross section (\cf\cref{eq:swaveanncrosssectionsttilde}) split up by the color representation of their two-particle final state $\hat{\mathbf{R}}_i$ and the 
\begin{equation}
    \label{eq:Sann0Ri}
    \expval{\Sannl{0}^{\scriptscriptstyle [\hat{\mathbf{R}}_i]}}\equiv\frac{x^{3/2}}{2\sqrt{\pi}}\int_0^\infty\dd{\vrel} \vrel^2e^{-\frac{x}{4}\vrel^2}\Sannl{0}^{\scriptscriptstyle [\hat{\mathbf{R}}_i]}
\end{equation}
denote the thermally averaged s-wave \gls{se} factors. This distinction of representations is necessary because particles in different initial representations (which have to match the final state representations due to color conservation) will experience different potentials. Therefore, also the \gls{se} factor for a Coulomb potential as given in \cref{eq:SEfactorannCoulomb} depends on $\hat{\mathbf{R}}_i$ through $\alpha^{\scriptscriptstyle S}_{g,\scriptscriptstyle{[\hat{\mathbf{R}}_i]}}$ . Mediator annihilations into the final states $Z+Z$, $h+h$, $W^++W^-$ can only happen in the singlet configuration due to the lack of color-charge in the final state. Therefore, they have $\alphagSs=4/3\,\alphasS$ with $\alphasS\equiv\alpha_s(k)$ being evaluated at $k$. Annihilations into $\bar{t}+t$ are also approximated color singlets because the gluon mediated interactions are p-wave suppressed. The final states $g+Z$ and $g+\gamma$ are color octets and therefore possess a coupling of $\alphagSo=-1/6\,\alphasS$. The \gls{se} factors for pure singlet and octet potentials are given by
\begin{equation}
     \Sannl{0}^{\scriptscriptstyle [\mathbf{1}]}=S_0\left(\frac{4}{3}\zeta\right),\quad \Sannl{0}^{\scriptscriptstyle [\mathbf{8}]}=S_0\left(-\frac{1}{6}\zeta\right)\,,
\end{equation}
with $\zeta\equiv\alphasS /\vrel$, here. For the annihilation process $\ttilde+\ttilde^*\to g+g$, both color configurations are possible. One has to correct here for the treatment of the initial states in the computation of the matrix elements that enter $\sigma_0^i$ by weighting the \gls{se} factors \cite{DeSimone:2014qkh,ElHedri:2016onc} 
\begin{equation}
    \label{eq:Sann0rep1+8}
    \Sannl{0}^{\scriptscriptstyle [\mathbf{1}]+[\mathbf{8}]}=\frac{2}{7}S_0\left(\frac{4}{3}\zeta\right)+\frac{5}{7}S_0\left(-\frac{1}{6}\zeta\right).
\end{equation}
The lepto-philic mediator model just features one \gls{se} factor for all contributions $\sigma_0^i$ (\cf\cref{eq:swaveanncrosssectiontautilde}) to the annihilation cross sections. However, in this case the \gls{se} factor has to be calculated numerically due to the Yukawa potentials from Higgs and $Z$ boson exchange (\cf\cref{eq:ttautildepotential}). To accomplish this, we employ the technique presented in \cref{subsubsec:SEannihilations} using the differential equation for the radial part of the s-wave scattering wave function (\cf\cref{eq:scatteringSchroedingerRadial})
\begin{equation}
    \tilde{\chi}_{\vec{k},0}''(y)+\left(1+\frac{2}{y}\left[\zeta_\gamma+\zeta_Ze^{-b_Zy}+\zeta_He^{-b_Hy}\right]\right) \tilde{\chi}_{\vec{k},0}(y)=0\,,
\end{equation}
with $y=kr$, $b_i=m_i/k$, where we employ the initial conditions $\tilde{\chi}_{\vec{k},0}(y)\xrightarrow{y\to0}y$, $\tilde{\chi}_{\vec{k},0}'(y)\xrightarrow{y\to0} 1-(\zeta_\gamma+\zeta_Z+\zeta_H) y$.\footnote{The same parametrization as in \cref{eq:scatteringSchroedingerRadial} is not feasible here, due to the different coupling strengths of the potentials. However, this choice does not affect \cref{eq:SEannYukawa}.}\newp
For \gls{bs} decays, we can approximate the thermal average $\Gammadecexp\approx \Gammadec$ for $x\gtrsim 10$. In the top-philic mediator model we require the \gls{bs} to be a color singlet for reasons explained above. Applying the decomposition given in \cref{eq:SEsplittingdec}, we then obtain 
\begin{equation}
    \Gammadec\approx \frac{g_{\ttilde}^2}{g_B}\Sdecnl{\{100\}}{0}\sum_i w_{[\mathbf{1}]}^i\sigma^i_{0}\,,
\end{equation}
with $\Sdecnl{\{100\}}{0}=\kappa^3/\pi=8\mttilde^3(\alphasBs)^3/(27\pi)$ being the \gls{se} factor for the ground state decay rate when applying a Coulomb potential and $\sigma^i_{0}$ contains all contributions to the annihilation cross section where a singlet final state is possible. The factor $g_{\ttilde}^2/g_B$ corrects for the difference in \gls{dof} between the scattering and the \gls{bs} in the cross section due to color-charge, while $w_{[\mathbf{1}]}^i$ is a singlet weighting factor, which yields $2/7$ for mixed $[\mathbf{1}]+[\mathbf{8}]$ contributions (\cf\cref{eq:Sann0rep1+8}), $1$ for the purely $[\mathbf{1}]$ final states and $0$ for the pure $[\mathbf{8}]$ representations. In the lepto-philic mediator model, where \gls{bs} are only mediated by the Coulomb potential (see \cref{app:potential_for_different_mediators}), the picture is very similar with $g_{\tautilde}^2/g_B=1$, $\Sdecnl{\{100\}}{0}=\kappa^3/\pi=\alpha_\gamma^3\mtautilde^3/(8\pi)$ (and no weighting factor).


\subsubsection{Bound state formation cross section and ionization rate}
\label{subsubsec:sWSEBSF}

\begin{figure}
    \centering
    \includegraphics[width=\textwidth]{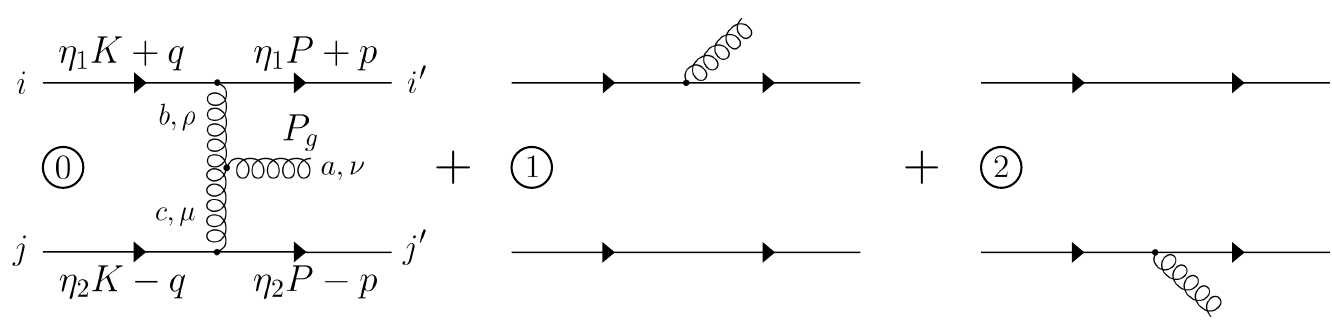}
    \caption[Leading order diagrams to $\Cfiveamp{g}$ in the top-philic mediator model.]{Leading order diagrams to $\Cfiveamp{g}$ in the top-philic mediator model. All three diagrams share the same color and Lorentz indices. The momenta of the incoming (outgoing) particles are displayed in terms of total and relative momenta of the system. In the lepto-philic mediator model (where we exchange the gluon with photon lines), the leftmost diagram is absent in the calculation of $\Cfiveamp{\gamma}$ due to the abelian nature of $U(1)_{\text{em}}$. Adapted from \eref\cite{Harz:2018csl}.}
    \label{fig:C5ampgluon}
\end{figure}

\begin{figure}
    \centering
    \includegraphics[width=\textwidth]{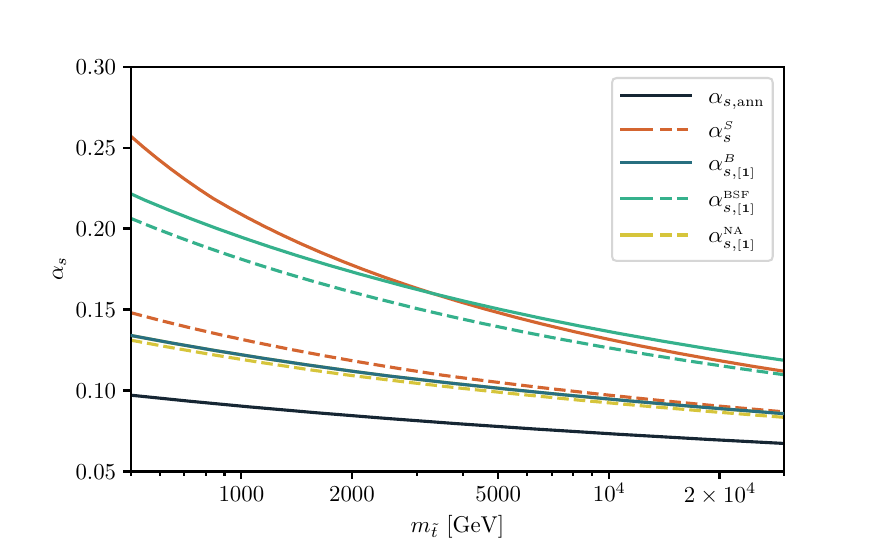}
    \caption[Mediator mass scaling of the strong couplings arising in the top-philic mediator model.]{Mediator mass scaling of the strong couplings arising in the top-philic mediator model as defined in the text. Velocity dependent quantities are evaluated at $\vrel=0.01$ ($\vrel=0.1$), which is indicated by the solid (dashed) lines. Note that scaling for the coupling $\alphasNA$ at $\vrel=0.01$ is not visible as it coincides with $\alphasBs$ for small velocities. The running of $\alpha_s$ has been implemented up to one loop order following the prescription in \eref\cite{Deur:2016tte}.}
    \label{fig:alphasQCD}
\end{figure}

To calculate the \gls{bsf} cross section for the top-philic mediator model, we will use the results obtained in \cref{subsubsec:BSFfromBetheSalpeter} (or \eref\cite{Petraki:2015hla}) and from there on revise the derivation in \eref\cite{Harz:2018csl} for non-abelian theories. First, we need to calculate the transition matrix element (\cf\cref{eq:Mtrans}) 
\begin{equation*}
	\left[\Mtrans(\vec{p},\vec{q})\right]^{\nu,a}_{ii',jj'}\equiv \mathcal{S}_0^{-1}(\vec{p},P)\mathcal{S}_0^{-1}(\vec{q},K)\int\frac{\dd{p^0}}{2\pi}\frac{\dd{q^0}}{2\pi}\left[\Cfiveamp{g}(q,p;K,P,P_g)\right]^{\nu,a}_{ii',jj'}
\end{equation*}
of the process $(\ttilde+\ttilde^*)_{[\mathbf{8}]}\rightarrow \mathscr{B}(\ttilde~\ttilde^*)_{[\mathbf{1}]}+g_{[\mathbf{8}]}$, where the color representation of the initial state has been fixed by the final state demands. $\Cfiveamp{g}$ denotes the sum of all connected diagrams contributing to the process 
\begin{equation}
    \chi_{1,i}(\eta_1K+q)+\chi_{2,j}(\eta_2K-q)\rightarrow \chi_{1,i'}(\eta_1P+p) +\chi_{2,j'}(\eta_2P-p) + g^a(P_g)
\end{equation}
where we relabelled $\ttilde\to\chi_1$, $\ttilde^*\to\chi_2$ for better comparability with \cref{subsec:nonperturbativeeffectsfromQFT}. The incoming and outgoing scalars $\chi_{1,2}$ are off-shell and carry color-charge indices $i$, $i'$, $j$ and $j'$. The radiated gluon with momentum $P_g$ and color-charge $a$ is on-shell and amputated, and the total and relative momenta $K$, $P$, $p$ and $q$ are defined equivalently to the ones in \cref{subsubsec:BSFfromBetheSalpeter}. To leading order in the strong coupling, the diagrams contributing to 
\begin{equation}
    \left[\Cfiveamp{g}\right]^{\nu,a}_{ii',jj'}=\left[\mathcal{C}_0\right]^{\nu,a}_{ii',jj'}+\left[\mathcal{C}_1\right]^{\nu,a}_{ii',jj'}+\left[\mathcal{C}_2\right]^{\nu,a}_{ii',jj'}
\end{equation}
are sketched in \cref{fig:C5ampgluon}. They yield 
\begin{align}
    i&\left[\mathcal{C}_0\right]^{\nu,a}_{ii',jj'}=\,\nonumber\\
    &\quad\tilde{S}_1(\eta_1P+p)[-i\gsna(T_1^b)_{ii'}(\eta_1K+\eta_1P+q+p)_\rho]\tilde{S}_1(\eta_1K+q)\frac{-i}{(\eta_1K+q-\eta_1P-p)^2} \nonumber\\
    &\quad\tilde{S}_2(\eta_2P-p)[-i\gsna(T_2^c)_{jj'}(\eta_2K+\eta_2P-q-p)_\mu]\tilde{S}_2(\eta_2K-q)\frac{-i}{(\eta_2K-q-\eta_2P+p)^2} \nonumber\\
    &\quad(-i\gsbsf f^{abc})\left\{g^{\rho\mu}[(\eta_1K+q-\eta_1P-p)-(\eta_2K-q-\eta_2P+p)]^\nu\right.\nonumber\\
    &\qquad\left.-g^{\nu\rho}[(\eta_1K+q-\eta_1P-p)+P_g]^\mu+g^{\mu\nu}[(\eta_2K-q-\eta_2P+p)+P_g]^\rho\right\}\,,
\end{align}
\begin{align}
    i\left[\mathcal{C}_1\right]^{\nu,a}_{ii',jj'}=\,&\tilde{S}_1(\eta_1P+p)[-i\gsbsf(T^a_1)_{ii'}(\eta_1K+\eta_1P+q+p)^\nu]\tilde{S}_2(\eta_1K+q)\tilde{S}_2(\eta_2K-q)\nonumber\\
    &\delta_{jj'}(2\pi)^4\delta^{(4)}(\eta_1K+q-\eta_1P-p-P_g)\,,\\
    i\left[\mathcal{C}_2\right]^{\nu,a}_{ii',jj'}=\,&\tilde{S}_1(\eta_2P-p)[-i\gsbsf(T^a_2)_{jj'}(\eta_2K+\eta_2P-q-p)^\nu]\tilde{S}_2(\eta_2K-q)\tilde{S}_1(\eta_1K+q)\nonumber\\
    &\delta_{ii'}(2\pi)^4\delta^{(4)}(\eta_2K-q-\eta_2P+p-P_g)\,,
\end{align}
where $\gsbsf$ refers to the coupling involved in gluon emission and $\gsna$ is the coupling between $\chi_{1,2}$ and an internal gluon in the first diagram.\footnote{One would expect $\mathcal{C}_0$ to be of higher order in the strong coupling. However, this is not the case due to the scaling of the relative momenta with $\alpha_s$ and $\vrel$ as will become clear in the following.} Following the approximations performed in \cref{subsubsec:BSFfromBetheSalpeter}, we can estimate the expressions above to leading order in $\alpha_s$ and $\vrel$ in the \gls{com} frame. 
To obtain the squared matrix element later, we need to consider the sum over the polarisation states of the radiated gluon
\begin{align}
    \label{eq:M2polarizationsumsimplification}
    \sum_{\text{pol.}}\abs{\epsilon_\mu\Mkn^\mu}^2&=\Mkn^{\mu\,*}\Mkn^\nu\sum_{\text{pol.}} \epsilon_\mu^*\epsilon_\nu=-g_{\mu\nu}\Mkn^{\mu\,*}\Mkn^\nu\nonumber\\
    &=\big|\Mknvec\big|^2-\big|\uvec{P}_g\cdot \Mknvec\big|^2.
\end{align}
The contribution coming from the $\mu,\nu=0$ component cancels the components parallel to $\uvec{P}_g$, such that we can express the final result in terms of the spatial entities only. Therefore, we can disregard the time-like $\nu=0$ component in the computation of $\Cfiveamp{g}$. The spatial parts simplify to 
\begin{align}
    \left[\vec{\mathcal{C}}_0\right]^{a}_{ii',jj'}\simeq& +if^{abc}(T_1^b)_{ii'}(T_2^c)_{jj'}~8\gsbsf (\gsna)^2 m \mu \frac{(\vec{q}-\vec{p})}{(\vec{q}-\vec{p})^4}S(q;K)S(p;P)\,,
\end{align}
\begin{align}
    \left[\vec{\mathcal{C}}_1\right]^{a}_{ii',jj'}\simeq&-\delta_{jj'}(T_1^a)_{ii'}\gsbsf(\vec{q}+\vec{p})S(q;K)\tilde{S}_1(\eta_1P+p)(2\pi)^4\delta^{(4)}(q-p-\eta_2P_g)\,,\\
    \left[\vec{\mathcal{C}}_2\right]^{a}_{ii',jj'}\simeq&+\delta_{ii'}(T_2^a)_{jj'}\gsbsf(\vec{q}+\vec{p})S(q;K)\tilde{S}_2(\eta_2P-p)(2\pi)^4\delta^{(4)}(q-p+\eta_1P_g)\,,
\end{align}
where we used the on-shell relation $K=P+P_g$ for the expressions in the $\delta$-distribution.\footnote{In $\vec{\mathcal{C}}_0$, the $g^{\rho\mu}$ term yields the dominant contribution. Also, the energy transfer along gluon propagators can be neglected for the same reasons as in \cref{app:SEderivation} using the instantaneous approximation.}
Using the approximations \cite{Petraki:2015hla}
\begin{align}
    \label{eq:propagatorintegralapprox1}
   \int\frac{\dd{p^0}}{2\pi}\frac{\dd{q^0}}{2\pi}S(q;K)\tilde{S}_1(\eta_1P+p)(2\pi)\delta(q^0-p^0-\eta_2P^0_g)\simeq 2m_2 \mathcal{S}_0(\vec{p},P)\mathcal{S}_0(\vec{q},K)\,,\\
   \label{eq:propagatorintegralapprox2}
   \int\frac{\dd{p^0}}{2\pi}\frac{\dd{q^0}}{2\pi}S(q;K)\tilde{S}_2(\eta_1P-p)(2\pi)\delta(q^0-p^0+\eta_1P^0_g)\simeq 2m_1 \mathcal{S}_0(\vec{p},P)\mathcal{S}_0(\vec{q},K)\,,
\end{align}
at $\vec{q}-\vec{p}-\eta_2\vec{P}_g=0$ (or $\vec{q}-\vec{p}+\eta_1\vec{P}_g=0$, respectively), the transition matrix element in the \gls{com} frame (\ie $\vec{K}=0$) can be written as 
\begin{align}
    \left[\Mtransvec(\vec{p},\vec{q})\right]^{a}_{ii',jj'}\simeq -4m \sqrt{4\pi\alphasBSFs} \Bigg\{&-if^{abc}(T_1^b)_{ii'}(T_2^c)_{jj'}8\pi\mu\alphasNA\frac{(\vec{q}-\vec{p})}{(\vec{q}-\vec{p})^4}\nonumber\\
    &+\eta_2(T_1^a)_{ii'}\delta_{jj'}~\vec{p}~(2\pi)^3\delta^{(3)}(\vec{q}-\vec{p}-\eta_2\vec{P}_g)\nonumber\\
    &-\eta_1\delta_{ii'}(T_2^a)_{jj'}~\vec{p}~(2\pi)^3\delta^{(3)}(\vec{q}-\vec{p}+\eta_1\vec{P}_g)\Bigg\}.
\end{align}
Note that we set $(\vec{q}+\vec{p})\delta^{(3)}(\vec{q}-\vec{p}\mp\eta_{2,1}\vec{P}_g)=(2\vec{p}\pm\eta_{2,1}\vec{P}_g)\delta^{(3)}(\vec{q}-\vec{p}\mp\eta_{2,1}\vec{P}_g)$ in the second and third term and neglected the $\vec{P}_g=\vec{P}$ contributions as they are of $\order{\alpha_s^2+\vrel^2}$. For a \gls{bs} in the singlet representation, the coupling strength $\alphasBSFs=\alpha_s(\omega)$ is evaluated at the gluon energy $\omega=\Ek-\En=\mu/2[\vrel^2+(\alphagBs/n)^2]$, whereas for $\alphasNA$, the average momentum transfer is $\abs{\vec{q}-\vec{p}}$, which yields $\alphasNA=\alpha_s\big(\mu\sqrt{\vrel^2+(\alphagBs)^2}\big)$ considering $\abs{\vec{q}}\sim k$ and $\abs{\vec{p}}\sim \kappa$. An overview on the different scaling of the strong couplings with the mediator mass adopted in this derivation, is given in \cref{fig:alphasQCD}. Employing \cref{eq:MknSEapprox}, the spatial matrix element in the non-relativistic limit is given by 
\begin{align}
    \left[\Mknvec\right]^a_{ii',jj'}=&-\sqrt{2^5\pi\alphasBSFs m^2/\mu}~\Big\{-if^{abc}(T_1^b)_{ii'}(T_2^c)_{jj'}\Ykn{n}\nonumber\\
    &+\eta_2(T_1^a)_{ii'}\delta_{jj'}\Jkn{n}(\eta_2\vec{P}_g)-\eta_1\delta_{ii'}(T_2^a)_{jj'}\Jkn{n}(-\eta_1\vec{P}_g)\big\}\,,
\end{align}
where we again neglected contributions from $\vec{p}^2$ and $\vec{q}^2$. The definition of the integrals $\Jkn{n}$ and $\Ykn{n}$, which compute different overlaps of the scattering and \gls{bs} wave functions over the momentum space, can be found in \cref{app:overlapintegrals}. These integrals will depend on the color representation of the incoming scattering and outgoing \gls{bs} through the different \gls{qcd} couplings involved. Since for $\Jkn{n}(\vec{b})$, $\abs{\vec{b}}\propto\abs{\vec{P}_g}$, the dominating contribution will be independent of $\vec{b}$. Therefore, we can set $\Jkn{n}\equiv\Jkn{n}(\vec{b}=0)$ in the following.\newp
To gradually match this so far rather general discussion with our model, we start by assuming that the initial state particles transform under conjugate color state representations, \ie $\mathbf{R}_1\equiv \mathbf{R}$ and $\mathbf{R}_2\equiv \mathbf{\bar{R}}$ (with $\mathbf{R}=\mathbf{3}$ in our case), such that $T_1^a=T^a$ and $T_2^a=-(T^a)^*$. Singling out the \gls{bs} singlet configuration by setting $i'=j'$, the matrix element yields
\begin{equation}
    \left[\Mknvec\right]^a_{ij}\equiv\delta_{i'j'}\left[\Mknvec\right]^a_{ii',jj'}=-\sqrt{\frac{2^5\pi\alphasBSFs m^2}{\mu}}\left[\Jkn{n}+\frac{C_2(\mathbf{G})}{2}\Ykn{n}\right]T_{ij}^a\,,
\end{equation}
where we employed the identity $f^{abc}T^bT^c=(i/2)C_2(\mathbf{G})T^a$ with $C_2(\mathbf{G})$ denoting the quadratic Casimir operator of the group $\mathbf{G}$ (taken to be $\mathbf{G}=SU(3)\equiv\mathbf{3}$ later). The squared, color-summed and averaged spatial matrix element is then given by
\begin{equation}
    \label{eq:spatialsquaredmatrixelement}
    \overline{\big|\Mknvec\big|^2}=\frac{1}{d_{\mathbf{R}}^2}\sum_{i,j,a}\abs{\left[\Mknvec\right]^a_{ij}}^2=\left(\frac{2^5\pi\alphasBSFs m^2}{\mu}\right)^2\frac{C_2(\mathbf{R})}{d_{\mathbf{R}}^2}\abs{\Jkn{n}+\frac{C_2(\mathbf{G})}{2}\Ykn{n}}^2\,,
\end{equation}
where $d_{\mathbf{R}}$ denotes the dimension and $C_2(\mathbf{R})\equiv T^a_{ij}T^a_{ji}$ is the quadratic Casimir operator of a given $\mathbf{R}$. Employing \cref{eq:M2polarizationsumsimplification}, the \gls{bsf} cross section reads (\cf\cref{eq:generalBSFcrosssection})
\begin{equation}
    \sigmabsf^{(n)}\vrel = \frac{(\Ek-\En)}{64\pi^2m^2\mu}\int\dd{\Omega}\left(\overline{\big|\Mknvec\big|^2}-\overline{\big|\uvec{P}_g\cdot \Mknvec\big|^2}\right).
\end{equation}
Considering just the ground state $n=\{100\}$ for which $\Ek-\mathcal{E}_{1}=\mu/2((\alphagBs)^2+\vrel^2)$, the overlap integrals in \cref{eq:spatialsquaredmatrixelement} are both proportional to $\uvec{k}$ at leading order in the expansion (\cf\cref{app:overlapintegrals}). Therefore, 
\begin{equation}
    \overline{\big|\vec{\mathcal{M}}_{\vec{k}\to\{100\}}\big|^2}-\overline{\big|\uvec{P}_g\cdot\vec{\mathcal{M}}_{\vec{k}\to\{100\}}\big|^2}=\overline{\big|\vec{\mathcal{M}}_{\vec{k}\to\{100\}}\big|^2}\sin^2\theta\,,
\end{equation}
where the squared spatial matrix element is independent of $\theta$, which denotes the angle between $\vec{k}$ and $\vec{P}_g$. Performing the orbital integration thus leaves us with 
\begin{equation}
    \sigmabsf^{\{100\},\ttilde}\vrel = \frac{(\alphagBs)^2+\vrel^2}{48\pi m^2}\overline{\big|\vec{\mathcal{M}}_{\vec{k}\to\{100\}}\big|^2}\,.
\end{equation}
Calculating $\Jkn{\{100\}}$, $\Ykn{\{100\}}$ to leading order (see \cref{app:overlapintegrals}) and defining $\zetas\equiv \alphagSo/\vrel$, $\zetab\equiv \alphagBs/\vrel$, the \gls{bsf} cross section into the ground state yields
\begin{align}
    \label{eq:sigmabsfvrel100ttilde}
    \sigmabsf^{\{100\},\ttilde}\vrel=&\frac{\pi\alphasBSFs\alphagBs}{\mu^2}\frac{2^7C_2(\mathbf{R})}{3d^2_{\mathbf{R}}}\left[1+\frac{C_2(\mathbf{G})}{2C_2(\mathbf{R})}\left(\frac{\alphasNA}{\alphasBs}\right)\right]^2\SBSF^{\ttilde}(\zetas,\zetab)\,,\\
    \implies\quad\sigmabsf^{\{100\},\ttilde}\vrel=&\frac{2^717^2}{3^5}\frac{\pi\alphasBSFs\alphasBs}{\mttilde^2}\SBSF^{\ttilde}(\zetas,\zetab)\,,
\end{align}
where we inserted the color factors $d_{\mathbf{R}}=3$, $C_2(\mathbf{R})=4/3$, $C_2(\mathbf{G})=3$ as well as the masses $\eta_1=\eta_2=1/2$, $\mu=\mttilde/2$ in the last step and approximated $\alphasNA\approx\alphasBs$ in our regime of interest ($\vrel\ll 1$; see \cref{fig:alphasQCD}). We also defined a \gls{se} factor for the \gls{bsf} process incorporating all terms which are $\zeta$-dependent, namely
\begin{equation}
	\SBSF^{\ttilde}(\zetas,\zetab)\equiv S_0(\zetas)\frac{(1+\zetas^2)\zetab^4}{(1+\zetab^2)^3}e^{-4\zetas\arccot(\zetab)}.
\end{equation}
As expected for $\zetas,\zetab \ll 1$, $\SBSF^{\ttilde}\to\zetab^4\ll 1$ is inefficient at large velocities. Moreover, at small velocities the \gls{bsf} cross section in the top-philic mediator model becomes also suppressed exponentially due to the repulsion of the scattering state, since $\zetas<0$ and thus $S_0(\zetas)\to 0$ for $\zetas\ll -1$.\newp
By taking the thermal average of the \gls{bsf} cross section, we can assume that the scattering state follows a Maxwell-Boltzmann distribution. However, since the gluon is relativistic, we need to take into account its statistical factor in the calculation. Therefore, the thermal average reads
\begin{align}
    \label{eq:avgBSFcrossection}
    \sigmabsfexp&=\frac{x^{3/2}}{2\sqrt{\pi}}\int_0^\infty \dd{\vrel}\vrel^2e^{-\frac{x}{4}\vrel^2}(1+f_g(\omega))\sigmabsf\vrel\nonumber\\
    &=\frac{x^{3/2}}{2\sqrt{\pi}}e^{\frac{x}{4}(\alphagBs)^2}\int_0^\infty \dd{\vrel}\vrel^2f_g(\omega)\sigmabsf\vrel\,,
\end{align}
where $f_g(\omega)=1/(e^{-\omega\,x/\mttilde}-1)$ is the gluon distribution function in the SM plasma. The ionization rate is characterized by the reverse process $ \mathscr{B}(\ttilde~\ttilde^*)_{[\mathbf{1}]}+g_{[\mathbf{8}]}\rightarrow (\ttilde+\ttilde^*)_{[\mathbf{8}]}$. It is related to the \gls{bsf} cross section through the principle of detailed balance which is in this context also referred to as Milne relation \cite{Harz:2018csl,Biondini:2023zcz}. The thermal average of the ionization rate can then be written as
\begin{equation}
    \label{eq:ionizationrate}
    \Gammaionexp= \frac{{\ndeq{\tilde{t}}}^2}{\ndeq{B}} \sigmabsfexp = \frac{g_{\ttilde}^2}{g_B}\frac{\mttilde^3}{(4\pi)^{3/2}}x^{-3/2}e^{-\frac{x}{4}(\alphagBs)^2}\sigmabsfexp.
\end{equation}\newp
The computation of the \gls{bsf} cross section in the lepto-philic mediator model will in principle be similar to the derivation shown above. Since the contributions from radiating a Higgs or $Z$ boson are highly suppressed (\cf\cref{app:potential_for_different_mediators}), we only take into account the process $\tautilde+\tautilde^*\rightarrow \mathscr{B}(\tautilde~\tautilde^*)+\gamma$, where the leading order contributions are structurally the same as in \cref{fig:C5ampgluon} with the first diagram missing due to the abelian nature of \gls{qed}. Instead of redoing the full calculation, we can simply modify \cref{eq:sigmabsfvrel100ttilde} in the following way to obtain the correct result: First, we take $\alphasNA\to 0$ to eliminate the non-abelian diagram, which is not present in the lepto-philic case. Then, we erase the effects of color-charge and the running of the strong coupling by neglecting the color-dependent factors $C_2(\mathbf{R}),d_{\mathbf{R}}\to 1$ and substituting $\alphasBSFs,\alphagBs\to \alpha_\gamma$ as well as $\zetas,\zetab\to\zetagamma$ with $\zetagamma=\alpha_\gamma/\vrel$. Eventually, the \gls{bsf} cross section in the lepto-philic mediator model yields
\begin{equation}
    \sigmabsf^{\{100\},\tautilde}\vrel=\frac{2^9}{3}\frac{\pi\alpha_\gamma^2}{\mtautilde^2}\SBSF^{\tautilde}(\zetagamma)\,,
\end{equation}
with 
\begin{equation}
    \SBSF^{\tautilde}(\zetagamma)\equiv S_0(\zetagamma)\frac{\zetagamma^4}{(1+\zetagamma^2)^2}e^{-4\zetagamma\arccot(\zetagamma)}.
\end{equation}
The behaviour of $\SBSF^{\tautilde}$ at large velocities is the same as for the top-philic mediator model. However, since $\zetagamma>0$, at small velocities the \gls{se} factor $\SBSF^{\tautilde}\propto 1/\vrel$ and thus the \gls{bsf} cross section will diverge in this limit. Since we are only considering thermally averaged quantities in our analysis, this does not pose a problem (see \cref{subsubsec:SEannihilations}). By performing this matching of the results from the top-philic to the lepto-philic mediator model, we did not consider the contributions of the Higgs and $Z$ boson within the scattering state wave function $\phik$, as they would require a numerical treatment. However, due to the mass suppression of the heavy bosons, these corrections to the $\Jkn{n}$ integral are small and can therefore be neglected. The thermal average to the \gls{bsf} cross section and ionisation rate has been computed in analogy to the top-philic mediator model.


\subsection{Dark matter abundance in the superWIMP mechanism} 
\label{subsec:DMabundancesW}

As discussed in \cref{subsec:nonthermalprodmechanisms}, the \gls{dm} abundance in the \gls{sw} mechanism is obtained via the mediator yield after \gls{fo}. The yield is governed by the \gls{be} (\cf\cref{eq:MFOwoSEwoBSF}), or, respectively, a system of coupled \gls{be} given in \cref{eq:BoltzmannYttbarnonpert,eq:BoltzmannYBnonpert} 
\begin{align*}
    \dv{\Yttbar}{x}&=-\frac{1}{2}\xi_1(x)\sigmaannexp\left(\Yttbar^2-{\Yeq{\tilde{t}\tilde{t}^*}}^2\right)-\frac{1}{2}\xi_1(x)	\sigmabsfexp\Yttbar^2+2\xi_2(x)\Gammaionexp\YB\,,\\
	\dv{\YB}{x}&=-\xi_2(x)\Gammadecexp\left(\YB-\Yeq{B}\right)+\frac{1}{4}\xi_1(x)\sigmabsfexp\Yttbar^2-\xi_2(x)\Gammaionexp\YB,
\end{align*}
if the non-perturbative effects outlined in \cref{subsec:nonpertubativeeffectsdmproduction} are taken into account. Since these types of equations lack analytical solutions, we must resort to numerical methods to evaluate the mediator yield. Solving the coupled system numerically, in particular around the \gls{fo} temperature near $x\sim 20$, poses a challenge. Fortunately, within this regime we can simplify our calculations significantly. We will employ in the following two distinct approaches to validate the accuracy of our results.\newp
The first idea has been proposed in \eref\cite{Ellis:2015vaa}. A more recent study has shown that its validity may persist even if its primary assumptions are relaxed \cite{Binder:2021vfo}. It extends the method to higher \gls{bs} and accounts for \gls{bs} transitions, which we will disregard in this analysis. We start by presuming that at high temperatures \gls{bsf} and ionization processes are very efficient, while at smaller temperatures \gls{bs} decays are dominant. In both regimes, the rates are sufficiently high to assume quasi-steady state conditions for the \gls{bs}, where $\dd{\YB}/\dd{x}\approx 0$. Utilizing \cref{eq:ionizationrate}, we can rewrite \cref{eq:BoltzmannYBnonpert} to express the bound state yield as
\begin{equation}
    \YB\approx\frac{\Gammadecexp+\Yttbar^2/{\Yeq{\tilde{t}\tilde{t}^*}}^2\Gammaionexp}{\Gammadecexp+\Gammaionexp}\Yeq{B}.
\end{equation}
Through substitution into \cref{eq:BoltzmannYttbarnonpert}, we can describe the evolution of the mediator particle using a single \gls{be} even in the presence of non-perturbative effects. The newly obtained equation is given by
\begin{align}
    \dv{\Yttbar}{x}&=-\frac{1}{2}\xi_1(x)\left[\sigmaannexp+\sigmabsfexp\frac{\Gammadecexp}{\Gammadecexp+\Gammaionexp}\right]\left(\Yttbar^2-{\Yeq{\tilde{t}\tilde{t}^*}}^2\right)\nonumber\\
    &\equiv -\frac{1}{2}\xi_1(x)\Big[\sigmaannexp+\sigmabsfeffexp\Big]\left(\Yttbar^2-{\Yeq{\tilde{t}\tilde{t}^*}}^2\right) \nonumber\\
    &\equiv -\frac{1}{2}\xi_1(x)\sigmaanntotexp\left(\Yttbar^2-{\Yeq{\tilde{t}\tilde{t}^*}}^2\right)\,,
    \label{eq:MFOwSEwBSF}
\end{align}
and turns out to be structurally equivalent to \cref{eq:MFOwoSEwoBSF}, where no non-perturbative effects are present. 
\begin{figure}
    \centering
    \includegraphics[width=\textwidth]{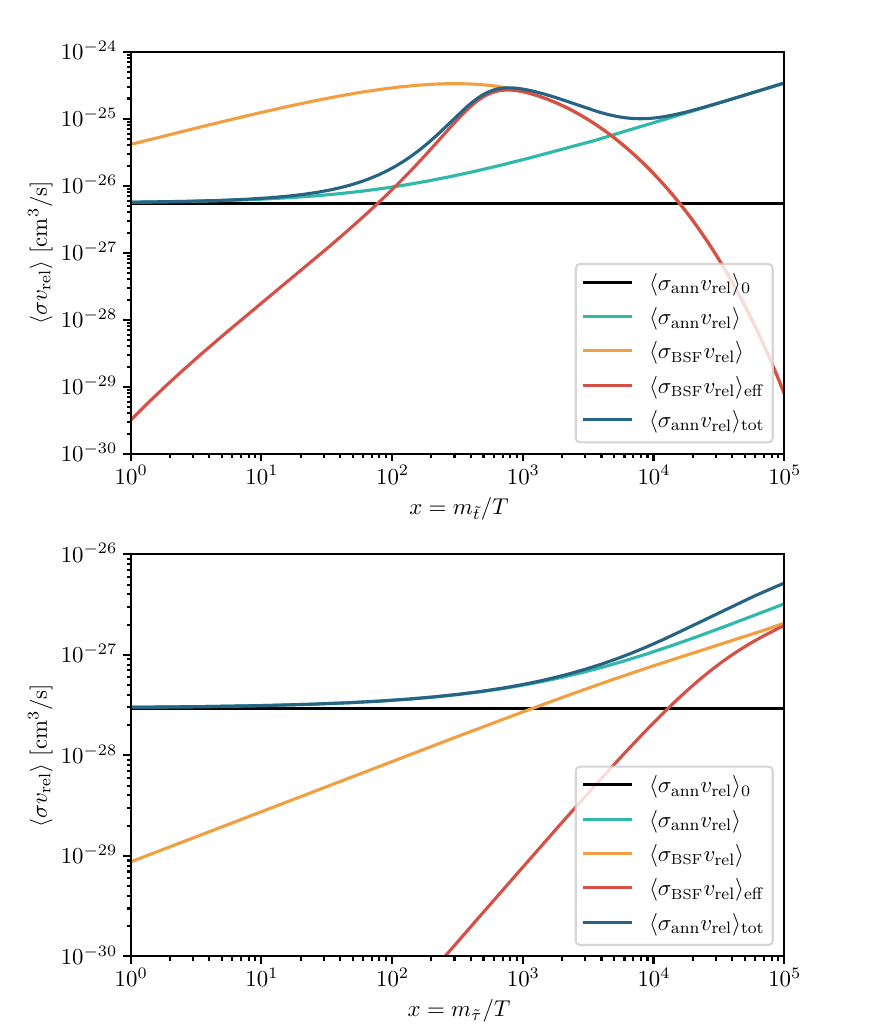}
    \caption[Contributions to the overall annihilation cross section when considering non-perturbative effects.]{Contributions to the overall annihilation cross section, defined in \cref{eq:MFOwSEwBSF}, when considering non-perturbative effects as a function of the mediator mass over temperature. Also displayed is the perturbative cross section $\sigmaannexp_0$ (black) for comparison. The results are presented for a top-philic (top) and a lepto-philic (bottom) mediator model at an exemplary mediator mass of $\SI{5}{\TeV}$.}
    \label{fig:avgcrosssectionsnonth}
\end{figure}
Under this approximation, the inclusion of \gls{se} factors and the presence of \gls{bs} thus leads to a modification of the thermally averaged cross section within the framework of an ordinary \gls{be} describing \gls{fo}. The \textit{total annihilation cross section} $\sigmaanntotexp$ consists of the sum of \gls{se} corrected annihilation and \gls{bsf} contributions. The latter is weighted by a factor which accounts for the depletion of \gls{bs} into the dark sector through ionization. This term suppresses the \textit{effective \gls{bsf} cross section}, denoted as $\sigmabsfeffexp$, particularly at high temperatures when ionization processes are rapid, and tends to unity when the temperature drops below the energy of the \gls{bs}. This behaviour can be observed in \cref{fig:avgcrosssectionsnonth}, where we compare the different contributions to the total annihilation cross section for the top-philic (top) and lepto-philic (bottom) mediator model at an exemplary mediator mass of $\SI{5}{\TeV}$. In the top-philic mediator model, both \gls{se} and \gls{bs} corrections noticeably increase the overall cross section shortly after decoupling, potentially exerting a significant impact on the mediator yield through non-perturbative effects. In the lepto-philic mediator model, we anticipate a sizeable impact from \gls{se} corrections, whereas contributions from \gls{bsf} are expected to be insignificant as they occur much later. \newp
Another approach to simplify the coupled system of \glspl{be} is to assume Saha equilibrium (see \eg\eref\cite{Peter:2013avv} and \eref\cite{saha1921physical} for the original paper) at the time, where the mediator starts to deviate from its equilibrium value around $x\sim 20$. This assumption posits that ionization processes are rapid enough to maintain chemical equilibrium between bound and scattering states, ensuring that their relative abundances remain constant, \ie $n_B/\ndeq{B}=n_{\tilde{t}}^2/{\ndeq{\tilde{t}}}^2$ or $\YB=R(x)s(x)\Yttbar^2$ with $R(x)\equiv \ndeq{B}/{\ndeq{\tilde{t}}}^2$. We checked this condition to hold up to $x\lesssim 30$. Defining further $\bar{x}\equiv\YB/\Ytot$ as the fraction of the \gls{bs} yield compared to the total yield of the free mediator and \gls{bs} $\Ytot\equiv \Yttbar+\YB$, we can use the Saha equilibrium condition to rewrite the \gls{be} for the total yield
\begin{equation}
    \label{eq:BSFBoltzmannSaha}
    \dv{\Ytot}{x}=-\frac{1}{2}\xi_1(x)\expval{\sigma_{\text{eff}}\vrel}\left((1-\bar{x})^2\Ytot^2-{\Yeq{\tilde{t}\tilde{t}^*}}^2\right)\,,
\end{equation}
through a combination of \cref{eq:BoltzmannYttbarnonpert,eq:BoltzmannYBnonpert} with $\expval{\sigma_{\text{eff}}\vrel}\equiv\sigmaannexp+2R(x)\Gammadecexp$. At early times, when $\bar{x}\ll 1$, we can safely approximate $(1-\bar{x})^2\approx 1$ and take $\Ytot(x_0)=\Yeq{\tilde{t}\tilde{t}^*}(x_0)$ at $x_0=1$ as an initial condition. We solve \cref{eq:BSFBoltzmannSaha} up to $x_1=30$ and subsequently transition to solving the coupled system of \glspl{be} (\cref{eq:BoltzmannYttbarnonpert,eq:BoltzmannYBnonpert}). The initial conditions for the full system are taken to be $\Yttbar(x_1)=\Ytot(x_1)$ and $\YB(x_1)$ being defined through the Saha relation. The final yields obtained through this method closely match those obtained using the steady-state approximation $\dd{\YB}/\dd{x}\approx 0$, with differences of less than $1\%$ within our range of interest. Hence, we will use the method proposed in \eref\cite{Harz:2018csl} going forward, as it offers a simpler applicability.\newp
\begin{figure}[ht]
\centering
    \includegraphics[width=\textwidth]{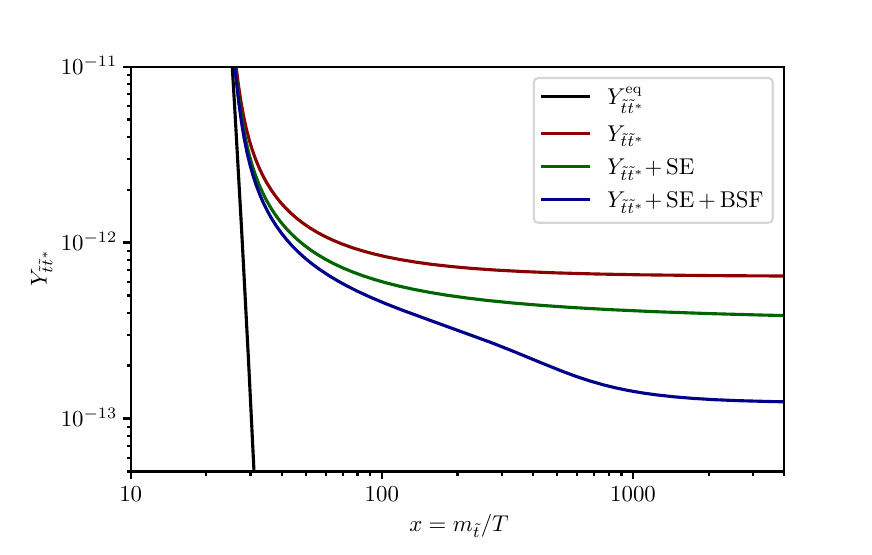}
    \caption[The evolution of the mediator yield in the top-philic mediator model as a function of $x\equiv\mttilde/T$.]{The evolution of the mediator yield in the top-philic mediator model as a function of $x=\mttilde/T$ for an exemplary mass of $\mttilde=\SI{5}{\TeV}$ and $\lH=0$. For comparison, the yield of the perturbative computation (red) is displayed alongside an only \gls{se} corrected (green) and an \gls{se} and \gls{bsf} corrected result (blue). The equilibrium yield is shown in black. Adapted from \eref\cite{Bollig:2021psb}.}
    \label{fig:foexample}
\end{figure}
We have presented in \cref{fig:foexample} the evolution of the mediator yield in the top-philic mediator model for a benchmark point of $\mttilde=\SI{5}{\TeV}$ and $\lH=0$. Three different scenarios are compared here: 1) the solution to \cref{eq:MFOwoSEwoBSF}  using only the perturbative \gls{dm} cross section $\sigmaannexp_{0}\approx\sigma_0$ (red), 2) the solution taking into account the \gls{se} factors in the annihilation cross section $\sigmaannexp$ (green), and 3) the solution to \cref{eq:MFOwSEwBSF} using $\sigmaanntotexp$ with \gls{se} and \gls{bsf} effects present (blue). This example illustrates the significant impact of \gls{se} and \gls{bsf} effects on the mediator yield. While \gls{se} effects become significant shortly after deviation from equilibrium, \gls{bsf} effects start to influence the yield at $x\sim\order{10^2-10^3}$. This behavior aligns with our expectations when comparing it to \cref{fig:avgcrosssectionsnonth}. Both effects lead to a significant increase in the effective cross section around $x\sim 100$ which in turn depletes the yield, a behavior which has been observed previously \eg in coannihilation scenarios \cite{Harz:2018csl,Biondini:2018pwp,Biondini:2018ovz}. Considering only \gls{bsf} into the ground state, the mediator has frozen out at $x_p\gtrsim 1000$ which we will take as a reference point to define $\Ysw{\chi,\infty}$ (\cf\cref{subsec:BEfiandfo,subsec:nonthermalprodmechanisms}).\footnote{This calculation will serve as a lower limit to \gls{bsf} effects on the mediator yield since it has been shown \eg in \eref\cite{Binder:2023ckj}, that the inclusion of higher \gls{bs} as well as \gls{bs} transitions will deplete the abundance even further. However, this might lead to a scenario where \gls{fo} and decay of the mediator are no longer separable, a scenario which is beyond the scope of this work.} By utilizing \cref{eq:DMabundance}, we can exclude combinations of $\mchi$, $\mttilde$, and $\lH$ that result in abundances $\Omegadm^{\text{sW}}>\Omegadmt$, given that later modifications to the cosmological history of the Universe are absent (\eg late entropy production \cite{Co:2015pka,Calibbi:2021fld}). Abundances resulting from \gls{sw} production yielding $\Omegadm^{\text{sW}}<\Omegadmt$ are acceptable as they can be supplemented with contributions from \gls{fi} production. Assuming that these two mechanism make up for the entire \gls{dm} abundance we observe today, we can define the required \gls{fi} fraction for a given \gls{sw} abundance as $p\equiv 1-\Omegadm^{\text{sW}}/\Omegadmt$. Given $\mttilde$, $\lH$, and a fixed $p$, the required \gls{dm} mass can be calculated via (\cf\cref{eq:DMabundance}) 
\begin{equation}
    \label{eq:mchiofmtt}
     \mchi=\frac{(1-p)\Omegadmt\rhocrit}{s_0\Ysw{\chi,\infty}}\,.
\end{equation}
We can easily see that it is largest, if all \gls{dm} is produced through the \gls{sw} mechanism and decreases linearly with $p>0$. Conversely, if a \gls{dm} mass $\mchi$ (alongside $\mttilde$ and $\lH$) is given, we are able to determine $p$, which also fixes $\lambda_\chi$ via
\begin{equation}
    \label{eq:lambdachi}
     \lchi=\sqrt{\frac{p}{(1-p)}\frac{\Ysw{\chi,\infty}}{\Yfi{\chi,\infty}}}\,.
\end{equation}
As anticipated from \cref{fig:avgcrosssectionsnonth}, in the lepto-philic mediator model \gls{bsf} plays a subordinate role compared to \gls{se} in determining the mediator yield due to the much smaller binding energy of electromagnetic interactions. Consequently, the overall corrections due to non-perturbative effects are less than $15\%$, and therefore, they are not explicitly shown.


\subsection{Model constraints from colliders and cosmological observations} 
\label{subsec:constraintsnth}

\begin{figure}
    \centering
    \includegraphics[width=\textwidth]{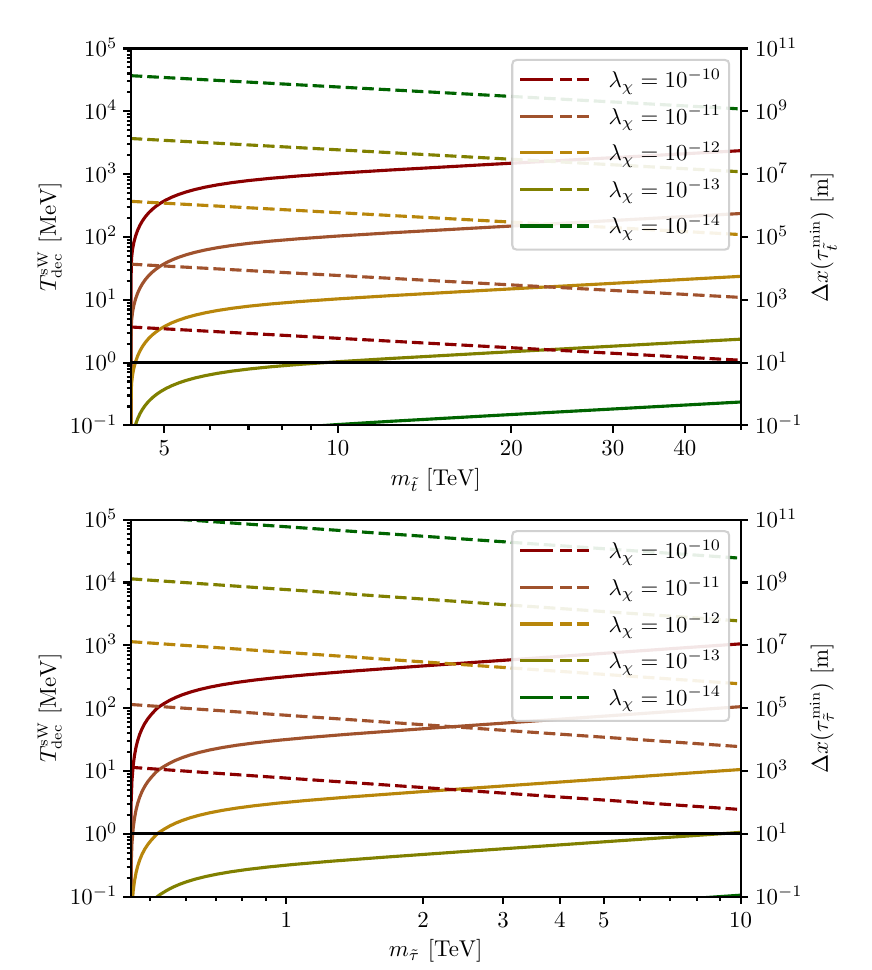}
    \caption[Lower limits on the decay temperature and decay length of the mediator as a function of its mass in the top-philic and lepto-philic mediator model for various couplings $\lchi$.]{Lower limits on the decay temperature (solid lines) and decay length (dashed lines) of the mediator as a function of its mass in the top-philic (top) and lepto-philic (bottom) mediator model for various couplings $\lchi$. The decay temperature $T^{\text{\scriptsize{sW}}}_{\text{dec}}$ has been calculated via \cref{eq:decaytemperature} assuming an \gls{sw} contribution only and can be compared to the bath temperature $T_{\text{\tiny{BBN}}}$ at the onset of \gls{bbn} (black line). The decay length after production at the \gls{lhc} is given via \cref{eq:decaywidthcollider} for a minimal mediator lifetime (see text) and is compared to a detector length of $\Delta x_{\text{\tiny{DL}}}=\SI{10}{\metre}$ (black line).}
    \label{fig:constraintsDeltaxTdec}
\end{figure}

Apart from overclosing the Universe (\ie $\Omegadm^{\text{sW}}>\Omegadmt$), there are a few more observations which will constrain the parameter space of our models. We will start with limits from collider searches. If the masses of the mediators are small enough, they can be produced \eg in proton-proton collisions through Drell-Yann-like processes due to their \gls{sm} gauge couplings. Production through Higgs interactions are also possible but suppressed even for sizeable values of $\lH$ \cite{Hessler:2014ssa}. Especially the top-philic mediator has a large production cross section of $\sigma_{\ttilde\ttilde^*}\simeq\SI{1}{\femto\barn}$ for $\mttilde\sim\SI{1300}{\GeV}$ which make various classes of \gls{dm} searches at the \gls{lhc} sensitive to it.\footnote{Correspondingly, for the lepto-philic mediator, a cross section of $\sigma_{\tautilde\tautilde^*}\simeq\SI{0.1}{\femto\barn}$ for a mediator mass of $\mtautilde\sim\SI{500}{\GeV}$ can be estimated.} Moreover, a small coupling of $\lchi\ll 10^{-8}$ as estimated from \cref{eq:estimate_non_thermal}, corresponds to a long lifetime of the mediator, which can render it long-lived or even stable on collider scales. To test this presumption, we estimate the mediator decay length within a collider experiment via
\begin{equation}
    \label{eq:decaywidthcollider}
	\Delta x \simeq \beta\gamma~\tau_{\ttilde}\,,
\end{equation}
with $\tau_{\ttilde} \equiv 1/\Gamma_{\ttilde}$ its lifetime and $\beta$ the velocity of the decaying particle (in natural units), boosted by a Lorentz factor $\gamma$. For the velocity factors, we choose a typical value of $\beta\gamma=\sqrt{1-\mttilde^2/E_{\ttilde}^2}\sim 0.3$ for heavy particles produced at the \gls{lhc}.\footnote{We checked the validity of this choice by estimating the range of $\beta\gamma$, applying a simplified $t$-channel model from MicrOMEGAs \cite{Belanger:2001fz,Belanger:2013oya} to a \gls{dm} production simulation performed in CalcHEP \cite{Belyaev:2012qa}.} Considering only the $1\to 2$ process as the dominant contribution and neglecting the top mass, we can set a lower limit on the lifetime of the mediator, yielding $\tau_{\ttilde}\geq 32\pi/(\mttilde\lchi^2)\equiv \tau_{\ttilde}^{\text{min}}$ for  $\mchi\to0$ (and respectively larger for $\mchi>0$). For mediator masses of $\order{\SI{1}{\TeV}}$, we observe in \cref{fig:constraintsDeltaxTdec} (dashed lines) that within our regime of interest ($\lchi\ll 10^{-8}$), we always obtain $\Delta x \gg \SI{10}{\metre}$ in both models. 
For the values of $\lchi$ preferred by \gls{fi} and a realistic distribution of $\beta\gamma$ factors, we checked that even $\Delta x \geq \SI{100}{\metre}$ throughout our parameter space, implying that we can treat the mediator to be stable on collider scales. \newp
A heavy, non-relativistic, stable and charged mediator will leave distinct signatures at the \gls{lhc} experiments ATLAS and CMS. These are characterized by an unusual energy loss in the calorimeters and a low velocity measured in the tracking system \cite{ATLAS:2019gqq,CMS-PAS-EXO-16-036}. We will apply in the following the most stringent limits on a stop-like R-hadron (for a top-philic mediator) as well as on a directly produced stau (for a lepto-philic mediator), imposed by an ATLAS search \cite{ATLAS:2019gqq}, which is based on an integrated luminosity of $\SI{36.1}{\femto\barn^{-1}}$ at a collider energy of $\sqrt{s}=\SI{13}{\TeV}$. The measurements for the top-philic mediator exclude masses up to $\SI{1345}{\GeV}$ at $95\%$ \gls{cl}, while those for the lepto-philic mediator rule out masses above $\mtautilde\gtrsim \SI{430}{\GeV}$. We expect these limits to strengthen in the future up to $\mttilde\lesssim\SI{1600}{\GeV}$ and $\mtautilde\lesssim \SI{600}{\GeV}$, respectively, after the \gls{hl} upgrade of the \gls{lhc}, assuming a similar efficiency as today and an integrated luminosity of $\SI{3000}{\femto\barn^{-1}}$. The supersymmetric theoretical models employed in the analysis of \eref\cite{ATLAS:2019gqq} are very similar to our own models, although not identical. Particularly, one might worry that contributions from supersymmetric partners, which are not present in our model, lead to different results. Therefore, we performed a reevaluation of the theoretical mediator production cross section using our model with CalcHEP \cite{Belyaev:2012qa} and compared it with the predictions in \eref\cite{ATLAS:2019gqq}. The results agree well, which suggests a limited effect of the supersymmetric model on these observables. This validates the use of these results for our objectives.\newp
Other than collider experiments, there are several constraints emerging from cosmological observations. In our parameter regime of interest, the most important ones stem from \gls{bbn}. A sufficiently long-lived mediator could decay during the formation of the light elements and inject energy into the \gls{sm} plasma, or form bound states with \gls{sm} particles triggering unexpected nuclear reactions \cite{Jedamzik:2007qk,Kawasaki:2017bqm}. If sufficiently effective, both processes will spoil the delicate balance of the nuclear reactions and alter the primordial abundances of the light elements. In order to avoid these kinds of complications, one usually demands the vast majority of the mediator particles to decay before the relevant stage of \gls{bbn}, such that the remnants are too dilute to affect the nuclear reactions. We will therefore require the lifetime of the mediator to be shorter than $\tau_{\ttilde}=1/\Gamma_{\ttilde}\lesssim \SI{10}{\second}$ which corresponds to a bath temperature of $T_{\text{\tiny{BBN}}}\sim\SI{1}{\MeV}$ at the time of decay \cite{Kawasaki:2017bqm}. Utilizing the time-temperature relation in a radiation dominated Universe as given in \cref{eq:timetemperaturerelation} (neglecting the slight differences in values of $\geff$, $\heff$, and $\gstar$, fixing them to $\geff=10.75$ at $\SI{1}{\MeV}$), we can define a corresponding decay temperature of the mediator as a function of the decay width
\begin{equation}
    \label{eq:decaytemperature}
	T_{\text{dec}}\simeq \sqrt{0.301\geff^{-1/2}\MPl\Gamma_{\ttilde}(\mttilde,\mchi,\lchi)}\,,
\end{equation}
which can be compared to $T_{\text{\tiny{BBN}}}$. If kinematically allowed, the two-body decay $\ttilde \rightarrow t_R + \chi$ (or $\tautilde \rightarrow \tau^+_R + \chi$) will dominate the decay width and usually lead to a sufficiently fast decay. We have displayed its impact on the decay temperature in \cref{fig:constraintsDeltaxTdec} (solid lines) for a variety of $\lchi$ assuming \gls{sw} production only ($p=0$), which will serve us as a lower limit. An \gls{fi} contribution ($p>0$) will result in lower \gls{dm} masses (\cf\cref{eq:mchiofmtt}) which leads to higher decay rates. As we can see, for couplings $\lchi\gtrsim 10^{-12}$, bounds from \gls{bbn} can be naturally avoided. If two-body decays are not kinematically allowed in the top-philic mediator model, we will consider the dominant three-body decay $\ttilde\rightarrow W^+ +b+\chi$, which vastly increases the lifetime of the mediator for a fixed $\lambda_\chi$. This process has to be calculated numerically and we have included it in our analysis below the threshold of the two-body decay. For the lepto-philic mediator, this procedure is unnecessary for reasons explained in \cref{subsec:parameterspacenthprod}.\newp
Another category of cosmological constraints can become important, when the velocity of \gls{dm} after decoupling is high enough to affect structure formation. This occurs, when the closely related \gls{dm} free-streaming length (in this context, the average distance a \gls{dm} particle can travel without scattering after matter-radiation equality) becomes larger than the size of primordial density fluctuations, such that \gls{dm} can escape overdensities and prevents structure on these scales to grow. In the literature, these so-called \gls{wdm} bounds are presented in terms of the \gls{dm} mass $\mwdm$ assuming a thermal velocity distribution. However, for comparison with our model predictions, it will be more convenient to work with the \gls{rms} velocity of \gls{dm} today, which can be related to the mass limit via \cite{Bode:2000gq,Barkana:2001gr}
\begin{align}
     \vrms\approx 0.04 \left(\frac{\Omegadmt h^2}{0.12}\right)^{1/3}  \left(\frac{\mwdm}{\SI{1}{\keV}}\right)^{-4/3}\frac{\si{\km}}{\si{\s}}\,.
\end{align}
Numerous astrophysical observations can be used to constrain $\mwdm$, typically confining it to the $\si{\keV}$ regime (see \eg\erefs\cite{Irsic:2017ixq,Dekker:2021scf,Hsueh:2019ynk,Gilman:2019nap}). We will employ in the following a limit from \eref\cite{Irsic:2017ixq} of $\mwdm \geq \SI{3.5}{\keV}$ or $\vrms\leq\SI{7.5}{\m/\s}$, respectively, which is based on a \textit{Lyman-$\alpha$ forest} data analysis.\newp
The mean velocity of a (non-relativistic) \gls{dm} particle today is given by
\begin{align}
    \label{eq:v0WDMbound}
   v_0=\frac{p_0}{\mchi}=\frac{p_{\text{prod}}}{\mchi}\frac{a_{\text{prod}}}{a_0}=\frac{p_{\text{prod}}}{\mchi}\left(\frac{\heff(T_0)}{\heff(\Tprod)}\right)^{1/3}\frac{T_0}{\Tprod}\,,
\end{align}
with $p$ its momentum and the subscript $0$ ($\text{prod}$) refers to the values of the corresponding entities today (at production). Note that we assumed no efficient (self-)interaction processes of \gls{dm} after its production and applied entropy conservation in the last step. Given these assumptions, \cref{eq:v0WDMbound} holds irrespective of the underlying \gls{dm} model. For simplicity, we will use in the following $v_0$ instead of the \gls{rms} velocity when calculating the \gls{wdm} mass bound, since a conversion of $v_0$ to $\vrms$ would only yield a $\order{10\%}$ correction. Considering \gls{dm} produced by \gls{fi}, the mean \gls{dm} momentum at production is simply $\Tprod/2$ (for a mediator decaying at rest) times an $\order{1}$ factor, given that $\mttilde\gg \mchi+m_t$. Taking the proportionality factor to be $2.5$ \cite{Heeck:2017xbu} and estimating $\heff(\Tprod)\sim 100$, the mean velocity of \gls{dm} from \gls{fi} is approximately
\begin{align}
    v^{\text{FI}}_{0} \approx 30\left(\frac{\SI{1}{\keV}}{m_\chi}\right) \frac{\mbox{m}}{\mbox{s}}\,,
\end{align}
which can be translated to a lower \gls{dm} mass limit of $\mchi\gtrsim\SI{4}{\keV}$.\footnote{For comparison, a limit of $m_\chi \gtrsim \SI{15}{\keV}$ has been stated in \cite{Decant:2021mhj} by performing a full modeling of the transfer function for \gls{fi}.}\newp
For \gls{dm} produced via the \gls{sw} mechanism, the \gls{dm} momentum is given by the kinematics of the non-relativistic frozen out mediator at its decay. For simplicity, we will model the mediator decay as instantaneous, such that we can compute the production temperature through $\Gamma_{\ttilde}=H|_{\Tprod}$ (see \eg\eref\cite{Jedamzik:2005sx}). Determining $\mchi$ through \cref{eq:mchiofmtt} with $p=0$ (\gls{sw} production only), we can employ \cref{eq:v0WDMbound} with $p_{\text{prod}}\approx(\mttilde^2-m_\chi^2)/(2\mttilde)$, $\heff(\Tprod)\sim 10$ and $v_0=\SI{7.5}{\m/\s}$ to obtain a lower limit on $\lchi$ as a function of the mediator mass. For our regime of interest, which encompasses mediator masses in the $\si{\TeV}$ and multi-$\si{\TeV}$ regime, we find $\lambda_\chi\gtrsim 10^{-15}-10^{-14}$ for both models, too low to provide compatible constraints on the parameter space.


\subsection{The parameter space for non-thermal production} 
\label{subsec:parameterspacenthprod}

We will now combine our findings regarding the non-perturbative corrections to the \gls{dm} abundance (\cf\cref{subsec:DMabundancesW}) with the constraints presented in \cref{subsec:constraintsnth} to obtain a complete picture of the parameter space of our models which allows for non-thermal production. With our precise knowledge of $\Omegadmt h^2=0.1200\pm 0.0012$ \cite{Planck:2018vyg}, we can fix the Yukawa-like coupling $\lchi$ while keeping the \gls{dm} and mediator masses alongside $\lH$ as free parameters. We expect the latter to be of no importance for non-perturbative effects from \gls{bsf} for reasons explained above, but it will modify effects from \gls{se} due to its impact on the perturbative annihilation cross sections and the scattering state wave function in the lepto-philic mediator model. Therefore, we will compare in the following the two scenarios of $\lH=0$, \ie no Higgs portal, and a sizeable value of $\lH=0.3$ to investigate its significance in the corresponding models. We are then left with a parameter scan over the mediator mass $\mttilde$ ($\mtautilde$) as well as the \gls{dm} mass, which we express through $\Delta m = \mttilde-\mchi$ ($\Delta m = \mtautilde-\mchi$) to enhance the visibility of our model constraints.\newp
\begin{figure}
    \centering
    \includegraphics[width=\textwidth]{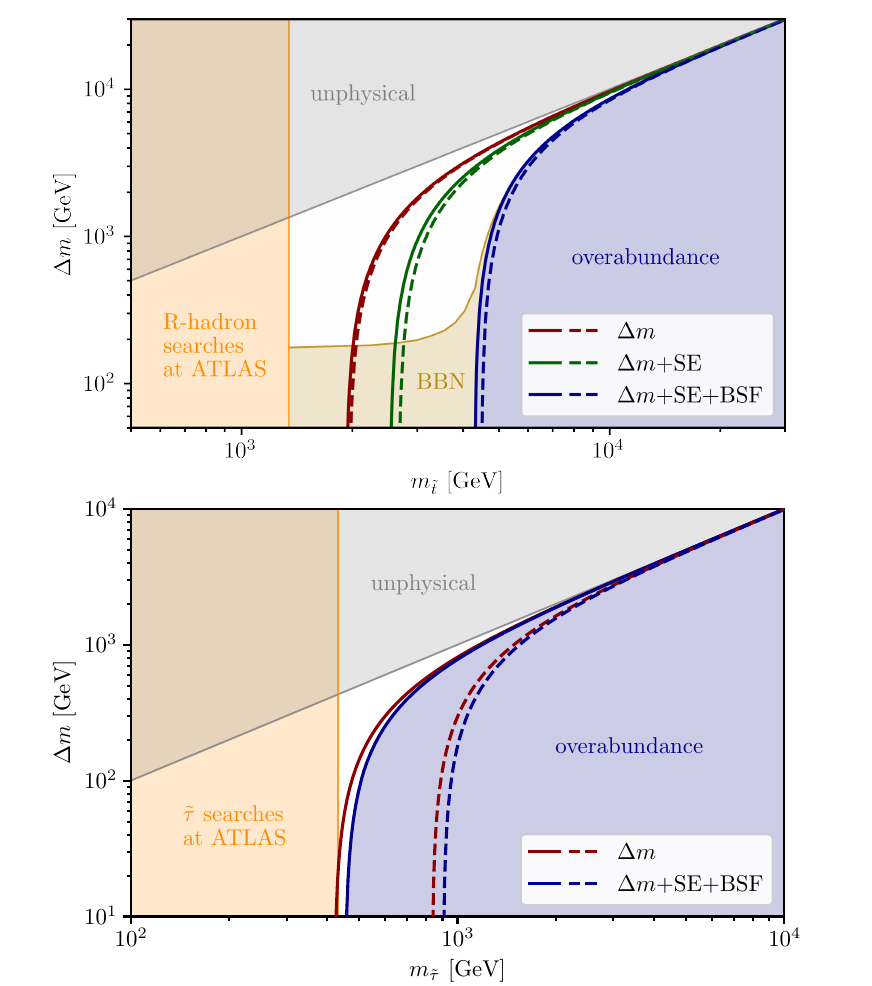}
    \caption[The parameter space of non-thermal production in the top-philic and lepto-philic mediator model.]{The parameter space of non-thermal production in the top-philic (top) and lepto-philic (bottom) mediator model spanned by the mediator mass $\mttilde$ ($\mtautilde$) and the mass difference $\Delta m = \mttilde-\mchi$ ($\Delta m = \mtautilde-\mchi$). For the calculation of $\mchi$ considering \gls{sw} production only (\cf\cref{eq:mchiofmtt}), the perturbative (red), SE corrected (green) and SE+BSF corrected (blue) yields of the mediator after \gls{fo} have been used, considering a Higgs portal coupling of $\lH=0$ (solid lines) and $\lH=0.3$ (dashed lines). Excluded regions are shaded (refer to text for additional information). Adapted from \eref\cite{Bollig:2021psb}, where a former mistake in the calculation of the \gls{bbn} constraints has been corrected.}
    \label{fig:parameterspacenth}
\end{figure}
The impact of non-perturbative effects on the parameter space of the top- and lepto-philic mediator model is presented in \cref{fig:parameterspacenth}. In addition, \cref{fig:heatplots} highlights the computed values of $\lchi$ for each combination of masses, along with the \gls{fi} fraction to the total \gls{dm} abundance $p$ (\cf\cref{eq:lambdachi}). From above, the parameter space is constrained by the trivial condition that $\Delta m$ cannot be larger than the mediator mass (gray). Very close to this bound lies another bound stemming from \gls{wdm} constraints on \gls{fi} production. However, as this constraint only applies to \gls{dm} masses below $\order{\SI{10}{\keV}}$, it is not visible in the plots. As evident from \cref{fig:heatplots} (bottom), even the lowest values of $\lchi$, in the regime where \gls{dm} production is dominated by the \gls{sw} mechanism, are well above the corresponding \gls{wdm} constraints of $\lambda_\chi\gtrsim 10^{-15}-10^{-14}$ calculated in the previous section. Therefore, a pure \gls{sw} scenario is possible for all mediator masses considered here. \gls{lhc} searches for heavy stable particles constrain the parameter space from the left (orange), since they give a lower bound on the allowed mediator mass. We have checked that the mediator is stable on collider scales, such that the limit is independent of $\mchi$ and thus appears as a vertical line in both figures.\newp
From the right, the parameter space is constrained from an \textit{overabundance} bound on \gls{dm} (blue). Since \gls{sw} production is independent of $\lchi$, and the frozen out mediator yield scales proportional to its mass, we expect for high mediator masses an \gls{sw} contribution to the total \gls{dm} abundance which exceeds $\Omegadmt$, \ie the \gls{sw} mechanism alone overproduces \gls{dm} (\cf\cref{eq:DMabundance}). In order to avoid this bound, one needs for a fixed mediator mass a smaller value of $\mchi$ or a larger value of $\Delta m$, respectively. Above this threshold, the \gls{sw} mechanism alone is insufficient to explain the entire \gls{dm} abundance. This creates space for \gls{fi} production to contribute to the remaining portion. How much production from \gls{fi} is required for each configuration of masses has been displayed in \cref{fig:heatplots} (top).\newp 
As anticipated from our discussion in \cref{subsec:DMabundancesW} and evident from \cref{fig:parameterspacenth}, non-perturbative corrections have a considerable impact on the overabundance limit. We illustrate this by displaying the surfaces where $\Omegadmt=\Omegadm^{\text{sW}}$, for $\lH=0$ (solid) and $\lH=0.3$ (dashed), calculated using the perturbative cross section as well as the enhanced ones from non-perturbative effects, while retaining the same color coding as in \cref{fig:foexample}. In case of the top-philic mediator model, the highest allowed mediator mass for the lowest possible $\Delta m$, increases from $\SI{2}{\TeV}$ for a perturbative cross section to over $\SI{4}{\TeV}$ if \gls{se} and \gls{bs} effects are taken into account. Neglecting \gls{bs} effects, we still observe an enhancement of $\sim 40\%$. The influence of $\lH$ in the top-philic mediator model is small as expected but it still allows for a shift of the overabundance limit of $\order{\SI{100}{\GeV}}$. In the context of the lepto-philic mediator model, the situation differs. Due to the significantly lower \gls{qed} couplings, \gls{bsf} has no real impact on the mediator yield after \gls{fo}, which is why the difference between the \gls{se} only and \gls{se}+\gls{bsf} scenario in both figures cannot be resolved (and is therefore not displayed). The difference in the overabundance limit using a perturbative and an \gls{se} corrected cross section is of $\order{10\%}$. We want to note here that for a $\lH=0$, the current collider limits from the \gls{lhc} are already very close to the cosmological ones for the lowest $\Delta m$ and have the prospects to rule out large \gls{dm} masses within this scenario after the \gls{hl} upgrade of the \gls{lhc}. However, unlike the top-philic mediator model, a non-zero Higgs coupling in the lepto-philic mediator model has a large impact on the allowed parameter space, as discussed at the beginning of this section. Hence, a full exclusion of this regime is not foreseeable.\newp
\begin{figure}[ht]
    \centering
    \includegraphics[width=\textwidth]{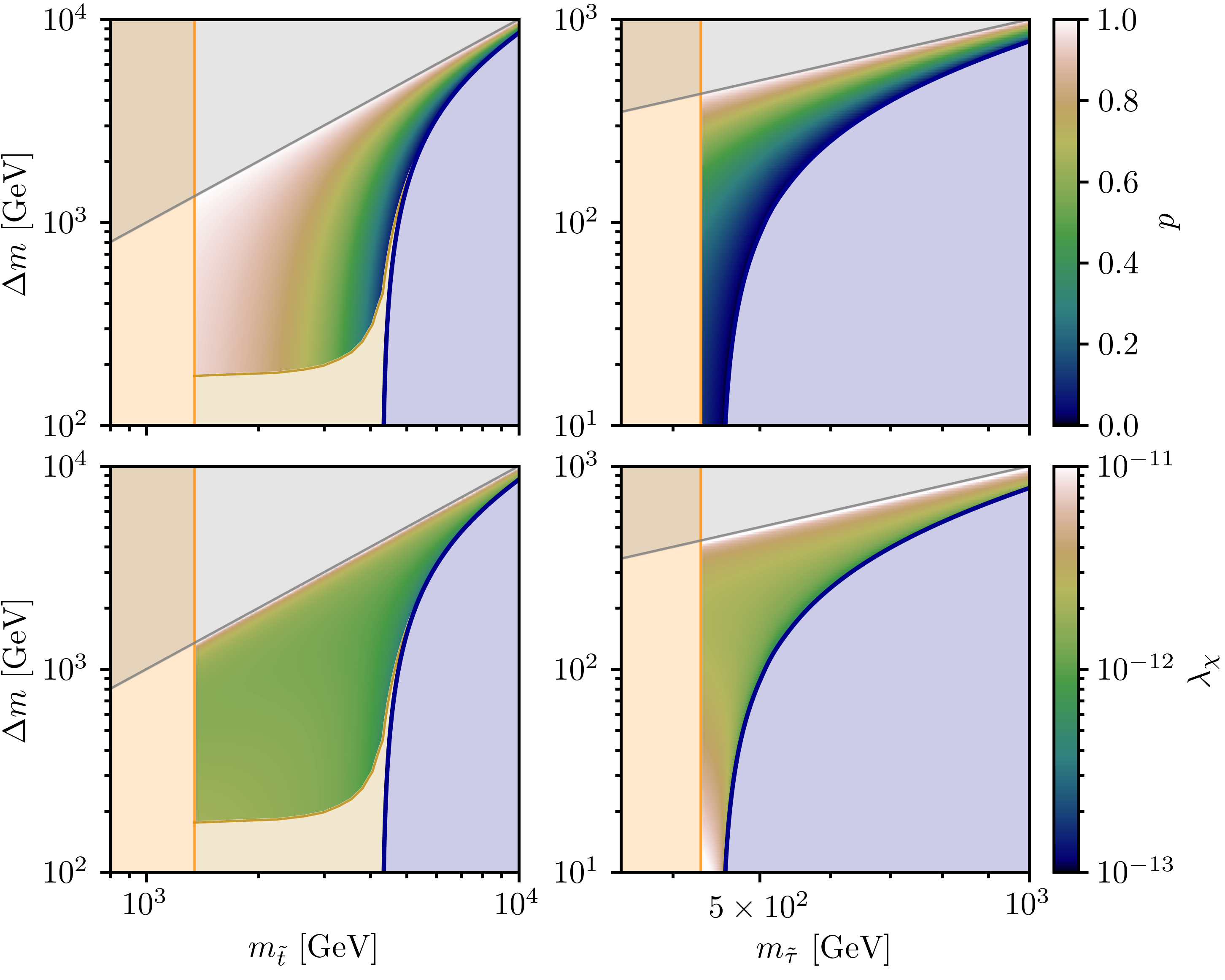}
    \caption[The parameter space for non-thermal production in the top-philic and lepto-philic mediator model, where the values of the Yukawa-like coupling $\lchi$ as well as the FI fraction $p$ have been highlighted.]{The parameter space of non-thermal production in the top-philic (left) and lepto-philic (right) mediator model, where the values of the Yukawa-like coupling $\lchi$ (bottom) as well as the \gls{fi} fraction $p$ (top) in the viable parameter space considering non-perturbative corrections at $\lH=0$ have been highlighted. For information on the labels and shaded areas, see \cref{fig:parameterspacenth}. Adapted from \eref\cite{Bollig:2021psb}.}
    \label{fig:heatplots}
\end{figure}
The last bound we take into account, and which constrains the parameter space from below, stems from \gls{bbn} (beige). In the top-philic mediator model, for most mediator masses, this limit is approximately independent of $\mttilde$ and solely excludes mass differences below the top mass. This is expected as it coincides with the kinematic threshold of the two body decay, which leads to a drastic increase in the mediator lifetime beyond $\Delta m \lesssim m_t$ for a constant coupling. Being close to this threshold will also lower the contributions from the two-body decay channel to \gls{fi} production. This will, in turn, lead to a preference for larger $\lambda_\chi$ couplings to compensate for this effect (see \cref{fig:heatplots}, bottom left). Due to sizeable contributions from the $2\to2$ production channels (\cf\cref{fig:fi_example}) in this regime, this effect will, however, not be sufficient to balance the suppression from the three-body decay, such that the limit remains close to the top mass in the regime where \gls{fi} dominates. As we approach the regime where \gls{sw} production becomes dominant (\cf\cref{fig:fi_example}, top left), the \gls{bbn} constraints are strengthened due to overall smaller values of $\lchi$ which will boost the lifetime of the mediator. A similar exclusion limit exists in the lepto-philic mediator model around the mass of the tau lepton. However, this is outside our primary regime of interest and is therefore not displayed.\newp 
We observe the predictions of $\lchi$ in both models to be roughly constant in the regime, where both \gls{fi} and the \gls{sw} mechanism play a role in \gls{dm} production (see \cref{fig:heatplots}, bottom). This suggests that both regimes can be well separated and dominate \gls{dm} production for $\lchi\gtrsim 10^{-12}$ and $\lchi\lesssim 10^{-12}$, respectively. For the largest mass splittings, \gls{dm} is solely produced via \gls{fi}. This can be important, since within this regime \gls{sw} production might enter a phase where the mediator \gls{fo} and decay are no longer separable due to the increased mediator lifetime as a consequence of larger $\lchi$. Fortunately, this will have no impact on the \gls{dm} abundance due to the negligible \gls{sw} contribution in this regime.\newp
Overall, we showed that a perturbative calculation of the non-thermal \gls{dm} abundance within the top-philic mediator model drastically underestimates the parameter space allowed by cosmological constraints. We expect a moderate improvement up to $\approx \SI{1.6}{\TeV}$ on the exclusion limits by collider searches after the \gls{hl} upgrade of the \gls{lhc}, leaving us doubtful that the parameter space of this model can be tested in the foreseeable future. In the lepto-philic mediator model, where non-perturbative effects are of lesser importance, the upgrade could be sufficient to rule out large mass splittings assuming $\lH=0$. However, if a sizable Higgs portal is allowed, testing the viable regime remains a challenge for far-future colliders probing \gls{bsm} physics at significantly higher energies.

\section{Non-perturbative effects in dark matter indirect detection}
\label{sec:indirectdetection}
\fancyhead[RO]{\uppercase{Non-perturbative effects in DM indirect detection}}

Another intriguing avenue to explore the impact of non-perturbative effects on \gls{dm} models apart from \gls{dm} production is to focus on its detection properties. Among the different approaches to search for \gls{dm} introduced in \cref{subsec:detectionmethods}, indirect detection distinguishes itself as being the most directly linked to the production process. Given that the \gls{cta}, a cutting-edge gamma-ray telescope \cite{CTAConsortium:2017dvg}, has entered the building phase, it is now the right time to reevaluate constraints on \gls{dm} imposed by current indirect detection experiments and juxtapose these bounds with the prospects forecasted by the \gls{cta} collaboration.\newp
Having already explored non-thermal production of \gls{dm} in the previous section, we will focus in this section on a typical thermal \gls{fo} scenario for the \gls{dm} candidate. However, in contrast to the common \gls{wimp} paradigm, \gls{dm} annihilations here will not proceed into \gls{sm} particles but rather into states beyond the \gls{sm}. These types of models go under the name of \textit{secluded \gls{dm}} \cite{Pospelov:2007mp,Pospelov:2008jd} and they are a particularly interesting scenario for indirect detection as it facilitates the decoupling of the relic density from other observables, thereby circumventing stringent constraints imposed by direct detection experiments \cite{PandaX-4T:2021bab,LZ:2022lsv,XENON:2023cxc} and collider searches \cite{ATLAS:2019wdu,ATLAS:2021kxv,CMS:2017jdm,CMS:2018mgb}. If the new states are light compared to the \gls{dm} mass, long-range self-interactions can occur, leading to \gls{se} and, if the interactions are attractive, to the formation of \gls{bs} (\cf\cref{sec:non-perturbative-effects}).\newp
Over recent years, the integration of \gls{se} has become a common practice in both calculating the relic density of \gls{dm} and establishing indirect detection limits \cite{Hisano:2004ds,Feng:2010zp,Slatyer:2011kg,Abazajian:2011ak,Lu:2017jrh,Ando:2021jvn}, while accounting for \gls{bsf} remains less established and is still under ongoing methodological development. When considering \gls{bs}, most studies either focus on the \gls{dm} relic density \cite{vonHarling:2014kha,Petraki:2015hla,Beneke:2016ync,Ellis:2015vna,Liew:2016hqo,Mitridate:2017izz,Cirelli:2016rnw,Beneke:2016jpw,Harz:2018csl,Biondini:2018pwp,Oncala:2019yvj,Oncala:2021tkz,Biondini:2021ycj,Garny:2021qsr,Binder:2019erp,Binder:2021vfo} \textit{or} on experimental signatures \cite{March-Russell:2008klu,Laha:2015yoa,Asadi:2016ybp,Coskuner:2018are,Chu:2018faw,Bottaro:2021srh}, whereas only a limited number of works consistently integrate the influence of non-perturbative effects on both the cross sections for thermal \gls{fo} and indirect detection. Relevant examples have so far scrutinized models with a vector mediator, \ie a dark photon, which undergoes kinetic mixing with the $U(1)_Y$ gauge group of the \gls{sm} \cite{An:2016gad,Pearce:2015zca,Cirelli:2016rnw,Baldes:2020hwx}. In the study of this section, our objective is to address the gap concerning a category of \gls{dm} models, where the self-interaction is mediated by a scalar particle. Specifically, we investigate both scalar and pseudo-scalar interactions of the mediator with a fermionic \gls{dm} candidate. Employing \gls{pnreft}, we derive the cross sections, which are crucial for determining the \gls{dm} relic density as well as \gls{dm} annihilations in astrophysical environments, such as the \gls{gc} and \glspl{dsph}. Subsequently, we utilize these results to delineate the cosmologically viable parameter space for the model, which we then integrate with existing and prospective constraints from the \gls{cmb} as well as gamma-ray telescopes, namely \gls{fermi} and \gls{cta}, respectively.\newp
In \cref{subsec:dmmodelid}, we introduce the \gls{dm} model and its pertinent energy scales. Subsequently, in \cref{subsec:dmproductionpnreft}, we present the cross sections and decay widths relevant for \gls{dm} production within the framework of \gls{pnreft} and perform the relic density calculation including non-perturbative effects. \Cref{subsec:indirectdetection} is dedicated to scrutinize the impact of \gls{se} and \gls{bsf} on \gls{dm} annihilation in various astrophysical and cosmological environments, including discussions on current bounds and future prospects, such as those offered by the next-generation detector \gls{cta}. These constraints are complemented in \cref{subsec:other_exp_limit} by various other searches. The parameter space currently constrained by existing bounds as well as those that can be tested by future experiments is laid out in \cref{subsec:parameterspacethermaldm}.\newp
The work presented in this section has been published in \eref\cite{Biondini:2023ksj}. For transparency, we want to note that the analytic computation of the observables presented in \cref{subsec:dmproductionpnreft} using \gls{pnreft} has been carried out by Simone Biondini and is added to this thesis for the sake of completeness and consistency. However, the obtained results have been cross-checked by the author through a rederivation, employing the approach introduced in \cref{subsec:nonperturbativeeffectsfromQFT}. Due to perfect agreement, these computations have been omitted in the publication but can be found in \cref{app:BSF_thermalfo}. 


\subsection{The dark matter model and energy scales}
\label{subsec:dmmodelid}

We will start by introducing the \gls{dm} model employed in this section and discuss the relevant energy scales of our system, as we will use \gls{pnreft} (\cf\cref{subsec:nonperturbativeeffectsfromEFT}) to compute our observables. Our \gls{dm} candidate is a Dirac fermion $\chi$ which features no direct interactions with \gls{sm} particles and is thus naturally a singlet under the \gls{sm} gauge group. It does, however, possess Yukawa-like interactions with a real scalar mediator $\phi$ of an extended dark sector, which in turn couples to the \gls{sm} via mixing with the Higgs boson. The Lagrangian density of the two-particle dark sector is then given by
\begin{equation}
    \label{eq:LmodID}
    \mathcal{L}_{\text{\scriptsize{DS}}}= \bar{\chi} \left(i \gamma_\mu\partial^\mu -\mchi\right) \chi + \frac{1}{2} \partial_\mu \phi \, \partial^\mu \phi -\frac{1}{2}\mphi^2 \phi^2 -  \bar{\chi} (g + ig_5 \gamma_5)  \chi \phi - \frac{\lambda_\phi}{4!} \phi^4  +\mathcal{L}_{\text{\scriptsize{portal}}} \,,
\end{equation}
where $\mchi$, $\mphi$ denote the \gls{dm} and mediator masses and $g$, $g_5$ are the scalar and pseudo-scalar Yukawa-type couplings between the dark sector particles. The mediator self coupling $\lambda_\phi$ is chosen to be negligible such that it plays no role in the subsequent discussion. To allow for the study of non-perturbative effects, the mediator mass is chosen to be considerably smaller than the \gls{dm} mass. We further require $\alpha \equiv g^2/(4 \pi) \gg \alpha_5 \equiv g_5^2/(4 \pi)$ to ensure that the non-perturbative effects predominantly arise from scalar-type interactions, which induce an attractive mediator potential (see \cref{subsec:potentialssecondmodel}). This allows us to neglect mixed and pure pseudo-scalar contributions to \gls{se} and \gls{bsf} effects (see \cref{subsec:nonperturbativeeffectsfromEFT} and \erefs\cite{Kahlhoefer:2017umn,Biondini:2021ycj}).\newp
Since we are chiefly interested in investigating the impact of non-perturbative effects on indirect detection, we opt for a minimal \gls{dm} model \cite{Kaplinghat:2013yxa,DeSimone:2016fbz}. Consequently, the mass parameters and couplings in \cref{eq:LmodID} are chosen freely, without making any assumptions about a specific \gls{uv} completion.\footnote{However, it is not challenging to devise a mechanism for generating masses in our model. Assuming, for instance, a non-zero \acrshort{vev} for the scalar mediator, we would obtain $\mchi\simeq g v_\phi$, $\mphi\simeq \sqrt{\lambda_\phi}v_\phi$, allowing for an arbitrary splitting between both dark sector masses $\mphi/\mchi\simeq \sqrt{\lambda_\phi}/g<1$, since $\lambda_\phi$ can be tuned freely. Interestingly, this choice would also maintain a negligible thermal contribution to the mediator mass $\mphi^{\text{th}}/\mphi\simeq g T/\mchi$ at \gls{fo} (since $T\ll \mchi$) largely independent of $\lambda_\phi$.} We refer to \erefs\cite{Kahlhoefer:2015bea,Duerr:2016tmh} as instances of similar simplified models, which feature gauge symmetries and spontaneous symmetry breaking in the dark sector. Technically, the mediator sector can be extended to include a richer set of interactions, as discussed in works such as \erefs\cite{Wise:2014jva,Kahlhoefer:2017umn,Oncala:2018bvl}. For instance, an interaction of the form $\rho_\phi \phi^3$ with a dimensionful coupling could exist, thereby facilitating additional \gls{bsf} processes \cite{Oncala:2018bvl,Oncala:2019yvj}.\newp
The Higgs sector containing the Higgs portal interactions with the dark sector mediator is given by
\begin{equation}
    \label{eq:LHiggsportal}
	\mathcal{L}_{\text{\scriptsize portal}}=-\mu^2 H^\dagger H -\lambda (H^\dagger H)^2-\mu_{\phi h}\phi\left(H^\dagger H -\frac{v^2}{2}\right)-\frac{1}{2}\lambda_{\phi h}\phi^2\left(H^\dagger H -\frac{v^2}{2}\right) \, ,
\end{equation}
where $H$ denotes the \gls{sm} Higgs doublet with $\mu$, $\lambda$ its \gls{sm} potential couplings and $v^2\equiv -\mu^2/\lambda$ the usual Higgs \gls{vev}. Note that we have applied a shift to the scalar mediator field such that it does not develop a \gls{vev} on its own. The quartic coupling $\lambda_{\phi h}$ is taken to be negligible in the following such that only the dimensionful coupling $\mu_{\phi h}$ induces a mixing between the mediator and the Higgs.\footnote{By this choice, we naturally avoid sizeable contributions to the Higgs mass through mediator loops. Thermal contact between the \gls{sm} and the dark sector can still be maintained with relatively small values of $\lambda_{\phi h}$ (see \cref{subsubsec:thermalizationlimits}).} We made this choice, along with $v_\phi=0$, to minimize the number of relevant free model parameters. An alternative choice could have been to set $\mu_{\phi h}=0$ and retain the quartic mixing $\lambda_{\phi h}$ alongside the quartic scalar coupling $\lambda_\phi$. Inevitably, this would introduce a \gls{vev} for the scalar ($v_\phi>0$) analogous to the \gls{sm} Higgs sector. Nonetheless, the definition of the mixing angle $\delta$ and the mixing term in the Lagrangian after \gls{ewsb} remain the same at leading order when substituting $\mu_{\phi h}\to v_\phi \lambda_{\phi h}$. We have verified that the phenomenology of both scenarios, including the resulting dark scalar and Higgs boson masses, is mostly equivalent up to $\order{\sin^2\delta}$, which we will neglect in the following due to the small value of $\delta$. However, a sizeable self-interaction of the mediator would introduce a thermal mass for the mediator $\mphi^{\text{th}}=T\sqrt{\lambda_\phi/12}$, a complication which is not pursued further here. A discussion of the scalar mixing and its implications for indirect detection is postponed to \cref{subsubsec:photonspectrum}.\newp
The \gls{dm} relic density in this model is fixed via thermal \gls{fo} (\cf\cref{subsec:BEfiandfo}). The essential process that activates this mechanism and plays a crucial role in generating astrophysical signals for indirect detection is the annihilation of \gls{dm} pairs into mediator particles, $\bar{\chi}+ \chi \to \phi + \phi$. Both \gls{dm} production and detection processes occur in the non-relativistic regime, where $\vrel\ll 1$. To study non-perturbative effects in this model, we reintroduce the dimensionless parameters from \cref{eq:zetaandxi}
\begin{equation*}
    \zeta \equiv\frac{\alpha}{\vrel} \, \quad \text{and} \, \quad  \xi\equiv\frac{\mchi \alpha}{2 \mphi}\,,
\end{equation*}
where the criteria for large \gls{se} ($\zeta\gtrsim 1$) and \gls{bs} ($\xi\gg 1$) effects have already been discussed in \cref{subsec:SEannihilationBSdecay}. In addition to the \gls{dm} mass $\mchi$, there are two dynamical energy scales at $\vrel\sim\alpha$ that will play a major role in the subsequent discussion (\cf\cref{subsec:nonperturbativeeffectsfromEFT}): 1) the typical momentum transfer in $\bar{\chi}\chi$-scattering which is proportional to the Bohr momentum $\kappa\propto \mchi\alpha$ of the \gls{bs} and 2) the \gls{bs} energy scale $\mchi \alpha^2$. With decreasing $\alpha$, the separation between the scales becomes more pronounced, with $\mchi\gg \mchi\alpha \gg \mchi\alpha^2$. As discussed in \cref{subsec:nonperturbativeeffectsfromEFT}, we will refer to them as the hard, soft, and ultra-soft energy scales, respectively.\newp
The mediator mass $\mphi$ is another pertinent energy scale. We consider it to be smaller than the \gls{dm} mass $\mchi$ and, at most, as large as the soft scale $\mchi \alpha$. A key distinction from massless mediators is the screening of \gls{dm} self-interactions when the soft scale $\mchi \alpha$ is of the order of the mediator mass $\mphi$.\footnote{One might wonder, if the mediator induced \gls{dm} self coupling places any bounds on the parameter space of our model. This does seem to be the case, as \eg\eref\cite{Kahlhoefer:2017umn} found limits for a very similar model only at \gls{dm} and mediator masses lower than considered in this study.} In the case of scattering states subject to an attractive potential as produced by the scalar coupling, the \gls{se} typically levels off at low velocities (instead of further increasing $\propto \vrel^{-1}$). Furthermore, a characteristic resonance structure emerges due to the Yukawa potential, which depends on the ratio of mediator to \gls{dm} mass characterized by $\xi$ \cite{Hisano:2003ec,Hisano:2004ds,Arkani-Hamed:2008hhe}. The Coulomb limit is regained for velocities significantly larger than $\alpha$, \ie $\zeta \ll 1$, and/or in the massless mediator limit $\xi \to \infty$. Phenomenologically, the latter does not occur since our primary interest lies in the regime where $\mphi\geq 2 m_{\pi^{0}}$. This is motivated by the observation that lighter mediators do not produce a substantial amount of gamma-rays, which constitute the primary observational signature examined in this study. Nonetheless, lighter mediators can impact other observables such as the positron flux or the distribution of \gls{dm} mass in dense halos (see \eg\erefs\cite{Kahlhoefer:2017umn,Kainulainen:2015sva}).\newp
\begin{figure}
    \centering
    \includegraphics[width=\textwidth]{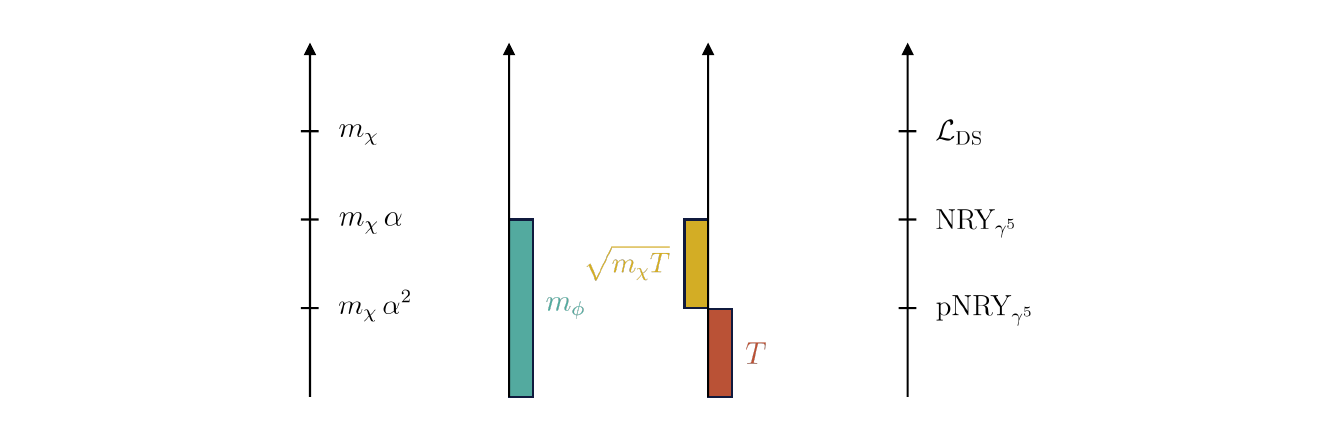}
    \caption[Hierarchy of the relevant energy scales in the model employed in \cref{sec:indirectdetection}.]{Hierarchy of the relevant energy scales in the model employed in \cref{sec:indirectdetection}. Besides the two mass scales, dynamically generated as well as thermal scales are depicted. Effective field theories considered in this work (\cf\cref{subsec:nonperturbativeeffectsfromEFT}) for the \gls{dm} model defined in \cref{eq:LmodID} are matched with their corresponding scales. Adapted from \eref\cite{Biondini:2023ksj}.}
    \label{fig:energyscales}
\end{figure}
Eventually, during the \gls{fo} of \gls{dm} particles in the early Universe, thermal scales can become relevant. The first one to note is the temperature $T$ of the thermal plasma, which is already $T\ll\mchi$ at the onset of \gls{fo}. Nevertheless, it can still be of the order of the soft or ultra-soft scale, which one has to account for. In this study, we incorporate thermal effects due to the medium, presuming the temperature to be approximately at the ultrasoft scale $\mchi \alpha^2$ or smaller. This assumption implies that thermal effects do not affect the non-relativistic potential, which we can then still assume to be the in-vacuum Yukawa one. For \gls{dm} fermions in thermal equilibrium, the relative velocity is approximately $\vrel \sim \sqrt{T/\mchi}$, resulting in a typical momentum of $\mchi \vrel \sim \sqrt{\mchi T} \lesssim \mchi \alpha$, indicating $\vrel \lesssim \alpha$. Additionally, within the temperature range considered in this work, we have $\sqrt{\mchi T}\gtrsim \mchi\alpha^2$. These conditions classify $\mchi \vrel$ as a soft scale, and $\mchi\vrel^2 \sim T$ as an ultrasoft scale (\cf\cref{subsec:nonperturbativeeffectsfromEFT}). Visualized in \cref{fig:energyscales}, we summarize the hierarchy of the energy scales as 
\begin{equation}
\mchi \gg \mchi\alpha \gtrsim \sqrt{\mchi T} \gg \mchi\alpha^2 \gtrsim T\,,
\label{eq:hierarchyofenergyscales}
\end{equation}
where the mediator mass is assumed to be $\mphi \lesssim \mchi \alpha$. For reasons explained above, thermal masses are negligible in this setup and do not enter the scale hierarchy.


\subsection{Dark matter production from pNREFT}
\label{subsec:dmproductionpnreft}

Before we can study the indirect detection properties of our \gls{dm} model, we first need to determine how much \gls{dm} is produced during \gls{fo}. As for the model discussed in \cref{sec:DMproduction}, we will subsequently calculate the annihilation cross section with \gls{se}, the \gls{bsf} cross section as well as the \gls{bs} ionization and decay rate. Contrary to before, we will employ \gls{pnreft} to perform the computations, as it allows for a better control of the different scales within the model.


\subsubsection{Heavy pair annihilation and decay}
\label{subsubsec:heavypairannihilationanddecay}

\begin{figure}
    \centering
    \includegraphics[width=\textwidth]{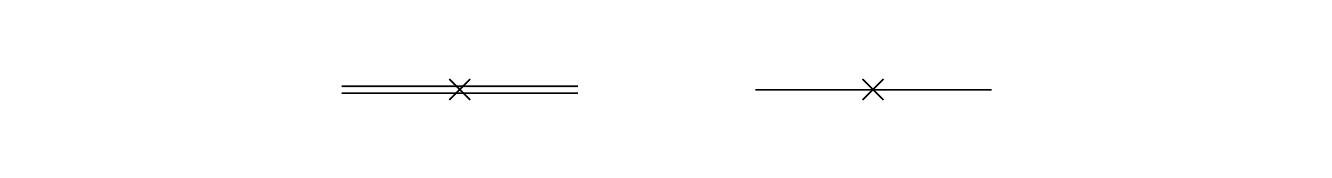}
    \caption[The localized scattering state and BS self energy diagrams in $\pnry$.]{The localized ($\propto\delta^{(3)}(\vec{r})$) contributions to the scattering state (left) and \gls{bs} (right) self energy diagrams in $\pnry$. Their imaginary parts, given in the Lagrangian of \cref{eq:LannpNRY}, can be matched to the $\bar{\chi}\chi$ annihilation cross section and the \gls{bs} decay rate, respectively, where a double solid line corresponds to a scattering state and a single solid line to a \gls{bs}. The crosses denote the interaction points.}
    \label{fig:selfenergyann}
\end{figure}
We will start by calculating the annihilation cross section of the heavy $\bar{\chi}\chi$ pair. This can be achieved by relating the cross section via the optical theorem to the imaginary contribution of the scattering pair self-energy as visualized in \cref{fig:selfenergyann} (left). In $\nry$, annihilation processes are encoded in local 4-fermion operators. This translates to $\pnry$, where these operators generate local terms ($\propto\delta^{(3)}(\vec{r})$) in the imaginary part of the potential. The matching coefficients required for the subsequent computation, are thus the ones given in \cref{eq:LannpNRY}, which we will collectively denote as $V_{\text{ann}}$ in the following. The annihilation cross section can be written as
\begin{equation}
    \label{eq:anncrosssectionVann}
    \sigmaann\vrel(\vec{p})=\frac{1}{4}\langle\vec{p},0|\int\dd[3]{r}\varphi^\dagger(\vec{r},\vec{R},t)\left[2\Im{-V_{\text{ann}}(\vec{r},\vec{p},\vec{\sigma}_1,\vec{\sigma}_2)}\right]\varphi(\vec{r},\vec{R},t)|\vec{p},0\rangle,
\end{equation}
where we have considered the scattering amplitude from an initial to a final state $|\vec{p},0\rangle$ in the \gls{com} frame (with $\vec{P}=0$). We sum here implicitly over all possible spin configurations of $|\vec{p},0\rangle$ and average over the initial spins which results in a prefactor $1/4$. In order to project out the spin factors, it is advantageous to insert a $\mathds{1}=\sum_{S,m_S}|S,m_S\rangle\langle S,m_S|$, where $\sum_{m_S}\langle 1,m_S|S^iS^j|1,m_S\rangle=2\delta^{ij}$ and all singlet projections vanish. After inserting the definitions of $\varphi^{(S)}(\vec{r},\vec{R},t)$ given in \cref{eq:bilocalfielddef} and decomposing the scattering state wave function $\phiq{p}(\vec{r})$ into partial waves (\cf\cref{eq:scatteringwavefunctiondec}), the cross section yields 
\begin{align}
    \sigmaann\vrel (\vec{p})~=~& \frac{\Im{f(\spec{1}{S}{0})}}{\mchi^2}|\mathcal{R}_{\vec{p},0}(r)|^2_{r=0}~+~\frac{\Im{f(\spec{3}{P}{0})}+5\Im{f(\spec{3}{P}{2})}}{3\mchi^4}|\mathcal{R}'_{\vec{p},1}(r)|^2_{r=0}\nonumber\\
    &-~\frac{\Im{g(\spec{1}{S}{0})}}{\mchi^4}\Re{\mathcal{R}_{\vec{p},0}^*(r)\vec{\nabla}_{\vec{r}}^2\mathcal{R}_{\vec{p},0}^{}(r)}_{r=0}\,.
\end{align}
The $\mathcal{R}_{\vec{p},l}(r)\equiv (2l+1)\chiqs{p}{l}(\kappa r)/(\kappa r)$ denote the radial wave functions, which are related to the \gls{se} factors via $|\mathcal{R}_{\vec{p},0}(r)|^2_{r=0}=\Sannl{0}(\zeta,\xi)$ and $|\mathcal{R}'_{\vec{p},1}(r)|^2_{r=0}=p^2\Sannl{1}(\zeta,\xi)$ (\cf\cref{eq:SEannihilation}), with $p=\mchi\vrel/2$.\footnote{We want to note here that the corresponding radial $l$-wave functions are projected out from the decomposition of $\phiq{p}(\vec{r})$ because the $\nry$ matching coefficients have a concrete angular momentum dependence (and thus come with angular projection operators), which has been left implicit in \cref{eq:LannpNRY}.} The expression $\Re{\mathcal{R}_{\vec{p},0}^*(r)\vec{\nabla}_{\vec{r}}^2\mathcal{R}_{\vec{p},0}^{}(r)}$ formally diverges for $r=0$, however, we can exchange this term with the finite expression $-p^2|\mathcal{R}_{\vec{p},0}(r)|^2_{r=0}$ as explained in \eref\cite{Brambilla:2002nu}. As we can see, by computing the annihilation cross section in $\pnry$, the resummation of soft exchanges (\ie \gls{se}) is naturally included because $\pnry$ is a theory of interacting pairs. The hard dynamics is encoded in the matching coefficients of $\nry$, whereas soft dynamics are contained in the radial wave functions. The annihilation cross section corrected for \gls{se} is then given by
\begin{align}
    \label{eq:annihilationcrosssectionID}
    \sigmaann\vrel =~& \frac{1}{\mchi^2} \left\lbrace  \left(\Im{f(\spec{1}{S}{0})} + \Im{g(\spec{1}{S}{0})} \frac{\vrel^2}{4}\right) \Sannl{0}(\zeta, \xi) \right.\nonumber \\
    &\left.\hspace{1cm} +~\left(\frac{\Im{f(\spec{3}{P}{0})}+5\Im{f(\spec{3}{P}{2})}}{3} \right) \frac{\vrel^2}{4}\Sannl{1}(\zeta, \xi)  \right\rbrace \,, 
\end{align}
where the \gls{se} factors have to be computed numerically with the methods introduced in \cref{subsubsec:SEannihilations}.
\begin{figure}
    \centering
    \includegraphics[width=\textwidth]{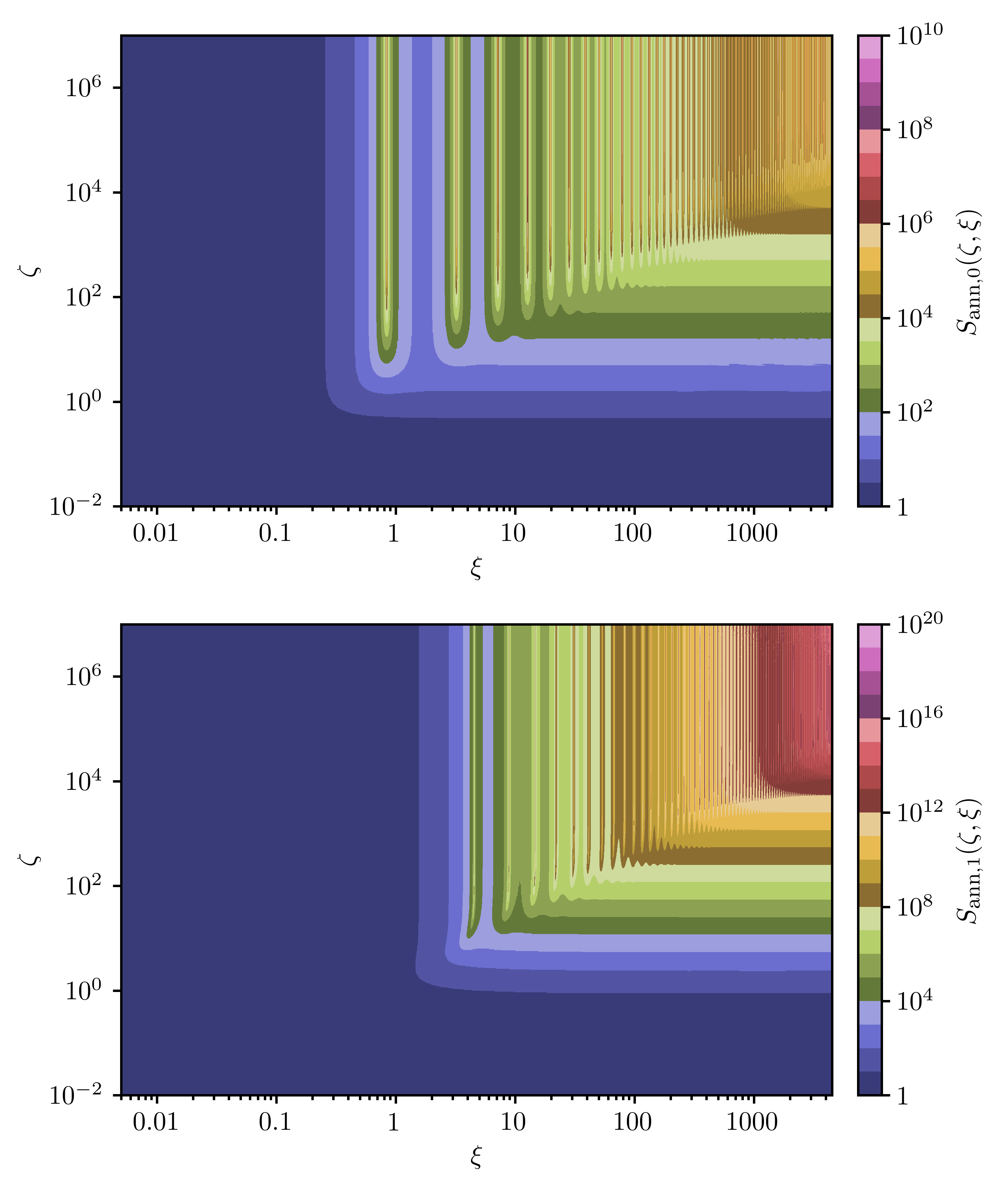}
    \caption[The SE factors of s- and p-wave scattering states in $\bar{\chi}\chi$ pair annihilation for finite mediator masses.]{The \gls{se} factors of s-wave (top) and p-wave (bottom) scattering states in $\bar{\chi}\chi$ pair annihilation for a finite mediator mass as a function of the dimensionless variables $\zeta$ and $\xi$, encompassing the \gls{dm} to mediator mass ratio, the scalar coupling strength and the relative \gls{dm} velocity.}
    \label{fig:Sannlspectrum} 
\end{figure}
We have displayed them in \cref{fig:Sannlspectrum}, where we observe the aforementioned resonance structure along the $\xi$ axis. These resonances are caused by the Yukawa potential in the Schrödinger equation, which for some mediator masses exactly cancels the kinetic and angular contributions resulting in a zero energy \gls{bs}. The resonance peaks vary for the displayed partial waves due to the different angular dependencies (\ie the $l(l+1)/x^2$ terms in the Schrödinger equation) \cite{Beneke:2024iev}. We cross-checked \cref{eq:annihilationcrosssectionID} utilizing \cref{eq:SEsplittingann}, where we calculated the s- and p-wave contributions to the hard annihilation cross section directly from the \gls{qft} matrix element of the full theory.\newp
The decay rate of a \gls{bs} can be calculated equivalently to the annihilation cross section by relating it to its self-energy via the optical theorem (see the right diagram in \cref{fig:selfenergyann}). We employ again the imaginary part of the matching coefficients $V_{\text{ann}}$ and project it onto the \gls{bs} $|n,0\rangle$ in the \gls{com} frame, yielding
\begin{equation}
    \Gammadecn=\frac{1}{2J+1}\langle n,0|\int\dd[3]{r}\varphi^\dagger(\vec{r},\vec{R},t)\left[2\Im{-V_{\text{ann}}(\vec{r},\vec{p},\vec{\sigma}_1,\vec{\sigma}_2)}\right]\varphi(\vec{r},\vec{R},t)|n,0\rangle\,,
\end{equation}
where we average over the degenerate states with the same total angular quantum number $J$.\footnote{Note that a sum over the degenerate quantum states with different $m$ is not necessary, because only the $m=0$ state contributes due to the azimuthal symmetry of the potential (\cf\cref{eq:SEdecay}).}
We can already see from the spin and angular momentum structure in \cref{eq:LannpNRY}, that an $S$-state $|nS\rangle\equiv|n00\rangle$ will project out a spin singlet, whereas for a $P$-state $|nP\rangle\equiv|n1m\rangle$ only a spin triplet contribution will remain. Performing similar steps as before, we obtain
\begin{align}
    \label{eq:boundstatedecaynSID}
    \Gamma_{\text{dec}}^{(nS)} &= \frac{\Im{f(\spec{1}{S}{0})}}{\pi \mchi^2}|\mathcal{R}_{n0}(r)|_{r=0}^2=\frac{4\Im{f(\spec{1}{S}{0})}}{\mchi^2} \Sdecnl{(nS)}{0}(\xi),\\
    \Gamma_{\text{dec}}^{(nP_J)} &= \frac{\Im{f(\spec{3}{P}{J})}}{\pi \mchi^4} |\mathcal{R}'_{n1}(r)|_{r=0}^2 = \frac{\Im{f(\spec{3}{P}{J})}}{3 \mchi^2}\Sdecnl{(nP)}{1}(\xi),  
    \label{eq:boundstatedecaynPJID}
\end{align}
where $\mathcal{R}_{nl}(r)\equiv\kappa^{3/2}\chinl(\kappa r)/(\kappa r)$ is the radial wave function of the \gls{bs} (\cf\cref{eq:boundstatewavefunctiondec}), which translates to the decay \gls{se} factors (\cf\cref{eq:SEdecay}) via $|\mathcal{R}_{n0}(r)|_{r=0}^2=4\pi\,\Sdecnl{(nS)}{0}(\xi)$ and $|\mathcal{R}'_{n1}(r)|_{r=0}^2=\pi/3\,\mchi^2\,\Sdecnl{(nP)}{1}(\xi)$. Methods to compute the \gls{se} factors for \gls{bs} decay have been discussed in \cref{subsubsec:SEboundstatedecays}. For reasons explained later, we will only consider the decay of the $1S$ state in the following.


\subsubsection{Bound state formation and ionization}
\label{subsubsec:BSFandIionID}

\begin{figure}
    \centering
    \includegraphics[width=\textwidth]{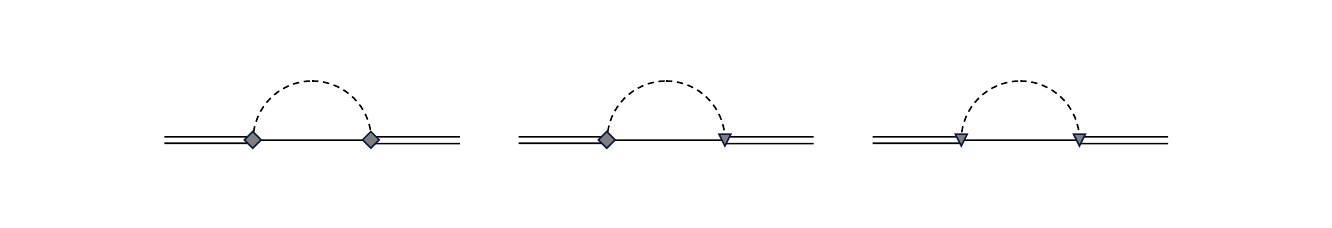}
    \caption[The leading order self-energy contributions to the scattering state of a heavy $\bar{\chi}\chi$ pair from quadrupole and derivative interactions in $\pnry$.]{The leading order self-energy contributions to the scattering state (double solid line) of a heavy $\bar{\chi}\chi$ pair from quadrupole (diamond vertex) and derivative (triangle vertex) interactions in $\pnry$ (\cf\cref{eq:pNRY}). Their imaginary parts can be matched to the \gls{bsf} cross section. The internal propagators correspond to a \gls{bs} (single solid line) and a mediator (dashed line).}
    \label{fig:selfenergyBSF}
\end{figure}

We will continue by calculating the \gls{bsf} cross section as well as the \gls{bs} ionization rate. The latter will only be important for \gls{fo} because ionization requires a thermal population of mediator particles, which is not present in the late Universe. For \gls{bsf}, we will restrict ourselves to the leading order ultra-soft transitions between a scattering and a \gls{bs}, which radiate off a single scalar mediator particle, namely $\bar{\chi}+\chi\to\mathscr{B}(\bar{\chi}\chi)+\phi$.\footnote{Additional \gls{bsf} processes can emerge from $2\to 2$ inelastic scatterings between \gls{sm} particles and the $\bar{\chi}\chi$ pair (see \eg\erefs\cite{Biondini:2018pwp,Binder:2019erp,Biondini:2017ufr}). However, as they are induced in our case by the mixing between the scalar and the \gls{sm} Higgs, they are strongly suppressed due to the small mixing angle and, therefore, will be neglected. Inelastic scattering processes involving the mediator are absent because we assume the trilinear mediator coupling $\rho_\phi\phi^3$ to vanish.} These are given in the $\pnry$ Lagrangian by the terms in the second line of \cref{eq:pNRY}. The \gls{bsf} cross section at finite temperature can then be calculated from the imaginary parts of the respective contributions to the self-energy of the scattering state, which are illustrated in \cref{fig:selfenergyBSF}. Note that the monopole diagrams are absent because they vanish due to the orthogonality of the scattering and \gls{bs} wave functions. Therefore, the \gls{bsf} processes are driven by the quadrupole as well as the derivative term. In principle, the \gls{bsf} cross section in $\pnry$ is given by an expression similar to \cref{eq:anncrosssectionVann}. However, since none of the interaction vertices are spin dependent, we can drastically simplify the notation in this case, yielding
\begin{equation}
    \sigmabsf\vrel\big\vert_{T}=\langle \vec{k}|2\Im{-\Sigma{\text{\tiny{BSF}}}}|\vec{k}\rangle\,,
\end{equation}
with $|\vec{k}\rangle$ a scattering state in the \gls{com} frame and $\Sigma{\text{\tiny{BSF}}}$ the self energy contributions from the three diagrams in \cref{fig:selfenergyBSF}. The scattering and \gls{bs} wave functions are then simply given by $\langle \vec{r}|\vec{k}\rangle=\phik(\vec{r})$ and $\langle \vec{r}|n\rangle=\psin(\vec{r})$.\newp
As an example, we will proceed by calculating the self-energy from the leftmost diagram, corresponding to the pure quadrupole contribution. In dimensional regularization with $D=4-2\epsilon$, the self-energy contribution with purely quadrupole interactions is given by \cite{Biondini:2021ycj}
\begin{align}
    \label{eq:selfenergyBSFquadrupole}
    \Sigma^{\mathcal{Q}} = &-i \frac{\pi\alpha}{4}\mu^{4-D} r^ir^j\int\frac{\dd[D]{q}}{(2\pi)^D}\frac{i}{K^0-q^0-H+i\eta}q^iq^jq^mq^n\nonumber\\
    &\hspace{2.8cm}\left[\frac{i}{(q^0)^2-q^2-\mphi^2+i\eta}+2\pi\delta\left((q^0)^2-q^2-\mphi^2\right)f_\phi\left(\abs{q^0}\right)\right]r^mr^n,
\end{align}
where $\mu=\mchi/2$ is the reduced mass of the system, $K^0=\Ek$ denotes the energy of the scattering state, the $q^i$ correspond to the $\nabla^i_{\vec{R}}$ in momentum space and the $r^i$ are understood as operators which act on the bound and scattering states. The terms in the brackets correspond to the scalar propagator, which splits up into an in-vacuum and a thermal part at finite temperature, where $f_\phi(x)=1/(e^{x/T}-1)$ is the Bose-Einstein equilibrium distribution of the scalar.\footnote{The scalar propagator at finite temperature has been derived in the \textit{real-time formalism} and can be found \eg in \eref\cite{Biondini:2021ycj}; see \erefs\cite{Bellac:2011kqa,Kapusta:1989tk} for further details on the construction.} Note that the thermal non-relativistic propagator of a fermion antifermion pair interacting through a potential $V(r)$ as given \eg in \eref\cite{Brambilla:2008cx}, can be replaced by its in-vacuum form, as displayed in the first line of \cref{eq:selfenergyBSFquadrupole}, when considering physical amplitudes ($H$ denoting the Hamiltonian). To ensure that the internal loop propagator describes \gls{bs}, we insert a complete set of \gls{bs}
\begin{equation}
    \frac{i}{K^0-q^0-H+i\eta}=\sum_n\frac{i}{\Ekn{n}-q^0+i\eta}|n\rangle\langle n|\,,
\end{equation}
where $H|n\rangle=\En|n\rangle$ projects out the \gls{bs} energy and 
\begin{equation}
    \Ekn{n}\equiv \Ek-\En = \frac{k^2}{2\mu}+\gammanl^2(\xi)\frac{\kappa^2}{2\mu}=\frac{\mchi\alpha^2}{4}\left(\frac{1+\gammanl^2(\xi)\zeta^2}{\zeta^2}\right)
\end{equation}
denotes the energy difference between the scattering and the \gls{bs}. Note that $\gammanl$ for the massive mediator scenario must be determined simultaneously with the \gls{bs} wave function (see \cref{subsubsec:SEboundstatedecays}). By applying the usual cutting rules $1/(x-q^0+i\eta)\to -\pi i\delta(x-q^0)$ and $1/((q^0)^2-x^2+i\eta)\to -2\pi i\delta((q^0)^2-x^2)$, as outlined in \eref\cite{Peskin:1995ev}, the imaginary part of the self-energy is obtained, yielding
\begin{equation}
    \label{eq:ImselfenergyBSFQ}
    \Im{\Sigma^{\mathcal{Q}}}=-\frac{\alpha}{240}\sum_n\left[(\Ekn{n})^2-\mphi^2\right]^{5/2}\hspace{-0.1cm}\left(\vec{r}^2|n\rangle\langle n|\vec{r}^2+2r^ir^j|n\rangle\langle n|r^ir^j\right)\hspace{-0.1cm}\left[1+f_\phi(\Ekn{n})\right].
\end{equation}\newp
The self-energy contributions from the other two diagrams in \cref{fig:selfenergyBSF} can be computed in a similar manner. As evident from \cref{eq:ImselfenergyBSFQ} (and also holds in general to this order), the $T$ dependence of the \gls{bsf} cross section into a \gls{bs} state $(n)$ can be factored out
\begin{equation}
    \label{eq:BSFcrosssectionFT}
    \sigmabsf^{(n)}\vrel\big|_T=\sigmabsf^{(n)}\vrel\left[1+f_\phi(\Ekn{n})\right]\,,
\end{equation}
with the in-vacuum cross section at $T=0$ given by 
\begin{align}
    \label{eq:BSFcrosssectionID}
    \sigmabsf^{(n)}\vrel= &~\frac{\alpha}{120 } \ \left[ (\Ekn{n}) ^2-\mphi^2\right]^\frac{5}{2} \left[ |\langle\vec{k} | \vec{r}^2 |n  \rangle|^2 + 2 |\langle\vec{k} | r^i r^j |n  \rangle|^2\right] \nonumber\\
    &- \frac{\alpha}{3\mchi^2} \left[(\Ekn{n})^2-\mphi^2\right]^\frac{3}{2}\Re{\langle\vec{k}|\nabla^2_{\vec{r}}|n\rangle \langle n| \vec{r}^2|\vec{k}\rangle}\nonumber\\
    & + \frac{2\alpha}{\mchi^4} \left[(\Ekn{n})^2-\mphi^2\right]^\frac{1}{2}|\langle\vec{k}|\nabla^2_{\vec{r}}|n\rangle|^{2}\, .
\end{align}
The quantum mechanical matrix elements, sometimes also referred to as \textit{overlap integrals}, for an operator $\mathcal{O}$ are defined as 
\begin{equation}
    \label{eq:expvalueBSF}
    \langle\vec{k}|\mathcal{O}|n\rangle\equiv \int \dd[3]{r}\phik^*(\vec{r})\mathcal{O}\psin(\vec{r})
\end{equation}
and have to be evaluated numerically due to the finite mediator mass. To do so, we employed the leading order Yukawa potential in the Schrödinger equations (\cf\cref{eq:scatteringSchroedinger,eq:boundstateSchroedinger}) to derive the scattering and \gls{bs} wave functions with the methods introduced in \cref{subsec:SEannihilationBSdecay}. The total \gls{bsf} cross section contains in principle the sum over all possible \gls{bs} $\sigmabsf\vrel=\sum_n\sigmabsf^{(n)}\vrel$ which are kinematically allowed (\ie for which $\Ekn{n}>\mphi$). However, we will consider in the following only \gls{bsf} into the ground state $(1S)=\{100\}$ and estimate corrections of higher states in \cref{subsubsec:relicdensityid}. By factoring out $\pi\alpha^4/\mchi^2$, the \gls{bsf} cross section can be expressed as
\begin{equation}
    \label{eq:BSFcrosssectionID2}
    \sigmabsf\vrel=\frac{\pi\alpha^4}{\mchi^2}\SBSF(\zeta,\xi)\,,
\end{equation}
where we have suppressed the $(1S)$ superscript and defined the dimensionless \gls{se} factor for \gls{bsf} into the ground state as
\begin{align}
    \label{eq:SBSFID}
    \SBSF(\zeta,\xi&)\equiv ~\frac{\sqrt{\mathscr{P}_{10}(\zeta,\xi)}}{4\pi} \left(\frac{1+\zeta^2\gamma_{10}^2(\xi)}{\zeta^2}\right)\nonumber\\
    &\Bigg\{\frac{\kappa^7}{240}\mathscr{P}_{10}^2(\zeta,\xi)\left(\frac{1+\zeta^2\gamma_{10}^2(\xi)}{\zeta^2}\right)^4\left[|\langle\vec{k} | \vec{r}^2 |1S  \rangle|^2 + 2 |\langle\vec{k} | r^i r^j |1S  \rangle|^2\right]\nonumber\\
    &~~-\frac{\kappa^3}{6}\mathscr{P}_{10}(\zeta,\xi)\left(\frac{1+\zeta^2\gamma_{10}^2(\xi)}{\zeta^2}\right)^2\Re{\langle\vec{k}|\nabla^2_{\vec{r}}|1S\rangle\langle 1S| \vec{r}^2|\vec{k}\rangle}+\frac{1}{\kappa}|\langle\vec{k}|\nabla^2_{\vec{r}}|1S\rangle|^{2}\Bigg\}.
\end{align}
The quantity $\mathscr{P}_{10}(\zeta,\xi)$ denotes the phase space suppression factor and is given by 
\begin{equation}
    \label{eq:pssfactorgroundstate}
    \mathscr{P}_{10}(\zeta,\xi)\equiv 1-\frac{\mphi^2}{\big(\Ekn{1S}\big)^2}=1-\frac{4\zeta^4}{\alpha^2\xi^2(1+\zeta^2\gamma_{10}^2(\xi))^2}\,,
\end{equation}
yielding $\mathscr{P}_{10}(\zeta,\xi)\to 1$ in the Coulomb limit. In \cref{app:BSF_thermalfo}, we verify \cref{eq:SBSFID} by recalculating the \gls{bsf} cross section using the approach introduced in \cref{subsec:nonperturbativeeffectsfromQFT}.
\begin{figure}
    \centering
    \includegraphics[width=\textwidth]{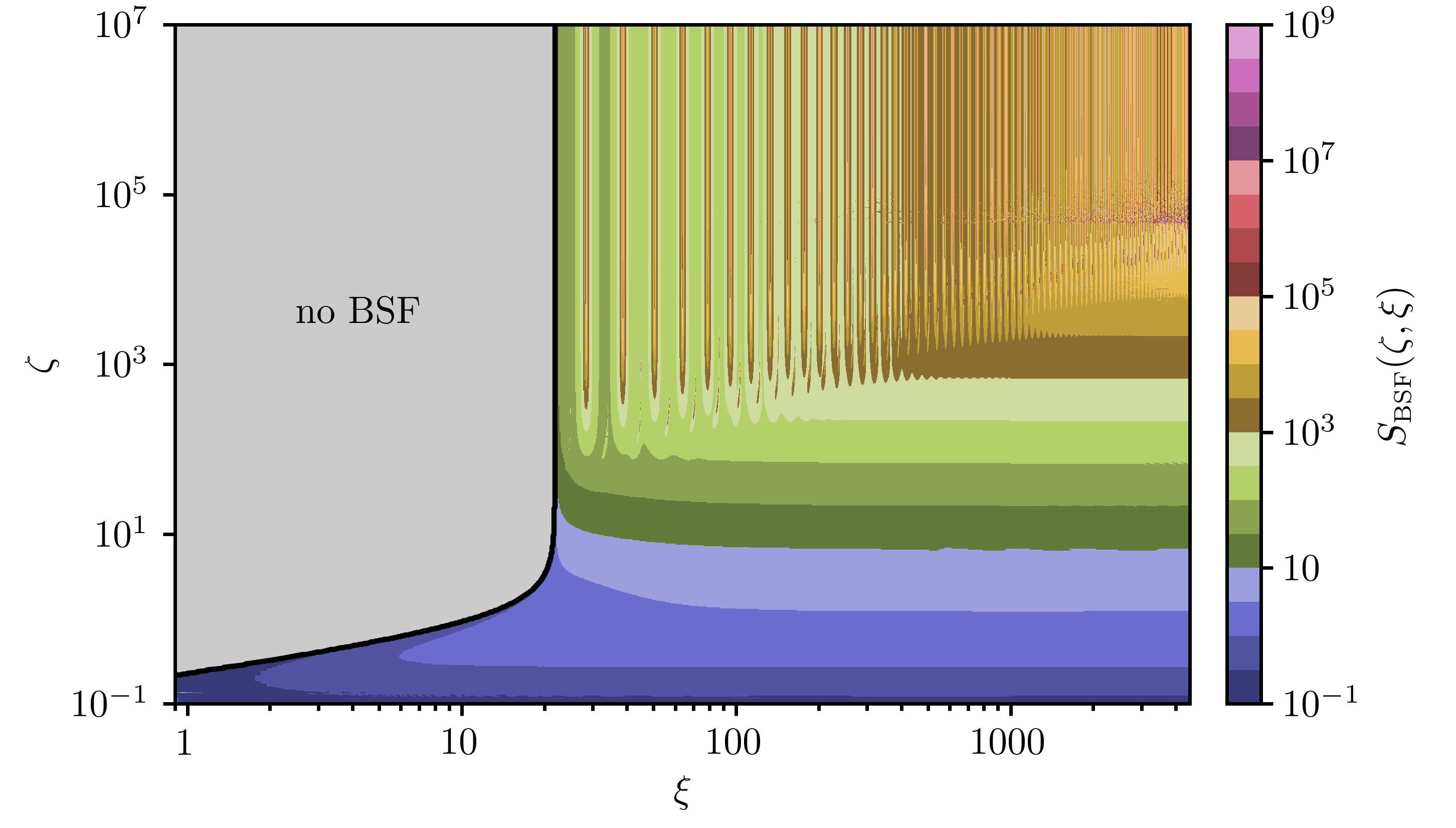}
    \caption[The SE factor for BSF into the ground state for a finite mediator mass.]{The \gls{se} factor for \gls{bsf} into the ground state for a finite mediator mass and $\alpha=0.1$ as a function of the dimensionless variables $\zeta$ and $\xi$, encompassing the \gls{dm} to mediator mass ratio, the scalar coupling strength and the relative \gls{dm} velocity. The gray area denotes the region of parameter space, where \gls{bsf} is not possible.}
    \label{fig:SBSFID}
\end{figure}
\begin{figure}
    \centering
    \includegraphics[width=\textwidth]{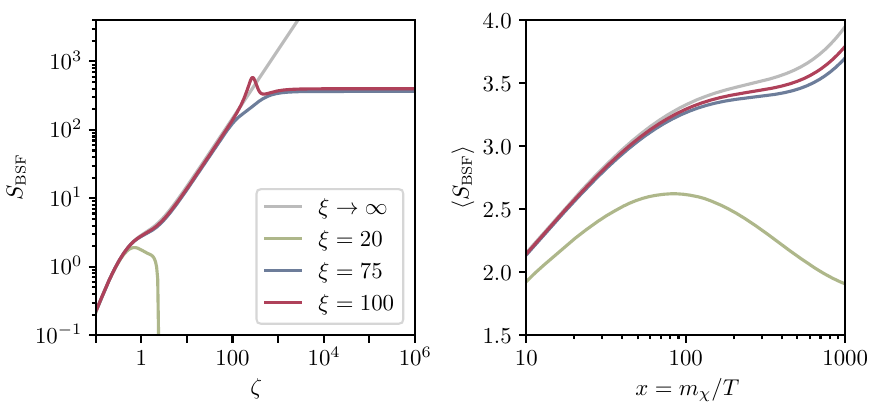}
    \caption[The SE factor for BSF into the ground state as well as its thermal average for selected values of $\xi$.]{The \gls{se} factor for \gls{bsf} into the ground state (left) as well as its thermal average (right) for $\alpha=0.1$ as a function of $\zeta$ or, respectively, $x\equiv\mchi/T$ for selected values of $\xi$ (identical for both plots). The case $\xi\to\infty$ (or $\mphi\to 0$) corresponds to the Coulomb limit. Adapted from \eref\cite{Biondini:2023ksj}.}
    \label{fig:SBSF_comparison}
\end{figure}\newp
As evident from \cref{fig:SBSFID}, the resonance structure in the $\xi$ direction is also present in the \gls{bsf} cross section, similar to the \gls{se} factors observed in \gls{dm} annihilations. Additionally, the gray regions in the parameter space indicate where \gls{bsf} is kinematically forbidden, \ie where $\Ekn{1S}<\mphi$. When $\xi>2/(\alpha\gamma_{10}^2(\xi))$, \gls{bsf} can always occur (see \cref{eq:bsfexistencecriterium}). For $\alpha=0.1$, this condition is approximately satisfied at $\xi_{\text{min}}\approx 22$. To better illustrate the dependence of the \gls{bsf} cross section on $\zeta$, we sliced through the parameter space at different fixed values of $\xi$ in \cref{fig:SBSF_comparison} (left) and also included the Coulomb case ($\xi\to\infty$). We observe a very different behaviour for various mediator to \gls{dm} mass ratios. As expected, when $\xi=20$, the \gls{se} factor drops to zero as soon as the $\bar{\chi}\chi$ pair no longer has sufficient kinetic energy to form a \gls{bs}. In the cases where \gls{bsf} is always possible, such as for $\xi=75$ and $\xi=100$, the curves flatten out towards large $\zeta$ or small $\vrel$, respectively. This behavior is attributed to the screening effect of the finite mediator mass. For $\xi=100$, a resonance structure appears.\newp
If we assume the non-relativistic scattering state follows a Maxwell-Boltzmann distribution, the thermal average of the \gls{bsf} cross section 
\begin{equation}
    \label{eq:avgBSFcrossectionID}
    \sigmabsfexp\equiv\frac{x^{3/2}}{2\sqrt{\pi}}\int_0^\infty \dd{\vrel}\vrel^2e^{-\frac{x}{4}\vrel^2}(1+f_\phi(\Ekn{1S}))\sigmabsf\vrel
\end{equation}
with $x\equiv \mchi/T$ is essentially the same as in \cref{eq:avgBSFcrossection}. Note that the statistical factor for the potentially relativistic mediator arises in $\pnry$ naturally when considering finite temperature effects (\cf\cref{eq:BSFcrosssectionFT}). In the right panel of \cref{fig:SBSF_comparison}, we displayed the thermal average of the \gls{se} factor of \gls{bsf}, defined via $\sigmabsfexp\equiv \pi\alpha^4/\mchi^2\,\langle\SBSF\rangle$, for the same choices of $\xi$ as in the left panel. We observe that the resonance structures observable for $\xi=100$ in the left panel are washed out and not visible any longer. The same is true with the drop off for $\xi=20$, such that we find smooth enhancement factors at \gls{fo} for all $\xi$.\newp
The computation of the \gls{bs} ionization rate can be performed in analogy to the \gls{bsf} cross section. The self-energy diagrams, which must be projected onto $|n\rangle$, are the same as in \cref{fig:selfenergyBSF} with the bound and scattering state lines exchanged (see \eg\erefs\cite{Biondini:2021ycj,Brambilla:2008cx}). However, the ionization cross section can be also just inferred from the \gls{bsf} cross section utilizing the principle of detailed balance, usually referred as Milne relation in this field of application \cite{Harz:2018csl,Biondini:2023zcz}
\begin{equation}
    \sigma_{\text{ion}}(\omega)=\frac{g_\chi^2}{g_B g_\phi}\frac{\mchi^2\vrel^2}{4\omega^2}\sigmabsf\,,
\end{equation}
with $\omega=\Ekn{n}$ the energy of the scalar mediator and $g_\chi=2$, $g_\phi=1$, $g_B$ the internal \gls{dof} of the \gls{dm} candidate, the mediator and the corresponding \gls{bs} ($g_{1S}=1$). The thermally averaged ionization rate is then given by
\begin{equation}
    \label{eq:GammaionexpID}
    \Gammaionexp\equiv g_\phi\int_{\omega_{\text{min}}}^\infty\frac{\dd{\omega}}{2\pi^2}\omega^2f_\phi(\omega)\sigma_{\text{ion}}(\omega)\,,
\end{equation}
where $\omega_{\text{min}}=\max\{\abs{\En},\mphi\}$ represents the minimum energy required for the scalar to ionize a \gls{bs}. Note that for $\mphi\to 0$ we recover the structural form of \cref{eq:ionizationrate}. For the calculation of the \gls{dm} abundance we will in the following only consider \gls{bs} ionization of the ground state, which is why we have suppressed the superscript $(n)$ in the Milne relation and in the definition of $\Gammaionexp$.\newp
So far we have not considered the running of the scalar coupling $\alpha$, on which we want to catch up on now. The scalar coupling strength appears in the hard matching coefficients of \cref{eq:IMmatch_coeff_1,eq:IMmatch_coeff_2,eq:IMmatch_coeff_3,eq:IMmatch_coeff_4}, in the \gls{se} factors of \cref{eq:annihilationcrosssectionID,eq:boundstatedecaynSID} describing soft processes, as well as in the \gls{bsf} cross section in \cref{eq:BSFcrosssectionID} generated via ultra-soft vertices. It therefore has to be evaluated at the hard, soft, and ultrasoft scale, resulting in different physical values one has to keep track of if its running is considered. We used two different approaches to study the running of $\alpha$: 1) Taking into account \gls{dm} fermion, scalar self energy, as well as vertex one-loop diagrams of the full theory and expanding the integrals according to their soft and ultra-soft momenta with respect to the mediator and \gls{dm} mass scales. 2) Performing the calculations directly in $\nry$ or $\pnry$. Both methods show that there is no significant running induced at scales below $\mchi$, thus $\alpha$ remains approximately frozen at the hard scale. The running of $\alpha_5$ is neglected because it is smaller than $\alpha$ and does not contribute at leading order to soft and ultra-soft processes.


\subsubsection{The relic density}
\label{subsubsec:relicdensityid}

To calculate the relic density of \gls{dm}, we employ the same approach as in \cref{subsec:DMabundancesW} utilizing a set of \glspl{be} for \gls{fo}. We first assume a quasi steady-state for the \gls{bs} yield, \ie $\dd{Y_B}/\dd{x}\approx 0$, holding if the processes governing the rate at which the \gls{bs} yield changes are faster than the Hubble rate. The \gls{dm} yield as a function of $x\equiv\mchi/T$ incorporating non-perturbative effects can then be summarized in a single equation (\cf\cref{eq:MFOwSEwBSF})
\begin{equation}
    \dv{Y_\chi}{x} = -\frac{1}{2}\xi_1(x)\sigmaanntotexp\left(Y_\chi^2-{\Yeq{\chi}}^2\right)\,,
\end{equation}
where $\Yeq{\chi}$ is the equilibrium yield given in \cref{eq:equilibriumyield}, $\xi_1(x)\equiv\sqrt{\pi/45}M_{\text{Pl}}\mchi\gstar^{1/2}(x) x^{-2}$ and the factor $1/2$ accounts for the Dirac nature of the \gls{dm} candidate. The total thermally averaged cross section (neglecting bound-to-bound transitions) is given by
\begin{equation}
    \label{eq:totalannihilationcrosssectionID}
    \sigmaanntotexp\equiv\sigmaannexp +\sum_n \frac{\big\langle\Gammadec^{(n)}\big\rangle}{\big\langle\Gammaion^{(n)}\big\rangle+\big\langle\Gammadec^{(n)}\big\rangle}\big\langle\sigmabsf^{(n)}\vrel\big\rangle\,,
\end{equation}
where $\sigmaannexp$ is the thermal average of \gls{se} corrected annihilation cross section given in \cref{eq:annihilationcrosssectionID} (\cf also \cref{eq:thermallyaveragedcrosssectionII}) and we sum over \gls{bs} contributions with different quantum numbers $(n)$. The factor in front of the \gls{bsf} cross section serves as a weight that determines which percentage of the produced \gls{bs} will actually decay and therefore effectively deplete the \gls{dm} abundance. For indirect detection, the prefactor becomes $1$ because the ionization rate goes to zero due to the lack of thermal mediators in the late Universe. As previously stated, we will only include the $1S$ \gls{bs} in the following. Therefore, $\sigmabsfexp$ and $\Gammaionexp$ are given by \cref{eq:avgBSFcrossectionID,eq:GammaionexpID}, respectively. The thermal average of the ground state decay rate of \cref{eq:boundstatedecaynSID} can be calculated from \cref{eq:Gammadecexp}.\newp
\begin{figure}[ht]
    \centering
    \includegraphics[width=\textwidth]{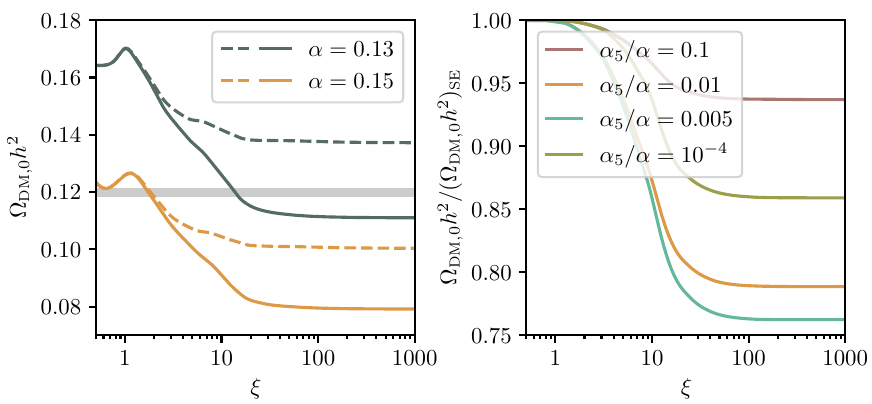}
    \caption[The DM relic abundance in a FO scenario considering non-perturbative effects for different benchmark points.]{\textit{Left}: The \gls{dm} relic abundance observed today as a function of $\xi$ for a benchmark point $\mchi=\SI{2}{\TeV}$, $\alpha_5/\alpha=0.01$ and two different values of $\alpha$. Solid lines correspond to the case where both, \gls{se} and \gls{bs} effects are taken into account, whereas for the dashed lines only \gls{se} has been included. The gray area corresponds to the observed relic abundance today $\Omegadmt h^2=0.1200\pm 0.0012$ \cite{Planck:2018vyg} with a $1\sigma$ uncertainty band. \textit{Right}: The ratio of the \gls{dm} abundance taking into account both non-perturbative effects with respect to the \gls{dm} abundance obtained by only considering \gls{se} for a benchmark point $\mchi=\SI{2}{\TeV}$, $\alpha=0.15$ and different values of $\alpha_5/\alpha$. Adapted from \eref\cite{Biondini:2023ksj}.}
    \label{fig:OmegaDM}
\end{figure}
In \cref{fig:OmegaDM} (left), we displayed the \gls{dm} relic abundance computed from \cref{eq:DMabundance} as a function of $\xi$ for the benchmark point $\mchi=\SI{2}{\TeV}$, $\alpha_5/\alpha=0.01$ and two different values for $\alpha$. Dashed lines correspond to the case where \gls{bsf} effects are not taken into account, \ie $\sigmaanntotexp=\sigmaannexp$, whereas for solid lines both non-perturbative effects contribute. We can see that for large $\xi$, the curves flatten as the solution converges to the Coulomb limit, where non-perturbative effects become maximal (\cf also \cref{fig:SBSF_comparison}). Whereas \gls{bs} effects become sizeable for $\xi\gtrsim 5$ considering this benchmark point, for $\xi\lesssim 3$ they are absent due to phase space suppression effects of \gls{bsf} and the overall existence limit of the ground state $\xi> 0.84$ (\cf\cref{subsubsec:SEboundstatedecays}). For $\alpha=0.13$, \gls{bs} effects are even crucial to avoid overclosing of the Universe.\newp
In the right panel of \cref{fig:OmegaDM}, we show the ratio of the \gls{dm} relic density when including or disregarding \gls{bs} effects. We fixed $\mchi=\SI{2}{\TeV}$ and $\alpha=0.15$, and then varied the ratio $\alpha_5/\alpha$ between the scalar and pseudo-scalar couplings over a range of values. While decreasing the coupling ratio an interesting trend can be observed. The effects from \gls{bs} first become stronger for smaller pseudo-scalar couplings as visible by the orange ($\alpha_5/\alpha=0.01$) and blue line ($\alpha_5/\alpha=0.005$) in comparison to the red graph ($\alpha_5/\alpha=0.1$). The reason for this is the linear dependence of the velocity-independent s-wave part of the annihilation cross section on the pseudo-scalar coupling (\cf\cref{eq:annihilationcrosssectionID}), whereas \gls{bsf} is independent of $\alpha_5$. For very small ratios, as for the green line ($\alpha_5/\alpha=10^{-4}$), the trend reverses because $\alpha_5$ also affects the decay rate of the ground state, therefore decreasing the weighting factor of the \gls{bsf} contribution.\footnote{In fact, for $\alpha_5\to 0$ the ground states cannot decay and $\Gammadecexp/(\Gammadecexp+\Gammaionexp)\to 0$, resulting in an absence of \gls{bs} effects. Allowing for three body decays, \ie diagrams with three scalar vertices or a trilinear coupling vertex from a $\rho_\phi\phi^3$ term in the Lagrangian, would alleviate the problem by making the decay width of the ground state independent of $\alpha_5$.} This behaviour as well as the importance of \gls{bs} effects in general is rather independent of the \gls{dm} mass.\newp
As mentioned earlier, for the relic density calculations as well as for indirect detection we will only include the ground state $1S$ in our estimate of \gls{bs} effects on the total annihilation cross section. In the light of recent works about the high relevance of excited states for the \gls{fo} of \gls{dm} \cite{Garny:2021qsr,Binder:2021vfo,Binder:2023ckj}, we need to justify this decision. The cited studies focus on models with massless vector mediators from an unbroken gauge symmetry. They find a large logarithmic enhancement of the total \gls{bsf} cross section (\ie in the right term in \cref{eq:totalannihilationcrosssectionID}) considering a large number of Coulomb \gls{bs}. Partly, this is caused by the enormous increase of \gls{bs} with $l\leq n-1$ for large $n$. The model in our work comprises of a massive scalar mediator. The resulting Yukawa potential only features a finite number of \gls{bs}, such that a large enhancement from summing up to large $n$ is not expected.\footnote{We can impose a conservative limit on the number of \gls{bs} supported by a Yukawa potential by requiring that the binding energy of the \gls{bs} exceeds the mediator mass. For the parameter space of interest in our model, this leads to the constraint of $n\leq 22$.} To estimate effects from the lowest excited states we have included in the Coulomb limit the four $n=2$ states, namely the $2S$ singlet and $2P$ triplet, for which we have calculated the \gls{se} \gls{bsf} factors in \cref{app:higherBS}. From \cref{fig:higherBS} in the same appendix, it can be observed that the inclusion of states with $n>2$ will not substantially alter this estimate. For the $n=2$ states we found that the corrections to the relic density always stay below $10\%$ within our parameter regime of interest. The dominant contribution to these corrections arises from the $2S$ states for two reasons: 1) The weighting factor for the $2P$ state is significantly smaller than for the $2S$ state, because the decay rate is strongly suppressed (\cf\cref{eq:boundstatedecaynPJID}). 2) Due to the missing dipole moment in the ultra-soft vertices of $\pnry$ (\cf\cref{eq:pNRY}), only $\Delta l=0,2$ transitions between \gls{bs} are allowed \cite{Oncala:2018bvl,Biondini:2021ccr,Biondini:2021ycj}. Therefore,  $2P$ states cannot transition into $2S$ states effectively. A more thorough study of the effects of higher \gls{bs} in finite mass mediator models goes beyond the scope of this work.


\subsection{Indirect detection}
\label{subsec:indirectdetection}

\begin{figure}
    \centering
    \includegraphics[width=\textwidth]{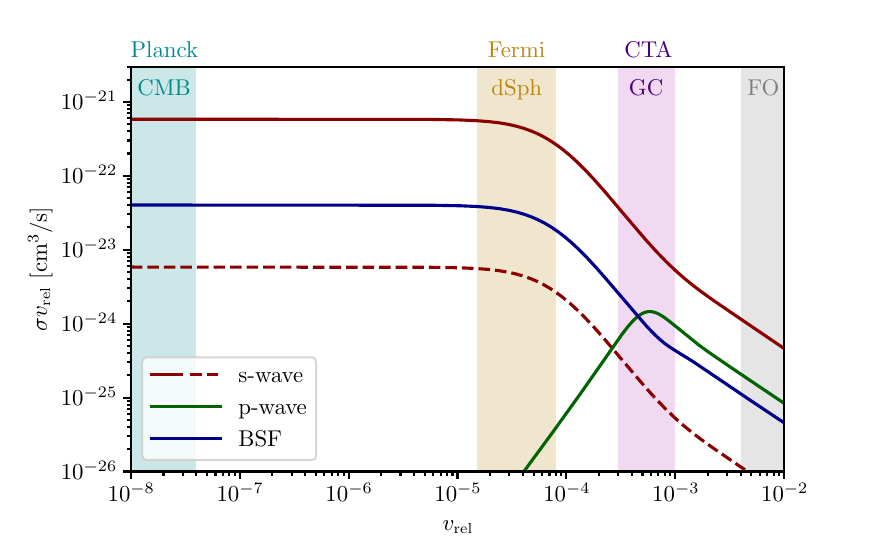}
    \caption[Velocity dependence of the leading order contributions to the total DM annihilation cross section from s-wave and p-wave annihilation as well as BSF.]{Velocity dependence of the leading order contributions to the total \gls{dm} annihilation cross section from s-wave and p-wave annihilation as well as BSF into the ground state for a benchmark point of $\mchi=\SI{1}{\TeV}$, $\mphi= \SI{1}{\GeV}$ and $\alpha=0.1$. For the pseudo-scalar coupling, two different values of $\alpha_5=10^{-3}$ (solid lines) and $\alpha_5=10^{-5}$ (dashed lines) have been chosen and are depicted for s-wave annihilation only, since the $\alpha_5$ dependence on the p-wave (\gls{bsf}) contribution is subdominant (absent). We have shaded regions of typical DM velocities during \gls{fo}, at times of the \gls{cmb} decoupling as well as for \glspl{dsph} and the \gls{gc}. Adapted from \eref\cite{Biondini:2023ksj}.}
    \label{fig:crossectionvrel}
\end{figure}

Within our model, \gls{dm} annihilations at late times are still possible (especially in regions with a high \gls{dm} density) but happen far too rarely to significantly affect the total \gls{dm} abundance on cosmological scales. However, their presence can be detected by methods of indirect detection, \eg through their effects on the flux of various cosmic rays. We will primarily focus on present and future gamma-ray telescopes in the following discussion, as they have the prospects to provide the most stringent limits in our parameter regime of interest. Specifically, we will look at data by \gls{fermi} on \glspl{dsph} as well as prospects from the \gls{cta} collaboration of a future survey of the \gls{gc} of the Milky Way (\cf\cref{subsec:detectionmethods}). Additionally, we will take into account bounds on late time energy injection during \gls{cmb} decoupling derived by the Planck collaboration.\newp
As sketched in \cref{fig:crossectionvrel} (shaded regions), all three environments under consideration, namely the \gls{cmb}, \glspl{dsph} and the \gls{gc}, feature a different velocity distribution of \gls{dm}. While the characteristic \gls{dm} velocities in \glspl{dsph} and the \gls{gc} are of $\order{10^{-5}}$ and $\order{10^{-3}}$, respectively, the \gls{dm} velocity at times around the \gls{cmb} is typically of $\order{10^{-8}}$. \Cref{fig:crossectionvrel} depicts the leading order s- and p-wave contributions as well as the \gls{bsf} cross section into the ground state for an exemplary configuration of masses and couplings, respectively.\footnote{Due to the considerably lower velocities relevant for indirect detection as compared to \gls{fo}, all higher order velocity contributions to the corresponding cross sections can be safely neglected.} It is apparent that the typical velocity dependence of $\propto 1/\vrel$, present for all contributions at the time of \gls{fo} (\cf\cref{subsubsec:SEannihilations,subsubsec:sWSEBSF}), morphs into a non-trivial behaviour at lower velocities due to the finite mass of the mediator. These low-velocity regimes are relevant for indirect detection. Therefore, it is imperative to discern the dominant contributions within the aforementioned astrophysical environments. For \glspl{dsph} and the \gls{gc}, the cross sections associated with s-wave annihilation and \gls{bsf} are enhanced due to their scaling with low velocities. This is also true for the p-wave cross section in the region $\vrel \gg \mphi/\mchi$, where \gls{se} is capable of compensating for the p-wave velocity suppression. The \gls{cmb} is mostly influenced by the s-wave and \gls{bsf} cross sections whereas their relative strength is governed by the value of $\alpha_5$. Within this regime, both contributions have flattened due to the finite size of the potential and thus their \gls{se} factors have been rendered constant. For all configurations of masses and couplings considered in this work, the p-wave annihilation cross section will have no relevance for the \gls{cmb} due to its $\vrel^2$ velocity suppression.\newp
In order to take into account the non-trivial velocity dependence of the annihilation cross section in the phenomenology of indirect detection, a few adjustments to the conventional approaches are necessary. Considering the \gls{cmb}, a redefinition of the thermally averaged cross section is sufficient, on which we will comment later. In case of galactic environments, such as \glspl{dsph} and the \gls{gc}, the matter is more intricate. In the context of gamma-ray signals from these objects, the pivotal observable is the photon flux. The contribution to the differential photon flux from (Dirac) \gls{dm} annihilations reads (see \eg\eref\cite{Ferrer:2013cla})
\begin{equation}
    \label{eq:photonfluxgeneral}
    \dv{\Phi_\gamma}{E_\gamma}=\frac{1}{16\pi \mchi^2}\dv{N_\gamma}{E_\gamma}  \int_{\Delta\Omega}\dd{\Omega}\!\int_0^\infty\dd{\psi}\!\int \dd[3]{v_1}\!\int\dd[3]{v_2}f_\chi(r(\psi,\Omega),\vec{v}_1) f_{\bar{\chi}}(r(\psi,\Omega),\vec{v}_2) \, \sigma\vrel.
\end{equation} 
Here, the \textit{photon spectrum} $\dd{N_\gamma}/\dd{E_{\gamma}}\equiv\sum_f\mathcal{B}_f\dd{N^{(f)}_\gamma}/\dd{E_{\gamma}}$ denotes the number of photons that get produced per photon energy $E_\gamma$ per annihilation, where the sum runs over all possible annihilation final states $f$ weighted by their branching fraction $\mathcal{B}_f$. The (in general velocity dependent) annihilation cross section $\sigma\vrel$ is first integrated over the velocities of the initial state \gls{dm} particles together with the \gls{dm} phase space \glspl{df} $f_\chi=f_{\bar{\chi}}$.\footnote{This cross section includes contributions from s- and p-wave annihilations as well as \gls{bsf}. The \glspl{df} here are position \text{and} velocity dependent because the astrophysical environment is neither homogeneous nor isotropic.} The remaining integration is then performed over the field of view of the gamma-ray telescope, which is usually given in galactic coordinates, namely the galactic longitude $l$ and latitude $b$ as well as the \gls{los} distance $\psi$. For the galactic coordinates parametrization, the angular measure is given by $\dd{\Omega}=\cos b \dd{b}\dd{l}$, and the distance of a DM particle from the center of the galaxy is determined by $r(\psi,\Omega)=\sqrt{D^2+\psi^2-2D\psi\cos b\cos l}$, where $D$ represents the distance of our sun to the center of the galaxy ($D=\SI{8.5}{\kpc}$ for the \gls{gc} and $\order{10-100\,\si{\kpc}}$ for the different \glspl{dsph} under consideration \cite{Boddy:2017vpe}).\newp
In case that $\sigma\vrel$ is velocity independent, as for the leading order s-wave contribution without \gls{se}, it can be pulled out of the integration, such that the remainder yields the usual $J$-factor, which is defined by (see \eg\cite{Bertone:2010zza})
\begin{align}
    \label{eq:J0}
    J_0\equiv\int\dd{\Omega}\int_0^\infty\dd{\psi}\rho_\chi(r(\psi,\Omega))^2\,,
\end{align}
where $\rho_\chi(r)\equiv\int \dd[3]{v} f_\chi(r,\vec{v})$ is the \gls{dm} energy density. In the following, we assume that the \glspl{df} and, consequently, the energy density are spherically symmetric, which is a reasonable assumption for galactic environments and will greatly simplify our subsequent calculations. If the cross section is p-wave dominated or non-perturbative effects play a role, the non-trivial velocity dependence of $\sigma \vrel$ demands a full calculation of the velocity average which in turn requires knowledge about the behaviour of the \glspl{df} $f_\chi$. To start with, we will in analogy to the thermally averaged total annihilation cross section in \cref{eq:totalannihilationcrosssectionID} assume that we can split up the total annihilation cross section into 
\begin{equation}
    \label{eq:crosssectionid}
    \sigma \vrel=\sum_{l=0}^{\infty}\sigma_{l}\vrel^{2l}\Sann(\zeta,\xi)+\sum_{n}\sigmabsf^{(n)}\vrel\,,
\end{equation} 
where any weighting factors in front of the $\sigmabsf^{(n)}\vrel$ can be neglected since the ionization rate is negligible. Since we are only considering \gls{bsf} into the ground state for the same reasons as in \cref{subsubsec:relicdensityid}, we will drop the sum over $n$ in the second term and define $\sigmabsf\vrel\equiv\sigmabsf^0\SBSF$ with $\sigmabsf^0\equiv \pi\alpha^4/\mchi^2$ (\cf\cref{eq:BSFcrosssectionID2}). In the first term, we only consider the leading order contributions to s- and p-wave annihilation, where the perturbative cross sections, which have been stripped from their angular and velocity dependence (\cf\cref{eq:annihilationcrosssectionID}), are given by 
\begin{equation}
    \sigma_0\equiv\frac{\Im{f(\spec{1}{S}{0})}}{\mchi^2},\quad \sigma_1\equiv\frac{\Im{f(\spec{3}{P}{0})}+5\Im{f(\spec{3}{P}{2})}}{12\mchi^2}\,.
\end{equation}
This splitting enables us to define a generalized $J$-factor for annihilation (and an analogous factor for \gls{bsf})
\begin{equation}
    \label{eq:Jsannl}
    \Jsann{l}(\xi)\equiv\int_{\Delta\Omega}\!\dd{\Omega}\!\int_0^\infty\!\dd{\psi}\!\int\!\dd[3]{v_1}f_\chi(r(\psi,\Omega),\vec{v}_1)\!\int\!\dd[3]{v_2}f_\chi(r(\psi,\Omega),\vec{v}_2)\,\vrel^{2l}\,\Sann(\zeta,\xi)\,,
\end{equation}
which only depends on the \gls{dm} and mediator masses through $\xi$ as well as explicitly on the scalar coupling $\alpha$. We can therefore write \cref{eq:photonfluxgeneral} as 
\begin{equation}
    \label{eq:photonflux_Jfactors}
    \dv{\Phi_\gamma}{E_\gamma}=\frac{1}{16\pi \mchi^2}\dv{N_\gamma}{E_\gamma}\left(\sigma_0\Jsann{0}(\xi) +\sigma_1 \Jsann{1}(\xi)+\sigmabsf^0\Jsbsf(\xi)\right)\,.
\end{equation}
In the subsequent sections, we will derive the generalized $J$-factors and the photon spectra $\dd{N_\gamma}/\dd{E_{\gamma}}$ separately. For the prior, however, we must first determine the distribution functions (\glspl{df}) of DM in galactic environments, with which we will start.


\subsubsection[The dark matter distribution function \texorpdfstring{$f_{\chi}$}{???}]{\protect\boldmath The dark matter distribution function \texorpdfstring{$f_{\chi}$}{???}}
\label{subsubsec:fDM}

The \gls{df} of \gls{dm} in galaxies is not directly observable through astrophysical measurements. Consequently, it is necessary to incorporate additional information to ascertain a form of $f_\chi$ that ensures consistency with observations of measurable quantities deduced from it. In the following, we will primarily employ methods developed for spherically symmetric distributions of \gls{dm} in galaxies, which are applicable to collisionless systems in quasi-static equilibrium. A detailed discussion of these techniques can be found in \eref\cite{binney:2011}, whereas \erefs\cite{Ferrer:2013cla,Boddy:2017vpe,Lacroix:2018qqh} focus on their application to galactic \gls{dm} structures. An alternative strategy that we do not use here is the direct sampling of the \gls{df} from numerical simulations of \gls{dm} halos, as has been done \eg in \eref\cite{Board:2021bwj,McKeown:2021sob}.\newp
Within the \glspl{dsph} under consideration (or the \gls{gc}), we assume a static and spherically symmetric \gls{dm} density function $\rho_\chi$. For a sufficiently isolated system, this allows us to determine the gravitational potential of the galaxy $\Psi$ through Poisson's equation
\begin{align}
    \label{eq:Poissonequation}
    \Delta \Psi = - 4 \pi G(\rho_\chi+\rho_b)\,,
\end{align}
with $G$ the Newtonian constant and $\rho_b$ the local baryon density within the galaxy. We can then determine a unique and ergodic \gls{df} for \gls{dm} supported by this potential via the Eddington inversion method \cite{binney:2011}
\begin{equation}
    \label{eq:EddingtonEquation}
    f_\chi(\epsilon)=\frac{1}{\sqrt{8}\pi^2}\left(\int_0^\epsilon\dv[2]{\rho_\chi}{\Psi}\frac{\dd{\Psi}}{\sqrt{\epsilon-\Psi}}-\frac{1}{\sqrt{-\epsilon}}\left.\dv{\rho_\chi}{\Psi}\right|_{\Psi=0}\right), 
\end{equation}
where $\epsilon=\Psi(r)-v^2/2$ denotes the \gls{dm} energy per unit mass. Following from the isolation criterion mentioned above, we will set the gravitational potential $\Psi(r)$ to $0$ at $r\to\infty$, which causes the second term to disappear.\footnote{Technically, setting $\Psi(r)=0$ at $r\to\infty$ is unphysical because the gravitational potential of a galaxy is naturally bounded by its neighbors. However, it was shown in \eref\cite{Lacroix:2018qqh} that the effect of a large but finite boundary compared to an infinite one is small such that we will neglect it in the following.} Since $\Psi$ is monotonic in $r$, we can rewrite \cref{eq:EddingtonEquation} directly as a function of $r$ and $v$ \cite{Ferrer:2013cla}
\begin{equation}
    \label{eq:distribtutionfunctionID}
    f_\chi(r,v)=\frac{1}{\sqrt{8}\pi^2} \int_{r_{\text{min}}}^\infty \frac{\dd{r'}}{\sqrt{\Psi(r)-\Psi(r')-\frac{v^2}{2}}}\left(\dv{\Psi}{{r'}}\right)^{-1}\left[\dv[2]{\rho_\chi}{{r'}}-\left(\dv{\Psi}{{r'}}\right)^{-1}\dv[2]{\Psi}{{r'}}\dv{\rho_\chi}{{r'}}\right]\,,
\end{equation}
where $r_{\text{min}}$ is determined by the solution to the equation $\Psi(r)-\Psi(r_{\text{min}})-v^2/2=0$.\newp
A range of different parametrizations for the energy density $\rho_\chi$ of a \gls{dm} halo have been considered in the literature (see \eg \eref\cite{Cardone:2005ne} for an overview). As for \glspl{dsph}, we will assume a \gls{nfw} profile \cite{Navarro:1995iw,Navarro:1996gj,Moore:1999nt} in the following. It is given by
\begin{equation}
    \rho_{\text{\tiny NFW}}= \frac{\rho_0}{\frac{r}{r_s}\left(1+\frac{r}{r_s}\right)^2}\,,
\end{equation}
where $r_s$ is the \textit{scale radius} and $\rho_0$ fixes the normalization. For a better comparability with other galactic models as well as astrophysical measurements, it is advisable to establish a frame of reference for the model parameters. Following \eg\eref\cite{Cardone:2005ne}, we will define the scale radius via the logarithmic slope
\begin{equation}
    \label{eq:logslope}
    \left.\dv{\log \rho_\chi}{\log r}\right|_{r=r_s}\equiv-2,\qquad \rho_\chi(r_s)=\rho_s\,,
\end{equation}
where $\rho_s$ denotes the \textit{scale density}, which is given by $\rho_s=\rho_0/4$ for the \gls{nfw} profile. For our \gls{dm} model of the \gls{gc}, we employ an \textit{Einasto profile} \cite{Navarro:2003ew,Graham:2005xx} to stay comparable to the sensitivity study performed by the \gls{cta} collaboration \cite{CTA:2020qlo}. It reads
\begin{equation}
    \label{eq:Einastoprofile}
    \rho_{\text{\tiny Ein}}= \rho_0 \,\exp\left[-\frac{2}{\gamma}\left[\left(\frac{r}{r_s}\right)^\gamma-1\right]\right]\,,
\end{equation}
with $\gamma=0.17$ and $\rho_s=\rho_0$. Note that although we will perform our study on indirect detection in \glspl{dsph} and the \gls{gc} using the designated \gls{dm} density profiles, we will employ the respective other profile for cross-verification and to estimate potential errors. For our analysis of \glspl{dsph} we will calculate the best-fit values for ($r_s,\rho_s)$ from the parameters $(r_{\text{max}},V_{\text{max}})$ given in Table 1 of \eref\cite{Boddy:2017vpe}, which denote the maximal circular velocity $v_{\text{cir}}(r_{\text{max}})=V_{\text{max}}$ and its radius. The circular velocity (\cf\cref{eq:circularvelocity}) is defined as  $v_{\text{cir}}\equiv \sqrt{G M(r)/r}$, with $M(r)=4\pi\int_0^r \dd{r'} (r')^2\rho_\chi(r')$ being the enclosed \gls{dm} mass at radius $r$. The relations of the two parameter tuples for the considered density profiles are given by $r_s=r_{\text{max}}/C_1$ and $\rho_s=(V_{\text{max}}/C_2)^2/(16\pi G r_s^2)$ with $(C_{1}^{\text{\tiny NFW}},C_{2}^{\text{\tiny NFW}})=(2.163,0.465)$ for the \gls{nfw} and $(C_{1}^{\text{\tiny Ein}},C_{2}^{\text{\tiny Ein}})=(2.204,0.473)$ for the Einasto profile \cite{Boddy:2017vpe}. For the \gls{gc}, we set $r_s= \SI{20}{\kpc}$, $\rho_s=\SI{0.081}{\GeV/\cm^3}$ as suggested by the \gls{cta} prospect analysis \cite{CTA:2020qlo}.\newp
To define the properties of a \gls{dm} halo which can in turn be used to further enhance the comparability of different density models, we will define two additional parameters: the \textit{virial radius} $\rvir$ and the \textit{virial mass} $\Mvir$. The virial radius of a system is a radius within which the virial theorem can be applied (\cf\cref{subsec:dmexpevidence}). Roughly speaking, this is the distance at which the galaxy, including its \gls{dm} halo, can still be regarded as a stable system of discrete particles bound by a gravitational force. In our context, the virial radius measures the extent of the \gls{dm} halo and the virial mass is defined as the \gls{dm} mass encapsulated by it, \ie $\Mvir\equiv M(\rvir)$. Vice versa, equating the virial mass of a spherical \gls{dm} halo with the mass of a sphere of the same size but a flat background times an overdensity parameter $\Delta_c$ (typically set to $\Delta_c=200$), defines the virial radius
\begin{equation}
    \label{eq:virialradius}
    4\pi\int_0^{\rvir} \dd{r} r^2\rho_\chi(r)\equiv\frac{4}{3}\pi \rvir^3\Delta_c \rhocrit\,.
\end{equation}
Using \cref{eq:virialradius}, one can calculate $\rvir$ and $\Mvir$ from other scaling parameters such as $(r_s,\rho_s)$. We have displayed them for the \glspl{dsph} under consideration as well as for the \gls{gc} in \cref{tab:DMhaloparameters} considering both \gls{dm} density profiles.\newp
\begin{table}
	\centering
	\caption[Astrophysical parameters of the DM halos for dSphs and the GC.]{Astrophysical parameters of the \gls{dm} halos for \glspl{dsph} and the \gls{gc} used for subsequent computations, namely the distance $D$ to the sun (in $\si{\kpc}$), the concentration parameter $c$ as well as the virial radius $\rvir$ (in $\si{\kpc}$) and the virial mass $\Mvir$ in units of the solar mass $M_\odot$. All model dependent parameters have been calculated for an underlying \gls{nfw} and Einasto density profile from data provided in \erefs\cite{Boddy:2017vpe,CTA:2020qlo}.}
	\begin{tabular}{r@{\hskip 0.3cm}c@{\hskip 0.3cm}c@{\hskip 0.3cm}c@{\hskip 0.3cm}c@{\hskip 0.3cm}c@{\hskip 0.3cm}c@{\hskip 0.3cm}c}
		\toprule \\[-0.4cm]
		galaxy & $D~[\si{\kpc}]$& $c^{\text{\tiny{NFW}}}$ & $c^{\text{\tiny{Ein}}}$ & $\rvir^{\text{\tiny{NFW}}}~[\si{\kpc}]$ & $\rvir^{\text{\tiny{Ein}}}~[\si{\kpc}]$ & $\Mvir^{\text{\tiny{NFW}}}~[M_{\odot}]$ & $\Mvir^{\text{\tiny{Ein}}}~[M_{\odot}]$\\
		[0.1cm]\midrule \\[-0.2cm]
        Coma B. & $44$ & $44.84$ & $44.32$ & $7.879$ & $7.640$ & $\SI{5.17e7}{}$ & $\SI{4.71e7}{}$ \\[0.05cm]
        Ursa Minor & $76$ & $34.55$ & $34.59$ & $21.09$ & $20.71$ & $\SI{9.91e8}{}$ & $\SI{9.39e8}{}$ \\[0.05cm]
        Draco & $76$ & $37.83$ & $37.72$ & $15.04$ & $14.71$ & $\SI{3.60e8}{}$ & $\SI{3.37e8}{}$ \\[0.05cm]
        Sergue 1 & $23$ & $38.85$ & $38.68$ & $13.65$ & $13.34$ & $\SI{2.69e8}{}$ & $\SI{2.50e8}{}$ \\
		[0.3cm]\midrule \\[-0.2cm]
        GC & $8.5$ & $11.83$ & $12.17$ & $236.6$ & $243.4$ & $\SI{1.40e12}{}$ & $\SI{1.52e12}{}$\\[0.2cm]
		\bottomrule
	\end{tabular}
	\label{tab:DMhaloparameters}
\end{table}
It will be more convenient in the following to switch to dimensionless coordinates \cite{Ferrer:2013cla}. We will define $x\equiv r/\rvir$ as a fractional measure of distance from the center and $c\equiv \rvir/r_s$ a concentration parameter which can be calculated using \cref{eq:virialradius} (also displayed in \cref{tab:DMhaloparameters}). The density functions can be written as $\rho_\chi\equiv \rho_0\rhot(x,c;\dots)$ with $\rhot$ its dimensionless part. We can express $\rho_0$ through the virial parameters $\Mvir=4\pi\rvir^3\rho_0 g(c)$, where we further define $g(c;\dots)\equiv\int_0^\infty x^2 \rhot(x,c;\dots)$. For the \gls{nfw} and Einasto profile this reparametrization reads
\begin{equation}
    \rho_\chi=\frac{\Mvir}{4\pi\rvir^3}\frac{1}{g(c)}\rhot(x,c)\equiv\frac{\Mvir}{4\pi\rvir^3}
    \begin{cases}
        \frac{1}{\gNFW(c)} \frac{1}{cx(1+cx)^2} & \text{NFW}\\
        \frac{1}{\gEin(c;\gamma)} \exp\left[-\frac{2}{\gamma}\left[(cx)^\gamma-1\right]\right]& \text{Einasto}\\
    \end{cases}\,,
\end{equation}
with \cite{Ferrer:2013cla}
\begin{align}
    \gNFW(c)&=\frac{1}{c^3}\left[\log(1+c)-\frac{c}{1+c}\right]\,,\\
    \gEin(c;\gamma)&=\frac{e^{2/\gamma}}{c^3\gamma}\left(\frac{2}{\gamma}\right)^{-3/\gamma}\left[\Gamma\left(\frac{3}{\gamma}\right)-\Gamma\left(\frac{2}{\gamma},\frac{2}{\gamma}c^\gamma\right)\right],
\end{align}
where $\Gamma$ denotes the (incomplete) gamma function (\cf\cref{eq:gammafunction}). Likewise, all other parameters within \cref{eq:distribtutionfunctionID} can be substituted with their dimensionless counterparts
\begin{equation}
    \Psi\equiv\frac{G\Mvir}{\rvir}~\psit,\quad \epsilon\equiv\frac{G\Mvir}{\rvir}~\et, \quad v^2\equiv \frac{G\Mvir}{\rvir}\vt^2,\quad 
\end{equation}
such that $\et=\psit-\vt^2/2$. The dimensionless \gls{df} then follows as
\begin{equation}
    \label{eq:ftildeDM}
    \fchit(x,\vt)=\int_{x_{\text{min}}}^\infty \frac{\dd{x'}}{\sqrt{\psit(x)-\psit(x')-\frac{\vt^2}{2}}}\left(\dv{\psit}{x'}\right)^{-1}\left[\dv[2]{\rhot}{{x'}}-\left(\dv{\psit}{x'}\right)^{-1}\dv[2]{\psit}{{x'}}\dv{\rhot}{{x'}}\right] \, ,
\end{equation}
with $x_{\text{min}}$ defined analogously to  $r_{\text{min}}$. The relation to $f_\chi$ is given by
\begin{equation}
    f_\chi(r,v)=\frac{1}{8\pi^3\sqrt{2G^3\rvir^3\Mvir}}\frac{1}{g(c)}\fchit(x,\vt).
\end{equation}
In the following we mostly make use of \glspl{df} which are normalized by their \gls{dm} density distribution and are defined via 
\begin{equation}
    P_r(v)\dd[3]v\equiv\frac{f_\chi(r,v)}{\rho_\chi(r)}\dd[3]v=\frac{1}{\sqrt{8}\pi^2}\frac{\fchit(x,\vt)}{\rhot(x,c)}\dd[3]{\vt}\equiv P_x(\vt)\dd[3]{\vt} \, ,
\end{equation}
such that $4\pi\int_0^{\vesc}\dd{v} v^2P_r(v)=4\pi\int_0^{\vesct} \dd{v} \vt^2P_x(\vt)=1$, where $\vesc=\sqrt{2\Psi(r)}$ (and analogously $\vesct$) denotes the \textit{escape velocity}, \ie the maximal velocity a particle within a gravitationally bound system can possess.\newp
The (dimensionless) gravitational potential arising from the \gls{dm} halo can be computed analytically from \cref{eq:Poissonequation}, yielding \cite{Ferrer:2013cla}
\begin{align}
    \psit_\chi^{\text{\tiny{NFW}}}(x) &= \frac{\log(1+cx)}{c^3\gNFW(c)x}\,, \\
    \psit_\chi^{\text{\tiny{Ein}}}(x) &= \frac{1}{x}\frac{\Gamma\left(\frac{3}{\gamma}\right)-\Gamma\left(\frac{3}{\gamma},\frac{2}{\gamma}(cx)^\gamma\right)}{\Gamma\left(\frac{3}{\gamma}\right)-\Gamma\left(\frac{3}{\gamma},\frac{2}{\gamma}c^\gamma\right)}+c\left(\frac{2}{\gamma}\right)^{1/\gamma}\frac{\Gamma\left(\frac{2}{\gamma},\frac{2}{\gamma}(cx)^\gamma\right)}{\Gamma\left(\frac{3}{\gamma}\right)-\Gamma\left(\frac{3}{\gamma},\frac{2}{\gamma}c^\gamma\right)}\,,
\end{align}
for the two different density profiles. In the case of \glspl{dsph}, the gravitational contribution from baryonic matter can be neglected due to the dominance of \gls{dm} in the dwarf galaxies under consideration \cite{Battaglia:2013wqa,Walker:2013}. In the \gls{gc}, however, baryonic matter plays a significant role and its influence on the overall gravitational potential must be considered. We will only model the baryonic bulge and the stellar disk in the following, since they give the largest contributions. In order to apply Eddington inversion, the gravitational potential of the baryons must also be spherically symmetric, which is certainly not the case. However, following \eref\cite{Ferrer:2013cla}, we can approximate the baryonic potentials using symmetrized models which enclose the same mass as the actual profiles. They read 
\begin{equation}
    \psit_{\text{bulge}}(x)=\frac{M_{\text{bulge}}}{\Mvir}\frac{1}{x+\frac{c_0}{\rvir}},\qquad \psit_{\text{disk}}(x)=\frac{M_{\text{disk}}}{\Mvir}\frac{1-\exp\left[-\frac{\rvir}{b_{\text{disk}}}x\right]}{x} \, ,
\end{equation}
where $M_{\text{bulge}}=\SI{1.5e10}{}M_\odot$ and $M_{\text{disk}}=\SI{5e10}{}M_\odot$ denote the bulge and disk masses, respectively, whereas $b_{\text{disk}}=\SI{4}{\kpc}$ and $c_0=\SI{0.6}{\kpc}$ are model parameters encoding the spatial extent of these objects.


\subsubsection[Generalized \texorpdfstring{$J$}{???}-factors]{\protect\boldmath Generalized \texorpdfstring{$J$}{???}-factors}
\label{subsubsec:Jsfactor}

After having established a technique to compute the \glspl{df} $f_\chi$, we will now determine the generalized $J$-factors as given in \cref{eq:Jsannl}
\begin{equation*}
    \Jsann{l}(\xi)=\int_{\Delta\Omega}\dd{\Omega}\!\int_0^\infty\dd\psi\!\int \dd[3]{v_1}f_\chi(r(\psi,\Omega),\vec{v}_1)\!\int\dd[3]{v_2}f_\chi(r(\psi,\Omega),\vec{v}_2)\,\vrel^{2l}\,\Sann(\zeta,\xi).
\end{equation*}
It is convenient to start with the velocity integrals in order to reduce complexity. Due to the spherical symmetry of the system we can perform three out of the six integrals trivially $\dd[3]{v_1}\dd[3]{v_2}=4\pi v_1^2 \dd{v_1} 4\pi v_2^2 \dd{v_2} \frac{1}{2}\dd{\!\cos\phi}$, where $\phi$ denotes the angle between $\vec{v}_1$ and $\vec{v}_2$. In order to carry out the remaining integrals, we perform a coordinate transformation $\{v_1,v_2,\cos\phi\}\to\{\vrel,\vcm,z\}$, where $\vrel=\abs{\vec{v}_1-\vec{v}_2}$ denotes the relative velocity as usual, $\vcm=\abs{\vec{v}_1+\vec{v}_2}/2$ is the \gls{com} velocity and $z=\cos\theta$, with $\theta$ being the angle between $\vec{v}_{\text{cm}}$ and $\vec{v}_{\text{rel}}$. We find $\vec{v}_{1,2}=\vec{v}_{\text{cm}}\pm\vec{v}_{\text{rel}}/2$, such that 
\begin{equation}
    v_{1,2}=\sqrt{\vcm^2+\frac{\vrel^2}{4}\pm \vcm\vrel z}, \quad \cos\phi=\frac{\vcm^2-\frac{\vrel^2}{4}}{\sqrt{\vcm^2+\frac{\vrel^2}{4}+\vcm\vrel z}\sqrt{\vcm^2+\frac{\vrel^2}{4}-\vcm\vrel z}},
\end{equation}
with $\abs{J}=\vcm^2\vrel^2/(v_1^2v_2^2)$ the Jacobian of the transformation. Therefore, we can identify $8\pi^2 v_1^2v_2^2 \dd{v_1}\dd{v_2}\dd{\!\cos\phi}=8\pi^2\vcm^2\vrel^2\dd{\vcm}\dd{\vrel}\dd{z}$. The limits $0\leq v_{1,2}<\vesc$ translate to $\vcm\leq\vesc$, $\vrel\leq 2\vesc$ and $\abs{z}\leq z_0$ with $z_0\equiv (\vesc^2-\vcm^2-\vrel^2/4)/(\vcm/\vrel)$. Employing the symmetry $z\to -z$ to restrict our integration to $0\leq z\leq z_0$, we need to consider that for $0\leq\vcm<\vesc-\vrel/2$, $z$ would exceed $1$, which is why we split up the integral and use $0\leq z< 1$ in this regime. Moreover, for $\vcm>\sqrt{\vesc^2-\vrel^2/4}$ we would obtain $z<0$, such that we use this as an upper bound for $\vcm$ instead. In the new coordinates, we can write \cref{eq:Jsannl} as
\begin{align}
     \Jsann{l}(\xi)&=\int_{\Delta\Omega}\dd\Omega\int_0^\infty\dd\psi \rho_\chi(x)^2\Sigma^{-2l}\int_0^{2\vesct}\dd{\vrelt}\Pxrel(\vrelt)\,\vrelt^{2l}\,\Sann(\Sigma\alpha/\vrelt,\xi)\,,
\end{align}
where we switched to a dimensionless coordinate representation and introduced a unit conversion constant $\Sigma\equiv c_\gamma\sqrt{\rvir/(G\Mvir)}$ with $c_\gamma$ the speed of light.\footnote{The $c_\gamma$ is needed here because the term $\vrel^{2l}\Sann(\alpha/\vrel,\xi)$ has been given in natural units.} The integrals over $\vcmt$ and $z$ have been absorbed in the definition of
\begin{align}
    \label{eq:Pxrel}
    \Pxrel(\vrelt)\equiv&\frac{2\vrelt^2}{\pi^2\rhot(x)^2}\left\{\int_0^{\vesct-\frac{\vrelt}{2}}\dd{\vcmt}\vcmt^2\int_0^1\dd{z}+\int_{\vesct-\frac{\vrelt}{2}}^{\sqrt{\vesc^2-\frac{\vrelt^2}{4}}}\dd{\vcmt}\vcmt^2\int_0^{z_0}\dd{z}\right\}\nonumber\\
    & \fchit\left(x,\sqrt{\vcmt^2+\vrelt^2/4+\vcmt\vrelt z}\right)\fchit\left(x,\sqrt{\vcmt^2+\vrelt^2/4-\vcmt\vrelt z}\right)\,.
\end{align}
In slight abuse of notation, we further define a velocity averaged \gls{se} factor
\begin{equation}
    \left\langle\Sann^{(\alpha)}\right\rangle(x;\xi)\equiv \Sigma^{-2l}\int_0^{2\vesct}\dd{\vrelt} \Pxrel(\vrelt)\,\vrelt^{2l}\,\Sann^{(l)}(\Sigma\alpha/\vrelt,\xi)\,,
\end{equation}
which approaches $1$ for the s-wave contribution of a heavy mediator (\ie no \gls{se}). The generalized $J$-factors for annihilation can then be written as
\begin{equation}
     \Jsann{l}(\xi)=\int_{\Delta\Omega}\dd\Omega\int_0^\infty\dd\psi \rho_\chi(x(\psi,\Omega))^2 \left\langle\Sann^{(\alpha)}\right\rangle(x(\psi,\Omega);\xi)\,,
\end{equation}
and analogously for $\Jsbsf(\xi)$.\newp
\begin{figure}
    \centering
    \includegraphics[width=\textwidth]{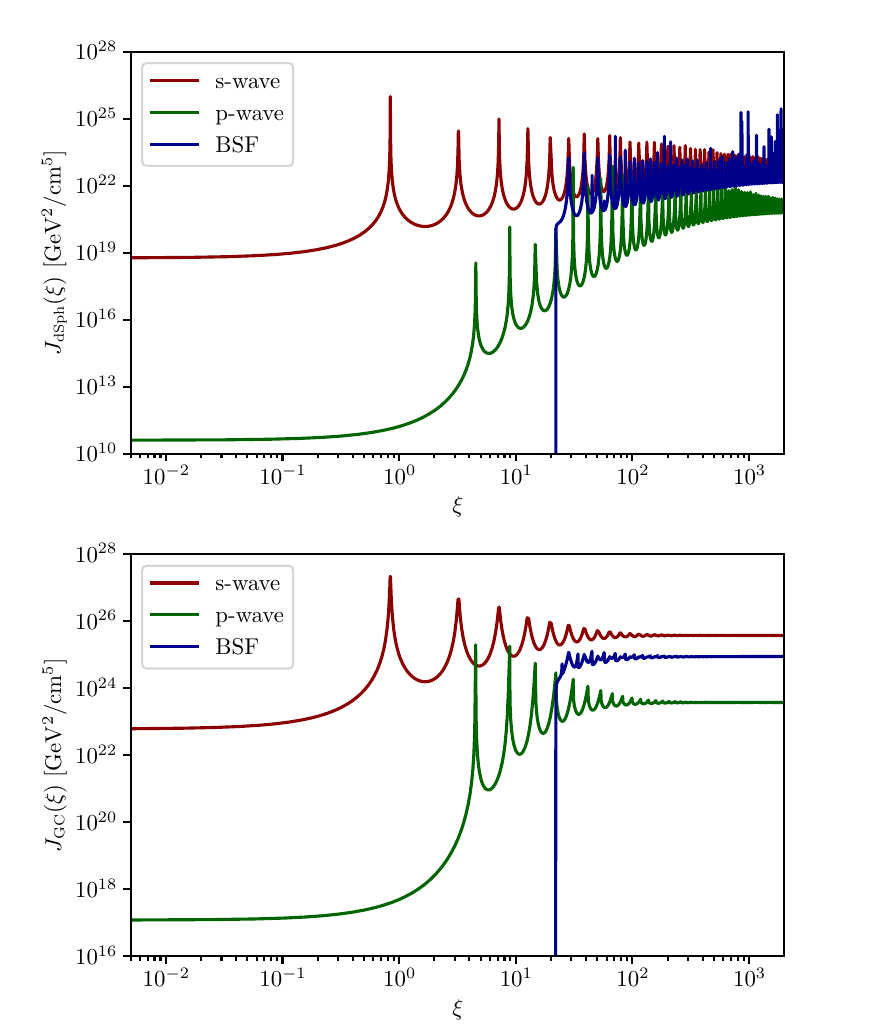}
    \caption[Generalized $J$-factors for the leading order s-wave and p-wave contributions as well as for BSF into the ground state.]{Generalized $J$-factors for the leading order s-wave (red) and p-wave (green) contributions as well as for \gls{bsf} into the ground state with a scalar coupling $\alpha=0.1$ as a function of $\xi$ for the exemplary \gls{dsph} \textit{Draco} (top) and the \gls{gc} (bottom), including a baryonic contribution to the gravitational potential for the latter. Adapted from \eref\cite{Biondini:2023ksj}.}
    \label{fig:J-factors}
\end{figure}
We are now left with an integration over the field of view. Regarding \glspl{dsph}, the solid angle observed by telescopes is usually large enough to cover the entire \gls{dm} halo and the distance between the galaxy and us is much larger than its extent, \ie $D\gg\rvir$. This enables us to substitute the integration over the cone representing the field of view by a sphere centered around the galaxy, \ie $\int\dd{\Omega}\int\dd{\psi}\to 4\pi/D^2\int r^2\dd{r}$. The \glspl{dsph} considered in the subsequent computation are given in \cref{tab:DMhaloparameters}. Assuming an underlying \gls{nfw} profile, we calculated their generalized $J$-factors for annihilation and \gls{bsf} and displayed them for an exemplary galaxy in \cref{fig:J-factors} (top) for a scalar coupling $\alpha=0.1$. We observe two limits, in which the $J$-factors are rendered constant. One occurs for $\xi\to 0$, for which the mediator mass becomes so large that the mediator potential becomes short-ranged and the \gls{se} becomes ineffective. In this limit we recover the $J_0$-factor for s-wave annihilation (\cf\cref{eq:J0}) and its p-wave equivalent. The other plateau exists for $\xi\to\infty$ where we approach the Coulomb limit. In between these limits we observe a series of peaks inherited from the \gls{se} factors of a Yukawa potential as expected (\cf\cref{subsec:dmproductionpnreft}). The enhancement becomes very pronounced for large $\xi$ and the natural hierarchy between s-wave and the p-wave contributions is significantly reduced. For $\mphi \lesssim |E_{1S}|$, which happens around $\xi\approx 22$ for $\alpha=0.1$, also \gls{bsf} is allowed (\cf\cref{eq:bsfexistencecriterium}). For larger $\xi$, the \gls{bsf} $J$-factor rises steeply and settles slightly below the s-wave and well above the p-wave contribution. This behaviour has already been anticipated from \cref{fig:crossectionvrel}. We would like to remind the reader at this point, that due to the different behaviour of the velocity-independent factors $\sigma_l$ and $\sigmabsf^0$ on the masses and couplings, the $J$-factors alone are not sufficient to assess the relative importance of the annihilation and \gls{bsf} contributions to the expected \gls{dm} flux (\cf\cref{eq:photonflux_Jfactors}).\newp
For the \gls{gc}, the simplification on calculating the generalized $J$-factors as done for the \glspl{dsph} is not possible due to $D\ll \rvir$. We consider in the following a region of interest of $l,b\in[-6^{\circ},6^{\circ}] $ motivated by the sensitivity study performed by the \gls{cta} collaboration \cite{CTA:2020qlo}. The generalized $J$-factors for the \gls{gc} are presented in \cref{fig:J-factors} (bottom) for the same $\alpha=0.1$, where we employed an Einasto profile including the impact of baryonic matter on the gravitational potential. The overall structure (as well as its origin) is comparable to the \glspl{dsph} and we will only comment on the main differences. First, the absolute values of the $J$-factors are larger due to the higher concentration of \gls{dm} in the \gls{gc} and thus a much larger virial \gls{dm} mass. Second, the resonance structure is more washed out due to the overall higher average velocities in comparison with \glspl{dsph} (see again \cref{fig:crossectionvrel}).


\subsubsection{The photon spectrum}
\label{subsubsec:photonspectrum}

We define the photon spectrum $\dd N_\gamma/\dd E_\gamma$ as the number of photons $N_\gamma$ which are produced by annihilating or decaying \gls{dm} per photon energy $E_\gamma$. In our case, we will deal with annihilating \gls{dm} producing a pair of mediators $\bar{\chi}+\chi\to 2\phi$ at first. The mediators eventually decay into pairs of \gls{sm} particles $\phi\to \bar{f}f$, which undergo further decay or hadronization processes, ultimately leaving only particles stable over astrophysical distances (\eg photons) in the resulting spectrum. Since the decay of $\phi$ is mediated through mixing with the \gls{sm} Higgs boson, the mediator inherits all kinematically allowed Higgs decay modes leaving us with a plethora of possible \gls{sm} final states. By looking at
\begin{equation}
    \dv{N_\gamma}{E_\gamma}=\sum_f\mathcal{B}_f\dv{N_\gamma^{(f)}}{E_\gamma}\,,
\end{equation}
we can see that essentially two tasks have to be performed: 1) Determining the relevant branching ratios $\mathcal{B}_f\equiv\text{BR}(\phi\to\bar{f}f)$ of the mediator decaying into \gls{sm} final states and 2) calculating the photon spectrum produced by a given \gls{sm} final state while taking into account the large boosts originating from the mediator decays in the galactic rest frame. For the latter we will in the following consider two mass regimes of the mediator, namely $\mphi \geq \SI{10}{\GeV}$, denoted as \textit{high mass range} and $ 2 m_{\pi^0} \leq \mphi \leq \, \SI{1}{\GeV}$, which we label the \textit{low mass range}). The reasons for this splitting and the avoidance for masses in between are explained below.

\paragraph{\protect\boldmath 1) Branching ratios for $\phi\to\bar{f}+f$}$~$\newpp
After \gls{ewsb} the Lagrangian containing couplings and mass terms of the physical Higgs field $h$ and mediator $\phi$ reads (\cf\cref{eq:LmodID,eq:LHiggsportal})
\begin{equation}
    \mathcal{L}\supset -\frac{1}{2}\mphi^2\phi^2-\muphih v\,\phi\,h -\frac{1}{2}m_h^2h^2-\frac{1}{2}\muphih\,\phi\,h^2 -\frac{m_h^2}{2v}h^3-\frac{m_h^2}{8v^2}h^4,
\end{equation}
where we have neglected the quartic couplings $\lambda_{\phi h}$ and $\lambda_\phi$. The \gls{vev} of the Higgs field is given by $v^2=-\mu^2/\lambda$ as usual and the Higgs mass is consistently defined as $m_h^2\equiv 2\lambda v^2$. After rotating into the mass eigenbasis $\phi\to \cos\delta \,\phi'-\sin\delta\, h'$, $h\to \sin\delta\,\phi' +\cos\delta\,h'$ and expanding around $\sin\delta$, we obtain the Lagrangian
\begin{align}
    \mathcal{L}\supset& -\frac{1}{2}\mphi^2\phi'^2 -\frac{1}{2}m_h^2h'^2 -\frac{m_h^2}{2v}h'^3-\frac{m_h^2}{8v^2}h'^4 \nonumber\\
    & +\left(- \frac{m_h^2}{2v^2}\phi'h'^3 -\frac{2m_h^2+\mphi^2}{2 v} \phi' h'^2\right)\sin\delta +\order{\sin^2\delta}\,,
\end{align}
where fields that correspond to mass eigenstates are highlighted here by $\,^\prime$ for illustration purposes. This indication is dropped in the following. We observe, that if we choose a very small mixing angle $\delta$, which is connected to the trilinear coupling via 
\begin{equation}
    \label{eq:sindeltaapprox}
 \sin\delta\approx \frac{\mu_{\phi h} v}{|m_h^2 -\mphi^2|}\,,
\end{equation} 
we can neglect the corrections to the mediator and Higgs masses as well as the Higgs self-interactions because they only arise at $\order{\sin^2\delta}$. However, we do obtain novel interactions between the mediator and the Higgs boson which are only suppressed by $\sin\delta$ and can give rise to $\phi\to hh$ and $\phi\to hhh$ decay channels if kinematically allowed. Likewise, the mediator will inherit all Higgs couplings to \gls{sm} fermions and bosons, equally suppressed by a factor of $\sin\delta$ at leading order. Therefore, its decay rates will mimic the decay rates of a Higgs boson with equal mass and the branching ratios $\mathcal{B}_f$ will be largely independent of the mixing angle. Moreover, since also for small mixing angles the mediator decays instantly on astrophysical scales, its absolute value is irrelevant for indirect detection.\newp
\begin{figure}[ht]
    \centering
    \includegraphics[width=\textwidth]{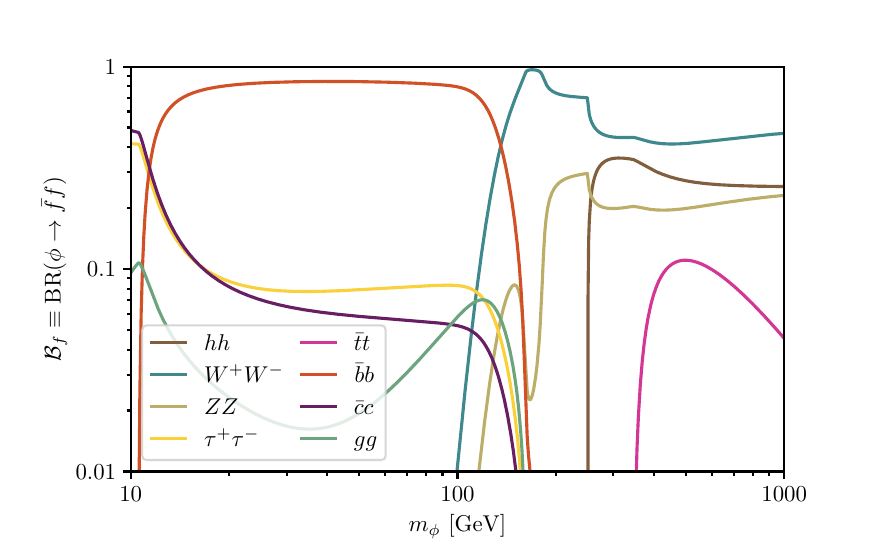}
    \caption[Branching ratios of the mediator $\phi$ decaying into SM particles in the high mediator mass range.]{Branching ratios of the mediator $\phi$ decaying into \gls{sm} particles in the high mediator mass range (see text). Displayed are all channels which contribute at least $1\%$ to the total decay width in the regime of interest. Below mediator masses of $\mphi<2m_h$, the branching ratios of the mediator are equivalent to the ones of a Higgs boson with equal mass. Adapted from \eref\cite{Biondini:2023ksj}.}
    \label{fig:mediator_decay}
\end{figure}
Leading order approximations of the decay rates of a scalar \gls{sm} Higgs boson with variable masses above $\gtrsim\SI{2}{\GeV}$ are taken from \eref\cite{Djouadi:2005gi}, where we have included the running of the masses and couplings through \eref\cite{Chetyrkin:2000yt}.\footnote{Older computations from an era where the Higgs mass was still unknown have been performed in \eg\erefs\cite{Gunion:1989we,Spira:1995rr}. For an in-depth analysis and numerical results on each channel, the reader is directed to the combined studies of the \textit{Higgs Cross Section Working Group}, exemplarily stating \erefs\cite{Denner:2011mq,Dittmaier:2012vm} and references therein.} Below the kinematic threshold of $\phi\to hh$, the branching ratios $\mathcal{B}_f$ of the mediator will be equivalent to the branching ratios of a Higgs boson with equal mass. For $\mphi>2m_h$ and $\mphi>3m_h$, respectively, decay channels into $2$ and $3$ Higgs bosons open, but only the $\phi\to hh$ contribution changes the branching ratios of the mediator enough within this regime (it actually becomes comparable to the $W$ and $Z$ contributions) to take it into account. Its decay rate is given by
\begin{equation}
	\Gamma_{\phi\to hh}=\frac{\left(v^2+2m_h^2\right)^2}{32\pi \mphi v^2}\sqrt{1-\frac{4m_h^2}{\mphi^2}}\sin^2\delta.
\end{equation}
We display the branching ratios of all relevant channels contributing more than $1\%$ to the total decay rate in \cref{fig:mediator_decay} for a mediator mass range between $\SI{10}{}-\SI{1000}{\GeV}$. In our analysis of the high mediator mass range we consider all of the depicted channels, namely $\phi\to\{hh,\,W^+W^-,\,ZZ,\,gg,\,\bar{t}t,\,\bar{b}b,\,\bar{c}c,\,\tau^+\tau^-\}$.\newp
For quark and gluon final states, all methods outlined above to calculate the decay rates break down close to the confinement scale. Hence, in the low mediator mass range $\mphi\lesssim\SI{1}{\GeV}$ we employ the branching ratios into hadrons derived in \eref\cite{Winkler:2018qyg}. Within this regime, we notice that below the kaon threshold around $\SI{1}{\GeV}$, most mediators ($\gtrsim 80\%$) will decay into pairs of pions with minor contributions from the muon, electron and photon decay channels. Since $\text{BR}(\pi^0\to 2\gamma)\approx 99\%$, a large fraction of the energy in this mass range goes directly to hard gamma-rays. Therefore, it is sufficient to solely focus on the $\pi^0$ decay channel to derive limits and prospects.

\paragraph{2) Cascade annihilations}$~$\newpp
The photon spectrum $\dd{N^{(f)}_\gamma}/\dd{E_\gamma}$ of \gls{dm} annihilation through a single mediator decay channel can be characterized by a two step process: 1) \gls{dm} annihilation into two mediator particles $\bar{\chi}+\chi\to \phi +\phi $ and 2) the decay of each mediator into \gls{sm} particles which in turn cascade into $N$ photons and other stable particles $\phi \to f +\bar{f} \to\ldots\to N~\gamma+X$. The quantity $N^{(f)}_\gamma(E_\gamma)$ in this context is then defined as the average number of photons with an energy $E_\gamma$ produced per \gls{dm} annihilation process.\newp
For mediator masses above a few $\si{\GeV}$ (\ie well above the \gls{qcd} confinement scale), the second process can and has already been simulated utilizing event generators for high-energy physics collisions like Pythia \cite{Bierlich:2022pfr} or Herwig \cite{Bellm:2015jjp}. In fact, simulated data on photon spectra for \gls{dm} annihilations into all relevant \gls{sm} final states has been published by several groups including \erefs\cite{Belanger:2001fz,Bringmann:2018lay} as well as \eref\cite{Cirelli:2010xx}, which we will use in the following. This public data source provides us with $\dd{N^{(f)}_\gamma}/\dd(\log_{10}x)$ as a function of an annihilating \gls{dm} mass $m_{\text{\tiny DM}}=\sqrt{s}/2$ taken to be half the \gls{com} energy, as well as the energy of the photons produced in the \gls{com} frame of the annihilating system denoted by $E_\gamma$ and parametrized as $x\equiv E_\gamma/m_{\text{\tiny DM}}$. In our case, $\chi$ does not annihilate directly to \gls{sm} particles  but through a one-step cascade producing mediator particles. Therefore, the data on $\dd{N^{(f)}_\gamma}/\dd(\log_{10}x_0)$ refers here to the number of photons produced for a decaying mediator particle in the rest frame of the mediator. The \gls{com} energy is then just $\sqrt{s}=\mphi$ such that $m_{\text{\tiny DM}}\equiv \mphi/2$ and $x_0\equiv 2E_0/\mphi$, where the index $0$ emphasizes that the corresponding quantity is in the mediator rest frame. Overall, a mediator mass range of $m_{\phi}\in [10,2\times10^5]\,\si{\GeV}$ is covered by the data. \newp
We are then left with translating the photon spectra of \eref\cite{Cirelli:2010xx}, given in the rest frame of the mediator, into the galactic rest frame, which is equivalent to the \gls{com} frame of the annihilating \gls{dm} particles. Since the scalar mediator decays isotropically, the spectrum in the \gls{com} frame is given by \cite{Mardon:2009rc}
\begin{equation}
    \label{eq:onestepcascadephotonspectra}
	\dv{N^{(f)}_\gamma}{E_\gamma}=\int_{-1}^{1}\dd{\!\cos\theta}\int_0^{\mphi/2}\dd{E_0}\dv{N^{(f)}_\gamma}{E_0}\delta(E_\gamma-E_\gamma^*(E_0))\,,
\end{equation}
with $E_\gamma^*(E_0)$ the Lorentz boost constraint which we will determine in the following. Assuming $\chi$ to be non-relativistic, the energy of the mediator in the \gls{com} frame is $E^*_\phi=E^*_\chi\approx \mchi$ with $E^*_\chi$ the energy of the annihilating \gls{dm} particles. Starting in turn from the energy of the mediator in its rest frame $E_\phi$, a boost to the \gls{com} frame yields $E_\phi^*=\gamma(E_\phi-\vec{v}\cdot\vec{k})=\gamma\mphi$ since $\vec{k}=0$ and $E_\phi=\mphi$. Equating these two formulas leaves us with a boost factor of $\gamma=1/\epsilon_\phi$ or $v=\sqrt{1-\epsilon_\phi^2}$, respectively, where we defined $\epsilon_\phi\equiv\mphi/\mchi$. The boost of the photon energy to the \gls{com} frame is then given by $E^*_\gamma(E_0)=\gamma(E_0+\vec{v}\cdot\vec{k}_0)=E_0/\epsilon_\phi\left(1+\sqrt{1-\epsilon_\phi^2}\cos\theta\right)$, where $\abs{\vec{k_0}}=E_0$ and $\theta$ denotes the angle between $\vec{v}$ and $\vec{k}_0$. Redefining $x\equiv E_\gamma/\mchi$, we can write \cref{eq:onestepcascadephotonspectra} as
\begin{equation}
	\dv{N^{(f)}_\gamma}{x}=2\int_{-1}^{1}\dd{\!\cos\theta}\int_0^{1}\dd{x_0}\,\delta\left(2x-x_0\left(1+\sqrt{1-\epsilon_\phi^2}\cos\theta\right)\right)\dv{N^{(f)}_\gamma}{x_0}\,,
\end{equation}
which leaves us with 
\begin{equation}
	\dv{N_\gamma}{x}=2\int_{t_{\text{min}}}^{t_{\text{max}}}\frac{\dd{x_0}}{x_0\sqrt{1-\epsilon_\phi^2}}\dv{N_\gamma}{x_0}\,,
\end{equation}
where $t_{\text{max}}=\min\left\{1,2x/\epsilon_\phi^2\left(1+\sqrt{1-\epsilon_\phi^2}\right)\right\}$ and $t_{\text{min}}=2x/\epsilon_\phi^2\left(1-\sqrt{1-\epsilon_\phi^2}\right)$.\newp
If $\mphi\lesssim\SI{10}{\GeV}$, the approach of calculating photon spectra via event simulation from high-energy physics collisions breaks down due to large corrections from non-perturbative \gls{qcd} effects. Attempts have been made to derive photon spectra for light vector mediators \cite{Plehn:2019jeo,Coogan:2022cdd}, however, they are not applicable for a scalar $\phi$. Therefore, we exclude the range $\SI{1}{\GeV} \leq \mphi \leq \SI{10}{\GeV}$ from our subsequent analysis. In contrast, for the light mediator mass range $2 m_{\pi^0} \leq  \mphi \leq 1\, \si{\GeV}$, the prominent mediator decay into pions once again enables a computation of the photon spectrum.\footnote{Below the pion threshold, mediators will decay predominantly into light lepton pairs. The most stringent bounds in this regime come from the positron flux \cite{AMS:2015tnn}, which has been interpreted in the context of \gls{dm} \eg in \eref\cite{Elor:2015bho}. However, since we are predominantly interested in the gamma-ray flux, we will not include these final states in the following. Moreover, in this mediator mass regime it will also become challenging to avoid \gls{bbn} and direct detection constraints, as we will see in \cref{subsec:other_exp_limit}.} The process under consideration $\bar{\chi}+\chi\to 2\phi\to 4\pi^0\to8\gamma$ is a two-step cascade annihilation with a branching fraction of $\approx 50\%$. The photon spectrum in the rest frame of the pion is simply given by $\dd{N^{(\pi)}_{\gamma}}/\dd{x_0}=2\delta(x_0-1)$ with $x_0\equiv 2E_0/m_{\pi^0}$. Defining further $x_1\equiv 2E_1/\mphi$ and $\epsilon_{\pi^0}=2m_{\pi^0}/\mphi$, we can use the methodology outlined in \eref\cite{Elor:2015tva} to obtain (after two boosts) the photon spectrum in the \gls{com} frame of the annihilating \gls{dm}
\begin{equation}
	\dv{N^{(\pi)}_\gamma}{x}=2 \int_{t_{2,\text{min}}}^{t_{2,\text{max}}}\frac{\dd{x_1}}{x_1\sqrt{1-\epsilon_\phi^2}}\left(2 \int_{t_{1,\text{min}}}^{t_{1,\text{max}}}\frac{\dd{x_0}}{x_0\sqrt{1-\epsilon_{\pi^0}^2}}\dv{N^{(\pi)}_\gamma}{x_0}\right)\,,
\end{equation} 
with the integration limits
\begin{align}
	t_{1,\text{min}}=~&\frac{2x_1}{\epsilon_{\pi^0}^2}\left(1-\sqrt{1-\epsilon_{\pi^0}^2}\right),\quad t_{1,\text{max}}=\min\left\{1,\frac{2x_1}{\epsilon_{\pi^0}^2}\left(1+\sqrt{1-\epsilon_{\pi^0}^2}\right)\right\} \nonumber\\ 
    t_{2,\text{min}}=~&\frac{2x}{\epsilon_\phi^2}\left(1-\sqrt{1-\epsilon_\phi^2}\right),\quad
    t_{2,\text{max}}=\min\left\{\frac{1}{2}\left(1+\sqrt{1-\epsilon_{\pi^0}^2}\right),\frac{2x}{\epsilon_\phi^2}\left(1+\sqrt{1-\epsilon_\phi^2}\right)\right\}.
\end{align} 
Performing the integrations over $x_0$ and $x_1$, we are left with an analytic result
\begin{equation}
	\dv{N_\gamma}{x}=\frac{8}{\sqrt{1-\epsilon_{\pi^0}^2}\sqrt{1-\epsilon_\phi^2}}
	\begin{cases}
		\ln\Big(\frac{x}{K_+^-}\Big) & K_+^-\leq x<\min\{K_+^+,K_-^-\} \\
		\ln\Big(\frac{\min\{K_+^+,K_-^-\}}{K_+^-}\Big) & \min\{K_+^+,K_-^-\}\leq x < \max\{K_+^+,K_-^-\} \\
		\ln\Big(\frac{K_-^+}{x}\Big) &  \max\{K_+^+,K_-^-\} \leq x<K_-^+\\
	\end{cases}
\end{equation}
and $0$ elsewhere, where we have defined for convenience
\begin{equation}
	K_\mp^\pm\equiv\frac{\left(1\pm\sqrt{1-\epsilon_{\pi^0}^2}\right)\epsilon_\phi^2}{4\left(1\mp\sqrt{1-\epsilon_\phi^2}\right)}\,.
\end{equation}
The photon spectrum of \gls{dm} annihilation through pion decay is very hard. Therefore, at sufficiently high energies a dedicated analysis searching for discrete spectral features may boost the sensitivity to astrophysical observation considerably \cite{Bringmann:2011ye,Bringmann:2012vr}. However, this will be not pursued further here and is left for future work. 


\subsubsection{Indirect detection bounds and prospects}
\label{subsubsec:boundsandprospects}

We have now all ingredients to calculate our model predictions of the \gls{dm} photon flux and compare them with observations. To achieve this, we will in the following match bounds from the \gls{fermi} collaboration as well as prospects for the \gls{cta}. Additionally, we will consider limits from \gls{cmb} anisotropies derived by the Planck collaboration.

\begin{figure}
    \centering
    \includegraphics[width=\textwidth]{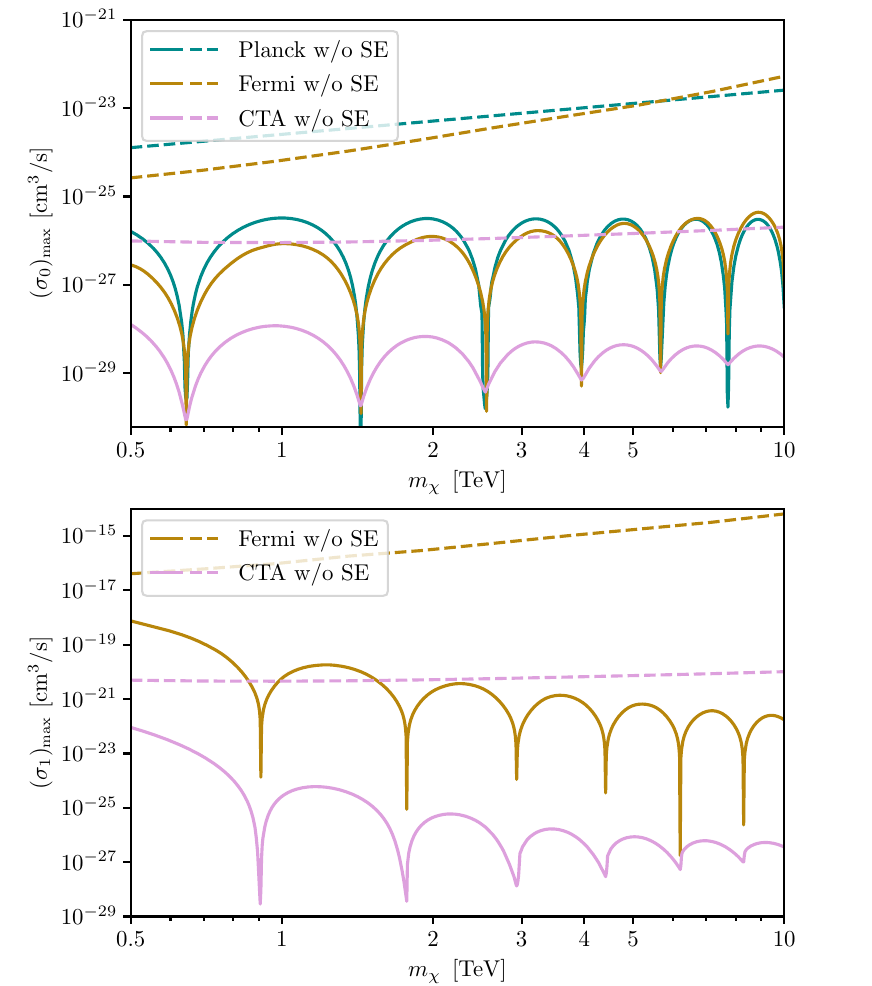}
    \caption[Indirect detection limits on the perturbative cross section in s-wave and p-wave dominated DM annihilation scenarios.]{Indirect detection limits on the perturbative cross section in s-wave (top) and p-wave (bottom) dominated \gls{dm} annihilation scenarios as a function of the DM mass $\mchi$ for a benchmark point of $\mphi=\SI{10}{\GeV}$ and $\alpha=0.1$. The turquoise lines represent current CMB limits based on Planck data (only for s-wave domination), the golden lines indicate limits from an analysis of \gls{fermi} data from \glspl{dsph}, and the pink lines denote prospects for upcoming \gls{cta} measurements in the \gls{gc}. Dotted lines depict the limits without accounting for \gls{se}, whereas solid lines incorporate \gls{se} into the analysis. Adapted from \eref\cite{Biondini:2023ksj}.}
    \label{fig:crosssectionlimit}
\end{figure}

\paragraph{Fermi-LAT limits}$~$\newpp
The Fermi collaboration has published a statistical analysis of a variety of different \glspl{dsph} on the basis of 6 years of \gls{fermi} data \cite{Fermi-LAT:2015att}. They evaluated the significance of a \gls{dm} hypothesis using the following \gls{ts} 
\begin{equation}
    \text{TS} = -2 \log\left(\frac{\mathfrak{L}(\vec{\mu}_0,\hat{\vec{\theta}}|\mathcal{D})}{\mathfrak{L}(\hat{\vec{\mu}},\hat{\vec{\theta}}|\mathcal{D})}\right),
\end{equation}
which is based on a negative log-likelihood ratio on a data set $\mathcal{D}$, where $\mathfrak{L}$ denotes the likelihood function, either of an individual galaxy or as a joint likelihood of all \glspl{dsph} under consideration. The $\vec{\mu}_0$ are the model parameters of the null hypothesis (no \gls{dm}) and $\hat{\vec{\mu}}$, $\hat{\vec{\theta}}$ represent, respectively, the best fit values of $\vec{\mu}, \vec{\theta}$, the model and nuisance parameters  under the \gls{dm} hypothesis.\footnote{Note that this definition of the \gls{ts} departs from the conventional one by fixing both $\hat{\vec{\theta}}$ to the best-fit values of the \gls{dm} hypothesis, where the signal is modelled by a $\dd{\Phi_\gamma}/\dd{E_\gamma}=-2$ power law (see \eref\cite{Fermi-LAT:2013sme}).} In order to derive upper limits on the \gls{dm} cross section as a function of the \gls{dm} mass, the collaboration performed a combined global fit of 15 \glspl{dsph} for a few selected channels into pure \gls{sm} final states, which is however not applicable to our situation. Fortunately, they also published the energy-bin by energy-bin likelihood as a function of the integrated photon flux $\expval{\Phi_\gamma}$ (as the only model parameter) for each \gls{dsph} individually \cite{Fermi_logL}. More concretely, they provided the \textit{delta-log-likelihood ratio} (see also \erefs\cite{Bartlett:1953,Rolke:2004mj})
\begin{equation}
    \Delta_{ij}(\expval{\Phi_\gamma}_i)=\log\left(\frac{\mathfrak{L}(\expval{\Phi_\gamma},\vec{\hat{\theta}}_j|\mathcal{D}_{ij})}{\mathfrak{L}(\widehat{\expval{\Phi_\gamma}},\vec{\hat{\theta}}_j|\mathcal{D}_{ij})}\right)\,,
\end{equation}
for each \gls{dsph} $j$ and energy bin $i$. We can derive our indirect detection limits easily from this data, by demanding that 
\begin{equation}
    \label{eq:IDcondition_dSph}
    \sum_{i,j}\Delta_{ij}(\expval{\Phi_\gamma}_i)+\frac{2.71}{2}>0\,,
\end{equation}
with $j=1,\dots,4$ (for our classical dwarfs Coma Berenices, Ursa Minor, Draco, and Sergue 1), where the maximally allowed deviation of $2.71/2$ from the extremum corresponds to a $95\%$ \gls{cl} with the factor of $1/2$ accounting for a one-sided statistical test \cite{Cowan:2013pha,Rolke:2004mj}. For this purpose, we integrate \cref{eq:photonflux_Jfactors} using the same binning as in the supplementary material given, defining 
\begin{equation}
   \expval{\Phi_\gamma}_i\equiv\int_{E_{\text{min},i}}^{E_{\text{max},i}} \dd{E_\gamma} E_\gamma \dv{\Phi_\gamma}{E_\gamma}\,, \qquad \expval{N_\gamma}_i\equiv\sum_f \mathcal{B}_f\int_{E_{\text{min},i}}^{E_{\text{max},i}} \dd{E_\gamma} E_\gamma\dv{N^{(f)}_\gamma}{E_\gamma}\,,
\end{equation}
such that we can write 
\begin{equation}
    \label{eq:intphotonflux_Jfactors}
    \expval{\Phi_\gamma}_i=\frac{1}{16\pi \mchi^2}\left(\sigma_0 \Jsann{0}(\xi) +\sigma_1 \Jsann{1}(\xi)+\sigmabsf^0 \Jsbsf(\xi)\right)\expval{N_\gamma}_i\,,
\end{equation}
and check the condition of \cref{eq:IDcondition_dSph} for each set of masses and couplings. For each \gls{dsph}, the globally maximized nuisance parameters $\hat{\vec{\theta}}_j$ used in the analysis of the Fermi collaboration, can be split up into best-fit parameters from the \gls{fermi} analysis $\hat{\vec{\alpha}}_j$ and the corresponding $J_0$-factors of the galaxy, where an \gls{nfw} profile has been assumed. Uncertainties on the derived limits due to a varying $J$-factor by e.g. assuming a different density profile have been estimated by the collaboration to yield not more than $\sim 40\%$, mostly caused by a change in the relative importance of each \gls{dsph}. We checked this for the case of generalized $J$-factors by recalculating them for an underlying Einasto profile and found a deviation of $\sim 30\%$. This difference is largely independent of the \gls{dm} velocity, in fact, the velocity dependence changes the ratio of generalized $J$-factors considering an NFW and Einasto profile by $\lesssim 5\%$ for the highest $\xi$. As a comprehensive revision of the statistical analysis for our particular model is beyond the scope of this work, this statement reaffirms the robustness of the limits when considering non-perturbative effects. As a cross-check, we also derived limits on the perturbative cross section for the benchmark channels analyzed in \eref\cite{Fermi-LAT:2015att}, where we used our reduced set of \glspl{dsph} as well as the $J_0$-factors given by the collaboration. To accomplish this, we changed $\Delta_{ij}(\expval{\Phi_\gamma}_i)\to\Delta_{ij}(\expval{\sigma\vrel}_0)$ and equated the relation of \cref{eq:IDcondition_dSph}. Good agreement with the official limits has been found.\footnote{Note that an exact agreement cannot be expected since the statistical treatment for the energy binned data is simplified compared to the full method used by the collaboration. Moreover, we have included less \glspl{dsph}.}\newp
Assuming that only one annihilation channel significantly contributes to the photon flux, we can employ the same approach to establish an upper limit (at $95\%$ \gls{cl}) on the individual perturbative cross section, incorporating non-perturbative effects. This involves considering all mediator decay channels along with the corresponding generalized $J$-factor. We have illustrated in \cref{fig:crosssectionlimit} two scenarios of s-wave (top) and p-wave domination (bottom) as a function of the \gls{dm} mass for a benchmark point with $\mphi=\SI{10}{\GeV}$ and $\alpha=0.1$. It is evident that the constraint on the s-wave cross section is relatively modest when the \gls{se} is absent. It falls within a similar range as the limits for pure \gls{sm} final states at the corresponding \gls{dm} masses. The incorporation of the \gls{se} significantly tightens the constraints by several orders of magnitude. As expected, the most substantial impact is observed around the resonances of the $J$-factors, potentially lowering the limit to values even below $\SI{e-29}{\cm^3/\s}$.  For the p-wave case in the lower panel, the limits are notably weaker due to the velocity suppression of the annihilation rate. In the absence of \gls{se}, the limits are far from the values relevant for a thermal relic. Even with \gls{se}, only the parts resonantly enhanced around specific masses dip below $\SI{e-23}{\cm^3/s}$. The overall limit improves with higher masses since we are maintaining $\mphi$ constant, leading to stronger \gls{se} as $\mchi$ is increased.

\paragraph{Planck limits}$~$\newpp
Measurements of \gls{cmb} properties can be quite constraining for \gls{dm} models with late time annihilation. This is due to the fact that the \gls{cmb} is susceptible to exotic energy injection during the cosmic dark ages (at redshifts of $z\approx600 -1000$). \gls{dm} annihilations into \gls{sm} particles around that time provide this additional energy injection into the \gls{igm}. The resulting increase of the residual ionization fraction will in turn broaden the last scattering surface and modify \gls{cmb} anisotropies and polarization \cite{Chen:2003gz,Padmanabhan:2005es,Galli:2011rz}. The Planck collaboration \cite{Planck:2018vyg} has established the most rigorous constraints on non-standard energy injection. Under the assumption that the redshift dependence of the energy injection rate is solely governed by the variation in the \gls{dm} density (implying a velocity-independent cross section) they have imposed an upper limit on 
\begin{equation}
    p_{\text{ann}}\equiv f_{\text{eff}} \frac{\sigmavrelavg}{\mchi}\leq \SI{3.5e-28}{\frac{\cm^3}{\GeV\,\s}}\,,
\end{equation}
with $f_{\text{eff}}$ an efficiency factor controlling how much of the energy released into the visible sector due to, \eg\gls{dm} annihilation, is absorbed by the \gls{igm}.\footnote{The underlying assumption here is that the energy which is deposited into the \gls{igm} is proportional to the energy injected at the same redshift. In all generality, this may not be the case, since the absorption from the \gls{igm} can also happen at a much later time. To account for this, one would need to define a general function $f(z)$, which depends on the redshift as well as the underlying \gls{dm} model. However, it has been shown in \eref\cite{Finkbeiner:2011dx} that the impact on the \gls{cmb} in contrast to employing a constant $ f_{\text{eff}}$ is negligible.} Based on the findings in \erefs\cite{Slatyer:2015jla,Slatyer:2015kla}, we will use in the following a conservative estimate of $f_{\text{eff}}\simeq 0.137$ for all annihilation channels. This yields a lower limit on
\begin{equation}
    \label{eq:CMBbound}
    \sigmavrelavg<\SI{2.5e-24}{\frac{\cm^3}{\s}}\left(\frac{\mchi}{\si{\TeV}}\right) \, ,
\end{equation}
where $\sigmavrelavg$ is understood as the sum of contributions to the thermally averaged annihilation cross section which scale as $\vrel^0$.
By looking at \cref{fig:crossectionvrel}, for typical \gls{dm} velocities of $\order{10^{-8}}$ we can see that this is the case in our model for the s-wave and \gls{bsf} cross section, which both become constant in this regime due to a finite mediator mass. Note that this limit cannot be applied for p-wave annihilating \gls{dm} due to its $\vrel^2$ scaling. However, p-wave contributions in this regime are highly suppressed and can be neglected.\newp
The limits on the perturbative s-wave cross section have been displayed in \cref{fig:crosssectionlimit} (left) for a representative set of parameters. The inclusion of the \gls{se} enhances the Planck limit on the cross section, resulting in an improvement of several orders of magnitude across the mass range. Notably, the Planck limit is relatively similar to the one derived from \gls{fermi} observations, and at high \gls{dm} masses, Planck even outperforms the gamma-ray limits. This qualitative observation can be explained based on the scaling of the relevant rates with $\mchi$. The \gls{fermi} sensitivity is tied to the photon flux, scaling as $\sigmavrelavg \rho_\chi^2/\mchi^2$, while the \gls{cmb} is sensitive to the energy release rate, which scales as $\mchi\sigmavrelavg \rho_\chi^2/\mchi^2$. Consequently, the dependence on the \gls{dm} mass is weaker in the \gls{cmb} case.

\paragraph{CTA prospects}$~$\newpp
The \textit{Cherenkov Telescope Array Consortium} has examined \gls{dm} prospects for \gls{cta} observations of the GC \cite{CTA:2020qlo}. Similar to the \gls{fermi} analysis, tabulated bin-by-bin likelihoods have been made publicly accessible \cite{bringmann_torsten_2020_4057987}. These tables enable us the estimation of the anticipated upper limit on the perturbative \gls{dm} annihilation cross section for any spectrum, analogous to the approach used for \glspl{dsph}. The binned likelihood presented in the data is given in terms of the  $(\text{TS})_i\equiv -2\Delta_i$, which has been constructed only slightly different compared to the \gls{fermi} analysis. It has been tabulated as a function of the energy flux $\dd{\Phi\gamma}/\dd{E_\gamma}|_{E_\gamma=\bar{E}_i}$ evaluated at the mean energy $\bar{E}_i$ of the corresponding bin. The potential upper limits on the perturbative cross section are also outlined in \cref{fig:crosssectionlimit}. As anticipated, the \gls{cta} is likely capable to meet or exceed both \gls{fermi} and Planck limits by orders of magnitude within the specified mass range.\newp
Before moving on, we briefly address the influence of the halo profile on the limits and prospects we have derived from observations of the \gls{gc}. Given that the matter content in the \gls{gc} is dominated by baryons, the $J$-factor is not as constrained as for \glspl{dsph}, resulting in a less robust interpretation of (prospective) observations. Specifically, cuspy profiles tend to predict higher fluxes than cored ones. The Einasto profile, adopted in this study with a parameter choice of $\gamma=0.17$, falls between a cuspy and a cored profile. However, further investigation is warranted regarding the dependence of the prospects on this choice. To assess the impact of a more clearly cored profile, we adopt the suggestion of the \gls{cta} collaboration and examine an artificially cored Einasto profile \cite{CTA:2020qlo}. In this scenario, the density remains constant in the inner part of the halo below $\SI{1}{\kpc}$. Since this profile precludes Eddington inversion as the derivatives of $\rho_\chi$ would vanish there, we employ an alternative method to estimate the effect on the velocity-dependent cross section. Specifically, we calculate the standard velocity-independent $J_0$-factors for s-wave annihilations in both halos and assume that the ratio between the velocity-dependent $J$-factors remains consistent. Our analysis reveals that the cored $J$-factor is approximately $50\%$ smaller, resulting in a weaker limit on the cross section by the same ratio. This relatively small change in the $J$-factor, contrary to previous studies, has also been noted in \eref\cite{CTA:2020qlo}. This phenomenon can be attributed to the larger region of interest considered by the \gls{cta} analysis. To validate our approach, we conducted a similar analysis using an \gls{nfw} profile, which permits the use of the Eddington inversion method. The computation of the ratio of velocity-dependent $J$-factors using this method revealed that the ratio between the Einasto and the NFW $J$-factors is largely velocity-independent and agrees to within $10\%$ with the ratio estimated using the velocity-independent $J_0$-factors. Therefore, we assert the robustness of our outlined method.\footnote{It is worth noting that, besides the assumed \gls{dm} density profiles, various other uncertain parameters can affect the statistical limits, including the signal morphology and additional information about correlations in the energy bins. However, addressing these factors would require a specialized statistical analysis, which is beyond the scope of this work.}


\subsection{Complementary searches}
\label{subsec:other_exp_limit}

In addition to indirect detection constraints, other experimental limitations may exist that could restrict our parameter space. In the following, we will explore those that seem particularly significant from a preliminary perspective and could supplement the constraints imposed by indirect detection. These include potential restrictions arising from direct detection, \gls{bbn}, \glspl{edm}, and thermalization requirements.


\subsubsection{Direct detection and BBN limits}
\label{subsubsec:DD_BBN}

Our model allows for elastic scattering between \gls{dm} and \gls{sm} particles due to the mixing of the mediator with the Higgs. Direct detection experiments \cite{PandaX-4T:2021bab,LZ:2022lsv,XENON:2023cxc} impose stringent constraints on the rates of these processes, which are controlled by the mixing angle $\delta$. The scattering cross section for spin-independent interactions with a nucleon yields \cite{Duerr:2016tmh}
\begin{align}
    \sigma_{\text{SI}}= \frac{4\alpha\mu^2_{\chi N} m^2_N f_N^2}{v^2}\sin^2{\delta} \cos^2{\delta} \left(\frac{1}{m^2_\phi}-\frac{1}{m_h^2}
    \right)^2.
\end{align}
Here, $m_N$ represents the nucleon mass, $\mu_{\chi N}$ stands for the reduced mass of the \gls{dm}-nucleon system, and $f_N\approx 0.35$ denotes the effective coupling of the Higgs to the nucleon.\footnote{Pseudo-scalar interactions with \gls{sm} particles result in a momentum dependent direct detection cross section \cite{Anand:2013yka}. These interactions are constrained more loosely, allowing us to disregard processes involving $g_5$ for the subsequent discussion.} In the limit where direct detection constraints become relevant, \ie $\mphi\ll m_h$ and $\mchi \gg m_N$, we can infer an upper limit on the mixing angle 
\begin{equation}
    (\sin\delta)_{\text{max}} \simeq \frac{\mphi^2\,v}{2\, f_N \,m_N^2}\sqrt{\frac{\sigma_{\text{SI}}(\mchi)}{\alpha}}.
\end{equation}
At present, the most stringent upper bound on the spin-independent \gls{dm}-nucleus scattering cross section comes from the LZ experiment \cite{LZ:2022lsv}. Throughout this work we assume that $\sin \delta$ is sufficiently small to avoid conflict with the direct detection data. Since the exact value is inconsequential for indirect detection, direct detection constraints do not have any impact on the phenomenology discussed above.\newp
Nevertheless, direct detection constraints can be combined with limits from \gls{bbn} to rule out low-mass mediators \cite{Wise:2014jva}. To preserve the abundances of primordial elements, it is essential to guarantee that the mediators decay before the start of \gls{bbn} (\cf\cref{subsec:constraintsnth}). Therefore, we require that the lifetime of the mediator $\tau_\phi=1/\Gamma_{\phi}$ is shorter than the age of the Universe at the onset of \gls{bbn}, which we consider to be $T_{\text{\tiny{BBN}}}\sim\SI{1}{\MeV}$. Employing \cref{eq:decaytemperature}, we can see that \gls{bbn} sets a lower limit on $\sin\delta$, while direct detection establishes an upper limit, such that we can combine both arguments to exclude a range of mediator masses. Numerically, we observe that a kinematically allowed decay into muons is essential for ensuring a sufficiently short mediator lifetime. Consequently, masses of $\mphi$ that slightly exceed $2m_\mu$ (and lower) are ruled out.


\subsubsection{Electric dipole moments}
\label{subsubsec:EDM}

The pseudo-scalar interaction between the mediator and the \gls{dm} particle in \cref{eq:LmodID} introduces CP violation in the dark sector. Through portal interactions, this CP violation can be transferred to the \gls{sm} sector, which can induce \glspl{edm} \cite{Bernreuther:1990jx,Chupp:2017rkp,Alarcon:2022ero}. In our model, the mixing with the Higgs boson does not generate a pseudo-scalar interaction between the dark scalar $\phi$  and the \gls{sm} fermions at the leading order. Instead, only a scalar interaction of the form $\mathcal{L}_{\text{int}} \supset - \sin\delta\bar{f} f\phi$ is found. Consequently, there is no contribution to one-loop topologies. The initial contribution to the \gls{edm} of \gls{sm} fermions can emerge at the two-loop level, where one scalar and one pseudo-scalar vertex, along with a \gls{dm} fermion loop, are involved.\newp
The electron \gls{edm} places the most stringent experimental bound with $\abs{d_e}<\SI{1.1e-29}{\cm\,e}$ \cite{ACME:2018yjb}. We can estimate our two-loop contribution to it as
\begin{equation}
    d_e \approx \frac{e}{(4 \pi^2)^2} 4\pi \sqrt{\alpha \alpha_5} \; y_e^2 \; \sin^2 \delta \frac{1}{m_e} \times \left( 1, \frac{m_e^2}{\mphi^2},  \frac{m_e^2}{\mchi^2}\right), 
\end{equation}
where $m_e$ denotes the electron mass and $y_e$ the electron Yukawa coupling. Possibly relevant combinations from the scales running in the loop, have been included through the very right term with the electron mass indicating the smallest scale. Even by considering the least suppressed contribution possible with the following values for the parameters involved, $m_e \simeq \SI{0.5}{\MeV} \simeq \SI{2e10}{\cm^{-1}}$, $y_e \simeq\SI{2e-6}{}$, $\alpha=0.1$, $\alpha_5/\alpha=0.1$ and $\sin\delta=10^{-2}$, we obtain $\abs{d_e}\simeq\SI{e-30}{\cm\,e}$, which falls slightly below the experimental limit. However, we want to make two remarks on this conservative estimate, reasoning that the actual electron \gls{edm} will be much smaller: 1) The values for the mixing angle employed in this estimation are considerably larger than those considered in our study. A more realistic estimate would suppress $d_e$ by a factor of $\sim 10^{-4}$ or more. 2) The presence of heavy scales in the problem is anticipated to drive the \gls{edm} to much smaller values. Therefore, we conclude that the \gls{edm} bound from the electron is irrelevant for the parameter space of this model.\newp
In the calculation of the \gls{edm} of the muon, a larger Yukawa coupling enters, which enhances the theoretical prediction by a factor of $\sim 200$ after compensating for the higher muon mass. However, this is not sufficient to counter the $\order{10^{10}}$ weaker experimental constraint.


\subsubsection{Thermalization of the dark sector}
\label{subsubsec:thermalizationlimits}

\begin{figure}[t!]
    \centering
    \includegraphics[width=\textwidth]{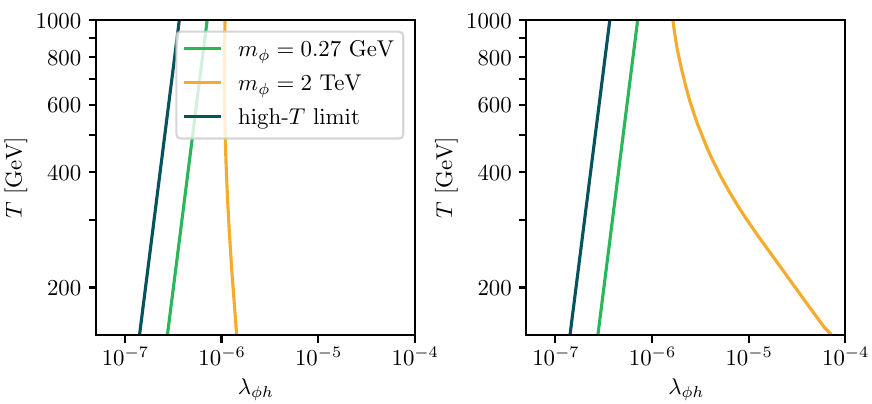}
    \caption[Thermalization limits on the quartic $\lambda_{\phi h}$ coupling for different scenarios.]{Thermalization limits on the quartic $\lambda_{\phi h}$ coupling using the condition $\mathscr{C}_{2 \to 2} = 3H \ndeq{\phi}$ for the $2\to 2$ processes $\phi H \leftrightarrow \phi H$ (left) and $\phi \phi \leftrightarrow H H^\dagger$ (right) at temperatures above \gls{ewsb}. The green and orange curves (in both plots) fulfill this condition for the two most extreme values for the mediator mass $\mphi$ employed in this context. The blue lines display the approximate result of \cref{eq:lambdaphihlimit}. Adapted from \eref\cite{Biondini:2023ksj}.}
    \label{fig:thermalizationrequirements}
  \end{figure}

The calculation of the relic density performed in \cref{subsubsec:relicdensityid} assumes that the \gls{sm} and the dark sector share a common temperature and that thermal contact is maintained during \gls{fo}. Although a deviation from this assumption is not inherently problematic, a more sophisticated analysis of the \gls{fo} process would be in order, and the accuracy of the relic density prediction will depend on the specifics of the thermal decoupling between the sectors. Hence, it is essential to verify the validity of our assumptions. To ensure an efficient thermal contact between the two sectors, we follow standard arguments in the literature, as outlined \eg in \eref\cite{Lebedev:2021xey}.\newp
Before \gls{ewsb}, the predominant processes are $2\to 2$ scatterings of the form $\phi H \leftrightarrow \phi H$ and $\phi \phi \leftrightarrow H H^\dagger$. Thermal contact is maintained, when these processes are efficient, \ie the interaction rates surpass the expansion rate of the Universe. As usual, this can be rephrased as $\mathscr{C}_{2 \to 2} \geq 3  H \ndeq{\phi}$, where $H$ is the Hubble rate during radiation domination (\cf\cref{eq:Hubble}), $n^{\text{eq}}_{\phi}$ denotes the equilibrium number density of the mediator, and $\mathscr{C}_{2 \to 2}$ represents the rate density (\ie the integrated collision term without back-reaction as defined in \cref{eq:integratedcollisionterm}). For a rough estimate, it suffices to consider the leading $T$ dependence entering the \gls{be}. This is captured by $\mathscr{C}_{2 \to 2}^{\text{high-}T} \propto \lambda^2_{\phi h}T^4/(32 \pi^2)$ at high temperatures.
Further approximating $\ndeq{\phi}\approx T^3/\pi^2$, we can estimate
\begin{equation}
    \label{eq:lambdaphihlimit}
    \lambda_{\phi h} \gtrsim 12.6~\geff^{1/4}(T) \sqrt{\frac{T}{\MPl}} \,.
\end{equation}
For $T \in [150, 1000]\,\si{\GeV}$, we find $\lambda_{\phi h} \gtrsim [\SI{e-7}{}, \SI{5e-7}{}]$. In \cref{fig:thermalizationrequirements}, we present limits on $\lambda_{\phi h}$ obtained from a more sophisticated analysis based on the full thermally averaged cross section of the processes $\phi H \leftrightarrow \phi H$ (left) and $\phi \phi \leftrightarrow H H^\dagger$ (right), which also accounts for the finite mass of the mediator. The values displayed have been calculated for the two most extreme benchmark points $\mphi=\SI{0.27}{\GeV}$ (green) and $\mphi=\SI{2}{\TeV}$ (orange), together with our approximation in \cref{eq:lambdaphihlimit} (blue). We can see that the elastic scattering $\phi H \leftrightarrow \phi H$ is more efficient. Across the range of scalar masses considered throughout our work, $\lambda_{\phi h} \simeq 10^{-6}$ is sufficient to maintain thermal contact.\newp
At temperatures below the \gls{ewsb} scale, the primary thermal connection between the \gls{sm} and the dark sector is facilitated by the mixing between the Higgs and the mediator. In this scenario, the most relevant processes for establishing thermal contact involve $f \phi \leftrightarrow f V$ scattering processes, where $f$ denotes an \gls{sm} fermion and $V$ is a gauge boson (see \eg\app\,A.2 of \eref\cite{Evans:2017kti} for a detailed discussion). The size of the mixing angle, needed to ensure thermal contact between the sectors as a function of the \gls{fo} temperature $T_\text{\tiny FO}$, can be extracted from Fig.\,2 of \eref\cite{Evans:2017kti}. In our region of interest, \eg $T_{\text{\tiny FO}}\gtrsim \SI{25}{\GeV}$, we find that $\sin \delta \gtrsim 10^{-6} - 10^{-7}$ is required to establish thermal contact. For most of the parameter space explored in this study, these values remain well below the experimental limits. Nevertheless, for very light scalars with $\mphi\lesssim \SI{500}{\MeV}$, the constraints from direct detection become highly restrictive and start to rule out the mixing angles necessary to sustain thermal contact. This specific region, however, largely overlaps with the region excluded by the combined constraints from direct detection and \gls{bbn}, so we will not display it separately.


\subsection{The parameter space of thermal dark matter}
\label{subsec:parameterspacethermaldm}

One of our main interests in this work is to gain insight about the status of our indirect detection limits and prospects when compared to the theoretical predictions for thermally produced \gls{dm}. Hence, we direct our focus to the parameter space of our model that enables a full production of the relic density through the \gls{fo} mechanism. There are five free parameters which characterize our model: the \gls{dm} and mediator masses (denoted as $\mchi$ and $\mphi$), the scalar and pseudo-scalar interaction strengths in the dark sector ($\alpha$ and $\alpha_5$), and the mixing angle between the mediator $\phi$ and the \gls{sm} Higgs. Provided the mixing angle remains sufficiently small to evade constraints from direct detection and large enough not to trouble \gls{bbn}, its precise value is irrelevant to phenomenology (\cf\cref{subsubsec:DD_BBN}). This reduces the number of relevant parameters to four. The relic density, which is accurately measured at the percent level \cite{Planck:2018vyg}, enables a further reduction of free parameters by requiring the correct \gls{dm} abundance in the early Universe after \gls{fo}. We choose to fix $\alpha$ in this manner, allowing for the variation of the two masses and $\alpha_5$. With only three parameters in play, we can illustrate slices through the parameter space for two of them by keeping the third fixed. Since the phenomenology is primarily influenced by the masses of the involved particles, our analysis considers $\mchi$ and $\mphi$ as variables, showcasing fixed ratios of $\alpha_5/\alpha$ to explore this direction.\newp
\begin{figure}
    \centering
    \includegraphics[width=\textwidth]{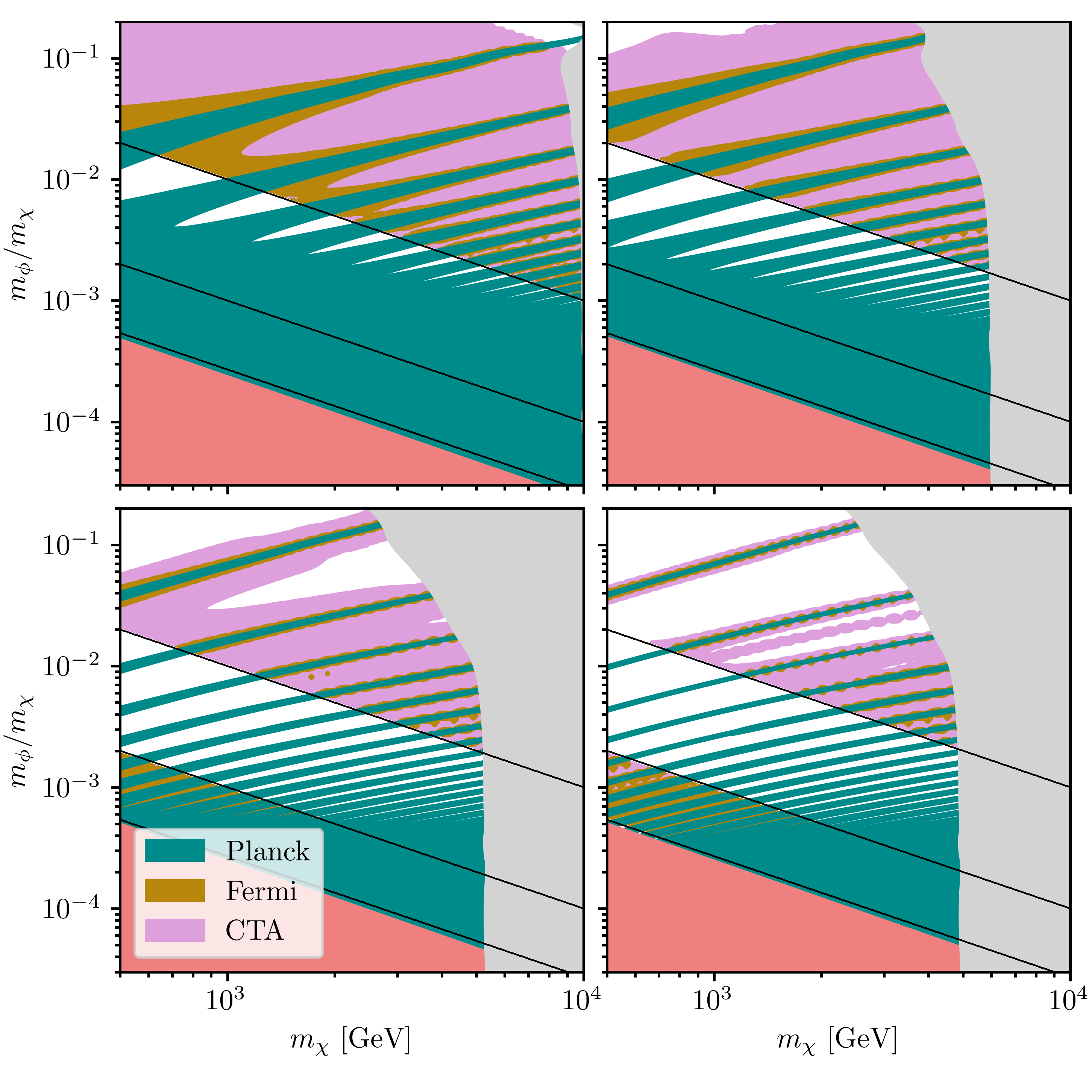}
    \caption[Slices through the cosmologically preferred parameter space in the $\mchi$-$\mphi/\mchi$ plane for four benchmark values of $\alpha_5/\alpha$.]{Slices through the cosmologically preferred parameter space in the $\mchi$-$\mphi/\mchi$ plane for four benchmark values of $\alpha_5/\alpha\in\{10^{-1},10^{-2},10^{-3},10^{-4}\}$ (left-to-right, top-to-bottom). The value of $\alpha$ has been fixed by the relic density requirement. Exclusion limits from indirect detection have been overlayed in the following order (starting with the top layer): Regions excluded by Planck are shown in turquoise and the \gls{fermi} exclusion limits are presented in gold, whereas prospects for \gls{cta} are depicted in pink. Mediator mass regions, which are excluded by the combined \gls{bbn} and direct detection limits on the mixing angle, are shaded in red. Regions where relic density requires $\alpha>0.25$ are shown in gray. Black lines indicate characteristic values of $\mphi=2 m_{\pi^0}$, $\SI{1}{\GeV}$ and $\SI{10}{\GeV}$ that separate the different gamma-ray production regimes defined in \cref{subsubsec:photonspectrum}. Adapted from \eref\cite{Biondini:2023ksj}.}
    \label{fig:parameterspaceID}
\end{figure}
\Cref{fig:parameterspaceID} shows representative examples of slices through the cosmologically preferred parameter space. The analysis is confined to the region $\SI{0.5}{\TeV} \leq \mchi \leq \SI{10}{\TeV}$. We also require $ \mphi \lesssim \alpha \mchi$ as an upper limit to ensure the applicability of the \gls{nreft} used in deriving the cross sections and non-perturbative effects from long-range interactions. Constraints and prospects related to gamma-rays are assessed for $\mphi \geq 2 m_{\pi^0}$ (\cf\cref{subsubsec:photonspectrum}) while \gls{cmb} limits are not constrained by this condition. However, the joint limit imposed by \gls{bbn} and direct detection searches excludes mediator masses slightly below this threshold anyways (shaded in red). In regions where relic density computations suggest values of $\alpha \gtrsim 0.25$ (shaded in gray), certain corrections in our derivations, such as $\alpha$ and $\alpha_5$ corrections to the matching coefficients in \cref{eq:IMmatch_coeff_1,eq:IMmatch_coeff_2,eq:IMmatch_coeff_3,eq:IMmatch_coeff_4} and corrections to the binding energy of \gls{bs}, become significant. However, accounting for these corrections is very challenging and clearly beyond the scope of this work. Hence, we will exclude this region from our discussion.\newp
We observe that indirect detection limits can be highly restrictive when $\alpha_5$ is not significantly smaller than $\alpha$. This is because larger values of $\alpha_5$ amplify the s-wave cross section, which is not velocity suppressed at any $\mphi$. Further including \gls{se}, this essentially excludes configurations with $\mphi \lesssim \SI{3}{\GeV}$ by Planck measurements for $\alpha_5/\alpha= 0.1$. At higher mediator masses, \gls{cmb} exclusions rely on resonances in the generalized $J$-factors due to \gls{se}, resulting in excluded parameter space stripes which extend to very high values of $\mchi$ and $\mphi$. These stripes become thinner with a decreasing total cross section for an increasing \gls{dm} mass, as the resonance condition must be fulfilled with higher precision. Gamma-ray telescopes can provide additional information in this regime. The \gls{fermi} limits are mostly comparable to the \gls{cmb} ones, however, due to Fermi's higher sensitivity, they result in broader stripes that encompass the \gls{cmb} ones for low $\mchi$. At the higher end of the mass range, this effect diminishes due to the sensitivity scaling differently with the \gls{dm} mass. Therefore, \gls{fermi} does not contribute significantly to \gls{cmb} bounds on the edge of our parameter space for $\mchi \approx \SI{10}{\TeV}$. Remarkably, \gls{cta} presents a significant advancement, offering stronger exclusion prospects without relying on resonance features. This effectively closes gaps in between the stripes and enables a probe of nearly the entire parameter space. Only very high \gls{dm} masses and large $\mphi$ can potentially evade these prospects. It is noteworthy that gamma-ray searches also contribute at low $\mphi$, where the low mediator mass regime ($2 m_{\pi^0} \leq \mphi \leq \SI{1}{\GeV}$) is excluded by \gls{fermi}. These limits are intentionally overlaid below the \gls{cmb} ones to enhance the readability.\newp
For lower $\alpha_5/\alpha$ ratios, the significance of p-wave and \gls{bsf} contributions increases. For $\alpha_5/\alpha=10^{-2}$, the overall qualitative picture remains similar, with new features becoming visible for $\alpha_5/\alpha=10^{-3}$. In this regime, bounds from Planck decrease substantially in the low mediator mass regime, causing configurations with $\mphi \lesssim \SI{1}{\GeV}$ to no longer be entirely constrained by CMB limits alone. However, \gls{fermi} limits remain significant in this region due to the substantial number of energetic photons produced in the pion cascade regime. As a result, a significant portion of this parameter space is still excluded by current experiments. For $\mphi > \SI{10}{\GeV}$, only narrow stripes with a strong resonant enhancement are excluded by existing experiments. \gls{cta} prospects remain excellent, and most gaps between enhanced stripes can be closed if the \gls{cta} sensitivity aligns with our expectations. It is important to note that the granularity of the gamma-ray limits, evident compared to the Planck limits in this figure, is a numerical artifact arising from the lower resolution of \gls{fermi} and \gls{cta} scans. Determining the generalized $J$-factors required for gamma-ray limits and prospects is computationally expensive. Therefore, we chose a coarser scanning grid given the available computing power.\newp
At $\alpha_5/\alpha=10^{-4}$, signals are predominantly influenced by p-wave and \gls{bsf}. Existing limits are relatively weak, and \gls{cta} is crucial in closing gaps in the low mediator mass region. For $\mphi\geq \SI{10}{\GeV}$, most of the parameter space is presently unconstrained. While the \gls{cta} remains influential, for $\mphi/\mchi\gtrsim 10^{-2}$, it also depends on resonances in the \gls{se} factors. Additional strips are resolved in this figure, driven by resonances of p-wave \gls{se} factors, which do not coincide with those of the s-wave (\cf\cref{fig:Sannlspectrum}). Overall, indirect detection proves highly effective in testing the model where s-wave annihilation is relevant. In the p-wave and \gls{bsf}-dominated regime, some promising limits and good prospects exist for parts of the parameter space. However, a definitive test of the model for low $\alpha_5/\alpha$ ratios remains challenging even with near-future experiments like the \gls{cta}.
\section{Conclusion}
\label{sec:conclusion}

Over the past two decades, researchers in the \gls{dm} community have increasingly focused on non-perturbative effects in \gls{dm} interactions. These effects have gained attention for their importance in accurately predicting \gls{dm} properties in various models. Particularly, the exploration of \gls{bs} alongside \gls{se} has become an active area of research in recent years. This thesis, centered around two publications \cite{Bollig:2021psb,Biondini:2023ksj}, has been dedicated to investigating these phenomena and their impact on the production and detection of \gls{dm}.\newp
In the first study, we focused on \gls{dm} models with non-thermal production, comprising a \gls{dm} candidate with feeble couplings to the visible sector and a mediator with \gls{sm} quantum numbers. Such models present markedly different experimental signatures compared to the prevalent \gls{wimp} paradigm, rendering them intriguing subjects for study. Additionally, due to the mediator interactions with the \gls{sm}, significant portions of the models' parameter space are accessible to collider searches at the \gls{lhc}. This is particular interesting, considering the impending \gls{hl}-\gls{lhc} upgrade. Given that gauge interactions wiht the mediator may also lead to substantial non-perturbative corrections for \gls{dm} production in the early Universe, a comprehensive study of the cosmologically favored parameter space must account for these effects to provide a realistic evaluation of experimental capabilities.\newp
In \cref{sec:impactonnonthermalDMproduction}, we conducted a comprehensive analysis of non-thermal \gls{dm} production within a specific class of simplified models. These models feature a fermionic singlet \gls{dm} candidate and a scalar (color-)charged mediator. We accounted for both \gls{fi} and \gls{sw} production of \gls{dm} and factored in corrections from \gls{se} and \gls{bs} effects. Comparing our findings with perturbative calculations, we observed a significant discrepancy in the case of a color-charged mediator, where neglecting non-perturbative effects led to an overestimation of the \gls{sw} contribution to the \gls{dm} abundance by an order of magnitude. This underscores the necessity of accounting for such corrections and reveals a much broader parameter space conducive to non-thermal \gls{dm} production than previously anticipated. In contrast, for a lepto-philic mediator, the non-perturbative effects were more subdued, with a $\sim15\%$ correction to the \gls{dm} yield from \gls{sw} production. However, this scenario is sensitive to the coupling between the mediator and the Higgs field, where even modest changes can result in considerable alterations to the \gls{sw} contribution.\newp
By combining constraints from \gls{lhc} searches, cosmological bounds from \gls{bbn}, and predictions from \gls{dm} production in the early Universe, we scrutinized the viable parameter space of both models. We found that while a small mass gap between \gls{dm} and the mediator is feasible for relatively low mediator masses in the range of a few $\si{TeV}$, larger mediator masses necessitate a substantial mass separation to prevent a \gls{dm} overabundance. Therefore, overall larger \gls{dm} masses are preferred. Non-perturbative effects significantly expand this viable parameter space. Altogether, our findings indicate that collider tests of these models pose greater challenges than initially anticipated. The increased luminosity at the \gls{hl}-\gls{lhc} may enhance experimental reach, but a collider with higher energy levels would be highly advantageous for probing these type of \gls{dm} models effectively.\newp
The focus of the second study was to investigate how non-perturbative effects impact the detectability of \gls{dm} in cosmological searches. For this purpose, we utilized a thermal \gls{dm} model with annihilations occurring within the dark sector. This setup naturally circumvents direct detection and collider constraints, singling out indirect detection as the preferred avenue for \gls{dm} searches. This is particularly pertinent given the imminent availability of new, more sensitive instruments like the \gls{cta}, which will provide new data in the near future. Thus, it is crucial to understand the sensitivity of the \gls{cta} with respect to thermal \gls{dm} and compare its abilities with existing constraints. Given the wide range of potential observables, we selected robust limits from Planck and \gls{fermi} as our benchmarks.\newp
In \cref{sec:indirectdetection}, we therefore concentrated on a model featuring fermionic \gls{dm}, which possesses (pseudo)-scalar couplings to a massive scalar mediator. The mediator in turn interacts with the \gls{sm} through mixing with the Higgs. When the mass of the mediator is relatively light compared to the \gls{dm} mass and the potential created by the mediator is attractive, \gls{se} can become sizeable in \gls{dm} observables and \gls{bs} can form. As a result of these non-perturbative effects, the cross sections of interest, specifically for \gls{dm} indirect detection, exhibit a pronounced and intricate dependence on the relative \gls{dm} velocity. This contradicts a conventional assumption in indirect detection, which suggests that only the leading velocity-independent contribution to s-wave annihilation is detectable in a realistic experiment. We accounted for this in our computation of appropriate velocity-averaged $J$-factors, which necessitates position-dependent velocity distributions for the targets under consideration. These have been derived using Eddington inversion from density profiles for four \glspl{dsph} and the \gls{gc} of the Milky Way. This enabled us to predict the photon flux from \gls{dm} annihilation based on particle physics parameters and establish upper limits (or identify prospects).\newp
Combining this with our predictions for the relic density allowed us to evaluate the status of thermal \gls{dm} in this model. If the suppression of the coupling parameter $\alpha_5$ relative to $\alpha$ is modest, s-wave annihilations play a significant role, leading to stringent constraints on light mediators regardless of the \gls{dm} mass. At higher mediator masses, the resonant enhancement due to \gls{se} factors becomes significant, enabling exclusion of regions near a resonance with Planck and \gls{fermi} observations. Interestingly, remaining gaps in the testable parameter space can be addressed by the \gls{cta} in this scenario. Testing the model becomes more challenging for smaller $\alpha_5/\alpha$, where the sensitivity of Planck and \gls{fermi} to the cosmologically favored parameter space diminishes, making \gls{cta} crucial, as it will cover large regions of the parameter space for the first time.\newp
We hope to have convinced the reader that research on non-perturbative effects in \gls{dm} model building is to equal terms an important and exciting field, with the potential to study the nature of \gls{dm} to an unprecedented level of precision. The challenges ahead are manifold, such as the consistent and automatised inclusion of higher-order bound states or the consideration of the \gls{sm} bath as a thermal background, to name only a few. With the rapid improvement of our theoretical toolkit, we are confident that our field will overcome all these challenges, eventually providing the most robust predictions on \gls{dm} observables for experimental exploration.

\pagestyle{appendix}
\appendix
\appendixpage
\newpage
\thispagestyle{empty}
\mbox{}
\newpage
\renewcommand{\theequation}{\thesection.\arabic{equation}}
\section{Special functions}
\label{app:special_functions}

In this appendix, we list all special functions, along with their properties and identities, that we will use throughout this work. The information provided below is by no means exhaustive, as the properties are well-known. Therefore, we will refrain from citations or proofs.

\paragraph{Gamma function}$~$\newpp
The \textit{gamma function} is an extension of the factorial to the complex plane and can be defined in the regime $z\in\mathbb{C}\setminus(-\mathbb{N})$. It copies the properties of the factorial function $\Gamma(z+1)=z\Gamma(z)$, and reduces to it $\Gamma(n+1)=n!$ for $n\in\mathbb{N}$. For $\Re{z}>0$, there also exists an integral representation
\begin{equation}
    \label{eq:gammafunction}
    \Gamma(z,x)\equiv\int_x^\infty \dd{t}t^{z-1}e^{-t}\,, \quad  \Gamma(z)\equiv \Gamma(z,0)\,,
\end{equation}
where (for $x\geq 0$) the former quantity is often called \textit{incomplete gamma function}. A property, which will be used often throughout this work is
\begin{equation}
    \label{eq:GammaS0prod}
    \abs{\Gamma(1+l-i\zeta)}^2=\frac{2\pi\zeta}{1-e^{-2\pi\zeta}}e^{-\pi\zeta}\prod_{r=1}^l(r^2+\zeta^2)\quad\text{for}\quad l\in\mathbb{N}\,,\quad \zeta\in\mathbb{R}\,.
\end{equation}

\paragraph{Bessel functions}$~$\newpp
\textit{Bessel functions} are canonical solutions $y(x)$ to Bessel's differential equation
\begin{equation}
    \left[x^2\dv[2]{}{x}+x\dv{}{x}+(x^2-\alpha^2)\right]y(x)=0
\end{equation}
for an arbitrary complex number $\alpha\in\mathbb{C}$. For real $x$, there are two linearly independent solutions 
\begin{align}
    J_\alpha(x)&\equiv\sum_{m=0}^\infty \frac{(-1)^m}{m!\Gamma(m+\alpha+1)}\left(\frac{x}{2}\right)^{2m+\alpha}\,,\\
    Y_\alpha(x)&\equiv\frac{J_\alpha(x)\cos(\alpha\pi)-J_{-\alpha}(x)}{\sin(\alpha\pi)}\,,
\end{align}
denoted as \textit{Bessel functions of first and second kind}, where for $\alpha=n$ an integer, the latter is defined as $Y_n(x)\equiv\lim_{\alpha\to n}Y_\alpha(x)$. For purely imaginary arguments $i x$, $x\in\mathbb{R}$, the solutions are called \textit{modified Bessel functions of first and second kind} and are connected to the real solutions via
\begin{align}
    I_\alpha(x)&\equiv i^{-\alpha}J_\alpha(ix)\,,\\
    K_\alpha(x)&\equiv \frac{\pi}{2}\frac{I_{-\alpha}(x)-I_\alpha(x)}{\sin(\alpha\pi)}\,.
\end{align}
Within this work, we are chiefly interested in $K_n(x)$, $n\in\mathbb{N}$, which possess an integral representation
\begin{equation}
    \label{eq:BesselKdef}
	K_n(x)\equiv\frac{\sqrt{\pi}}{\Gamma\left(n+\frac{1}{2}\right)}\left(\frac{x}{2}\right)^n\int_1^\infty \dd{y} e^{-xy}(y^2-1)^{n-\frac{1}{2}}\,,
\end{equation}
as well as a series expansion for large arguments
\begin{equation}
    \label{eq:BesselKexplargearguments}
	K_n(x)=\sqrt{\frac{\pi}{2x}}e^{-x}\sum_{k=0}^{\infty}\frac{1}{k!(8x)^k}\prod_{l=0}^{k-1}\left(4n^2-(2l+1)^2\right)\,.
\end{equation}
The two linearly independent solutions for $\alpha^2=l(l+1)$, $l\in\mathbb{Z}$, are called \textit{spherical Bessel and Neumann functions} and can be defined via
\begin{align}
    \label{eq:sphBesseljdefsum}
    \jl(x) & \equiv \sqrt{\frac{\pi}{2x}}J_{n+\frac{1}{2}}(x)=\sum_{s=0}^\infty \frac{(-1)^s x^{l+2s}}{2^s s!(2s+2l+1)!!}\,,\\
    \nl(x) & \equiv \sqrt{\frac{\pi}{2x}}Y_{n+\frac{1}{2}}(x)=(-1)^{n+1}\sqrt{\frac{\pi}{2x}}J_{-n-\frac{1}{2}}(x)\,.
\end{align}
Closely related are the \textit{spherical Hankel functions}, which are just linear combinations of the spherical Bessel and Neumann functions
\begin{equation}
    \label{eq:sphHankelfunctions}
    \hlone(x)\equiv\jl(x)+i\nl(x)\,,\quad \hltwo(x)\equiv\jl(x)-i\nl(x)\,.
\end{equation}

\paragraph{Laguerre polynomials}$~$\newpp
The \textit{associated Laguerre polynomials} are solutions to Laguerre's differential equation
\begin{equation}
    \left[x\dv[2]{}{x}+(\alpha+1-x)\dv{}{x}+n\right] \Laguerre{n}{\alpha}{x}=0\,,
\end{equation}
with $n\in\mathbb{N}$, $\alpha\in\mathbb{R}$ and reduce to the regular \textit{Laguerre polynomials} $L_n$ for $\alpha=0$. For this work, we only need the case where $\alpha\in\mathbb{N}$. They can be defined through a polynomial series 
\begin{equation}
    \label{eq:genLaguerrepolsumdef}
    \Laguerre{n}{\alpha}{x}=\sum_{k=1}^n \frac{(-1)^k(n+\alpha)!x^k}{(n-k)!(k+\alpha)!k!}\,,
\end{equation}
and thus expressed through a confluent hypergeometric function via
\begin{equation}
    \label{eq:LaguerreHGoneconnection}
    \Laguerre{n}{\alpha}{x}=\frac{(\alpha+1)_n}{n!}\HGone{-n}{\alpha+1}{x}\,,
\end{equation}
with $(\alpha+1)_n$ defined in \cref{eq:Pochhammersymbols}.

\paragraph{Legendre polynomials}$~$\newpp
The \textit{associated Legendre polynomials} are solutions to the differential equation
\begin{equation}
    \label{eq:diffeqLegendrePol}
    \left[\dv{}{x}\left((1-x^2)\dv{}{x}\right)+\left(l(l+1)-\frac{m^2}{1-x^2}\right)\right]P_l^m(x)=0\,,
\end{equation}
with $l,m\in\mathbb{N}$, where the special case $m=0$ yields the usual \textit{Legendre polynomials} $P_l(x)\equiv P_l^0(x)$. Identifying $x\equiv \cos\theta$, the latter are also eigenfunctions to the Laplace operator in spherical coordinates
\begin{equation}
    \label{eq:LegendreDEQ}
    \Delta_{\vec{r}}P_{l}(\cos\theta)=-\frac{l(l+1)}{r^2}P_{l}(\cos\theta)\,,
\end{equation}
which can easily be derived from \cref{eq:diffeqLegendrePol}. The Legendre polynomials form a complete and orthogonal system of polynomials. The standard condition $P_l(1)\equiv 1$ fixes their normalization (with respect to the $L^2$ norm) on the interval $-1\leq x\leq 1$. Their orthogonality as well as completeness relations are given by
\begin{align}
    \label{eq:Legendrenorm}
    \int_{-1}^1\dd{x}P_l(x)P_{l'}(x)=\frac{2}{2l+1}\delta_{ll'}\,,\\
    \label{eq:Legendrecomplete}
    \sum_{l=0}^\infty \frac{2l+1}{2}P_l(x)P_l(y)=\delta(x-y)\,.
\end{align}
We will make extensive use of the special case $x=1$, $y=\uvec{r}\cdot\uvec{r}'$ of \cref{eq:Legendrecomplete}, yielding
\begin{equation}
    \label{eq:Legendresum}
    \sum_{l=0}^\infty \frac{2l+1}{2}P_l(\uvec{r}\cdot\uvec{r}')=\delta(1-\uvec{r}\cdot\uvec{r}')\,.
\end{equation}
Due to the properties described above, we can use the Legendre polynomials as a basis for any analytic azimuthally symmetric function 
\begin{equation}
    \label{eq:LPdecomposition}
    f(r,\theta)=\sum_{l=0}^\infty f_l(r) P_l(\cos\theta)\,,
\end{equation}
with coefficients $f_l(r)$ solely depending on $r$. A special analytic function used in the following is the plane wave, given as
\begin{equation}
    \label{eq:planewave}
    e^{i\vec{k}\cdot\vec{r}}=\sum_{l=0}^\infty(2l+1)i^l \jl(kr) P_l(\cos\theta)\,,
\end{equation}
with $j_l$ the spherical Bessel functions and $\cos\theta\equiv \uvec{k}\cdot\uvec{r}$. The Legendre polynomials of a scalar product of unit vectors can, in turn, be expanded in terms of spherical harmonics
\begin{equation}
    \label{eq:sphericalharmonicsadditiontheorems}
    P_l(\uvec{r}\cdot\uvec{r}')=\frac{4\pi}{2l+1}\sum_{m=-l}^{l}Y^*_{lm}(\Omega_{\vec{r}})Y_{lm}(\Omega_{\vec{r}'})\,,
\end{equation}
from  which also the following property can be derived
\begin{equation}
    \label{eq:Legendreint}
    \int\dd{\Omega_{\vec{k}}}P_l(\uvec{k}\cdot\uvec{r})P_{l'}(\uvec{k}\cdot\uvec{r}')=\frac{4\pi}{2l+1}\delta_{ll'}P_l(\uvec{r}\cdot\uvec{r}').
\end{equation}
A useful identity concerning Legendre polynomials, which has been derived \eg in \app\,D of \eref\cite{Petraki:2015hla} and will be used in \cref{app:SEderivation} is 
\begin{align}
    \label{eq:SommerfeldPlidentity}
    \int\dd{\Omega_{\vec{p}}}P_{l'}&(\cos\theta_{\vec{p}})\int\frac{\dd[3]{q}}{(2\pi)^3}\tilde{f}(\vec{q})\abs{\vec{q}}^lP_l(\cos\theta_{\vec{q},\vec{p}})\nonumber\\
    &=\frac{\delta_{ll'}(2l+1)!!}{i^l(2l+1)l!}\left[\dv[l]{}{r}\int\dd{\Omega_{\vec{r}}}P_l(\cos\theta_{\vec{r}})f(\vec{r})\right]_{\vec{r}=0}\,,
\end{align}
for an arbitrary analytic function $f(\vec{r})$ with Fourier transform $\tilde{f}(\vec{q})$.

\paragraph{Spherical harmonics}$~$\newpp
\textit{Spherical harmonics} are a complete set of orthonormal functions defined on the surface of a sphere $Y_{lm}:S^2\to \mathbb{C}$ and, therefore, form an orthonormal basis of the Hilbert space of square-integrable (complex) functions $L^2_{\mathbb{C}}(S^2)$. They can be defined through the associated Legendre polynomials
\begin{equation}
    \label{eq:sphericalharmonicsdef}
    Y_{lm}(\Omega)\equiv(-1)^m\sqrt{\frac{2l+1}{4\pi}\frac{(l-m)!}{(l+m)!}} P_l^m(\cos\theta)e^{im\phi}\,,
\end{equation}
with $Y^*_{lm}(\Omega)=(-1)^m Y_{l-m}(\Omega)$, and their orthonormality and completeness conditions are given by
\begin{align}
    \label{eq:sphericalharmonicsnorm}
    \int \dd{\Omega_{\vec{r}}}Y_{lm}(\Omega)Y^*_{l'm'}(\Omega)&=\delta_{ll'}\delta_{mm'}\,,\\
    \label{eq:sphericalharmonicscomplete}
    \sum_{l=0}^{\infty}\sum_{m=-l}^{l}Y_{lm}^*(\Omega)Y_{lm}(\Omega')&=\delta(\cos\theta-\cos\theta')\delta(\phi-\phi')\,.
\end{align}
From \cref{eq:sphericalharmonicsadditiontheorems} we can derive Uns\"old's theorem
\begin{equation}
    \sum_{m=-l}^{l}Y_{lm}^*(\Omega)Y_{lm}(\Omega)=\frac{2l+1}{4\pi}\,,
\end{equation}
by setting $\uvec{r}=\uvec{r}'$. Spherical harmonics are also eigenfunctions of the Laplace operator in spherical coordinates, yielding
\begin{equation}
    \label{eq:sphericalharmonicsDEQ}
    \Delta_{\vec{r}}Y_{lm}(\Omega)=-\frac{l(l+1)}{r^2}Y_{lm}(\Omega)\,.
\end{equation}

\paragraph{Wigner-3j symbols}$~$\newpp
The \textit{Wigner 3j-symbols} can be considered an alternative formulation to \textit{Clebsch-Gordon coefficients} and are linked to them via
\begin{equation}
    \begin{pmatrix}l_1&l_2&l_3\\m_1&m_2&m_3\end{pmatrix}\equiv \frac{(-1)^{l_1-l_2-m_3}}{\sqrt{2l_3+1}}\langle l_1,m_1,l_2,m_2|l_3,-m_3\rangle\,,
\end{equation}
where $l_i$, $m_i$ are angular momentum quantum numbers with $m_i\in\{-l_i,-l_i+1,\dots,l_i-1,l_i\}$. They vanish unless, the following additional conditions are satisfied: 1) $m_1+m_2+m_3=0$, 2) $|l_1-l_2|\leq l_3\leq l_1+l_2$, and 3) $l_1+l_2+l_3\in \mathbb{N}$ (and also even, if $m_1=m_2=m_3=0$). We will need in the following the orthogonality condition
\begin{equation}
    \label{eq:Wigner3jorthogonality}
    (2l_3+1)\sum_{m_1,m_2}
    \begin{pmatrix}l_1&l_2&l_3\\m_1&m_2&m_3\end{pmatrix}
    \begin{pmatrix}l_1&l_2&l'_3\\m_1&m_2&m'_3\end{pmatrix}
    =\delta_{l_3l'_3}\delta_{m_3m'_3}\{l_1~l_2~l_3\}\,,
\end{equation}
where $\{l_1~l_2~l_3\}$ is $1$ when the triangular condition 2) is fulfilled and $0$ elsewise, and the connection of the Wigner 3j-symbols to the spherical harmonics is given by
\begin{align}
    \int\dd{\Omega} Y_{l_1m_1}(\Omega)&Y_{l_2m_2}(\Omega)Y_{l_3m_3}(\Omega)\nonumber\\
    &=\sqrt{\frac{(2l_1+1)(2l_2+1)(2l_3+1)}{4\pi}}
    \begin{pmatrix}
        l_1 & l_2 & l_3 \\
        0 & 0 & 0
    \end{pmatrix}
    \begin{pmatrix}
        l_1 & l_2 & l_3 \\
        m_1 & m_2 & m_3
    \end{pmatrix}\,.
    \label{eq:Wigner3jrelation}
\end{align}

\paragraph{Hypergeometric functions}$~$\newpp
A \textit{generalized hypergeometric function} is a convergent polynomial series, defined via
\begin{equation}
    _pF_q(a_1,\dots,a_p;b_1,\dots,b_q;z)\equiv\sum_{k=0}^\infty \frac{(a_1)_k\dotsb(a_p)_k}{(b_1)_k\dotsb(b_q)_k}\frac{z^k}{k!}\,,
\end{equation}
with $p,q\in\mathbb{N}$, where we introduced \textit{Pochhammer symbols}
\begin{equation}
    \label{eq:Pochhammersymbols}
    (x)_n\equiv\frac{\Gamma(x+n)}{\Gamma(x)}
\end{equation}
for a more compact notation. We employ in the main text two functions of these type. The first one is the \textit{confluent hypergeometric function}
\begin{equation}
    \label{eq:HGonesumdef}
    \HGone{a}{b}{x}\equiv\sum_{k=0}^{\infty}\frac{(a)_k}{(b)_k}\frac{z^k}{k!}\,,
\end{equation}
which has an integral representation 
\begin{equation}
    \HGone{a}{b}{x}=\frac{\Gamma(b)}{\Gamma(b-a)\Gamma(a)}\int_0^1\dd{t}e^{zt}t^{a-1}(1-t)^{b-a-1}\,,
\end{equation}
for $\Re{b}>\Re{a}>0$. The second one is the \textit{hypergeometric function}
\begin{equation}
    \HGtwo{a}{b}{c}{x}\equiv\sum_{k=0}^{\infty}\frac{(a)_k(b)_k}{(c)_k}\frac{z^k}{k!}\,,
\end{equation}
with an integral representation
\begin{equation}
    \HGtwo{a}{b}{c}{x}=\frac{\Gamma(c)}{\Gamma(c-b)\Gamma(b)}\int_0^1\dd{t}\frac{t^{b-1}(1-t)^{c-b-1}}{(1-tz)^a}\,,
\end{equation}
if $\Re{c}>\Re{b}>0$. For our calculations in this work, we will make use of the following identities
\begin{align}
    \label{eq:HGoneD0}
    e^x\HGone{a}{c}{-x} &= \HGone{c-a}{c}{x}\,,\\
    \label{eq:HGoneD1}
    \dv[n]{(\HGone{a}{b}{z})}{z}&=\frac{(a)_n}{(b)_n}\HGone{a+n}{b+n}{z}\,,\\
    \label{eq:HGoneD2}
    \dv[n]{(z^{b-1}\HGone{a}{b}{z})}{z}& =(b-n)_n z^{b-n-1}\HGone{a}{b-n}{z}\,,
\end{align}
alongside with (for $\Re{c}>0$)
\begin{equation}
    \label{eq:HGoneHGtwoIntegralidentity}
    \int_0^\infty \dd{t}e^{-\rho t} t^{c-1}\!\HGone{a}{c}{t}\!\HGone{b}{c}{\lambda t}=\frac{\Gamma(c)~\rho^{a+b-c}}{(\rho-1)^{a}(\rho-\lambda)^b}\HGtwo{a}{b}{c}{\frac{\lambda}{(\rho-1)(\rho-\lambda)}}\!.
\end{equation}

\paragraph{Other identities}$~$\newpp
There are three more identities used in this work, which do not fit into the classification above:
\begin{enumerate}
    \item The three-dimensional delta distribution in spherical coordinates assuming azimuthal symmetry, can be written for non-vanishing $\uvec{r}\cdot\uvec{r}'$ as 
    \begin{equation}
        \label{eq:deltaspherical}
        \delta^{(3)}\left(\vec{r}-\vec{r}'\right)=\frac{1}{r^2}\delta(r-r')\frac{1}{2\pi}\delta(1-\uvec{r}\cdot\uvec{r}')\,.
    \end{equation}
    \item The integral representation of the Heavyside step function is given by 
    \begin{equation}
        \label{eq:HeavysideIntegral}
        \Theta(z)= \lim_{\epsilon\to 0^+}\frac{i}{2\pi}\int_{-\infty}^{\infty}\dd{k}\frac{e^{-ikz}}{k+i\epsilon}=\begin{cases} 1 & z\geq 0 \\ 0 & z<0 \end{cases}\,.
    \end{equation}
    \item An identity, which will be used to compute the \gls{se} factors for \gls{bsf}, reads
    \begin{equation}
        \label{eq:arccotrelation}
        \abs{\left(\frac{n+i\zeta_2}{n-i\zeta_2}\right)^{-i\zeta_1}}^2=e^{-4\zeta_1\arccot{(\zeta_2/n)}}\,.
    \end{equation}
\end{enumerate}

\section{Partial wave analysis}
\label{app:partial_wave_analysis}

We will give here a brief insight into partial wave analysis, a method developed in scattering theory to calculate the wave functions and cross sections of interacting particles in the non-relativistic limit. For this purpose, we closely follow \eref\cite{griffiths:2018}.\newp
We can describe a scattering process by starting with an incoming plane wave in the $z$-direction $\phi_{\vec{k},\text{in}}\propto e^{ikz}$, with $k=\sqrt{2\mu\Ek}$ being the wave-number (see \cref{eq:scatteringSchroedinger}) and $z=r\cos\theta$. The plane wave eventually meets a potential where it gets scattered, producing an outgoing spherical wave $\phi_{\vec{k},\text{out}}\propto e^{ikr}/r$. If we are sufficiently far from the potential (\ie $r\to\infty$), the solution to the Schrödinger equation is of the form 
\begin{equation}
    \label{eq:largerwavefunctionscattering}
    \phik(\vec{r})\approx A\left(e^{ikz}+f(\Ek,\theta_r)\frac{e^{ikr}}{r}\right)\,,
\end{equation}
with $\abs{A}$ being the amplitude of the wave function and $f(\Ek,\theta_r)$ denotes the scattering amplitude, a measure of the probability of scattering in a given direction $\theta_r$ (assuming azimuthal symmetry in the following). The scattering amplitude is related to the cross section of the process via
\begin{equation}
    \dv{\sigma}{\Omega}=\abs{f(\Ek,\theta_r)}^2\,.
\end{equation}
Due to the azimuthal symmetry of the problem, we can expand $\phi_{\vec{k},\text{out}}$ in the basis of Legendre polynomials (\cref{eq:LPdecomposition})
\begin{equation}
    \label{eq:partialwaveexpansionphi}
    \phi_{\vec{k},\text{out}}(\vec{r})=\frac{1}{\sqrt{4\pi}}\sum_{l=0}^{\infty} \sqrt{2l+1}\frac{\ukl(r)}{r}P_l(\cos\theta_r)\,,
\end{equation}
with the prefactors chosen for later convenience. The Schrödinger equation in terms of $\ukl$ then reads 
\begin{equation}
    \ukl''(r)\underbrace{-\frac{l(l+1)}{r^2}\ukl(r)}_{\text{\quotes{centrifugal term}}}\,\underbrace{-\vphantom{\frac{1}{r^2}}2\mu V(r)\ukl(r)}_{\text{\quotes{potential term}}}=-k^2 \ukl(r)\,.
\end{equation}
For pedagogical reasons, we will now consider three situations and compare them to the outgoing spherical wave in the large $r$ limit: 1) where $r$ is so large that we can neglect the potential as well as the centrifugal term, 2) where $r$ is in the intermediate regime where we can only neglect the potential term and 3) where we are close to the interaction point and have to take into account the whole Schrödinger equation. Scenario 3) will also correspond to the situation where we will have a long-ranged potential as used often throughout this work.\newp
If we can neglect both the centrifugal and potential term in scenario 1), the general solution to Schrödinger equation is given by 
\begin{equation}
    \ukl(r)= Ce^{ikr}+De^{-ikr}\,,
\end{equation}
where $e^{ikr}$ represents an outgoing and $e^{-ikr}$ an incoming spherical wave. Comparing this solution to \cref{eq:largerwavefunctionscattering}, it is obvious that $D=0$.\newp
In scenario 2), where we can only neglect the potential term, the solution yields
\begin{equation}
    \ukl(r)= Ar\jl(kr)+Br\nl(kr)=\tilde{A}r\hlone(kr)+\tilde{B}r\hltwo(kr)\,,
\end{equation}
with $\jl(x)$, $\nl(x)$, $\hlone(x)$, $\hltwo(x)$ being the spherical Bessel, Neumann and Hankel functions of first and second kind as defined in \cref{app:special_functions}. Since $\hlone(kr)\to(-i)^{l+1}e^{ikr}/(kr)$ and $\hltwo(kr)\to i^{l+1}e^{-ikr}/(kr)$ for large $r$, we deduce $\tilde{B}=0$ such that $\ukl(r)/r\propto \hlone(kr)$. Inserting this proportionality requirement into \cref{eq:partialwaveexpansionphi}, the expression for the wave function is then given by
\begin{align}
    \label{eq:scatteringwavefunctionsecenariotwo}
    \phik(\vec{r})&= A\left(e^{ikz}+\frac{1}{4\pi}\sum_{l=0}^{\infty}\sqrt{2l+1}C_l \hlone(kr)P_l(\cos\theta_r)\right)\nonumber\\
    &= A\left(e^{ikz}+k\sum_{l=0}^{\infty}i^{l+1}(2l+1)a_l \hlone(kr)P_l(\cos\theta_r)\right)\,,
\end{align}
where we substituted $C_l\equiv i^{l+1}k\sqrt{4\pi(2l+1)}a_l$ in the second line. Expanding also the scattering amplitude in partial waves
\begin{equation}
    f(\Ek,\theta_r)=\sum_{l=0}^{\infty} (2l+1) a_lP_l(\cos\theta_r)\quad\Rightarrow\quad \sigma=4\pi\sum_{l=0}^{\infty}(2l+1)\abs{a_l}^2\,,
\end{equation}
we can easily match the $a_l$ in \cref{eq:scatteringwavefunctionsecenariotwo} to the expansion coefficients of $f(\Ek,\theta_r)$ if we compare the expression with \cref{eq:largerwavefunctionscattering} in the large $r$ limit. Using Rayleigh's formula (see \eg\eref\cite{arfken:2013}) for the incident plane wave (\cf\cref{eq:planewave})
\begin{equation}
    e^{ikz}=\sum_{l=0}^\infty \frac{i^l(2l+1)}{2} \left(\hlone(kr)+\hltwo(kr)\right)P_l(\cos\theta_r)\,,
\end{equation}
with $\cos\theta_r\equiv\uvec{k}\cdot\uvec{r}$ and where we have split $\jl(x)$ into an incoming and outgoing spherical wave part (\cf\cref{eq:sphHankelfunctions}), the wave function yields
\begin{equation}
    \label{eq:wavefunctionscatteringintermediate}
    \phik(\vec{r})=A\sum_{l=0}\frac{i^l(2l+1)}{2}\left[\left(1+2ika_l\right)\hlone(kr)+\hltwo(kr)\right]P_l(\cos\theta_r)\,.
\end{equation}
Thus, we have reduced the problem to the task of finding the expansion coefficients of the scattering amplitude $a_l$ for each partial wave.\newp
If we take into account the potential as in scenario 3), outgoing spherical wave functions receive corrections if they are sufficiently close to the interaction point (since they cannot be considered to be free waves anymore). Since the amplitude of the incoming and outgoing spherical waves has to be the same in order to conserve probability, the change can only be a phase, defined as $e^{2i\delta}$, which depends on the potential under consideration. Due to conservation of angular momentum in a spherically symmetric potential, all partial waves scatter independently from each other, which means that if we perform a partial wave expansion on the whole wave function, each partial wave $l$ will have an independent phase $\delta_l$. Therefore, we can write \cref{eq:largerwavefunctionscattering} as 
\begin{equation}
    \phik(\vec{r})=\sum_{l=0}^{\infty}\phik^{(l)}(\vec{r})\xrightarrow{r\to\infty} A\sum_{l=0}\frac{i^l(2l+1)}{2ikr}\left[(-i)^le^{2i\delta_l}e^{ikr}-i^le^{-ikr}\right]P_l(\cos\theta_r)\,,
\end{equation}
where we started with the expansion of the solution $\phik^0(\vec{r})=Ae^{ikz}$ for no potential present (\ie no scattering), took the asymptotic limit for $r\to\infty$ and added a phase $\delta_l$ to each outgoing spherical partial wave. Comparing this asymptotic result to \cref{eq:wavefunctionscatteringintermediate}, we can identify $a_l=(e^{2i\delta_l}-1)/(2ik)$ reducing the problem even further from having a complex $a_l$ to a single phase for each partial wave. The wave function can then also be written as 
\begin{equation}
    \label{eq:fullscatteringwavefunction}
    \phik(\vec{r})\approx\sum_{l=0}^{\infty}i^le^{i\delta_l}(2l+1)\frac{\tilde{u}_{\vec{k},l}(r)}{r}P_l(\cos\theta_r)\,,
\end{equation}
with
\begin{equation}
    \label{eq:scatteringwavefunctionsineapprox}
    \tilde{u}_{\vec{k},l}(r)\xrightarrow{r\to\infty}\frac{A}{k}\sin\left(kr-\frac{l\pi}{2}+\delta_l\right).
\end{equation}
The phase has to be computed by solving the full Schrödinger equation for the potential under consideration after choosing appropriate initial conditions. The amplitude can then in principle be calculated from the phase shift (\cf\eref\cite{calogero:1967}). However, since we are only interested in the difference of the wave functions with and without a potential present, we can normalize the result with $\abs{\phik^0(\vec{r})}^2=\abs{A}^2\equiv 1$, without loss of generality.
\section{The non-relativistic potential for different mediator interactions}
\label{app:potential_for_different_mediators}

\begin{figure}
    \centering
	\includegraphics[width=\textwidth]{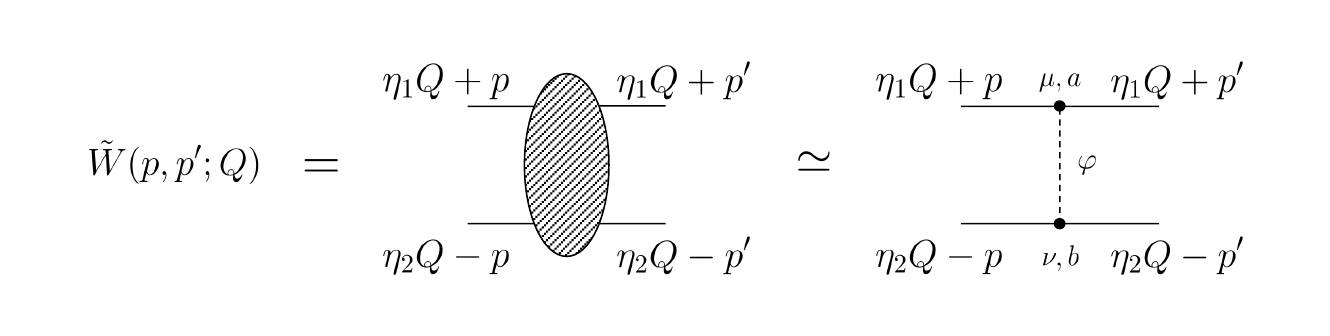}
	\caption[Interaction kernel of a $2\to 2$ scattering process exchanging an arbitrary number of mediators $\varphi$.]{Interaction kernel of a $2\to 2$ scattering process exchanging an arbitrary number of mediators $\varphi$. Its leading order approximation comprises a single boson exchange. The mediator can either be a scalar or a vector boson and the external scalar particles are allowed to carry color charge. Adapted from \eref\cite{Petraki:2015hla}.}
    \label{fig:interactionkernel}
\end{figure}

We will determine in the following the potential strength $\alpha$ alongside with the (potentially long-ranged) non-relativistic potential $V(\vec{r})$ arising from the interaction Lagrangian of two scalar particles and a mediator $\varphi$. We will do this by employing \cref{eq:potentialfrominteractionkernel}
\begin{equation*}
    V(\vec{r})\equiv-\frac{1}{i4m\mu}\int\frac{\dd[3]{p}}{(2\pi)^3}\mathcal{W}(\abs{\vec{p}})e^{i\vec{p}\cdot\vec{r}}\,,
\end{equation*}
where $m$ and $\mu$ denote the total and reduced mass of the two-particle system and $\mathcal{W}$ describes the interaction kernel $\tilde{W}$ after having taken the instantaneous approximation as described in \cref{subsubsec:BetheSalpeter}. We will consider in the following only interaction Lagrangians and particle types relevant for this thesis.\newp
As sketched in \cref{fig:interactionkernel}, where we used the same parametrization of total and relative momenta as in \cref{subsec:nonperturbativeeffectsfromQFT}, to leading order in the couplings the interaction kernel is given by a one particle exchange process of the mediator $\varphi$, which we take to be either a scalar or a vector boson. We further assume an interacting particle antiparticle system of scalars. The interaction kernel of a one boson exchange diagram in the non-relativistic regime is then given by 
\begin{equation}
	\label{eq:interactionkernel}
	\tilde{W}(p,p';Q)\simeq-\frac{i \epsilon_{\mu\nu}G^{ab}\Sigma^\mu_a(p,p';Q)\Sigma^\nu_b(p,p';Q)}{(p-p')^2-\mvphi^2}~\xrightarrow[\text{approx.}]{\text{inst.}}~\frac{i \Sigma}{(\vec{p}-\vec{p'})^2+\mvphi^2}=\mathcal{W}(\abs{\vec{p}-\vec{p'}}),
\end{equation}
where $\Sigma^{\mu}_{a}(p,p';Q)$, $\Sigma^{\nu}_{b}(p,p';Q)$ denote the vertex couplings at the corresponding vertices, $\epsilon_{\mu\nu}$ is the polarization sum of the mediator states and $G^{ab}$ describes the color matrix.\footnote{The polarization sum $\epsilon_{\mu\nu}$ only occurs if $\varphi$ is a vector boson and the color matrix enters solely for gluon exchange for which $G^{ab}\equiv\delta^{ab}$.} The product of $\epsilon_{\mu\nu}G^{ab}\Sigma^\mu_a(p,p';Q)\Sigma^\nu_b(p,p';Q)\simeq \Sigma$ can usually be expanded to leading order in the relative momenta such that $\Sigma$ is independent of $p,p'$ and $Q$. Inserting $\mathcal{W}(\abs{\vec{p}})$ into \cref{eq:potentialfrominteractionkernel} yields the Yukawa potential
\begin{align}
		V(\vec{r}) &= -\frac{\Sigma}{4m\mu}\int \frac{\dd[3]{p}}{(2\pi)^3}\frac{e^{i\vec{p}\cdot\vec{r}}}{\vec{p}^2+\mvphi^2}=-\frac{\Sigma}{4m\mu}\int_0^\infty \frac{\dd{p}}{(2\pi)^2}\frac{p^2}{p^2+\mvphi^2}\int_{-1}^{+1}\dd{\!\cos\theta}e^{ipr\cos\theta}\nonumber\\
		&= -\frac{\Sigma}{2m\mu r}\int_0^\infty \frac{\dd{p}}{(2\pi)^2}\frac{p\sin(pr)}{p^2+\mvphi^2}=-\frac{\Sigma}{16\pi m\mu r}e^{-\mvphi r}\equiv-\frac{\alpha}{r}e^{-\mvphi r},
\end{align}
where we labelled $p\equiv\abs{\vec{p}}$ within the calculation and defined the potential strength to be $\alpha\equiv \Sigma/(16\pi m\mu)$. As usual, the Yukawa potential simplifies to the Coulomb potential for massless mediators $\mvphi\to 0$. If $\Sigma>0$ the potential is attractive, in the other case it is considered repulsive. Our goal in the following is to determine $\alpha$ from the interaction kernels of the Lagrangians under consideration and single out the most important potentials for the corresponding model.


\subsection*{\protect\boldmath Non-relativistic potentials for a $t$-channel mediator model}
\label{subsec:potentialstchannel}
In the first model under consideration employed in \cref{sec:impactonnonthermalDMproduction}, we look at the three point interactions of an (un-)colored scalar mediator $\ttilde$ ($\tautilde$), neutral under $SU(2)_L$, with the \gls{sm} gauge fields after \gls{ewsb}
\begin{align}
	\mathcal{L}^{\ttilde}_{\text{int}}&\supset -ieQ^{\ttilde}_{\text{em}}\ttilde^*\partial^\mu \ttilde A_\mu+ ie\tan\theta_WQ^{\ttilde}_{\text{em}}\ttilde^*\partial^\mu \ttilde Z_\mu-ig_sT_a^{ij}\ttilde_i^*\partial^\mu \ttilde_j G^a_\mu+\lambda_H v \,\ttilde\,\ttilde^*h+h.c.\,,\\
    \mathcal{L}^{\tautilde}_{\text{int}}&\supset -ieQ^{\tautilde}_{\text{em}}\tautilde^*\partial^\mu \tautilde A_\mu+ ie\tan\theta_WQ^{\tautilde}_{\text{em}}\tautilde^*\partial^\mu \tautilde Z_\mu+\lambda_H v \,\tautilde\,\tautilde^*h+h.c.\,,
\end{align}
where $A_\mu, Z_\mu,G_\mu^a, h$ denote the photon, $Z$ boson, gluon and Higgs fields, respectively. The electromagnetic and strong coupling constants are given by $e$ and $g_s$, with $Q^{\ttilde}_{\text{em}}=2/3$ ($Q^{\tautilde}_{\text{em}}=-1$) the electromagnetic charge of the (un-)colored mediator and $T^{ij}_a$ the $SU(3)$ generators of the strong interaction in the fundamental representation. The variable $\lambda_H$ parameterizes the scalar Higgs coupling and $v$, $\theta_W$ denote the usual \gls{sm} Higgs \gls{vev} and Weinberg angle. 

\paragraph{Higgs boson exchange}$~$\newpp
The coupling strength for the Higgs boson exchange can be determined straight forwardly. In this case the interaction kernel in the instantaneous approximation is simply given by
\begin{equation}
	\mathcal{W}_h(p,p';Q)\simeq-\frac{i\lambda_H^2v^2}{(p-p')^2 -m_H^2} \quad\xrightarrow[\text{approx.}]{\text{inst.}} \quad \frac{i\lambda_H^2v^2}{(\vec{p}-\vec{p'})^2 +m_H^2}\,,
\end{equation} 
such that we identify $\mvphi=m_H$ and $\Sigma=\lambda_H^2 v^2$, leading to $\alpha_H=\lambda_H^2v^2/(16\pi M^2)$ with $M\equiv\mttilde$ ($M\equiv\mtautilde$) the mass of the (un-)colored mediator.

\paragraph{Photon exchange}$~$\newpp
The interaction kernel for the one-photon exchange is given by (see \eg\eref\cite{Bohm:2001yx})
\begin{equation}
	\mathcal{W}_\gamma(p,p';Q)\simeq\frac{ie^2Q_{\text{em},1}Q_{\text{em},2}}{(p-p')^2}g_{\mu\nu}(2\eta_a Q+p+p')^\mu(2\eta_b Q-p-p')^\nu\,.
\end{equation}
In the non-relativistic regime up to leading order in the relative momenta and setting $Q^2=m^2$, we can approximate this by
\begin{equation}
	\label{eq:photonexchangekernel}
	\mathcal{W}_\gamma(p,p';Q)\simeq\frac{4ie^2Q_{\text{em},1}Q_{\text{em},2}m_1m_2}{(p-p')^2}\quad\xrightarrow[\text{approx.}]{\text{inst.}} \quad -\frac{4ie^2Q_{\text{em},1}Q_{\text{em},2}m_1m_2}{(\vec{p}-\vec{p'})^2}\,,
\end{equation}
where we  identify $\mvphi=0$ and $\Sigma=-4e^2Q_{\text{em},1}Q_{\text{em},2}m_1m_2$. For $m_1=m_2=M$ and $Q_{\text{em},1}=-Q_{\text{em},2}\equiv Q_{\text{em}}$ we obtain $\alpha_\gamma = Q_{\text{em}}^2e^2/(4\pi)= Q_{\text{em}}^2 \alphaem$ with $\alphaem\equiv e^2/(4\pi)$ the typical electromagnetic potential strength.

\paragraph{\protect\boldmath $Z$ boson exchange}$~$\newpp
The case for the $Z$ boson can be treated very similarly to the photon exchange. We start with 
\begin{equation}
	\mathcal{W}_Z(p,p';Q)\simeq\frac{ie^2\tan^2\theta_W Q_{\text{em},1}Q_{\text{em},2}}{q^2-m_Z^2}\left(g_{\mu\nu}-\frac{q_\mu q_\nu}{m_Z^2}\right)(2\eta_a Q+p+p')^\mu(2\eta_b Q-p-p')^\nu\,,
\end{equation}
where $q=p-p'$. The $q_\mu q_\nu/m_Z^2$ term cancels completely (because we do not couple to the longitudinal Z mode), such that in the non-relativistic limit we obtain 
\begin{equation}
	\mathcal{W}_Z(p,p';Q)\simeq\frac{4ie^2\tan^2\theta_W Q_{\text{em},1}Q_{\text{em},2}m_1m_2}{(p-p'^2)-m_Z^2}~\xrightarrow[\text{approx.}]{\text{inst.}}~ -\frac{4ie^2\tan^2\theta_W Q_{\text{em},1}Q_{\text{em},2}m_1m_2}{(\vec{p}-\vec{p'})^2+m_Z^2}\,.
\end{equation}
We identify $\mvphi=m_Z$ and $\Sigma=-4e^2\tan^2\theta_WQ_{\text{em},1}Q_{\text{em},2}m_1m_2$. Using $m_1=m_2=M$ and $Q_{\text{em},1}=-Q_{\text{em},2}\equiv Q_{\text{em}}$ the potential strength for the $Z$ is given by $\alpha_Z = Q_{\text{em}}^2\tan^2\theta_We^2/(4\pi)= Q_{\text{em}}^2 \tan^2\theta_W\alphaem$. Due to the exponential suppression of the Yukawa interaction and the $\tan^2\theta_W$ term, the $Z$-potential is typically much weaker than the Coulomb potential for the photon exchange. 
\begin{table}
	\centering
	\caption[Potentials and fine structure constants from top-philic and lepto-philic mediator interactions with SM particles.]{Potentials and fine structure constants from top-philic (lepto-philic) mediator interactions $\ttilde~\ttilde^*$ ($\tautilde~\tautilde^*$) with \gls{sm} particles. The electric charge of the top-philic (lepto-philic) mediator is $Q^{\ttilde}_{\text{em}}=2/3$ ($Q^{\tautilde}_{\text{em}}=-1$). The gluon potential is only generated by the top-philic mediator, for which we consider $\mathbf{R_1}=\mathbf{3}$, $\mathbf{R_2}=\mathbf{\bar{3}}$ and $\mathbf{\hat{R}}\in\{\mathbf{1},\mathbf{8}$\}.}
	\begin{tabular}{c@{\hskip 0.8cm}c@{\hskip 0.8cm}c}
		\toprule
		gauge boson &$V(r)$ &$\alpha$ \\
		\midrule \\[-0.2cm]
		gluon & $V_g(r)=-\frac{\alpha_g}{r}$ & $\alpha_g=\frac{1}{2}\left[C_2(\mathbf{R_1})+C_2(\mathbf{R_2})-C_2(\mathbf{\hat{R}})\right]\alpha_s$\\[0.5cm]
		photon & $V_\gamma(r)=-\frac{\alpha_\gamma}{r}$ & $\alpha_\gamma=Q_{\text{em}}^2\alphaem$\\[0.5cm]
		$Z$ boson & $V_Z(r)=-\frac{\alpha_Z}{r} e^{-m_Zr}$ & $\alpha_Z=Q_{\text{em}}^2\tan^2\theta_W\alphaem$\\[0.5cm]
		Higgs boson& $V_H(r)=-\frac{\alpha_H}{r} e^{-m_Hr}$& $\alpha_H=\frac{\lambda_H^2v^2}{16\pi \mttilde^2}$ \\[0.3cm]
		\bottomrule
	\end{tabular}
	\label{tab:potentials}
\end{table}

\paragraph{Gluon exchange}$~$\newpp
The potential due to gluon exchange can only be generated for a $\ttilde~\ttilde^*$ pair. The kernel (in Feynman gauge) reads
\begin{equation}
	\mathcal{W}_g(p,p';Q)=\frac{ig_s^2T^a_{ii'}T^b_{jj'}\delta_{ab}}{(p-p')^2}g_{\mu\nu}(2\eta_a Q+p+p')^\mu(2\eta_b Q-p-p')^\nu\,.
\end{equation}
In the non-relativistic limit using the instantaneous approximation we arrive at
\begin{equation}
	\mathcal{W}_g(p,p';Q)\simeq\frac{4ig_s^2T^a_{ii'}T^a_{jj'}m_1m_2}{(p-p')^2}\quad\xrightarrow[\text{approx.}]{\text{inst.}} \quad -\frac{4ig_s^2T^a_{ii'}T^a_{jj'}m_1m_2}{(\vec{p}-\vec{p'})^2}\,,
\end{equation}
where we identify $\mvphi=0$ and $\Sigma=-4g_s^2T^a_{ii'}T^a_{jj'}m_1m_2$ (summing over $a$) with $i,j$ ($i'j'$) the initial (final) state color charges of the particles. We again take $m_1=m_2=\mttilde$ and set $\alpha_g=-T^{a}_{ii'}T^{a}_{jj'}g_s^2/(4\pi)= -T^{a}_{ii'}T^{a}_{jj'}\alpha_s$ with $\alpha_s= g_s^2/(4\pi)$ the fine structure constant of the strong interaction. It is also advantageous to rewrite $-T^{a}_{ii'}T^{a}_{jj'}=[C_2(\mathbf{R_1})+C_2(\mathbf{R_2})-C_2(\mathbf{\hat{R}})]/2$, with $C_2$ denoting the quadratic Casimir operator and $\mathbf{R_1}$, $\mathbf{R_2}$ ($\mathbf{\hat{R}}$) the initial (final) state representation(s).

\paragraph{Summary}$~$\newpp
A summary of the potentials and interaction strengths can be found in \cref{tab:potentials}. We can see that for a particle antiparticle pair, the potentials of the $\gamma$ and $Z$ are attractive. The Higgs potential is always attractive and the gluon potential can be either attractive or repulsive depending on the representation $\mathbf{\hat{R}}$ of the final state. Free colored particles are in the (anti-)fundamental representation $\mathbf{R_1}=\mathbf{3}$, $\mathbf{R_2}=\mathbf{\bar{3}}$, such that $\mathbf{\hat{R}}$ can be either the octet or singlet representation, since $\mathbf{3}\otimes\mathbf{\bar{3}}=\mathbf{8}\oplus\mathbf{1}$. With the quadratic Casimir operators $C_2(\mathbf{3})=C_2(\mathbf{\bar{3}})=4/3$, $C_2(\mathbf{8})=3$, and $C_2(\mathbf{1})=0$, we obtain two possible structure constants $\alpha_{g,[\mathbf{1}]}=4/3\,\alpha_s$ and $\alpha_{g,[\mathbf{8}]}=-1/6\,\alpha_s$, depending on the final state representation of the $\ttilde~\ttilde^*$ pair. We can immediately see, that the singlet final state potential is attractive and the octet one repulsive. \newp
Considering a colored mediator, $Q^{\ttilde}_{\text{em}}=2/3$, $\tan^2\theta_W=0.287$, $v=\SI{246}{\GeV}$, $\alphaem\sim 0.0078$, $\alpha_s\sim 0.15$, mediator masses $\mttilde\gtrsim \SI{1}{\TeV}$, and natural choices for $\lambda_H\sim \order{1}$, we can immediately see that the gluon potential always dominates, given that the temperature $T\gg\Lambda_{\text{QCD}}\sim\SI{400}{\MeV}$ remains well above the confinement scale of \gls{qcd} in the regime of interest, which we will assume throughout this work. Thus, we will assume a Coulomb potential for the colored mediator $\ttilde$ generated by gluon interactions and neglect all other contributions in the following.\newp
For an uncolored mediator $\tautilde$ with $Q^{\tautilde}_{\text{em}}=-1$, the $\gamma$, $Z$ and $h$ contributions are sizeable due to the absence of the gluon potential and have to be taken into account for the calculation of the \gls{se} factor in annihilations. A remaining open question is, if the potentials created by Higgs and $Z$ boson exchange are strong enough to support \gls{bs}. The answer follows from the discussion in \cref{subsubsec:SEboundstatedecays}, where we formulated the condition $\xi>\xi_c$ for \gls{bs} to exist. Only considering the ground state, for the $Z$-potential this bound can be translated to $\mtautilde\gtrsim \SI{68}{\TeV}$, which is well above our regime of interest. Employing a Higgs potential, even for sizable $\lH\sim\order{1}$ the bound leads to $\mtautilde\lesssim \SI{6}{\GeV}$, which has been ruled out by collider experiments (\cf\cref{subsec:constraintsnth}). Therefore, also in case of an uncolored mediator we can restrict ourselves to a Coulomb ($\gamma$-)potential when calculating the \gls{bsf} cross section and decay rate.


\subsection*{Non-relativistic potentials for a (pseudo-)scalar mediator model}
\label{subsec:potentialssecondmodel}

The model considered in \cref{sec:indirectdetection} involves the interaction Lagrangian 
\begin{equation}
    \mathcal{L}^\phi_{\text{int}}\supset -\bar{\chi}\left(g+ig_5\gamma^5\right)\chi\phi\,,
\end{equation}
where $\bar{\chi}\chi$ is a fermion antifermion pair and $\phi$ denotes a scalar mediator with $g$ ($g_5$) the interaction strength of the (pseudo-)scalar coupling. For a purely scalar interaction, the interaction kernel is given by 
\begin{equation}
	\mathcal{W}^{g^2}_\phi(p,p';Q)\simeq-\frac{ig^2\delta_{rr'}\delta_{ss'}}{(p-p')^2 -m_\phi^2} \quad\xrightarrow[\text{approx.}]{\text{inst.}} \quad \frac{ig^2}{(\vec{p}-\vec{p'})^2 +m_\phi^2}\,,
\end{equation} 
where $r^{(\prime)}$, $s^{(\prime)}$ denote the spin indices of the incoming (outgoing) particles, which are dropped in the non-relativistic limit. We identify $\mvphi=m_\phi$, $\Sigma=g^2$, and thus $\alpha\equiv\alpha_{g^2}=g^2/(4\pi)$ due to the extra factor of $4 m \mu$ in the definition of the non-relativistic potential for a fermionic system (\cf end of \cref{subsubsec:BetheSalpeter}).\newp
For interactions involving pseudo-scalar couplings, the simple modifications performed in \cref{subsubsec:BetheSalpeter} to obtain a non-relativistic potential for a fermionic system starting from a scalar one cannot be applied any longer. However, we argue that the strength of such potentials (apart from assuming $g_5<g$) will be suppressed by factors of $\vec{p}^2/(2\mu)$ (\cf\cref{subsubsec:pNREFT}). Therefore, the scalar potential will always dominate over the pseudo-scalar contributions, such that we will neglect the latter in our derivation of the Schrödinger wave functions.
\section{Sommerfeld enhancement from the Bethe-Salpeter approach}
\label{app:SEderivation}

\begin{figure}
	\centering
	\includegraphics[width=\textwidth]{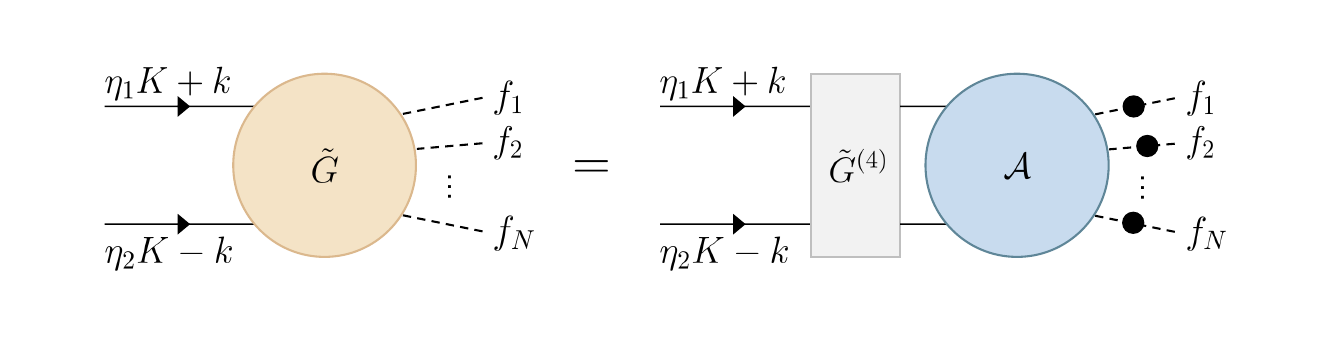}
	\caption[Decomposition of the two-particle annihilation / BS decay Green's function into N final states.]{Decomposition of the two-particle annihilation / \gls{bs} decay Green's function $\tilde{G}$ of a $\bar{\chi}\chi$ pair into N final states according to \cref{eq:5GreensdecompositionAnn}. The black filled dots refer to the full propagators of the final state particles, whereas the gray box on the right denotes the 4-point Green's function $\Gfourt$. The momenta of the incoming particles are also displayed in terms of total and relative momenta of the system. A definition of $\mathcal{A}$ is given in the text. Adapted from \eref\cite{Petraki:2015hla}.}
	\label{fig:GreensfunctiondecompositionAnn}
\end{figure}

Here, we want to derive the \gls{se} factors for annihilation and \gls{bs} decay that we used in \cref{subsubsec:SEannihilations,subsubsec:SEboundstatedecays} employing the tools introduced in \cref{subsec:nonperturbativeeffectsfromQFT}. The steps taken will be similar to the ones in \cref{subsubsec:BSFfromBetheSalpeter}. We start with the  corresponding S-matrix elements of a two-particle (bound) state annihilating (decaying) into $N$ final states, which are given by
\begin{align}
	\label{eq:SmatrixElementAnn}
	\prescript{}{\text{out}}{\left\langle f_1f_2\ldots f_N\middle|\mathcal{U}_{\vec{K},{\vec{k}}}\right\rangle_{\text{in}}}&=\left\langle f_1f_2\ldots f_N\middle|\mathsf{S}\middle|\mathcal{U}_{\vec{K},{\vec{k}}}\right\rangle\,,\\
	\prescript{}{\text{out}}{\left\langle f_1f_2\ldots f_N\middle|\mathcal{B}_{\vec{K},n}\right\rangle_{\text{in}}}&=\left\langle f_1f_2\ldots f_N\middle|\mathsf{S}\middle|\mathcal{B}_{\vec{K},n}\right\rangle\,,
\end{align}
and define the Green's function (which is structurally the same for both processes) as well as its Fourier transform
\begin{align}
	G(x_1,x_2;y_1,\ldots y_n)&\equiv\Vaccc T\varphi(X_\varphi)\chi_1(y_1)\ldots \chi_N(y_N)\chi_1^\dagger(x_1)\chi_2^\dagger(x_2)\Vac\,,\\
	\tilde{G}(k_1,k_2;p_1,\ldots p_N)&\equiv\prod_{j=1}^{N}\int \dd[4]{y_j} \dd[4]{x_1} \dd[4]{x_2} e^{-i(k_1x_1+k_2x_2-p_jx_j)}G(x_1,x_2;y_1,\ldots y_n).
\end{align}
With $\mathcal{A}$ denoting the sum of all connected and amputated diagrams with the respective in- and outgoing momenta $k_1,k_2$ and $p_j$, $j=1,\ldots,N$, we can decompose the Green's function into
\begin{align}
	\label{eq:5GreensdecompositionAnn}
	\tilde{G}(\eta_1K&+k,\eta_2K-k;p_1,\ldots p_N)\equiv\nonumber\\
	&\prod_{j=1}^{N} \tilde{S}_{f_j}(p_j)\int\frac{\dd[4]{k'}}{(2\pi)^4} \Gfourt(k,k';K)\mathcal{A}(\eta_1K+k',\eta_2K-k';p_1,\ldots p_N)\,,
\end{align} 
as sketched in \cref{fig:GreensfunctiondecompositionAnn} with $\tilde{S}_{f_j}(p_j)$ the propagators of the final state particles. Defining further the perturbative matrix element via
\begin{align}
	\mathcal{A}(\eta_1K+k,\eta_2K&-k;p_1,\ldots p_N)=\nonumber\\
	&i(2\pi)^4\delta^{(4)}\Bigg(K-\sum_{j=1}^{N} p_j\Bigg)\mathcal{M}_{\text{ann}}^{\text{pert}}(\eta_1K+k,\eta_2K-k;p_1,\ldots p_N)\,,
\end{align}
we can again use the \gls{lsz} reduction formula in a similar fashion as in \cref{subsubsec:BSFfromBetheSalpeter} to derive the two-particle annihilation and bound state decay matrix elements
\begin{align}
	\mathcal{M}_{\text{ann}}&=\prod_{j=1}^{N}\sqrt{Z_{f_j}(\vec{p}_j)}\int\frac{\dd[4]{q}}{(2\pi)^4}\PhiQBSt{K}{k}(q) \mathcal{M}_{\text{ann}}^{\text{pert}}(\eta_1K+q,\eta_2K-q;p_1,\ldots p_N)\\
	&\simeq\!\sqrt{2\ESSdiffq{K}{k}}\!\int\!\frac{\dd[3]{q}}{(2\pi)^3}\frac{\SEwavefSSt{k}(\vec{q})}{\sqrt{2\mathcal{N}_{\vec{K}}(\vec{q})}}\!\int\!\frac{\dd{q^0}}{2\pi}\frac{S(q;K)}{\mathcal{S}_0(\vec{q};K)}\mathcal{M}_{\text{ann}}^{\text{pert}}(\eta_1K+q,\eta_2K-q;p_1,\ldots p_N),
\end{align}
\begin{align}
	\mathcal{M}^{(n)}_{\text{dec}}&=\prod_{j=1}^{N}\sqrt{Z_{f_j}(\vec{p}_j)} \int\frac{\dd[4]{q}}{(2\pi)^4}\PsiQBSt{K}(q)\mathcal{M}_{\text{ann}}^{\text{pert}}(\eta_1K+q,\eta_2K-q;p_1,\ldots p_N)\\
	&\simeq\int\frac{\dd[3]{q}}{(2\pi)^3}\frac{\SEwavefBSt(\vec{q})}{\sqrt{2\mathcal{N}_{\vec{K}}(\vec{q})}}\int\frac{\dd{q^0}}{2\pi}\frac{S(q;K)}{\mathcal{S}_0(\vec{q};K)}\mathcal{M}_{\text{ann}}^{\text{pert}}(\eta_1K+q,\eta_2K-q;p_1,\ldots p_N),
\end{align}
where we have taken the instantaneous approximation (\cf\cref{eq:BSequationInstAppBS,eq:BSequationInstAppSS}) in the last step and set all renormalization factors $Z_{f_j}(\vec{p}_j)\simeq 1$ to leading order in $\alpha$. 
For all practical purposes, the diagrams contributing to the perturbative matrix element will be fully connected such that we can replace $\mathcal{M}_{\text{ann}}^{\text{pert}}(\eta_1K+q,\eta_2K-q;p_1,\ldots p_N) \simeq\hat{\mathcal{M}}_{\text{ann}}^{\text{pert}}(\eta_1\vec{K}+\vec{q},\eta_2\vec{K}-\vec{q};\vec{p}_1,\ldots \vec{p}_N)$ with the on-shell annihilation amplitude which is of order $\vec{q}^2$  (\cf \textit{on-shell approximation} in section 3.4 of \eref\cite{Petraki:2015hla}). Also approximating $\sqrt{2\mathcal{N}_{\vec{K}}(\vec{q})}\simeq \sqrt{2\mu}$ and $\ESSdiffq{K}{k}\simeq \mu$ to leading order in $\vec{q}^2$ and $\vec{k}^2$, we arrive at 
\begin{align}
        \label{eq:annMatrixelementonshellapprox}
		\mathcal{M}_{\text{ann}}&\simeq\int\frac{\dd[3]{q}}{(2\pi)^3}\SEwavefSSt{k}(\vec{q})\hat{\mathcal{M}}_{\text{ann}}^{\text{pert}}(\eta_1\vec{K}+\vec{q},\eta_2\vec{K}-\vec{q};\vec{p}_1,\ldots \vec{p}_N)\,,\\
        \label{eq:decMatrixelementonshellapprox}
		\mathcal{M}^{(n)}_{\text{dec}}&\simeq\frac{1}{\sqrt{2\mu}}\int\frac{\dd[3]{q}}{(2\pi)^3}\SEwavefBSt(\vec{q})\hat{\mathcal{M}}_{\text{ann}}^{\text{pert}}(\eta_1\vec{K}+\vec{q},\eta_2\vec{K}-\vec{q};\vec{p}_1,\ldots \vec{p}_N)\,,
\end{align}
for annihilation and decay of a $\bar{\chi}\chi$ pair into N arbitrary final states.\newp
Conforming to \cref{subsec:SEannihilationBSdecay}, we will now focus on particle antiparticle annihilation and bound state decay into two final state particles. Within the \gls{com} frame, the 3-momenta yield $\vec{K}=0$ and $\abs{\vec{p}_1}=\abs{\vec{p}_2}\equiv \abs{\vec{p}}$. The perturbative on-shell matrix element can then be expanded into partial waves like 
\begin{equation}
    \hat{\mathcal{M}}_{\text{ann}}^{\text{pert}}(\vec{q},-\vec{q};\vec{p},-\vec{p})=\sum_{l=0}^\infty \frac{\tilde{a}_l}{(m\mu)^l}\abs{\vec{p}}^l\abs{\vec{q}}^lP_l(\cos\theta_{\vec{q},\vec{p}})\,,
\end{equation}
where we can express the $\vec{q}$-dependence of $\tilde{a}_l(\vec{q})\simeq a_l+\mathcal{F}_l(\vec{q}^2,\vec{\epsilon}_A\cdot \vec{q})$ with $\vec{\epsilon}_A$ denoting the polarization vectors of possible vector boson final states and $\mathcal{F}_l\to 0$ for $\vec{q}\to 0$. In accordance with the approximations done so far, it will suffice to only consider $a_l$ in the following.\footnote{Note that $a_l$ and $\mathcal{F}_l$ can still depend on $\vec{p}$ through $\vec{p}^2$ and $\vec{p}\cdot \vec{\epsilon}_A$, as well as on $\vec{\epsilon}_A\cdot \vec{\epsilon}_B$ contributions.} We can also expand the full matrix elements $\mathcal{M}_{\text{ann}}$ in terms of partial waves
\begin{align}
    \mathcal{M}_{\text{ann}}(\Omega_{\vec{p}})&=\sum_{l=0}^{\infty}\frac{(2l+1)}{4\pi}P_l(\cos\theta_{\vec{p}})\mathcal{M}_{\text{ann},l}\,, \quad\text{where} \\
    \label{eq:lwavematrixelement}
    \mathcal{M}_{\text{ann},l}&\equiv \int \dd{\Omega_{\vec{p}}}P_l(\cos\theta_{\vec{p}})\mathcal{M}_{\text{ann}}(\Omega_{\vec{p}})
\end{align}
are its projected $l$-wave contributions, and equivalently for $\mathcal{M}^{(n)}_{\text{dec}}$. The partial wave expanded annihilation cross section and decay width are then given by (\cf\cref{eq:generalcrosssection,eq:1to2particledecayrate})
\begin{align}
    \label{eq:sigmaannpartialwaveexpanded}
    \sigmaann &= \frac{f_s}{64\pi^2 s}\frac{p_c^*}{p_a^*}\int\dd{\Omega_{\vec{p}}} \overline{\abs{\mathcal{M}_{\text{ann}}(\Omega_{\vec{p}})}^2}= \frac{f_s}{128\pi^2 m\mu \vrel}\sum_{l=0}^\infty \frac{2l+1}{4\pi}\overline{\abs{\mathcal{M}_{\text{ann},l}}^2}\,,\\
    \label{eq:Gammadecpartialwaveexpanded}
    \Gammadecn &=\frac{f_s}{32\pi^2}\frac{p^*}{M_n^2}\int\dd{\Omega_{\vec{p}}} \overline{\abs{\mathcal{M}_{\text{dec}}(\Omega_{\vec{p}})}^2}=\frac{f_s}{64\pi^2m}\sum_{l=0}^\infty \frac{2l+1}{4\pi}\overline{\abs{\mathcal{M}_{\text{dec},l}^{(n)}}^2}.
\end{align}
Note that we have in the second step also expanded $s$ and $p^*_a$ to leading order in $\vec{k}$, neglected the final state particle masses for simplicity and set $\abs{\vec{p}}\simeq(m+\Ek)/2\simeq m/2$ or $\abs{\vec{p}}\simeq M_n/2=(m+\En)/2\simeq m/2$ to their leading order values in $\vrel$ or $\alpha$, respectively.\newp
Inserting \cref{eq:annMatrixelementonshellapprox} (or \cref{eq:decMatrixelementonshellapprox}, respectively) into \cref{eq:lwavematrixelement} and using the identity \cref{eq:SommerfeldPlidentity} we arrive at 
\begin{align}
    \label{eq:Mannlal}
    \mathcal{M}_{\text{ann},l}&\simeq \frac{a_l\abs{\vec{p}}^l}{(m\mu)^l}\frac{(2l+1)!!}{i^l(2l+1)l!}\left[\dv[l]{}{r}\int\dd{\Omega_{\vec{r}}}P_l(\cos\theta_{\vec{r}})\phik(\vec{r})\right]_{\vec{r}=0}\,,\\
    \label{eq:Mdeclal}
    \mathcal{M}_{\text{dec},l}^{(n)}&\simeq\frac{a_l\abs{\vec{p}}^l}{\sqrt{2\mu}(m\mu)^l}\frac{(2l+1)!!}{i^l(2l+1)l!}\left[\dv[l]{}{r}\int\dd{\Omega_{\vec{r}}}P_l(\cos\theta_{\vec{r}})\psi_n(\vec{r})\right]_{\vec{r}=0}\,,
\end{align}
which can in turn be incorporated in \cref{eq:sigmaannpartialwaveexpanded,eq:Gammadecpartialwaveexpanded}. Defining further
\begin{align}
    \sigma_l &\equiv \frac{f_s}{32\pi m \mu}\frac{\abs{a_l}^2}{4^l(2l+1)}\,,\\
    \vrel^{2l}\Sannl{l}&\equiv\frac{\left[(2l+1)!/(l!)^2\right]^2}{4^{l+2}\pi^2 \mu^{2l}}\left|\dv[l]{}{r}\int\dd{\Omega_{\vec{r}}}P_l(\cos\theta_{\vec{r}})\phik(\vec{r})\right|^2_{\vec{r}=0}\,,\\
    \Sdec&\equiv\frac{\left[(2l+1)!/(l!)^2\right]^2}{4^{l+2}\pi^2 \mu^{2l}}\left|\dv[l]{}{r}\int\dd{\Omega_{\vec{r}}}P_l(\cos\theta_{\vec{r}})\psi_n(\vec{r})\right|^2_{\vec{r}=0}\,,
\end{align}
we can cast $\sigmaann\vrel$ and $\Gammadecn$ into the form given in \cref{eq:SEsplittingann,eq:SEsplittingdec}. Using the partial wave expansion for the Schrödinger wave functions as given in \cref{eq:scatteringwavefunctiondec,eq:boundstatewavefunctiondec}, we will recover the \gls{se} factors for two-particle annihilation and bound state decay as stated in \cref{eq:SEannihilation,eq:SEdecay}. To leading order in the momenta, the derivation of the \gls{se} factors is not affected by the spin of the interacting particles. Any differences between bosonic and fermionic particles arising in the non-relativistic limit can be absorbed in a redefinition of the $a_l$. 
\section{Overlap integrals}
\label{app:overlapintegrals}
We want to illustrate the computation of the following overlap integrals 
\begin{align}
    \label{eq:Ikndef}
	\Ikn{n}(\vec{b})&\equiv\int\frac{\dd[3]{p}}{(2\pi)^3}\SEwavefBSt^*(\vec{p})\SEwavefSSt{k}(\vec{p}+\vec{b})=\int\dd[3]{r}\psin^*(\vec{r})\phik(\vec{r})e^{-i\vec{b}\cdot\vec{r}}\,,\\
    \label{eq:Jkndef}
	\Jkn{n}(\vec{b})&\equiv\int\frac{\dd[3]{p}}{(2\pi)^3}\,\vec{p}\,\SEwavefBSt^*(\vec{p})\SEwavefSSt{k}(\vec{p}+\vec{b})=i\int\dd[3]{r}[\nabla\psin^*(\vec{r})]\phik(\vec{r})e^{-i\vec{b}\cdot\vec{r}}\,,\\
    \label{eq:Kkndef}
	\Kkn{n}(\vec{b})&\equiv\int\frac{\dd[3]{p}}{(2\pi)^3}\vec{p}^2\SEwavefBSt^*(\vec{p})\SEwavefSSt{k}(\vec{p}+\vec{b})=-\int\dd[3]{r}[\nabla^2\psin^*(\vec{r})]\phik(\vec{r})e^{-i\vec{b}\cdot\vec{r}}\,,\\
    \label{eq:Ykndef}
    \Ykn{n}&\equiv 8\pi\mu\alphasNA\int\frac{\dd[3]{p}}{(2\pi)^3}\frac{\dd[3]{q}}{(2\pi)^3}\frac{\vec{q}-\vec{p}}{(\vec{q}-\vec{p})^4}\SEwavefBSt^*(\vec{p})\SEwavefSSt{k}(\vec{q})=-i\mu\alphasNA\int\dd[3]{r}\psin^*(\vec{r})\phik(\vec{r})\uvec{r}\,,
\end{align}
for various purposes throughout this work. We note that the squared brackets within the definitions indicate that the derivatives are only taken with respect to $\psin^*(\vec{r})$. Their derivation has been adapted from the appendices of \erefs\cite{Petraki:2016cnz,Harz:2018csl}.

\paragraph{\protect\boldmath $\Jkn{\{100\}}$ and $ \Ykn{\{100\}}$ in the top-philic mediator model}$~$\newpp
The following derivation can also be found in \app\,B of \eref\cite{Harz:2018csl}. We will start here with a more convenient (but equivalent) definition of the Coulomb Schrödinger wave functions (\cf\cref{eq:scatteringwavefunctiondec,eq:boundstatewavefunctiondec,eq:ColumbscatteringwavefunctionRadial,eq:ColumbboundstatewavefunctionRadial}) 
\begin{align}
    \phik(\vec{r})~\equiv ~&\sqrt{S_0(\zetas)}\HGone{i\zetas}{1}{i(kr-\vec{k}\cdot\vec{r})}e^{i\vec{k}\cdot\vec{r}}\,,\\
    \psin(\vec{r})~\equiv ~&\kappa^{3/2}\sqrt{\frac{4(n-l-1)!}{n^4(n+l)!}}\left(\frac{2\kappa r}{n}\right)^{l} \Laguerre{n-l-1}{2l+1}{\frac{2\kappa r}{n}} e^{-\kappa r/n}Y_{lm}(\Omega_{\vec{r}})\,,
\end{align}
where $\HGone{a}{c}{x}$, $L_n^{(\alpha)}(x)$, and $Y_{lm}(\Omega_{\vec{r}})$ are defined in \cref{eq:HGonesumdef,eq:genLaguerrepolsumdef,eq:sphericalharmonicsdef}, respectively. Using $\uvec{r}e^{-i\vec{b}\cdot\vec{r}}=i \nabla_{\vec{b}}e^{-i\vec{b}\cdot\vec{r}}/r$ together with the identity \cite{Akhiezer:1996}
\begin{equation}
    \int \dd[3]{r}\frac{e^{i(\vec{k}-\vec{b})\cdot\vec{r}-\kappa r}}{4\pi r}\HGone{i\zetas}{1}{i(kr-\vec{k}\cdot\vec{r})}=\frac{\left[\vec{b}^2+(\kappa-ik)^2\right]^{-i\zetas}}{\left[(\vec{k}-\vec{b})^2+\kappa^2\right]^{1-i\zetas}}\equiv f_{\vec{k},\vec{b}}(\kappa)\,,
\end{equation}
we can write the overlap integrals for \gls{bsf} into the ground state as
\begin{align}
    \Jkn{\{100\}}(\vec{b})&=\mu\alphagBs\sqrt{16\pi\kappa^3 S_0(\zetas)}\left[\nabla_{\vec{b}} f_{\vec{k},\vec{b}}(\kappa)\right]\,, \\
    \Ykn{\{100\}}&=\mu\alphasNA\sqrt{16\pi\kappa^3 S_0(\zetas)}\left[\nabla_{\vec{b}} f_{\vec{k},\vec{b}}(\kappa)\right]_{\vec{b}=0}\,.
\end{align}
For the \gls{bsf} cross section in the top-philic mediator model, we only need $\Jkn{\{100\}}\equiv\Jkn{\{100\}}(0)$. In order to calculate the squared matrix element in \cref{eq:spatialsquaredmatrixelement} for the ground state, it is therefore sufficient to determine
\begin{equation}
    \frac{\Ykn{\{100\}}}{\Jkn{\{100\}}} = \frac{\alphasNA}{\alphagBs}\quad\text{and}\quad \abs{\Jkn{\{100\}}}^2 =\frac{2^6\pi}{k}S_0(\zetas)(1+\zetas^2)\frac{\zetab^5e^{-4\zetas\arccot{\zetab}}}{(1+\zetab^2)^4}\,,
\end{equation}
where for the latter quantity we made use of the identity stated in \cref{eq:arccotrelation}.

\paragraph{\protect\boldmath $\Ikn{\{nlm\}}$ and $ \Kkn{\{nlm\}}$ in the (pseudo-)scalar mediator model}$~$\newpp
The following derivation can also be found in \app\,B of \eref\cite{Petraki:2016cnz}. We first split up the Yukawa-type scattering and \gls{bs} wave functions into radial and angular parts (\cf\cref{eq:scatteringwavefunctiondec,eq:boundstatewavefunctiondec}), which we insert into the definitions of $\Ikn{\{nlm\}}$ and $ \Kkn{\{nlm\}}$. Expanding $e^{-i\vec{b}\cdot\vec{r}}$ in partial waves (\cf\cref{eq:planewave}) and rewriting the Legendre polynomials in terms of spherical harmonics (\cf\cref{eq:sphericalharmonicsadditiontheorems}), we can 
apply \cref{eq:Wigner3jrelation} to cast the overlap integrals into the following form
\begin{align}
    \Ikn{\{nlm\}}(\vec{b})=&\left(\frac{4\pi}{\kappa}\right)^{3/2}\sum_{s=0}^{\infty}\sum_{l_R=0}^{\infty}\left(\frac{b}{\kappa}\right)^{l_R+2s}\frac{(-1)^{l_R+s}\,i^{l_R}}{2^s s!(2l_R+2s+1)!!}\nonumber\\
    &\sum_{l_I=0}^{\infty}\sum_{m_R=-l_R}^{l_R}\sum_{m_I=-l_I}^{l_I}(-1)^{m_I}Y^*_{l_Im_I}(\Omega_{\vec{k}})Y_{l_Rm_R}(\Omega_{\vec{b}})\nonumber\\
    &\sqrt{(2l+1)(2l_R+1)(2l_I+1)}
    \begin{pmatrix}
        l & l_R & l_I \\
        0 & 0 & 0
    \end{pmatrix}
    \begin{pmatrix}
        l & l_R & l_I \\
        -m & -m_R & m_I
    \end{pmatrix}\nonumber\\
    &\int\dd{x}x^{l_R+2s}\chinl^*(x)\chiqs{k}{l_I}(x)\,,\\
    \Kkn{\{nlm\}}(\vec{b})=&\sqrt{(4\pi)^3\kappa}\sum_{s=0}^{\infty}\sum_{l_R=0}^{\infty}\left(\frac{b}{\kappa}\right)^{l_R+2s}\frac{(-1)^{l_R+s}\,i^{l_R}}{2^s s!(2l_R+2s+1)!!}\nonumber\\
    &\sum_{l_I=0}^{\infty}\sum_{m_R=-l_R}^{l_R}\sum_{m_I=-l_I}^{l_I}(-1)^{m_I}Y^*_{l_Im_I}(\Omega_{\vec{k}})Y_{l_Rm_R}(\Omega_{\vec{b}})\nonumber\\
    &\sqrt{(2l+1)(2l_R+1)(2l_I+1)}
    \begin{pmatrix}
        l & l_R & l_I \\
        0 & 0 & 0
    \end{pmatrix}
    \begin{pmatrix}
        l & l_R & l_I \\
        -m & -m_R & m_I
    \end{pmatrix}\nonumber\\
    &\int\dd{x}\left[-\gammanl^2(\xi)+\frac{2}{x}e^{-x/\xi}\right]x^{l_R+2s}\chinl^*(x)\chiqs{k}{l_I}(x)\,,
\end{align}
where we used the Schrödinger equation for \gls{bs} (\cf\cref{eq:boundstateSchroedinger}) in the last line to replace $\nabla^2\psin^*(\vec{r})$. The quantities in the third lines correspond to the Wigner-3j symbols (see \cref{app:special_functions}). Using their properties, we can write the overlap integrals for the S states (\ie $l,m=0$) as an expansion in $b/\kappa$
\begin{align}
    \label{eq:Ikn00expansion}
	\Ikn{\{n00\}}(\vec{b})=&-\sqrt{\frac{4\pi}{\kappa^3}}\left\{\left(\frac{b}{\kappa}\right)i P_1(\uvec{k}\cdot\uvec{b})\int_0^\infty \dd{x}x\chiij{n0}^*(x)\chiqs{k}{1}(x)\right.\nonumber\\
	& \hspace{1.7cm}+\left(\frac{b}{\kappa}\right)^2\left[\frac{1}{6}P_0(\uvec{k}\cdot\uvec{b})\int_0^\infty \dd{x}x^2\chiij{n0}^*(x)\chiqs{k}{0}(x)\right.\nonumber\\
    &\hspace{3.4cm}\left.\left.+\frac{1}{3}P_2(\uvec{k}\cdot\uvec{b})\int_0^\infty \dd{x}x^2\chiij{n0}^*(x)\chiqs{k}{2}(x)\right]\right\}+\order{(b/\kappa)^3}\\
    \label{eq:Kkn00expansion}
	\Kkn{\{n00\}}(\vec{b})=& \sqrt{4\pi\kappa}~P_0(y)\int_0^\infty\dd{x}\frac{2}{x}e^{-x/\xi}\chiij{n0}^*(x)\chiqs{k}{0}(x)+\order{(b/\kappa)}\,,
\end{align}
up to the corresponding order required in our work.\footnote{This expansion is essentially equivalent to an expansion in $\alpha$ and $\vrel\lesssim \alpha$, since in our work $b\propto\abs{\vec{P_\phi}}$, which is of $\order{\alpha^2+\vrel^2}$.} As we can see, the $(b/\kappa)^0$ term in $\Ikn{\{n00\}}$ as well as the $\gammanl^2$ term in $\Kkn{\{n00\}}$ vanish, due to the orthogonality relation
\begin{equation}
    \label{eq:orthogonalityradialwavefunctions}
	\int_0^\infty \dd{x}\chinl^*(x)\chikl(x)=0.
\end{equation}
\section{BSF in thermal freeze-out from the Bethe-Salpether approach}
\label{app:BSF_thermalfo}

\begin{figure}[ht]
    \centering
    \includegraphics[width=\textwidth]{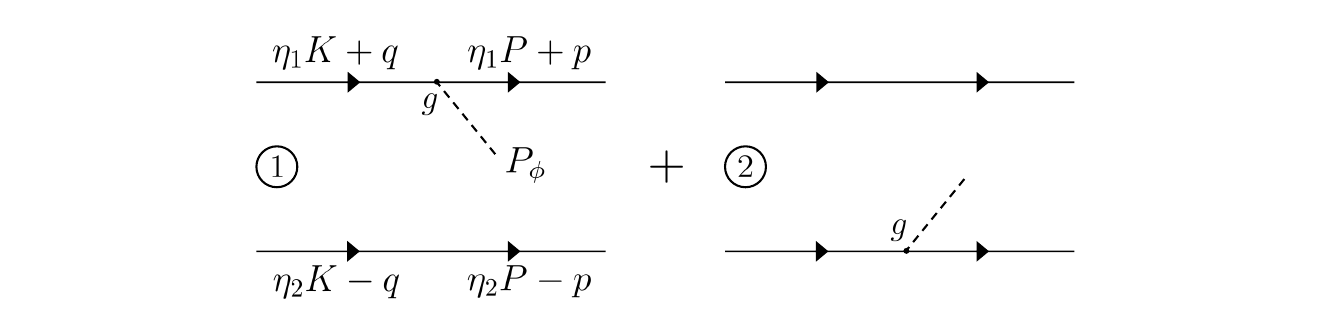}
    \caption[Leading order diagrams to $\Cfiveamp{\phi}$ in the (pseudo-)scalar mediator model.]{Leading order diagrams to $\Cfiveamp{\phi}$ in the (pseudo-)scalar mediator model. The momenta of the incoming (outgoing) particles are also displayed in terms of the total and relative momenta of the system. The vertices only feature scalar interactions and the spin indices are suppressed.}
    \label{fig:C5ampscalar}
\end{figure}

We will calculate in the following the \gls{bsf} cross section of thermal \gls{dm} within the model employed in \cref{sec:indirectdetection} using the methodology outlined in \cref{subsubsec:BSFfromBetheSalpeter}. The computation will be similar to the one performed for the non-thermal \gls{dm} model in \cref{subsubsec:sWSEBSF}. The process under consideration is given by $\bar{\chi} + \chi \to \mathscr{B}(\bar{\chi}\,\chi) + \phi$ with $\chi$ a (Dirac) fermionic \gls{dm} candidate and $\phi$ a scalar mediator. We will start again with the transition matrix element (\cf\cref{eq:Mtrans})
\begin{equation*}
	\Mtrans(\vec{p},\vec{q})\equiv (\mathcal{S}_0^f(\vec{p},P))^{-1}(\mathcal{S}_0^f(\vec{q},K))^{-1}\int\frac{\dd{p^0}}{2\pi}\frac{\dd{q^0}}{2\pi}\Cfiveamp{\phi}(q,p;K,P,P_\phi)\,,
\end{equation*}
where the superscript $f$ highlights the fermionic nature of the corresponding objects. As we will quickly change to a non-relativistic (scalar) interaction picture, following the prescription at the end of \cref{subsubsec:BetheSalpeter}, we subsequently suppress spin indices as well as spin matching conditions to enhance readability. The two processes contributing to $\mathcal{C}_{\phi-\text{amp}}=\mathcal{C}^{g}_1+\mathcal{C}^{g}_2$ at leading order in the couplings are displayed in \cref{fig:C5ampscalar}. As stated in \cref{subsec:dmmodelid}, all diagrams involving a $g_5$ vertex can be neglected at this order. The leading order contributions yield
\begin{align}
    i\mathcal{C}^g_1&=\tilde{S}^f_2(\eta_2 K -q)\tilde{S}^f_1(\eta_1K+q)(-ig)\tilde{S}^f_1(\eta_1P+p)(2\pi)^4\delta^{(4)}(\eta_1K+q-\eta_1P-p-P_\phi)\nonumber\\
    &\simeq (-ig)(2m_1)^2(2m_2)S(q;K)\tilde{S}_1(\eta_1P+p)(2\pi)^4\delta^{(4)}(q-p-\eta_2P_\phi)\,,\\
    i\mathcal{C}^g_2&=\tilde{S}^f_1(\eta_1 K +q)\tilde{S}^f_2(\eta_2K-q)(-ig)\tilde{S}^f_2(\eta_2P-p)(2\pi)^4\delta^{(4)}(\eta_2K-q-\eta_2P+p-P_\phi)\nonumber\\
    &\simeq(-ig)(2m_2)^2(2m_1)S(q;K)\tilde{S}_2(\eta_2P-p)(2\pi)^4\delta^{(4)}(q-p+\eta_1P_\phi)\,,
\end{align}
where we approximated the fermionic propagators $\tilde{S}^f_i\to 2m_i \tilde{S}_i$ with the scalar ones to leading order in the momenta and used four-momentum conservation $K=P+P_\phi$ within the $\delta$-distributions. Analogously, we transform $(\mathcal{S}_0^f)^{-1}\to (2m_1)^{-1}(2m_2)^{-1}(\mathcal{S}_0)^{-1}$ and use \cref{eq:propagatorintegralapprox1,eq:propagatorintegralapprox2} to approximate the integrals, such that the transition matrix element takes the simple form
\begin{equation}
    \Mtrans(\vec{p},\vec{q})\simeq (-ig)(2\pi)^3 \left[\delta^{(3)}(\vec{q}-\vec{p}-\eta_2\vec{P}_\phi)+\delta^{(3)}(\vec{q}-\vec{p}+\eta_1\vec{P}_\phi)\right].
\end{equation}
The fermionic wave functions can be mapped at leading order to the spin-agnostic Schrödinger wave functions via $(\SEwavefBSt^f(\vec{p}))^*\SEwavefSSt{k}^f(\vec{q})\to 4\mu m\,\SEwavefBSt^*(\vec{p})\SEwavefSSt{k}(\vec{q})$, such that we obtain the \gls{bsf} matrix element (\cf\cref{eq:MknSEapprox})
\begin{equation}
    \Mkn\simeq -4i g \mchi^{3/2}\left[\Ikn{n}(\vec{P}_\phi/2)+\Ikn{n}(-\vec{P}_\phi/2)-\frac{1}{2\mchi^2}\left(\Kkn{n}(\vec{P}_\phi/2)+\Kkn{n}(-\vec{P}_\phi/2)\right)\right]
\end{equation}
after identifying $m_1=m_2=\mchi$, $\eta_1=\eta_2=1/2$, and with the definitions of the overlap integrals $\Ikn{n}$ and $\Kkn{n}$ given in \cref{eq:Ikndef,eq:Kkndef}. 
We will only consider \gls{bsf} into the ground state in the following, \ie $n=\{100\}$. After expanding the overlap integrals in $\alpha$, one would usually expect the contributions from $\Kkn{\{100\}}$  to be of higher order, since they are naturally suppressed by a factor $\alpha^2$ (\cf\cref{eq:Ikn00expansion,eq:Kkn00expansion}). However, due to the vanishing $\alpha^0$ term and the anti-symmetry of the $\alpha^1$ term in the expansion of $\Ikn{\{100\}}$, both overlap integrals contribute at the same order in $\mathcal{M}_{\vec{k}\to \{100\}}$, which yields
\begin{equation}
    \mathcal{M}_{\vec{k}\to \{100\}}\simeq i\frac{32\sqrt{2}\pi}{\alpha}\left[\frac{\big|\vec{P}_\phi|^2}{2\kappa^2}\left(\frac{1}{6}P_0(\cos\theta)A_0+\frac{1}{3}P_2(\cos\theta)A_1\right)-\frac{\kappa^2}{m_\chi^2}P_0(\cos\theta)B(\xi)\right].
\end{equation}
This expression contains the following \textit{radial overlap integrals}
\begin{equation}
    \label{eq:radialoverlapintegrals}
    A_j\equiv\int_0^{\infty}\dd{x} x^2\chiij{10}^*(x)\chiqs{k}{2j}(x)\,,\qquad B(\xi)\equiv -\int_0^{\infty}\dd{x}\frac{2}{x}e^{-x/\xi}\chiij{10}^*(x)\chiqs{k}{0}(x)\,,
\end{equation}
which have to be determined numerically, and $\cos\theta=\uvec{P}_\phi\cdot\uvec{k}$. The cross section can then be calculated via \cref{eq:generalBSFcrosssection}, which yields for \gls{bsf} into the ground state $\sigmabsf\vrel=\pi\alpha^4/m_\chi^2\SBSF(\zeta,\xi)$ with
\begin{align}
    \label{eq:SBSFYukawa10}
	\SBSF(\zeta,\xi)\equiv\frac{\sqrt{\mathscr{P}_{10}(\zeta,\xi)}}{720}\left(\frac{1+\zeta^2\gamma_{10}^2(\xi)}{\zeta^2}\right)\left[\mathscr{P}_{10}^2(\zeta,\xi)\left(\frac{1+\zeta^2\gamma_{10}^2(\xi)}{\zeta^2}\right)^4\left(5A_0^2+4A_1^2\right)\right.\nonumber\\
	\left.-120\,\mathscr{P}_{10}(\zeta,\xi)\left(\frac{1+\zeta^2\gamma_{10}^2(\xi)}{\zeta^2}\right)^2\Re{A_0^*B(\xi)}+720 B(\xi)^2\right].
 \end{align}
Here, $\mathscr{P}_{10}(\zeta,\xi)$ denotes the phase space suppression factor defined in \cref{eq:pssfactorgroundstate}. We can see that this expression is equivalent to \cref{eq:SBSFID} when identifying
 \begin{equation}
     \langle 1S| \vec{r}^2 | \vec{k}  \rangle = \sqrt{\frac{4 \pi}{\kappa^7}} A_0,\quad\langle 1S \vert \nabla^2_{\vec{r}} \vert \vec{k} \rangle = \sqrt{ 4 \pi \kappa}B(\xi),\quad|\langle\vec{k} | r^i r^j |1S  \rangle|^2 = \frac{4 \pi}{\kappa^7} \left( \frac{1}{3} A_0^2 + \frac{2}{3} A_1^2 \right).
\end{equation}

\section{Higher bound states}
\label{app:higherBS}

We eventually want to estimate the effects of the inclusion of higher bound states on the total \gls{bsf} cross section for the (pseudo-)scalar mediator model in \cref{sec:indirectdetection}. For this purpose, we go to the Coulomb regime, which will provide us with an analytic solution and the attained cross section can serve as an upper bound on our Yukawa-type problem. In the Coulomb limit ($\xi\to \infty$ or $\mphi\to 0$), the \gls{bsf} cross section into an arbitrary $(n,l)$ bound state yields (\cf\cref{eq:BSFcrosssectionID})
\begin{align}
    \sigmabsf^{(nl)}\vrel(\zeta) = & \frac{\alpha}{120 }\sum_{n,l}(\Ekn{n})^5\left[ |\langle\vec{k} | \vec{r}^2 |nl \rangle|^2 + 2 |\langle\vec{k} | r^i r^j |nl \rangle|^2\right] \nonumber\\
    & + \frac{2 \alpha}{\mchi^2}  \sum_{n,l} \Ekn{n} \,  | \langle \vec{k} | \nabla^2_{\vec{r}} | nl  \rangle  |^{2} -  \frac{\alpha}{3\mchi^2}  \sum_{n,l} (\Ekn{n})^3 \,\Re{\langle \vec{k} | \vec{r}^2 | nl  \rangle  \langle nl | \nabla^2_{\vec{r}}  | \vec{k} \rangle} \, ,
\end{align}
where the summation over $m$ is implicit, \ie (\cf\cref{eq:expvalueBSF})
\begin{equation}
    \langle\vec{k} | \mathcal{O} |nl \rangle=\sum_m\langle\vec{k} |  \mathcal{O} |nlm \rangle=\sum_m\int \dd[3]{r}\phik^*(\vec{r})\,\mathcal{O}\,\psinlm(\vec{r})\,,
\end{equation}
for $\mathcal{O}$ denoting any of the operators above. We recall the decomposition of the scattering and \gls{bs} wave functions into radial and angular parts (\cf\cref{eq:scatteringwavefunctiondec,eq:boundstatewavefunctiondec})
\begin{align*}
	\phik(\vec{r})&=\sum_{l=0}^\infty (2l+1)\left[\frac{\chikl(\kappa r)}{\kappa r}\right] P_l(\uvec{k}\cdot\uvec{r})\,,\\
	\psinlm(\vec{r})&=\kappa^{3/2}\left[\frac{\chinl(\kappa r)}{\kappa r}\right]Y_{lm}(\Omega_{\vec{r}})\,,
\end{align*}
where $P_l(\uvec{k}\cdot\uvec{r})$ and $Y_{lm}(\Omega_{\vec{r}})$ denote the Legendre polynomials and spherical harmonics, respectively (\cf\cref{app:special_functions}). For later convenience, we will generalize the definition of the radial overlap integrals in \cref{eq:radialoverlapintegrals} to arbitrary angular states and apply them to the Coulomb case. They are given by
\begin{align}
     A_j^{(nl)}\equiv\int_0^{\infty}\dd{x} x^2\chiij{nl}^*(x)\chiqs{k}{l+2j}(x)\,,\qquad B_C^{(nl)}\equiv -\int_0^{\infty}\dd{x}\frac{2}{x}\chiij{nl}^*(x)\chiqs{k}{l}(x)\,,
\end{align}
with $x\equiv \kappa r$, where it is understood that $l+2j\geq 0$ for $j\in\{-1,0,1\}$. We will first perform the angular integrations, and calculate in a second step the radial overlap integrals in the Coulomb limit analytically. 

\paragraph{\protect\boldmath The first quadrupole term $|\langle\vec{k} | \vec{r}^2 |nl \rangle|^2$}$~$\newpp
Employing the definition of the spherical harmonics given in \cref{eq:sphericalharmonicsdef}, which relates $P_l(\cos\theta)=\sqrt{4\pi/(2l+1)}Y_{l0}(\Omega)$, and using the orthonormality condition of \cref{eq:sphericalharmonicsnorm}, we can immediately write down  
\begin{align}
    \langle\vec{k} | \vec{r}^2 |nlm \rangle =&\sqrt{\frac{4\pi}{\kappa^{7}}}\sqrt{2l+1}\,\delta_{m0}\int_0^\infty\dd{x}x^2\chiqs{k}{l}^*(x)\chiij{nl}(x)\,,\\
    |\langle\vec{k} | \vec{r}^2 |nl \rangle|^2=&\frac{4\pi}{\kappa^7}(2l+1)\left|A^{(nl)}_0\right|^2.
\end{align}

\paragraph{\protect\boldmath The derivative term $|\langle\vec{k} | \nabla_{\vec{r}}^2 |nl \rangle|^2$}$~$\newpp
Using the \gls{bs} Schrödinger equation (\cf\cref{eq:boundstateSchroedinger}), we can rewrite $\nabla^2\psinlm(\vec{r})=2\mu(V(\vec{r})-\Enl)\psinlm(\vec{r})$. The term with the potential contribution can be calculated analogously to the first quadrupole term, whereas the \gls{bs} energy term vanishes due to the orthogonality of the radial waves, given in \cref{eq:orthogonalityradialwavefunctions}. The computation yields 
\begin{align}
   \langle\vec{k}| \nabla^2_{\vec{r}} | nlm \rangle  &= \sqrt{4\pi\kappa}\sqrt{2l+1}\,\delta_{m0}\left(-\int_0^\infty\dd{x}\frac{2}{x}\chiqs{k}{l}^*(x)\chiij{nl}(x)\right)\,,\\
   |\langle\vec{k} | \nabla_{\vec{r}}^2 |nl \rangle|^2 &= 4\pi\kappa\,(2l+1)\left|B_C^{(nl)}\right|^2.
\end{align}
The mixed term between the first quadrupole and the derivative term is then given by
\begin{equation}
    \Re{ \langle \vec{k} | \vec{r}^2 | nl  \rangle  \langle nl | \nabla^2_{\vec{r}}  | \vec{k} \rangle}  =\frac{4\pi}{\kappa^3}(2l+1)\Re{\left(A^{(nl)}_0\right)^* B_C^{(nl)}}.
\end{equation}

\paragraph{\protect\boldmath The second quadrupole term $|\langle\vec{k} | r^ir^j |nl \rangle|^2$}$~$\newpp
Expressing again $P_l(\cos\theta)$ in terms of $Y_{l0}(\Omega)$, we can write the quadrupole contribution as
\begin{equation}
    \langle\vec{k} | r^i r^j |nlm \rangle=\sqrt{\frac{4\pi}{\kappa^{7}}}\sum_{l'=0}^{\infty}\sqrt{2l'+1}\int\dd{\Omega_r}\frac{r^ir^j}{r^2}Y_{l'0}(\Omega_r)Y_{lm}(\Omega_r)\int_0^\infty \dd{x} x^2 \chiqs{k}{l'}^*(x)\chiij{nl}(x).
\end{equation}
The quantity $r^ir^j/r^2$ is a rank $2$ tensor, which can be itself written in terms of spherical harmonics (see \eg\eref\cite{Merzbacher:1998}). The angular part of the squared quadrupole term then yields
\begin{align}
    &\left(\int\dd{\Omega_r}\frac{r^ir^j}{r^2}Y_{l'0}(\Omega_r)Y_{lm}(\Omega_r)\right)^*\!\left(\int\dd{\Omega_r}\frac{r^ir^j}{r^2}Y_{l''0}(\Omega_r)Y_{lm}(\Omega_r)\right)\nonumber\\
    &=\frac{4\pi}{15}\Bigg[5\left(\int\dd{\Omega_r}Y_{00}(\Omega_r)Y_{l'0}(\Omega_r)Y_{lm}(\Omega_r)\right)^*\!\left(\int\dd{\Omega_r}Y_{00}(\Omega_r)Y_{l''0}(\Omega_r)Y_{lm}(\Omega_r)\right)\nonumber\\
    &\hspace{0.4cm}+2\sum_{m'=-2}^{2}\left(\int\dd{\Omega_r}Y_{2m'}(\Omega_r)Y_{l'0}(\Omega_r)Y_{lm}(\Omega_r)\right)^*\!\left(\int\dd{\Omega_r}Y_{2m'}(\Omega_r)Y_{l''0}(\Omega_r)Y_{lm}(\Omega_r)\right)\Bigg].
\end{align}
The first line can be evaluated as usual utilizing the orthonormality condition of spherical harmonics (\cf\cref{eq:sphericalharmonicsnorm}), whereas the integrals in the second line can be connected to the Wigner-3j functions via \cref{eq:Wigner3jrelation}. Summing over $m$ and employing \cref{eq:Wigner3jorthogonality}, we obtain 
\begin{equation}
    |\langle\vec{k} | r^i r^j |nl \rangle|^2 =\,\frac{4\pi}{3\kappa^7}(2l+1)\left|A^{(nl)}_0\right|^2+\frac{8\pi}{3\kappa^7}(2l+1)\sum_{j=-1}^1K_{l+2j}\left|A^{(nl)}_j\right|^2\,,
\end{equation}
where we defined 
\begin{equation}
    K_{l'}\equiv (2l'+1)
    \begin{pmatrix}l&2&l'\\0&0&0\end{pmatrix}^2=
    \begin{cases}
        \frac{3}{2}\frac{(l-1)l}{(2l-1)(2l+1)}&\quad\text{for}\quad l'=l-2\\
        \hphantom{ \frac{3}{2}}\frac{l(l+1)}{(2l-1)(2l+3)}&\quad\text{for}\quad l'=l\\
        \frac{3}{2}\frac{(l+1)(l+2)}{(2l+1)(2l+3)}&\quad\text{for}\quad l'=l+2\\
    \end{cases}\,,
\end{equation}
and all other relations between $l'$ and $l$ vanish due to the Wigner-3j properties.\newp
Having performed all angular integrations, we can write the \gls{bsf} cross section as $\sigmabsf^{(nl)}\vrel(\zeta)=\pi\alpha^4/\mchi^2 \SBSF^{(nl)}(\zeta)$ with 
\begin{align}
    \label{eq:SBSFCoulombnl}
	\SBSF^{(nl)}(\zeta)\equiv\frac{2l+1}{720}\left(\frac{1+\zeta^2/n^2}{\zeta^2}\right)&\left[\left(\frac{1+\zeta^2/n^2}{\zeta^2}\right)^4\left(5\left|A^{(nl)}_0\right|^2+4\sum_{j=-1}^1K_{l+2j}\left|A^{(nl)}_j\right|^2\right)\right.\nonumber\\
	&\left.-120\left(\frac{1+\zeta^2/n^2}{\zeta^2}\right)^2\!\Re{\left(A^{(nl)}_0\right)^*\!B_C^{(nl)}}+720 \left|B_C^{(nl)}\right|^2\right].
 \end{align}
We can easily check that this is identical to \cref{eq:SBSFYukawa10} for $(nl)=(10)$, $\xi\to\infty$ and thus $\mathscr{P}_{10}, \gamma_{10} \to 1$. On the contrary, \cref{eq:SBSFCoulombnl} also holds for the Yukawa case with $\mphi>0$, when reinstating generalized phase space suppression factors $\mathscr{P}_{nl}$ in the logic of \cref{eq:SBSFYukawa10} and replacing $1/n^2\to \gammanl^2(\xi)$, $B_C^{(nl)}\to B^{(nl)}(\xi)$, all defined in analogy to $\mathscr{P}_{10}$, $\gamma_{10}$ and $B$. \newp
The next step is to calculate the radial overlap integrals for the Coulomb wave functions (\cf\cref{eq:ColumbscatteringwavefunctionRadial,eq:ColumbboundstatewavefunctionRadial})
\begin{align}
    \chiqs{k}{l'}(x)=&\sqrt{S_0(\zeta)\prod_{s=1}^{l'}(s^2+\zeta^2)}\frac{e^{i\delta_{l'}}}{(2l'+1)!}\left(\frac{2ix}{\zeta}\right)^{l'}xe^{-ix/\zeta}\HGone{1+l'+i\zeta}{2l'+2}{\frac{2ix}{\zeta}}\,,\\
    \chinl(x)=&\frac{1}{n(2l+1)!}\sqrt{\frac{(n+l)!}{(n-l-1)!}}\left(\frac{2x}{n}\right)^{l+1} e^{-x/n}\HGone{-n+l+1}{2l+2}{\frac{2x}{n}}\,,
\end{align}
where we employed \cref{eq:GammaS0prod,eq:LaguerreHGoneconnection} to attain this particular form, with the confluent hypergeometric functions $\HGone{a}{c}{x}$ defined in \cref{app:special_functions}. From there on, we can use \cref{eq:HGoneD0,eq:HGoneD1,eq:HGoneD2,eq:HGoneHGtwoIntegralidentity} to calculate the squared radial overlap integrals. The $\Delta l = 0$ integrals yield
\begin{align}
    \left|A^{(nl)}_0\right|^2=&\frac{2^{4(l+1)}n^{2l}(n+l)!}{(n-l-1)!((2l+1)!)^2}\frac{\zeta^{2(l+2)}}{(n^2+\zeta^2)^{2(l+1)}}\nonumber\\
    &\hspace{3cm}S_0(\zeta)\left(\prod_{s=1}^l(s^2+\zeta^2)\right)e^{-4\zeta\arccot(\zeta/n)}\abs{\HGtwo{a}{b}{c}{z}}^2\,,\\
    \left|B_C^{(nl)}\right|^2=&\frac{1}{16}\left(\frac{1+\zeta^2/n^2}{\zeta^2}\right)^4\left|A^{(nl)}_0\right|^2,\quad\Re{\left(A^{(nl)}_0\right)^*B_C^{(nl)}}=\frac{1}{4}\left(\frac{1+\zeta^2/n^2}{\zeta^2}\right)^2 \left|A^{(nl)}_0\right|^2\,,
\end{align}
with $a\equiv 1+l-n$, $b \equiv 1+l+i\zeta$, $c \equiv 2l+2$, $z \equiv 4in\zeta/(n+i\zeta)^2$ and $\HGtwo{a}{b}{c}{z}$ the hypergeometric function. This enables us to simplify the \gls{se} factor for \gls{bsf}
\begin{equation}
    \label{eq:SBSFnlCoulomb}
    \SBSF^{(nl)}(\zeta)=\frac{2l+1}{180}\left(\frac{1+\zeta^2/n^2}{\zeta^2}\right)^5\left(5\left|A^{(nl)}_0\right|^2+\sum_{j=-1}^1 K_{l+2j}\left|A^{(nl)}_j\right|^2\right).
\end{equation}
The $\Delta l=2 $ integrals are of the form
\begin{align}
    \left|A^{(nl)}_{\pm 1}\right|^2=&\frac{2^{4(l+1)}n^{2(l+2)}(n+l)!}{(l+1)^2(n-l-1)!((2l+1)!)^2}\frac{\zeta^{2(l+6)}}{(n^2+\zeta^2)^{2l+9}}\nonumber\\
    &S_0(\zeta)\left(\prod_{s=1}^{l}(s^2+\zeta^2)\right)\frac{e^{-4\zeta\arccot(\zeta/n)}}{(l^2+\zeta^2)}\nonumber\\
    &\quad\Big[K^1_\pm\abs{\HGtwo{a+1}{b-1}{c+1}{z}}^2+K^2_\pm\abs{\HGtwo{a+2}{b-1}{c+1}{z}}^2\nonumber\\
    &\quad +2K^{\text{Re}}_\pm\,\text{Re}\left\{\HGtwo{a+1}{b-1}{c+1}{z}^*\HGtwo{a+2}{b-1}{c+1}{z}\right\}\nonumber\\
    &\quad -2K^{\text{Im}}_\pm\,\text{Im}\left\{\HGtwo{a+1}{b-1}{c+1}{z}^*\HGtwo{a+2}{b-1}{c+1}{z}\right\}\Big]\,,
\end{align}
with $K_\pm$ (ratios of) polynomials of $n,l$ and $\zeta$, which are too extended to display here, and where it is understood that $|A^{(nl)}_{-1}|^2=0$ for $l<2$. 
\begin{figure}[ht]
    \centering
    \includegraphics[width=\textwidth]{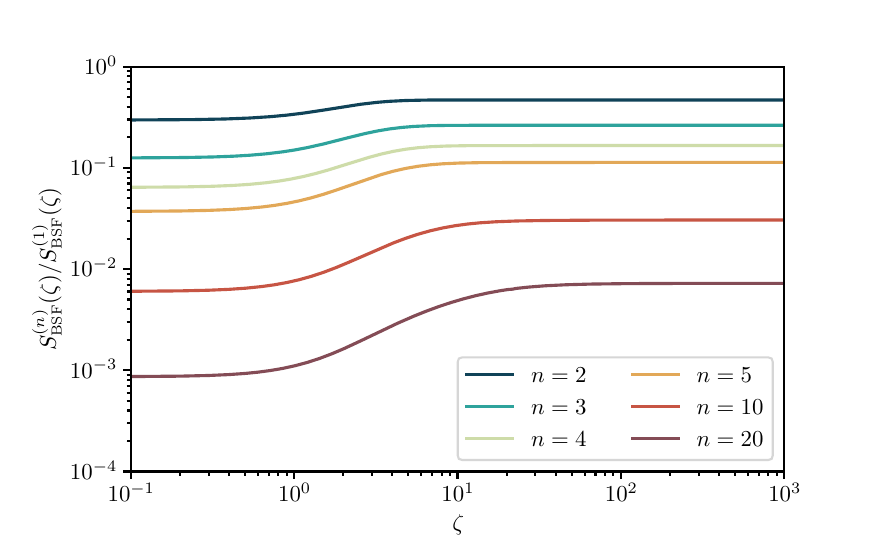}
    \caption[The SE factor of BSF into the $n$-th BS compared to the ground state as a function of $\zeta$ in the Coulomb limit.]{The \gls{se} factor of \gls{bsf} into the $n$-th \gls{bs} compared to the ground state as a function of $\zeta$ in the Coulomb limit. The sum over $l=0,\dots,n-1$ for each $n$ is left implicit.}
    \label{fig:higherBS}
\end{figure}
To obtain a first estimate on the effects of higher \gls{bs}, we have displayed in \cref{fig:higherBS} the relative size of the (\gls{se} factor of the) \gls{bsf} cross section into the $n$-th \gls{bs} with respect to the ground state (a sum over $l$ for each $n$ is implicit). We can see that the biggest contribution comes from the $n=2$ cross section, whereas the impact of higher \gls{bs} becomes quickly suppressed. As we are chiefly interested in finite mediator masses, the impact of higher \gls{bs} is expected to be even lower and the finite number of supported \gls{bs} by a Yukawa potential prevents a logarithmic growth due to an infinite number of \gls{bs}, as it has been observed in case of a Coulomb potential (see \eg\eref\cite{Binder:2023ckj}). For a realistic estimate, it is thus sufficient to include only the lowest excited \gls{bs} in the computations of \cref{subsubsec:relicdensityid}. The explicit \gls{se} factors for \gls{bsf} up to $n=2$ in the Coulomb limit from \cref{eq:SBSFnlCoulomb} are given by
\begin{align}
    \SBSF^{(10)}(\zeta)&=S_0(\zeta)\,\frac{2^6}{15}\,\frac{\zeta^2(7+3\zeta^2)}{(1+\zeta^2)^2}e^{-4\zeta\arccot(\zeta)}\,,\\
    \SBSF^{(20)}(\zeta)&=S_0(\zeta)\,\frac{2^5}{15}\,\frac{\zeta^2(448+528\zeta^2+100\zeta^4+15\zeta^6)}{(4+\zeta^2)^4}e^{-4\zeta\arccot(\zeta/2)}\,,\\
    \SBSF^{(21)}(\zeta)&=S_0(\zeta)\,\frac{2^9}{15}\,\frac{\zeta^4(1+\zeta^2)(36+5\zeta^2)}{(4+\zeta^2)^4}e^{-4\zeta\arccot(\zeta/2)}\,,
\end{align}
with $S_0(\zeta)=2\pi\zeta/(1-e^{-2\pi\zeta})$. These results agree well with the literature (see \eg\eref\cite{Biondini:2021ycj}).

\pagestyle{normal}

\printbibliography


\end{document}